\documentclass[10pt,a4paper,graphicx]{article}
\usepackage{slashed} 
\usepackage{braket}
\usepackage{graphicx}
\usepackage{epsfig}
\usepackage{multirow} 
\usepackage{epstopdf}
\usepackage{relsize}
\usepackage{amsmath,amssymb,mathtools}
\usepackage{caption}
\usepackage{enumitem}
\usepackage{jheppub}  
\usepackage{graphicx,color}
\usepackage{amsmath}
\usepackage{mathtools}
\usepackage{amsfonts}
\usepackage{hyperref}
\usepackage{ragged2e}
\usepackage{xr}
\usepackage{subfig}
\usepackage{xcolor}

\newcommand{\dis}[1]{\color{blue}\mathbold{#1}}
\newcommand{\alspi}{\ \frac{2\alpha_s}{3\pi}}
\newcommand{\LP}{leading power}
\newcommand{\NLP}{next-to-leading power}
\newcommand{\LL}{leading log }
\newcommand{\NLL}{next-to-leading log}
\newcommand{\B}{\color{blue}}
\newcommand{\R}{\color{red}}
\newcommand{\nn}{\nonumber}
\newcommand{\e}{\epsilon}
\newcommand{\ord}{{\cal O}}
\newcommand{\as}{\alpha_s}
\newcommand{\Ts}{\ta{1} \tb{2} \tc{3} \td{4} \te{5} \tg{6}}
\newcommand{\alg}{\,\,\,\, }

\newcommand{\ta}[1]{\mathbf{T}^{a}_{#1}} 
\newcommand{\tb}[1]{\mathbf{T}^{b}_{#1}}
\newcommand{\tc}[1]{\mathbf{T}^{c}_{#1}}
\newcommand{\td}[1]{\mathbf{T}^{d}_{#1}}
\newcommand{\tf}[1]{\mathbf{T}^{f}_{#1}} 
\newcommand{\tg}[1]{\mathbf{T}^{g}_{#1}}
\newcommand{\te}[1]{\mathbf{T}^{e}_{#1}}
\newcommand{\tj}[1]{\mathbf{T}^{j}_{#1}}
\newcommand{\thh}[1]{\mathbf{T}^{h}_{#1}}
\newcommand{\tkk}[1]{\mathbf{T}^{k}_{#1}}
\newcommand{\tm}[1]{\mathbf{T}^{m}_{#1}}
\newcommand{\tl}[1]{\mathbf{T}^{l}_{#1}}
\newcommand{\tn}[1]{\mathbf{T}^{n}_{#1}}
\newcommand{\tv}[1]{\mathbf{T}^{v}_{#1}}
\newcommand{\tu}[1]{\mathbf{T}^{u}_{#1}}
\newcommand{\tp}[1]{\mathbf{T}^{p}_{#1}}
\newcommand{\tq}[1]{\mathbf{T}^{q}_{#1}}
\newcommand{\trr}[1]{\mathbf{T}^{r}_{#1}}
\newcommand{\too}[1]{\mathbf{T}^{o}_{#1}}
\newcommand{\ts}[1]{\mathbf{T}^{s}_{#1}}
\newcommand{\tw}[1]{\mathbf{T}^{w}_{#1}}
\newcommand{\tx}[1]{\mathbf{T}^{x}_{#1}}
\newcommand{\ttt}[1]{\mathbf{T}^{t}_{#1}}
\newcommand{\tgg}[1]{\mathbf{T}^{gg}_{#1}}

\newcommand{\warning}[1]{\textcolor{red}{\textbf{#1}}}

\newcommand{\report}[1]{\textcolor{purple}{#1}}
\title{ Multiparton Cwebs at five loops} 
\author[a]{Shubham Mishra,}
\author[b]{Sourav Pal,}
\author[a]{Aditya Srivastav,}
\author[a]{Anurag Tripathi\,}

\affiliation[a]{Department of Physics, Indian Institute of Technology Hyderabad, Kandi, Sangareddy, Telangana State 502284, India.}
\affiliation[b] {Theoretical Physics Division, Physical Research Laboratory, Ahmedabad - 380009, \\ Gujarat, India.}

\emailAdd{shubhamhere82@gmail.com}
\emailAdd{sourav@prl.res.in}
\emailAdd{shrivastavadi333@gmail.com}
\emailAdd{tripathi@phy.iith.ac.in}

\abstract{Scattering amplitudes involving multiple partons are plagued with infrared singularities. The soft singularities of the amplitude are captured by the soft function which is defined as the vacuum expectation value of Wilson line correlators. Renormalization properties of soft function allows us to write it as an exponential of the finite soft anomalous dimension. 
An efficient way to study the soft function is through a set of Feynman diagrams known as Cwebs (webs). We present the mixing matrices and exponentiated colour factors (ECFs) for  the Cwebs at five loops that connect six Wilson  lines, except those that are related by relabeling of Wilson lines. Further, we express these ECFs in terms of 29 basis colour factors.
	We also find that this basis can be categorized into two colour structures.
	   Our results are the first key ingredients for the calculation of the soft anomalous dimension at five loops. 
}

\begin{document} 
\maketitle
\newcommand{\setfmfoptions}[0]{\fmfset{curly_len}{2mm}\fmfset{wiggly_len}{3mm}\fmfset{arrow_len}{3mm} \fmfset{dash_len}{2mm    }\fmfpen{thin}}
\pagebreak

\section{Introduction}
The study of the infrared (IR) structure of scattering amplitudes in perturbative quantum field theories that deal with massless gauge bosons has been a subject of interest for a century, with a long and rich history~\cite{Bloch:1937pw,Sudakov:1954sw,Yennie:1961ad,Kinoshita:1962ur,Lee:1964is,Grammer:1973db,Mueller:1979ih,Collins:1980ih,Sen:1981sd,
	Sen:1982bt,Korchemsky:1987wg,Korchemsky:1988hd,Magnea:1990zb,
	Dixon:2008gr,Gardi:2009qi,Becher:2009qa,Feige:2014wja}. A recent pedagogical review on this subject can be found in~\cite{Agarwal:2021ais}. The infrared structure of scattering amplitudes exhibit universality, that is they are independent of the relevant hard scattering processes. It is this universality property that helps one in studying IR singularities  to all orders in the perturbation theory without considering the details of the hard scattering processes. The structure of infrared singularities not only helps us in understanding gauge theories to all orders in the perturbation theory, they are also important for phenomenological applications. IR singularities always get canceled \cite{Kinoshita:1962ur,Lee:1964is} in an infrared safe observable, however leave their signature in certain kinds of logarithms. In certain kinematic regions these logarithms become large and resummation is needed for any sensible prediction that requires a thorough knowledge about these singularities. The universal structure of IR singularities also helps in the fixed order calculations where the cancellation of these IR singularities for complicated observables in colliders is very difficult in general. The use of universality property of the IR singularities helps one in developing several efficient subtraction procedures~\cite{GehrmannDeRidder:2005cm,Somogyi:2005xz,Catani:2007vq,
	Czakon:2010td,Boughezal:2015dva,Sborlini:2016hat,Caola:2017dug,
	Herzog:2018ily,Magnea:2018hab,Magnea:2018ebr,Capatti:2020xjc,Magnea:2022twu,Bertolotti:2022aih,Bertolotti:2022lcj}.        
    
The factorization property of QCD in the IR limit enables us in  studying the singular parts of scattering amplitudes  without calculating the  complicated hard part. The soft function, that controls the IR singular parts in a scattering process, can be expressed in terms of matrix elements of Wilson lines. The Wilson-line operator $\Phi ( \zeta )$  evaluated on smooth space-time contour $\zeta$ is defined as,
\begin{align}
	\Phi \left(  \zeta \right) \, \equiv \, \mathcal{P} \exp \left[ {\rm i} g \!
	\int_\zeta d x \cdot {\bf A} (x) \right] \, ,
	\label{genWL}
\end{align}
where ${\bf A}^\mu (x) = A^\mu_a (x) \, {\bf T}^a$ is a non-abelian gauge field, 
and ${\bf T}^a$ is a generator of the gauge algebra, which can be taken to belong
to any desired representation, and $ \mathcal{P} $ denotes path ordering of the gauge fields. The soft function for the scattering of $ n $ partons can be expressed as, 
\begin{align}
	{\cal S}_n \left( \zeta_i \right) \, \equiv \, \bra{0} \prod_{k = 1}^n
	\Phi \left(  \zeta_k \right) \ket{0} \, ,
	\label{genWLC}
\end{align}
here the semi-infinite Wilson lines originate from hard interaction vertex run along the direction of the hard particles. Thus, one can write the soft function as,
\begin{align}
	{\cal S}_n \Big( \beta_i \cdot \beta_j, \as (\mu^2), \e \Big) \, \equiv \, 
	\bra{0} \prod_{k = 1}^n \Phi_{\beta_k} \left( \infty, 0 \right) \ket{0} , 
\end{align}
where,
\begin{align}
	\Phi_\beta \left( \infty, 0 \right) \, \equiv \, \mathcal{P} \exp \left[ {\rm i} g \!
	\int_0^\infty d \lambda \, \beta \cdot {\bf A} (\lambda \beta) \right] .
	\label{softWLC}
\end{align}
$ \mathcal{S}_n$ suffers from both UV and IR (soft) singularities, and thus requires renormalization. In dimensional regularization $ d=4-2\epsilon $, $ \mathcal{S}_n $ is equal to zero, as it is made out of Wilson line correlators that involve only scaleless integrals. Thus, in a renormalized theory, $ \mathcal{S}_n $ contains pure UV counterterms. 

The renormalized soft function obeys a renormalization group equation, and solving this equation results in an exponentiation of the form, 
\begin{align}
	\mathcal{S}_n \Big( \beta_i \cdot \beta_j, \as (\mu^2), \e \Big) \, = \, 
	\mathcal{P} \exp \left[ - \frac{1}{2} \int_{0}^{\mu^2} \frac{d \lambda^2}  
	{\lambda^2} \, {\bf \Gamma}_n \Big( \beta_i \cdot \beta_j, \alpha_s (\lambda^2), 
	\e \Big) \right]  \, ,
	\label{softmatr}
\end{align}    
where $ {\bf \Gamma}_n $ is known as the soft anomalous dimension. As it is evident from the above equation, a knowledge of the soft anomalous dimension to any fixed order in perturbation theory enables us to predict sets of singular terms to all orders in perturbation theory,
thus, making it an object of immense interest and extensive study.
In case of processes involving multi-parton scatterings, the soft anomalous dimension is a matrix and is our main focus in this article. This renormalization group approach has been used for more than two decades to calculate the soft anomalous dimension. One-loop calculations for ${\bf \Gamma}_n$ were performed in \cite{ Kidonakis:1996aq, Kidonakis:1997gm,Kidonakis-1998,Korchemskaya:1994qp}, while two-loop calculations were done for both the massless case in \cite{Aybat:2006wq,Aybat:2006mz} and for the massive case in \cite{Mitov:2009sv, Ferroglia:2009ep, Ferroglia:2009ii, Kidonakis:2009ev, Kidonakis:2010tc, Kidonakis:2010ux, Kidonakis:2011wy,  Chien:2011wz}. The three-loop calculations were finally carried out for the massless case in \cite{Almelid:2015jia, Almelid:2017qju}. The calculation of soft anomalous dimension at four loops is an ongoing effort. Several studies in this direction are available in the literature in~\cite{Becher:2019avh, Kidonakis:2011wy, Falcioni:2020lvv,Falcioni:2021buo,Falcioni:2021ymu,Catani:2019nqv,Moch:2017uml,Ahrens:2012qz,
	Moch:2018wjh,Chetyrkin:2017bjc,
	Das:2019btv,Das:2020adl,vonManteuffel:2020vjv,Henn:2019swt,Duhr:2022cob, Kidonakis:2023lgc}.
The kinematic dependence of the soft anomalous dimension for scatterings, which involve only massless lines is restricted due to constraints discussed in  \cite{Gardi:2009qi,Gardi-Magnea,Becher:2009cu,Becher:2009qa}. However, these constraints do not hold true for scatterings involving massive particles. The state-of-the-art knowledge for soft anomalous dimension is known up to two loops for scatterings involving massive particle, and at three-loop for one massive Wilson line~\cite{Liu:2022elt}.

The soft function can be written as
\begin{align}
	{\cal S}_n \left( \gamma_i \right) \, = \, \exp \Big[ \sum{\cal W}_n  
	\Big]  \, ,
	\label{diaxp}
\end{align}
{where the sum is over different $ {\cal W}_n $ which are known as \textit{webs}, and can be directly computed using Feynman diagrams. Details are presented in section~\ref{sec:CwebProperties}}.  Webs were first defined as two-line irreducible diagrams for scattering involving two Wilson lines~\cite{Sterman-1981,
	Gatheral,Frenkel-1984}. In case of multi-parton scattering amplitude in a non-abelian gauge theory, a web at a given order in $ \alpha_s $  is defined as a set of diagrams that is closed under the permutation of gluon attachments on each Wilson line~\cite{Mitov:2010rp,Gardi:2010rn}. The kinematics and the colour factors of diagrams in a web~\cite{Gardi:2010rn} mix among themselves, through a web mixing matrix.

Cwebs --- a generalization of webs --- 
is a set of skeleton diagrams built out of connected gluon correlators attached to Wilson lines closed under shuffles of the gluon attachments to each Wilson line~\cite{Agarwal:2020nyc,Agarwal:2021him}. The state-of-the-art studies for massive multiparton webs at three loops~\cite{Gardi:2021gzz}, and for massless webs up to three loops~\cite{Gardi:2010rn,Gardi:2011yz,Gardi:2013ita,Gardi:2013saa} and at four loops~\cite{Agarwal:2020nyc,Agarwal:2021him} are available in the literature.

These web mixing matrices are crucial objects in the study of non-abelian exponentiation, and the general method to calculate these matrices is by applying a well-known replica trick algorithm \cite{Gardi:2010rn}. An alternative approach of generating functionals was developed in \cite{Vladimirov:2015fea,Vladimirov:2014wga,Vladimirov:2017ksc}. These web mixing matrices are combinatorial objects and are studied from the viewpoint of combinatorial mathematics using posets~\cite{Dukes:2013gea,Dukes:2013wa,Dukes:2016ger}. Further, a novel method of calculating the diagonal blocks --- using several new concepts such as Normal Ordering, Uniqueness theorem and Fused-Webs --- of the mixing matrices has been developed in~\cite{Agarwal:2022wyk} and has been applied for a certain classes of webs in~\cite{Agarwal:2022xec}.  The concepts introduced in that  study provided a systematic framework for simplifying and organizing the study of Cwebs at higher loop order. With the state-of-the-art of precision calculation reaching four loops~\cite{Becher:2019avh,Falcioni:2020lvv,Falcioni:2021buo,Falcioni:2021ymu,Catani:2019nqv,Moch:2017uml,Ahrens:2012qz,
	Moch:2018wjh,Chetyrkin:2017bjc,
	Das:2019btv,Das:2020adl,vonManteuffel:2020vjv,Henn:2019swt,Duhr:2022cob,Moch:2022frw,Duhr:2022yyp,Falcioni:2023luc} and five loops renormalization~\cite{ Luthe:2017ttg, Herzog:2017ohr} in perturbation theory, the calculation of colour structures present at five loops will be the first step for the calculation of soft anomalous dimension. 

In this article, we have enumerated all the unique Cwebs that appear at five loops and connect six Wilson lines using the algorithm introduced in~\cite{Agarwal:2020nyc}.
Following the method of replica trick and using the code developed in~\cite{Agarwal:2020nyc}, we have calculated the web mixing matrices and the exponentiated colour factors for all the Cwebs. 
Besides the explicit calculations, we have also shown that the concepts introduced in~\cite{Agarwal:2022wyk} predict mixing matrices for more than two thirds of the Cwebs without any calculation.

The rest of the paper is structured as follows. In section~\ref{sec:CwebProperties}, we review the known properties of mixing matrices. In section~\ref{sec:algorithm}, we review the details of the algorithm for generating Cwebs and enumerates the distinct Cwebs present at five loops connecting six Wilson lines. In section~\ref{sec:UNQ}, we  categorize these Cwebs into three groups and describe in detail the computations of mixing matrices and exponentiated colour factors of one Cweb from each group, 
and the details  for the remaining Cwebs are presented in the appendix~\ref{sec:appendix}. Section~\ref{sec:colourstru} lists down the colour structures that form a basis for five loops six lines Cwebs. Finally we conclude our findings in section~\ref{sec:conclusion}.

\section{Cwebs and properties of web mixing matrices}\label{sec:CwebProperties}

Cwebs~\cite{Agarwal:2020nyc,Agarwal:2021him} are all order quantities and thus
have their own perturbative expansion in powers of the strong coupling constant $g$. 
Throughout this article, we use the notation $W_n^{(c_2, \ldots , c_p)} (k_1, \ldots  , k_n)$ 
for a Cweb {contains precisely} $c_m$ $m$-point connected gluon correlators 
($m = 2, \ldots, p$), where $k_{i}$ denotes the number of attachments on $ i^{\text{th}} $ Wilson line.
Note that the 
perturbative expansion for an $m$-point connected gluon correlator starts 
at ${\cal O} (g^{m - 2})$, while each attachment to a Wilson line carries a 
further power of $g$; the perturbative expansion for a Cweb can thus be written as
\begin{equation}
	W_n^{(c_2, \ldots , c_p)} (k_1, \ldots  , k_n)  \, = \, 
	g^{\, \sum_{i = 1}^n k_i \, + \,  \sum_{r = 2}^p c_r (r - 2)} \, \sum_{j = 0}^\infty \,
	W_{n, \, j}^{(c_2, \ldots , c_p)} (k_1, \ldots  , k_n) \, g^{2 j} \, ,
	\label{pertCweb}
\end{equation}
which defines the perturbative coefficients $W_{n, \, j}^{(c_2, \ldots , c_p)} 
(k_1, \ldots  , k_n)$.  {For ease of notation we will write $W_{n, \, j}^{(c_2, \ldots , c_p)} 
(k_1, \ldots  , k_n)$ as $ W $ in the remaining parts of this section.}

Cwebs are the proper building blocks of the logarithm of soft function. Cwebs are very useful in the organization and counting of diagrammatic contributions at higher perturbative orders.
The logarithm of the soft function is a sum over all the Cwebs at each perturbative order:
\begin{align}
	{\cal S}_n \, = \, \exp \left[ \sum_{{W}}
	\sum_{d,d' \in  {W}} { K} (d) \, R_{{W}} (d, d') \, C (d')
	\right] \, .
	\label{Snwebs}
\end{align}
Here the perturbative order is $ g^{\, \sum_{i = 1}^n k_i \, + \,  \sum_{r = 2}^p c_r (r - 2)} $.
The $d$ here denotes a diagram in a  Cweb $W$ and its corresponding kinematic and colour factor are denoted by $ K (d) $ and $C(d)$. The action of web mixing matrix $R_{{W}}$ on the colour of a  diagram $ d $ generates its exponentiated colour factor $\widetilde{C}$,  
\begin{align}
	\widetilde{C} (d)  \, = \, \sum_{d'\in {{W}}} R_{{W}} (d, d') \, C(d') \, .
	\label{eq:ecf}
\end{align}
These mixing matrices have some general remarkable properties given in \cite{Gardi:2010rn, Gardi:2011yz, Gardi:2011wa, Gardi:2013ita,Agarwal:2022wyk} and are listed below.

\begin{enumerate}
	\item \textit{Idempotence} : $R^2=R$, thus the mixing matrix acts as a projection operator, \textit{i.e.} it has eigenvalues either $0$ or $1$.  
	The mixing matrix can be  diagonalized using a similarity transformation given by, 
	\begin{align}
	D=	Y \, R \, Y^{-1}= \text{diag}\,(\lambda_1, \lambda_2 \cdots \lambda_d),
		\label{eq:diagonalization}
	\end{align}
	where, 
\begin{equation}
	\lambda_n\, =
	\begin{cases}
		1 & 1  \leq n \leq r \\
		0 & r+1 \leq n \leq d \,.
	\end{cases}       
\end{equation}
Here $r$ and $d$ are the rank and dimension of the mixing matrix respectively.
This reduces $ D $ to the following form:
\begin{equation}\label{eq:DiagonalrankMatrix}
D=(\mathbf{1}_r,0)\,.
\end{equation}
From eq.~\eqref{Snwebs} the kinematic and colour contributions to a Cweb $ {W} $ can be rewritten as
		\begin{align}\label{eq:YC-1}
		  {W}  =   \left( K^T Y^{-1} \right) Y R Y^{-1}  
		\left( Y C \right) \,=\,  \sum_{j =  1}^{r} \left(  K^T Y^{-1}  \right)_j 
		\left( Y C \right)_j  \, ,
	\end{align}
	where, we have used the fact that $D_{ii} =1$ for $1 \leq i \leq r$ and the remaining  diagonal entries of $D$ vanish.
	\item \textit{Zero row-sum} : The mixing matrices follow zero row-sum property, which states that if we add all the elements of a given row they sum to zero, that is~\cite{Gardi:2010rn}
	\begin{align}
		\sum_{d'} R _{dd'}=0, \ \  \forall \  d\,.
	\end{align}
	%
	\item \textit{Weighted column-sum} : The web mixing matrices follow a conjectured weighted column-sum rule~\cite{Gardi:2011yz} of the form,
	\begin{align}
		\sum_d s(d) R_{{W}}(d,d')=0, \;\; \forall \  d^\prime\,.
	\end{align}
	where  $s(d)$ called the weight factor (or $ s $-factor) for a diagram $ d $ is defined as the number of different ways in which each of the gluon correlators 
	 can be {\it sequentially} shrunk to {the hard interaction vertex} (see appendix~\ref{sec:shrinking}). This conjectured property has been found to hold up to four loops~\cite{Agarwal:2020nyc,Agarwal:2021him}. {This conjecture is also a direct consequence of cancellation of leading divergences in webs, as discussed in~\cite{Gardi:2011yz}.  } 
	\item \textit{Uniqueness theorem} : This theorem~\cite{Agarwal:2022wyk} states that the mixing matrix is unique for a given column weight vector $S= \{s(d_i)\} \;\forall \  s(d_i) \neq 0$. 
	
\end{enumerate}
After having established the role of Cwebs in the determination of the soft function and enumerated the known properties of their mixing matrices 
now we proceed towards the main content of this article.

\section{Enumeration of Cwebs at five loops six lines  } \label{sec:algorithm}
In this article  we restrict ourselves to the Cwebs that  connect six  Wilson lines at five loops.  These Cwebs can be obtained by applying the three-step algorithm~\cite{Agarwal:2020nyc} on Cwebs present at four loops connecting five lines that were presented in~\cite{Agarwal:2020nyc}. Imagining that there is a supply of an infinite number of Wilson lines, the three steps of the algorithm are
\begin{enumerate}
	\item Add a two-gluon connected correlator  connecting any two Wilson lines
	(including Wilson lines that  had no attachments at lower orders).
	\item Connect an existing $m$-point correlator to any Wilson line (again, 
	including Wilson lines with no attachments at lower orders), turning  it into
	an  $(m+1)$-point correlator.
	\item Connect an existing $m$-point correlator to an existing $n$-point 
	correlator, resulting in an $(n+m)$-point correlator.
\end{enumerate}
The above steps generate a number of Cwebs that are related to one another by rearrangement of the Wilson lines. Since all these rearrangements gives the same colour structure we keep only one of them. This algorithm has been implemented in an in-house Mathematica code used to produce the results presented in this paper. 

\begin{figure}
	\centering
	\includegraphics[scale=0.5]{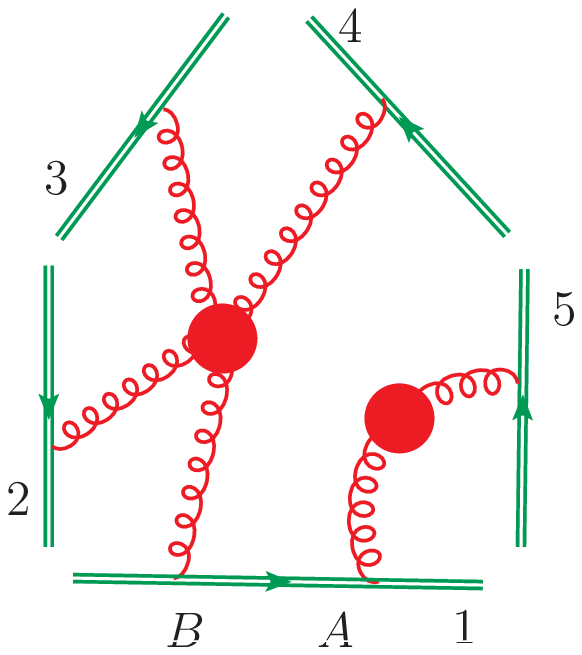}
	\caption{ Cweb $W^{(1,0,1)}_5 (2, 1, 1, 1, 1)$ at four loops and five lines.}
	\label{fig:4l5l1}
\end{figure}

\begin{figure}[t]
	\centering
	\vspace{-3mm}
	\hspace{-0.2cm}
	\subfloat[]{\hspace{0.3cm}\includegraphics[scale=0.32]{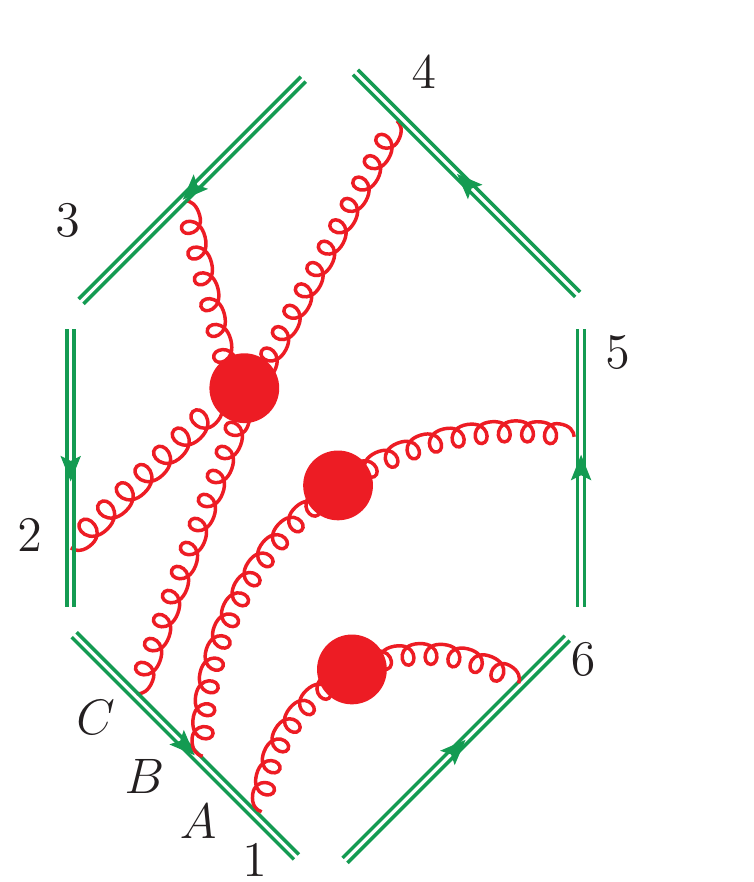} } 
 $ \; $
	\subfloat[]{\hspace{0.3cm}\includegraphics[scale=0.32]{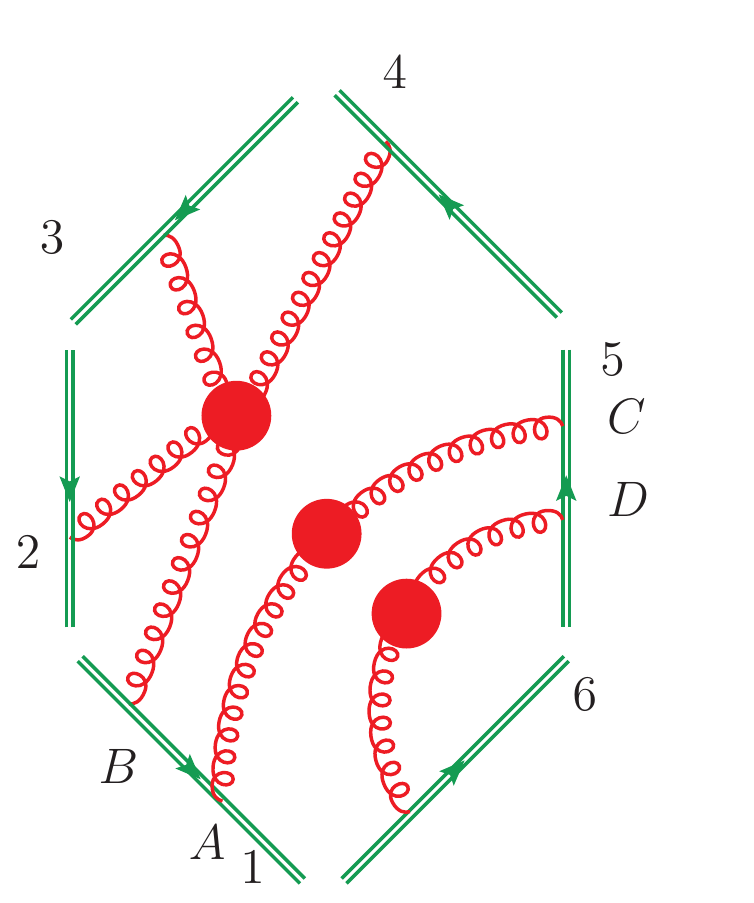} } 
    $ \; $
    \subfloat[]{\hspace{0.3cm}\includegraphics[scale=0.32]{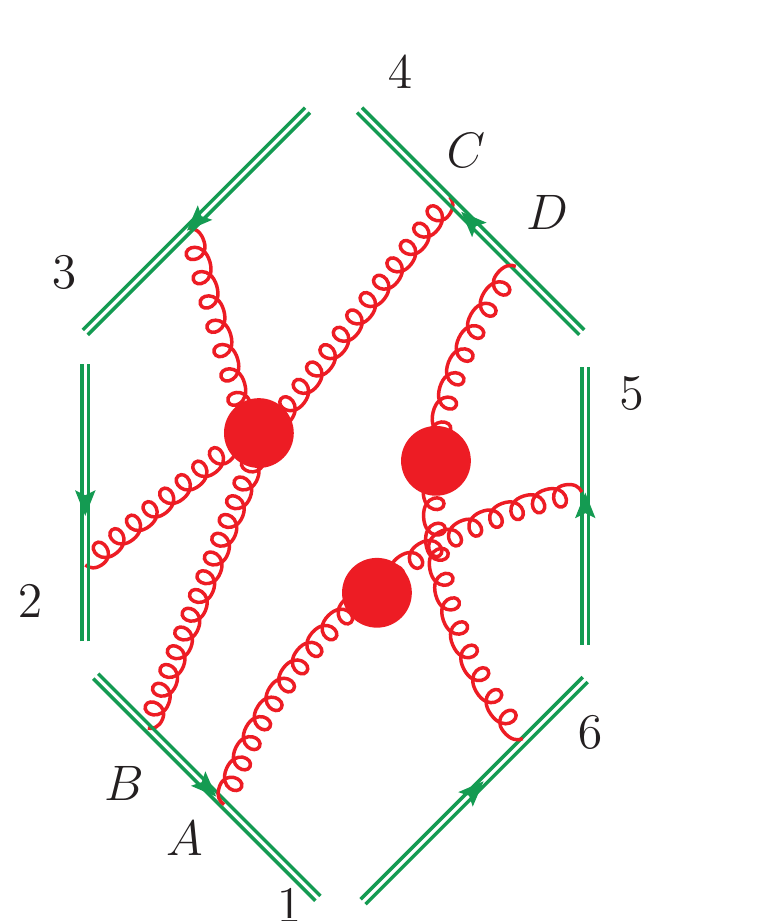} } 
	$ \; $
	\subfloat[]{\hspace{0.3cm}\includegraphics[scale=0.32]{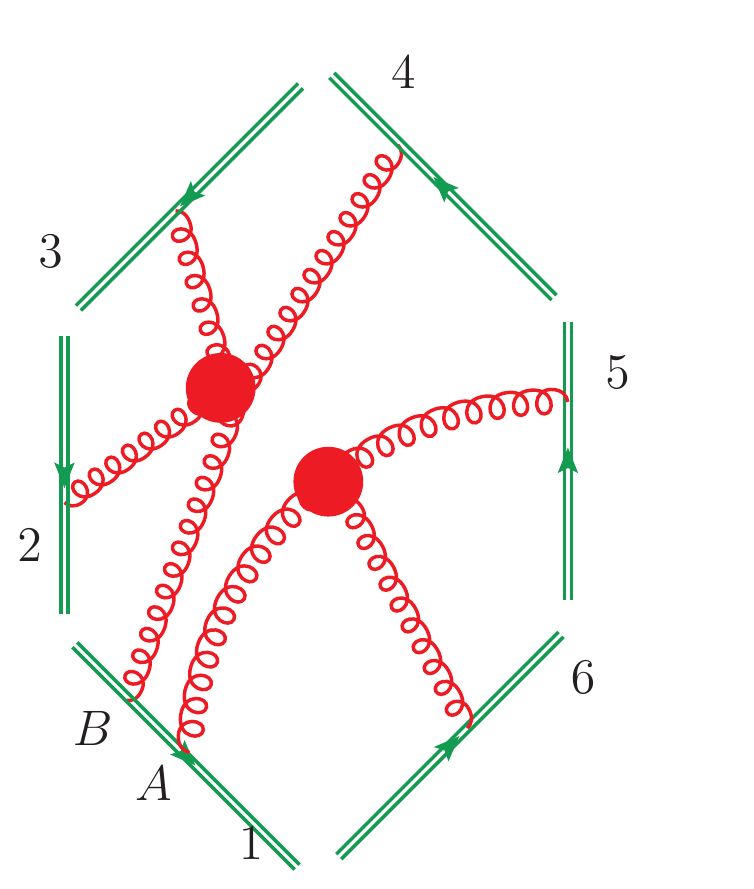} } 
	$ \; $
	\subfloat[]{\hspace{0.3cm}\includegraphics[scale=0.32]{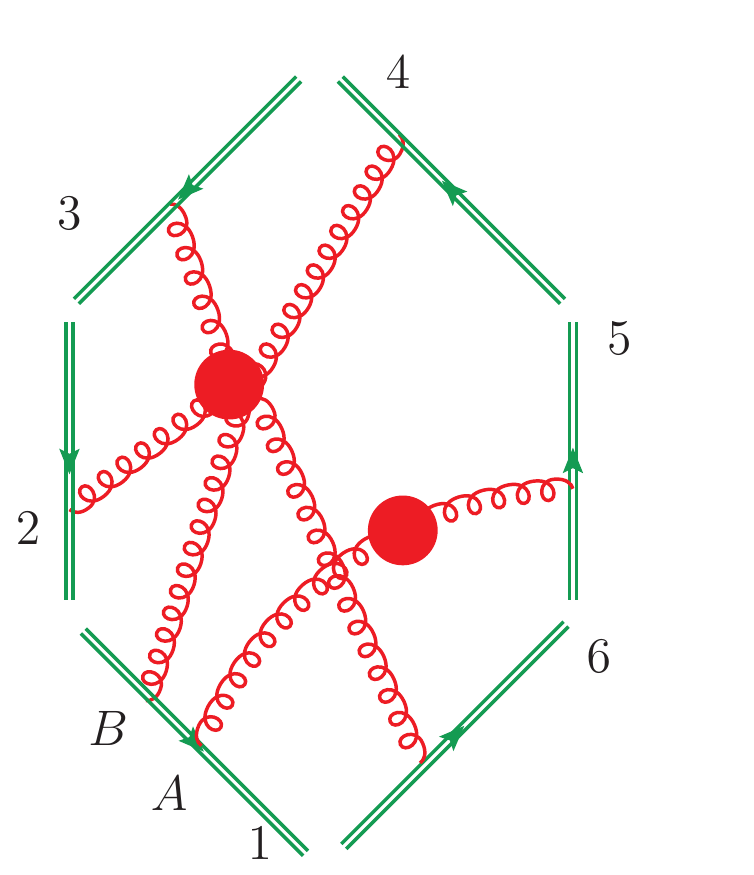} }
	\caption{Five Cwebs generated on applying the algorithm on Cweb $W^{(1,0,1)}_5 (2, 1, 1, 1, 1)$}
	\label{fig:Enumeration}
\end{figure}

To demonstrate the recursive algorithm for generating a five-loop Cweb, we consider the four loop Cweb $W^{(1,0,1)}_5 (2, 1, 1, 1, 1)$  shown in fig.~\ref{fig:4l5l1} that connects five Wilson lines.
We have put arrows on the Wilson lines and the tail ends of each of the Wilson lines meet at the hard interaction vertex.
Following step one, we can attach a two-point gluon correlator to a sixth line, which is absent at four loops. This generates 5 Cwebs, out of which only 3 are unique\footnote{Unique up to relabeling of Wilson lines}, which are shown in figs.~\ref{fig:Enumeration}\textcolor{blue}{a},~\ref{fig:Enumeration}\textcolor{blue}{b}, and~\ref{fig:Enumeration}\textcolor{blue}{c}.  
Step two can be applied in two different ways: by joining the two-point correlator with the sixth Wilson line turning it into a three-point correlator  we generate a new Cweb, shown in fig.~\ref{fig:Enumeration}\textcolor{blue}{d}; we can also join the four-point correlator with the sixth Wilson line turning it into a five-point correlator which generates only 1 Cweb, shown in fig.~\ref{fig:Enumeration}\textcolor{blue}{e}. 
The third step generates Cwebs at five loops and five lines, which are not considered in this article.
Thus starting from a $ \mathcal{O}(\alpha_s^4) $ Cweb in fig.~\ref{fig:4l5l1}, the recursive algorithm generates five unique Cwebs at $ \mathcal{O}(\alpha_s^5) $ shown in fig.~\ref{fig:Enumeration}. 

The application of these steps on all the eight Cwebs present at four loops connecting five lines~\cite{Agarwal:2020nyc} generates large number of Cwebs at five loops and six lines, out of which only 22 Cwebs are unique. In the next section, we present the mixing matrices and exponentiated colour factors of few of these 22 Cwebs, and the remaining will be presented in the appendix.

\section{Cwebs at five loops six lines} 
\label{sec:UNQ}
The Cwebs appearing up to four loops connecting highest number of Wilson lines fall into two classes.  The first class contains Cwebs made up of only two point gluon correlators known as \textit{basis Cwebs}~\cite{Agarwal:2022wyk};  and in the second class  the Cwebs   contain at least one higher point correlator  and these belong to existing families at lower orders that also contain basis Cwebs~\cite{Agarwal:2022wyk}. However, at five loops we see a new feature that is not present below this order ---  at this order we get a new category of the Cwebs that connect the highest number of Wilson lines which does not belong to any of the  two classes that we have mentioned, we call them as \textit{Orphan Cwebs}. Therefore, we classify
Cwebs present at five loops that connect six  Wilson  into three categories: Cwebs belonging to existing families, new basis Cwebs, and Orphan Cwebs. 
In the upcoming subsections, we present detailed calculation of one of the Cwebs from each of these three categories.
\subsection{Cwebs from existing families}

{In this section, we list down all the Cwebs that belong to one or the other existing family containing basis Cweb reported in~\cite{Agarwal:2022wyk}. Since the members of each family have the same column weight vector $S$, where none of the entries in the vector is 0, we get the same mixing matrix for each
 member of the family. This is the Uniqueness theorem as proven in~\cite{Agarwal:2022wyk}.}

\begin{figure}[t]
	\centering
	\subfloat[][]{\hspace{0.5cm}\includegraphics[scale=0.5]{C12N}}
	\qquad 
	\subfloat[][]{\hspace{0.5cm}\includegraphics[scale=0.5]{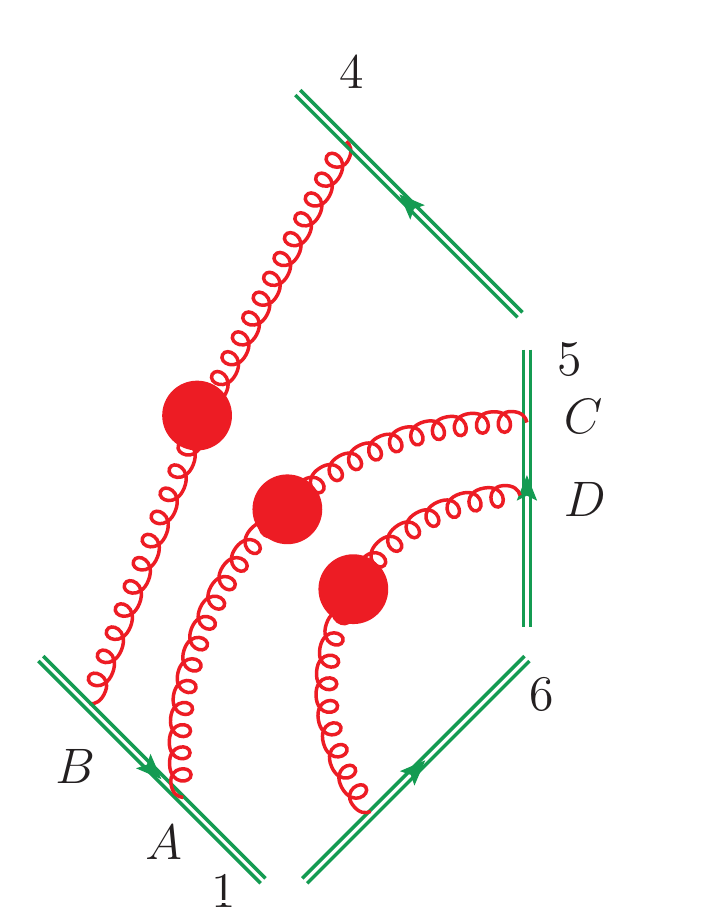}}    
	\caption{ Removing lines 2 and 3 from the five loops six lines Cweb $W^{(2,0,1)}_{6} (2, 1, 1, 1, 2, 1)$ in (a) reduces it to three loops four lines basis Cweb as shown in (b).}
	\label{fig:familyT}
\end{figure}

Let us take the  Cweb  $W^{(2,0,1)}_{6} (2, 1, 1, 1,2, 1)$, shown in fig.~\ref{fig:familyT}\textcolor{blue}{a} and use the Uniqueness theorem to determine its mixing matrix without any explicit computation. This Cweb has two attachments on  lines 1 and 5 whose shuffle generates four diagrams. We label the diagrams of the Cweb by the order of gluon attachments on Wilson lines. For the diagram $ d_1 $, we choose the order $ \{ \{AB\},\{CD\}\} $ as shown in fig.~\ref{fig:familyT}\textcolor{blue}{a}, where the letters 
are ordered opposite to the orientation of Wilson line arrows. The sequences for all the four diagrams are
\begin{align}
	\{d_1, d_2, d_3, d_4\} = 
	\big(\{AB\},\{CD\}\big), \big(\{AB\},\{DC\}\big), \big(\{BA\},\{CD\}\big), \big(\{BA\},\{DC\}\big).
\end{align}
Following the definition of $ s $-factors, the column weight vector (see properties 3 and 4 in section~\ref{sec:CwebProperties}) for this Cweb is given by, $ S=\{1,2,2,1\} $, { which is calculated using the method described in  appendix~\ref{sec:shrinking}}.
 Thus, we can associate this Cweb to the family of basis Cweb shown in fig.~\ref{fig:familyT}\textcolor{blue}{b}.
The mixing matrix of this basis Cweb was already calculated in~\cite{Gardi:2010rn,Gardi:2013ita} and was found to be
\begin{align}\label{eq:Family-mixingMatrix}
	R =\frac{1}{6} \left( 
	\begin{array}{cccc}
		\alg 2 & -2 & -2 & \alg 2 \\
		-1 & \alg 1 & \alg 1 & -1 \\
		-1 & \alg 1 & \alg 1 & -1 \\
		\alg 2 & -2 & -2 & \alg 2 \\
	\end{array}
\,\,	\right)\,.
\end{align}
Note that this matrix has rank one.
It follows from the Uniqueness theorem that this matrix is also the mixing matrix for the Cweb shown in fig.~\ref{fig:familyT}\textcolor{blue}{a}.
The diagonalizing matrix $ Y $ can be constructed with the right eigenvectors of $ R $ and is given by
\begin{align}
	Y =\left(
	\begin{array}{cccc}
		\alg 1 & -1 & -1 & \alg 1 \\
		-1 & \alg 0 & \alg 0 & \alg 1 \\
		\alg \frac{1}{2} & \alg 0 & \alg 1 & \alg 0 \\
		\alg \frac{1}{2} & \alg 1 & \alg 0 & \alg 0 \\
	\end{array}
	\right) \, , 
	\label{MixMat2}
\end{align}
 and the corresponding diagonal matrix $ D $ is given by, 
\begin{align}
	D=(\mathbf{1}_1,0)\,.
\end{align}
As the rank of $ R $ is one, this Cweb will have only one independent exponentiated colour factor (ECF). Putting the above matrices in eq.~\eqref{eq:YC-1} we determine the ECF of this Cweb which upon algebraic simplification using FORM \cite{Vermaseren:2000nd,Kuipers:2012rf} reads as
\begin{align}
	(YC)_1=- f^{abl}f^{cdl}f^{deh}f^{egk} \thh{1} \ta{2} \tb{3} \tc{4} \tkk{5} \tg{6} \, .
\end{align}
This ECF represents a fully connected diagram as predicted by the non-abelian exponentiation theorem~\cite{Gardi:2010rn}.
 In the similar manner, the calculation of column weight vectors for Cwebs present at five loops and six lines unveils that the explicit form of the mixing matrices for 15 of the 22 Cwebs can be obtained using the Uniqueness theorem.  
In table \ref{tab:UNQ}, we list down all the Cwebs whose mixing matrices are predicted using the Uniqueness theorem from the available results of lower order~\cite{Gardi:2013ita,Agarwal:2020nyc,Agarwal:2021him}.
\begin{table}[t]
	\begin{center}
		\begin{tabular}{|c|l|c|c|c|c|}
			\hline
			S. No.&	Name of the Cweb & Figure & No. of diagrams & $ r(R) $ & {Basis matrix}\\ 
			\hline 
			1 &	    $W^{(0,1,1)}_{6} (2, 1, 1, 1, 1, 1)$ & \footnotesize{~\ref{Diag:4}}&  2 & 1 & $ R(1_2) $\\
			\hline 
			2 &	    $W^{(1,0,0,1)}_{6} (2, 1, 1, 1, 1, 1)$ & \footnotesize{~\ref{Diag:5}} & 2 & 1 & $ R(1_2) $\\
			\hline 	
			3 &     $W^{(2,0,1)}_{6} (2, 1, 1, 1, 2, 1)$ & \footnotesize{~\ref{Diag:2}} & 4 & 1 & $ R(1_2,2_2) $\\
			\hline
			4 &    	$W^{(2,0,1)}_{6} (2, 1, 1, 2, 1, 1)$ & \footnotesize{~\ref{Diag:3}} & 4 & 1 & $ R(1_2,2_2) $\\
			\hline	
			5 &	    $W^{(1,2)}_{6} (2, 1, 1, 1, 2, 1)$ & \footnotesize{~\ref{Diag:7}} & 4 & 1 & $ R(1_2,2_2) $\\
			\hline 
			6 &     $W^{(1,2)}_{6} (2, 1, 1, 2, 1, 1)$ & \footnotesize{~\ref{Diag:15}} & 4 & 1& $ R(1_2,2_2) $ \\
			\hline
			7 &	    $W^{(1,2)}_{6} (3, 1, 1, 1, 1, 1)$ & \footnotesize{~\ref{Diag:6}} & 6 & 2 & $ R(1_6) $\\
			\hline 
			8 &	    $W^{(2,0,1)}_6 (3, 1, 1, 1, 1, 1)$ & \footnotesize{~\ref{Diag:1}} & 6 & 2 & $ R(1_6) $\\
			\hline 
			9 &	    $W^{(3,1)}_{6} (2, 2, 1, 1, 2, 1)$ & \footnotesize{~\ref{Diag:11}} & 8 & 1 & $ R(1_2,3_4,5_2) $\\
			\hline
			10 &	$W^{(3,1)}_{6} (2, 1, 1, 2, 2, 1)$ & \footnotesize{~\ref{Diag:14}} & 8 & 1 & $ R(1_2,3_4,5_2) $\\
			\hline 
			11 &    $W^{(3,1)}_{6} (3, 1, 1, 1, 2, 1)$ & \footnotesize{~\ref{Diag:9}} & 12 & 2 & $ R(1_4,2_4,3_4) $\\
			\hline
			12 &    $W^{(3,1)}_{6,\text{I}} (3, 1, 2, 1, 1, 1)$ & \footnotesize{~\ref{Diag:10}} & 12 & 2 & $ R(1_4,2_4,3_4) $\\
			\hline
			13 &    $W^{(3,1)}_{6,\text{II}} (3, 1, 2, 1, 1, 1)$ & \footnotesize{~\ref{Diag:21}} & 12 & 2 & $ R(1_4,2_4,3_4) $\\
			\hline
			14 &	$W^{(3,1)}_{6} (2, 1, 1, 3, 1, 1)$ & \footnotesize{~\ref{Diag:13}} & 12 & 2 & $ R(1_4,2_4,3_4) $\\
			\hline 
			15 &	$W^{(3,1)}_{6 }(4, 1, 1, 1, 1, 1)$ & \footnotesize{~\ref{Diag:8}} & 24 & 6 & $ R(1_{24}) $\\
			\hline 
		\end{tabular}
		\caption{Cwebs whose mixing matrices are predicted by the Uniqueness theorem appearing at five loops and six lines. The notation we have used for the basis matrix is introduced in~\cite{Agarwal:2022wyk}  where for a basis Cweb with $ S=\{s_1,s_2,s_3,\dots,s_n\} $ the notation of mixing matrix is $ R(S) $. If a given $ s_i $ is repeated multiple times then they are gathered together into the notation as $ S=\{s_1,s_1...,a \,\text{times},s_2,s_2,...,b \,\text{times},...\}\equiv \{{s_1}_a,{s_2}_b,...\} $.
			\label{tab:UNQ}}
	\end{center}
\end{table}\\

The explicit form of the mixing matrices, their diagonalizing matrices, diagonal matrices and corresponding ECFs for these 15 Cwebs are presented in appendix~\ref{sec:appdxFamily}. Note that using the uniqueness theorem we can easily predict the  mixing matrices of approximately  {two-thirds} of the Cwebs at five loops and six lines.  The calculation of remaining 7 Cwebs are described in the forthcoming sections.  

\subsection{New basis Cwebs}
\begin{figure}[b]
	\centering
	\includegraphics[scale=0.5]{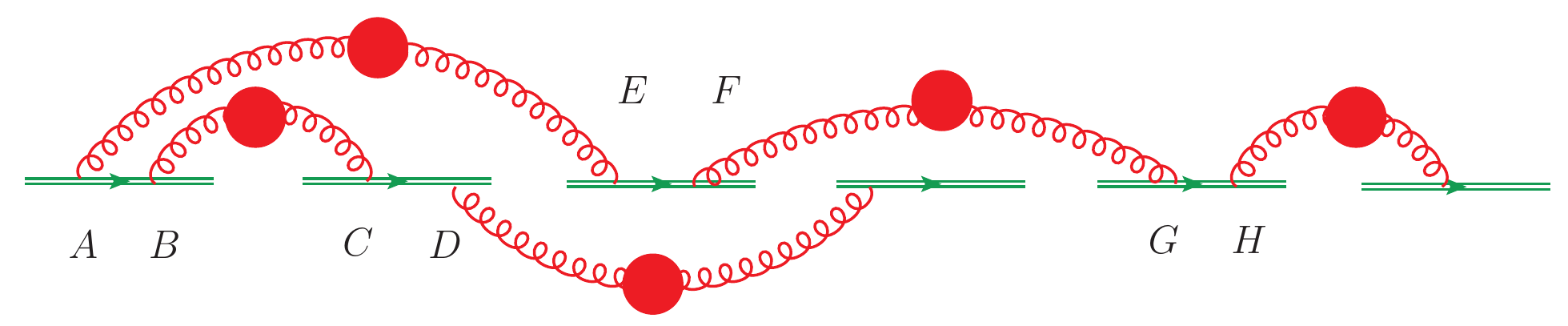}
	\captionof{figure}{ Basis Cweb appearing at five loops and connecting six lines $W^{(5)}_{6} (2, 2, 2, 1, 2, 1)$.}
	\label{fig:ElongBasis}	
\end{figure}

In section~\ref{sec:algorithm}, we have learnt that there are 22  Cwebs at five loops connecting six lines, out of which 15 Cwebs are member of the existing families and their mixing matrices were determined in the previous section. The six out of remaining 7 Cwebs are new basis Cwebs appearing at this perturbative order and their mixing matrices can not be predicted using the Uniqueness theorem. These are listed in table  \ref{tab:BASES}. 
\begin{table}[h]
	\begin{center}
		\begin{tabular}{|c|c|c|c|c|c|}
			\hline 
			S. No.&	Name of the Cweb & Figure & No. of diagrams & $ r(R) $ & Basis matrix\\ 
			\hline 
			1 &	$W^{(5)}_6 (2, 2, 2, 1, 2, 1)$ & \footnotesize{~\ref{Diag:16}} & 16 & 1 & $ R(1_2,4_4,6_2,9_4,11_2,16_2) $\\
			\hline
			2 &  $W^{(5)}_{6} (3, 2, 2, 1, 1, 1)$ & \footnotesize{~\ref{Diag:17}} & 24 & 2 & $ R(1_2,2_4,3_4,4_4,7_4,8_4,11_2) $\\
			\hline
			3 &	$W^{(5)}_{6} (2, 2, 3, 1, 1, 1)$ & \footnotesize{~\ref{Diag:18}} & 24 & 2 & $ R(1_4,3_4,4_4,6_4,7_4,9_4) $\\
			\hline
			4 &	$W^{(5)}_{6} (3, 1, 3, 1, 1, 1)$ & \footnotesize{~\ref{Diag:20}} & 36 & 4 & $ R(1_8,3_{16},4_4,6_8) $\\
			\hline
			5 &	$W^{(5)}_{6} (4, 1, 2, 1, 1, 1)$ & \footnotesize{~\ref{Diag:19}} & 48 & 6 & $ R(1_{12},2_{12},3_{12},4_{12}) $\\
			\hline
			6 &	$W^{(5)}_{6} (5, 1, 1, 1, 1, 1)$ & \footnotesize{~\ref{Diag:22}} & 120 & 24 & $ R(1_{120}) $\\
			\hline 
		\end{tabular}
		\caption{ New basis Cwebs appearing at five loops and six lines along with the basis matrix notation, rank and its dimension. 
			\label{tab:BASES}}
	\end{center}
\end{table}
We give the explicit results for one of the new basis Cwebs,  $W^{(5)}_{6} (2, 2, 2, 1, 2, 1)$  shown in fig.~\ref{fig:basisEXMPL} (whose conventional form~\cite{Agarwal:2022wyk} is given in fig.~\ref{fig:ElongBasis}). Here we have  two attachments on each of the lines $ 1, 2, 3, $ and $ 5, $ giving rise to sixteen diagrams, one of which is shown in fig.~\ref{fig:basisEXMPL}. 
The sequences of diagrams and corresponding $s$-factors are provided in table \ref{tab:basisExmpl}.

\vskip1cm

\begin{minipage}{0.45\textwidth}
\hspace{1.6cm}	\includegraphics[scale=0.5]{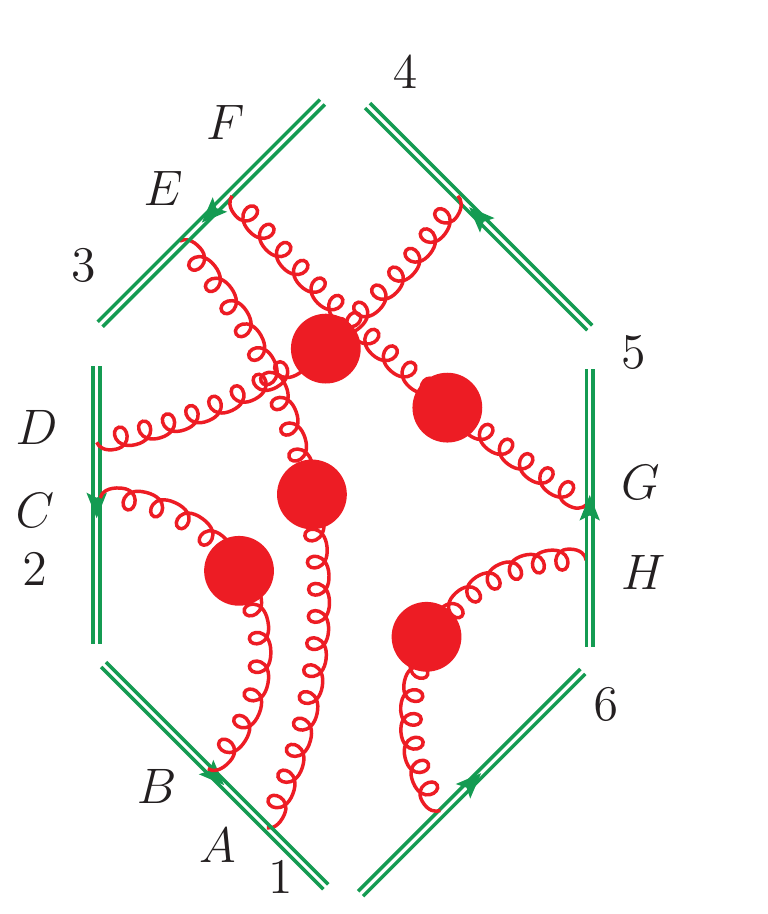}
	\captionof{figure}{$W^{(5)}_{6} (2, 2, 2, 1, 2, 1)$}
	\label{fig:basisEXMPL}	
\end{minipage}
\begin{minipage}{0.35\textwidth}
	\footnotesize{
		\begin{tabular}{ | c | c | c |}
			\hline
			\textbf{Diagrams} & \textbf{Sequences} & \textbf{$s$-factors} \\ \hline
			$d_{1}$ & $\{\{AB\},\{CD\},\{EF\},\{GH\}\}$  &  4\\ \hline
			$d_{2}$ & $\{\{AB\},\{CD\},\{EF\},\{HG\}\}$  &  11\\ \hline
			$d_{3}$ & $\{\{AB\},\{CD\},\{FE\},\{GH\}\}$  &  16\\ \hline
			$d_{4}$ & $\{\{AB\},\{CD\},\{FE\},\{HG\}\}$  &  9\\ \hline
			$d_{5}$ & $\{\{  AB\},\{DC\},\{EF\},\{GH\}\}$  & 1 \\ \hline
			$d_{6}$ & $\{\{  AB\},\{DC\},\{EF\},\{HG\}\}$  & 4 \\ \hline
			$d_{7}$ & $\{\{  AB\},\{DC\},\{FE\},\{GH\}\}$  &9  \\ \hline
			$d_{8}$ & $\{\{  AB\},\{DC\},\{FE\},\{HG\}\}$  & 6 \\ \hline
			$d_{9}$ & $\{\{  BA\},\{CD\},\{EF\},\{GH\}\}$  &  6\\ \hline
			$d_{10}$ & $\{\{  BA\},\{CD\},\{EF\},\{HG\}\}$  &  9\\ \hline
			$d_{11}$ & $\{\{  BA\},\{CD\},\{FE\},\{GH\}\}$  &  4\\ \hline
			$d_{12}$ & $\{\{  BA\},\{CD\},\{FE\},\{HG\}\}$  & 1 \\ \hline
			$d_{13}$ & $\{\{  BA\},\{DC\},\{EF\},\{GH\}\}$  & 9 \\ \hline
			$d_{14}$ & $\{\{  BA\},\{DC\},\{EF\},\{HG\}\}$  & 16 \\ \hline
			$d_{15}$ & $\{\{  BA\},\{DC\},\{FE\},\{GH\}\}$  &  11\\ \hline
			$d_{16}$ & $\{\{  BA\},\{DC\},\{FE\},\{HG\}\}$  &  4\\ \hline
		\end{tabular}
		\captionof{table}{Sequences and $s$-factors}
		\label{tab:basisExmpl}	}
\end{minipage} 

\vspace{1cm}

\noindent	The mixing matrix $R$, the diagonalizing matrix $Y$ and the diagonal matrix $D$ for this Cweb are given by, 

\begin{equation}
	R=\frac{1}{60} \left( \,\,\,
	\begin{array}{cccccccccccccccc}
		3 &\hspace{0.20cm} \hspace{-0.25cm}-3 &\hspace{0.20cm} \hspace{-0.25cm}-3 &\hspace{0.20cm} 3 &\hspace{0.20cm} \hspace{-0.25cm}-3 &\hspace{0.20cm} 3 &\hspace{0.20cm} 3 &\hspace{0.20cm} \hspace{-0.25cm}-3 &\hspace{0.20cm} \hspace{-0.25cm}-3 &\hspace{0.20cm} 3 &\hspace{0.20cm} 3 &\hspace{0.20cm} \hspace{-0.25cm}-3 &\hspace{0.20cm} 3 &\hspace{0.20cm} \hspace{-0.25cm}-3 &\hspace{0.20cm} \hspace{-0.25cm}-3 &\hspace{0.20cm} 3 \\
		\hspace{-0.25cm}-2 &\hspace{0.20cm} 2 &\hspace{0.20cm} 2 &\hspace{0.20cm} \hspace{-0.25cm}-2 &\hspace{0.20cm} 2 &\hspace{0.20cm} \hspace{-0.25cm}-2 &\hspace{0.20cm} \hspace{-0.25cm}-2 &\hspace{0.20cm} 2 &\hspace{0.20cm} 2 &\hspace{0.20cm} \hspace{-0.25cm}-2 &\hspace{0.20cm} \hspace{-0.25cm}-2 &\hspace{0.20cm} 2 &\hspace{0.20cm} \hspace{-0.25cm}-2 &\hspace{0.20cm} 2 &\hspace{0.20cm} 2 &\hspace{0.20cm} \hspace{-0.25cm}-2 \\
		\hspace{-0.25cm}-2 &\hspace{0.20cm} 2 &\hspace{0.20cm} 2 &\hspace{0.20cm} \hspace{-0.25cm}-2 &\hspace{0.20cm} 2 &\hspace{0.20cm} \hspace{-0.25cm}-2 &\hspace{0.20cm} \hspace{-0.25cm}-2 &\hspace{0.20cm} 2 &\hspace{0.20cm} 2 &\hspace{0.20cm} \hspace{-0.25cm}-2 &\hspace{0.20cm} \hspace{-0.25cm}-2 &\hspace{0.20cm} 2 &\hspace{0.20cm} \hspace{-0.25cm}-2 &\hspace{0.20cm} 2 &\hspace{0.20cm} 2 &\hspace{0.20cm} \hspace{-0.25cm}-2 \\
		3 &\hspace{0.20cm} \hspace{-0.25cm}-3 &\hspace{0.20cm} \hspace{-0.25cm}-3 &\hspace{0.20cm} 3 &\hspace{0.20cm} \hspace{-0.25cm}-3 &\hspace{0.20cm} 3 &\hspace{0.20cm} 3 &\hspace{0.20cm} \hspace{-0.25cm}-3 &\hspace{0.20cm} \hspace{-0.25cm}-3 &\hspace{0.20cm} 3 &\hspace{0.20cm} 3 &\hspace{0.20cm} \hspace{-0.25cm}-3 &\hspace{0.20cm} 3 &\hspace{0.20cm} \hspace{-0.25cm}-3 &\hspace{0.20cm} \hspace{-0.25cm}-3 &\hspace{0.20cm} 3 \\
		\hspace{-0.25cm}-12 &\hspace{0.20cm} 12 &\hspace{0.20cm} 12 &\hspace{0.20cm} \hspace{-0.25cm}-12 &\hspace{0.20cm} 12 &\hspace{0.20cm} \hspace{-0.25cm}-12 &\hspace{0.20cm} \hspace{-0.25cm}-12 &\hspace{0.20cm} 12 &\hspace{0.20cm} 12 &\hspace{0.20cm} \hspace{-0.25cm}-12 &\hspace{0.20cm} \hspace{-0.25cm}-12 &\hspace{0.20cm} 12 &\hspace{0.20cm} \hspace{-0.25cm}-12 &\hspace{0.20cm} 12 &\hspace{0.20cm} 12 &\hspace{0.20cm} \hspace{-0.25cm}-12 \\
		3 &\hspace{0.20cm} \hspace{-0.25cm}-3 &\hspace{0.20cm} \hspace{-0.25cm}-3 &\hspace{0.20cm} 3 &\hspace{0.20cm} \hspace{-0.25cm}-3 &\hspace{0.20cm} 3 &\hspace{0.20cm} 3 &\hspace{0.20cm} \hspace{-0.25cm}-3 &\hspace{0.20cm} \hspace{-0.25cm}-3 &\hspace{0.20cm} 3 &\hspace{0.20cm} 3 &\hspace{0.20cm} \hspace{-0.25cm}-3 &\hspace{0.20cm} 3 &\hspace{0.20cm} \hspace{-0.25cm}-3 &\hspace{0.20cm} \hspace{-0.25cm}-3 &\hspace{0.20cm} 3 \\
		3 &\hspace{0.20cm} \hspace{-0.25cm}-3 &\hspace{0.20cm} \hspace{-0.25cm}-3 &\hspace{0.20cm} 3 &\hspace{0.20cm} \hspace{-0.25cm}-3 &\hspace{0.20cm} 3 &\hspace{0.20cm} 3 &\hspace{0.20cm} \hspace{-0.25cm}-3 &\hspace{0.20cm} \hspace{-0.25cm}-3 &\hspace{0.20cm} 3 &\hspace{0.20cm} 3 &\hspace{0.20cm} \hspace{-0.25cm}-3 &\hspace{0.20cm} 3 &\hspace{0.20cm} \hspace{-0.25cm}-3 &\hspace{0.20cm} \hspace{-0.25cm}-3 &\hspace{0.20cm} 3 \\
		\hspace{-0.25cm}-2 &\hspace{0.20cm} 2 &\hspace{0.20cm} 2 &\hspace{0.20cm} \hspace{-0.25cm}-2 &\hspace{0.20cm} 2 &\hspace{0.20cm} \hspace{-0.25cm}-2 &\hspace{0.20cm} \hspace{-0.25cm}-2 &\hspace{0.20cm} 2 &\hspace{0.20cm} 2 &\hspace{0.20cm} \hspace{-0.25cm}-2 &\hspace{0.20cm} \hspace{-0.25cm}-2 &\hspace{0.20cm} 2 &\hspace{0.20cm} \hspace{-0.25cm}-2 &\hspace{0.20cm} 2 &\hspace{0.20cm} 2 &\hspace{0.20cm} \hspace{-0.25cm}-2 \\
		\hspace{-0.25cm}-2 &\hspace{0.20cm} 2 &\hspace{0.20cm} 2 &\hspace{0.20cm} \hspace{-0.25cm}-2 &\hspace{0.20cm} 2 &\hspace{0.20cm} \hspace{-0.25cm}-2 &\hspace{0.20cm} \hspace{-0.25cm}-2 &\hspace{0.20cm} 2 &\hspace{0.20cm} 2 &\hspace{0.20cm} \hspace{-0.25cm}-2 &\hspace{0.20cm} \hspace{-0.25cm}-2 &\hspace{0.20cm} 2 &\hspace{0.20cm} \hspace{-0.25cm}-2 &\hspace{0.20cm} 2 &\hspace{0.20cm} 2 &\hspace{0.20cm} \hspace{-0.25cm}-2 \\
		3 &\hspace{0.20cm} \hspace{-0.25cm}-3 &\hspace{0.20cm} \hspace{-0.25cm}-3 &\hspace{0.20cm} 3 &\hspace{0.20cm} \hspace{-0.25cm}-3 &\hspace{0.20cm} 3 &\hspace{0.20cm} 3 &\hspace{0.20cm} \hspace{-0.25cm}-3 &\hspace{0.20cm} \hspace{-0.25cm}-3 &\hspace{0.20cm} 3 &\hspace{0.20cm} 3 &\hspace{0.20cm} \hspace{-0.25cm}-3 &\hspace{0.20cm} 3 &\hspace{0.20cm} \hspace{-0.25cm}-3 &\hspace{0.20cm} \hspace{-0.25cm}-3 &\hspace{0.20cm} 3 \\
		3 &\hspace{0.20cm} \hspace{-0.25cm}-3 &\hspace{0.20cm} \hspace{-0.25cm}-3 &\hspace{0.20cm} 3 &\hspace{0.20cm} \hspace{-0.25cm}-3 &\hspace{0.20cm} 3 &\hspace{0.20cm} 3 &\hspace{0.20cm} \hspace{-0.25cm}-3 &\hspace{0.20cm} \hspace{-0.25cm}-3 &\hspace{0.20cm} 3 &\hspace{0.20cm} 3 &\hspace{0.20cm} \hspace{-0.25cm}-3 &\hspace{0.20cm} 3 &\hspace{0.20cm} \hspace{-0.25cm}-3 &\hspace{0.20cm} \hspace{-0.25cm}-3 &\hspace{0.20cm} 3 \\
		\hspace{-0.25cm}-12 &\hspace{0.20cm} 12 &\hspace{0.20cm} 12 &\hspace{0.20cm} \hspace{-0.25cm}-12 &\hspace{0.20cm} 12 &\hspace{0.20cm} \hspace{-0.25cm}-12 &\hspace{0.20cm} \hspace{-0.25cm}-12 &\hspace{0.20cm} 12 &\hspace{0.20cm} 12 &\hspace{0.20cm} \hspace{-0.25cm}-12 &\hspace{0.20cm} \hspace{-0.25cm}-12 &\hspace{0.20cm} 12 &\hspace{0.20cm} \hspace{-0.25cm}-12 &\hspace{0.20cm} 12 &\hspace{0.20cm} 12 &\hspace{0.20cm} \hspace{-0.25cm}-12 \\
		3 &\hspace{0.20cm} \hspace{-0.25cm}-3 &\hspace{0.20cm} \hspace{-0.25cm}-3 &\hspace{0.20cm} 3 &\hspace{0.20cm} \hspace{-0.25cm}-3 &\hspace{0.20cm} 3 &\hspace{0.20cm} 3 &\hspace{0.20cm} \hspace{-0.25cm}-3 &\hspace{0.20cm} \hspace{-0.25cm}-3 &\hspace{0.20cm} 3 &\hspace{0.20cm} 3 &\hspace{0.20cm} \hspace{-0.25cm}-3 &\hspace{0.20cm} 3 &\hspace{0.20cm} \hspace{-0.25cm}-3 &\hspace{0.20cm} \hspace{-0.25cm}-3 &\hspace{0.20cm} 3 \\
		\hspace{-0.25cm}-2 &\hspace{0.20cm} 2 &\hspace{0.20cm} 2 &\hspace{0.20cm} \hspace{-0.25cm}-2 &\hspace{0.20cm} 2 &\hspace{0.20cm} \hspace{-0.25cm}-2 &\hspace{0.20cm} \hspace{-0.25cm}-2 &\hspace{0.20cm} 2 &\hspace{0.20cm} 2 &\hspace{0.20cm} \hspace{-0.25cm}-2 &\hspace{0.20cm} \hspace{-0.25cm}-2 &\hspace{0.20cm} 2 &\hspace{0.20cm} \hspace{-0.25cm}-2 &\hspace{0.20cm} 2 &\hspace{0.20cm} 2 &\hspace{0.20cm} \hspace{-0.25cm}-2 \\
		\hspace{-0.25cm}-2 &\hspace{0.20cm} 2 &\hspace{0.20cm} 2 &\hspace{0.20cm} \hspace{-0.25cm}-2 &\hspace{0.20cm} 2 &\hspace{0.20cm} \hspace{-0.25cm}-2 &\hspace{0.20cm} \hspace{-0.25cm}-2 &\hspace{0.20cm} 2 &\hspace{0.20cm} 2 &\hspace{0.20cm} \hspace{-0.25cm}-2 &\hspace{0.20cm} \hspace{-0.25cm}-2 &\hspace{0.20cm} 2 &\hspace{0.20cm} \hspace{-0.25cm}-2 &\hspace{0.20cm} 2 &\hspace{0.20cm} 2 &\hspace{0.20cm} \hspace{-0.25cm}-2 \\
		3 &\hspace{0.20cm} \hspace{-0.25cm}-3 &\hspace{0.20cm} \hspace{-0.25cm}-3 &\hspace{0.20cm} 3 &\hspace{0.20cm} \hspace{-0.25cm}-3 &\hspace{0.20cm} 3 &\hspace{0.20cm} 3 &\hspace{0.20cm} \hspace{-0.25cm}-3 &\hspace{0.20cm} \hspace{-0.25cm}-3 &\hspace{0.20cm} 3 &\hspace{0.20cm} 3 &\hspace{0.20cm} \hspace{-0.25cm}-3 &\hspace{0.20cm} 3 &\hspace{0.20cm} \hspace{-0.25cm}-3 &\hspace{0.20cm} \hspace{-0.25cm}-3 &\hspace{0.20cm} 3. \\
	\end{array}
	\,\,\,\right),\nonumber
	\label{MM:basis}
\end{equation}
\vspace*{0.1cm}
\begin{equation}
	Y=\frac{1}{3}\left(\,\,\,
	\begin{array}{cccccccccccccccc}
		3 &\hspace{0.20cm} \hspace{-0.25cm}-3 &\hspace{0.20cm} \hspace{-0.25cm}-3 &\hspace{0.20cm} 3 &\hspace{0.20cm} \hspace{-0.25cm}-3 &\hspace{0.20cm} 3 &\hspace{0.20cm} 3 &\hspace{0.20cm} \hspace{-0.25cm}-3 &\hspace{0.20cm} \hspace{-0.25cm}-3 &\hspace{0.20cm} 3 &\hspace{0.20cm} 3 &\hspace{0.20cm} \hspace{-0.25cm}-3 &\hspace{0.20cm} 3 &\hspace{0.20cm} \hspace{-0.25cm}-3 &\hspace{0.20cm} \hspace{-0.25cm}-3 &\hspace{0.20cm} 3 \\
		\hspace{-0.25cm}-3 &\hspace{0.20cm} 0 &\hspace{0.20cm} 0 &\hspace{0.20cm} 0 &\hspace{0.20cm} 0 &\hspace{0.20cm} 0 &\hspace{0.20cm} 0 &\hspace{0.20cm} 0 &\hspace{0.20cm} 0 &\hspace{0.20cm} 0 &\hspace{0.20cm} 0 &\hspace{0.20cm} 0 &\hspace{0.20cm} 0 &\hspace{0.20cm} 0 &\hspace{0.20cm} 0 &\hspace{0.20cm} 3 \\
		2 &\hspace{0.20cm} 0 &\hspace{0.20cm} 0 &\hspace{0.20cm} 0 &\hspace{0.20cm} 0 &\hspace{0.20cm} 0 &\hspace{0.20cm} 0 &\hspace{0.20cm} 0 &\hspace{0.20cm} 0 &\hspace{0.20cm} 0 &\hspace{0.20cm} 0 &\hspace{0.20cm} 0 &\hspace{0.20cm} 0 &\hspace{0.20cm} 0 &\hspace{0.20cm} 3 &\hspace{0.20cm} 0 \\
		2 &\hspace{0.20cm} 0 &\hspace{0.20cm} 0 &\hspace{0.20cm} 0 &\hspace{0.20cm} 0 &\hspace{0.20cm} 0 &\hspace{0.20cm} 0 &\hspace{0.20cm} 0 &\hspace{0.20cm} 0 &\hspace{0.20cm} 0 &\hspace{0.20cm} 0 &\hspace{0.20cm} 0 &\hspace{0.20cm} 0 &\hspace{0.20cm} 3 &\hspace{0.20cm} 0 &\hspace{0.20cm} 0 \\
		\hspace{-0.25cm}-3 &\hspace{0.20cm} 0 &\hspace{0.20cm} 0 &\hspace{0.20cm} 0 &\hspace{0.20cm} 0 &\hspace{0.20cm} 0 &\hspace{0.20cm} 0 &\hspace{0.20cm} 0 &\hspace{0.20cm} 0 &\hspace{0.20cm} 0 &\hspace{0.20cm} 0 &\hspace{0.20cm} 0 &\hspace{0.20cm} 3 &\hspace{0.20cm} 0 &\hspace{0.20cm} 0 &\hspace{0.20cm} 0 \\
		12 &\hspace{0.20cm} 0 &\hspace{0.20cm} 0 &\hspace{0.20cm} 0 &\hspace{0.20cm} 0 &\hspace{0.20cm} 0 &\hspace{0.20cm} 0 &\hspace{0.20cm} 0 &\hspace{0.20cm} 0 &\hspace{0.20cm} 0 &\hspace{0.20cm} 0 &\hspace{0.20cm} 3 &\hspace{0.20cm} 0 &\hspace{0.20cm} 0 &\hspace{0.20cm} 0 &\hspace{0.20cm} 0 \\
		\hspace{-0.25cm}-3 &\hspace{0.20cm} 0 &\hspace{0.20cm} 0 &\hspace{0.20cm} 0 &\hspace{0.20cm} 0 &\hspace{0.20cm} 0 &\hspace{0.20cm} 0 &\hspace{0.20cm} 0 &\hspace{0.20cm} 0 &\hspace{0.20cm} 0 &\hspace{0.20cm} 3 &\hspace{0.20cm} 0 &\hspace{0.20cm} 0 &\hspace{0.20cm} 0 &\hspace{0.20cm} 0 &\hspace{0.20cm} 0 \\
		\hspace{-0.25cm}-3 &\hspace{0.20cm} 0 &\hspace{0.20cm} 0 &\hspace{0.20cm} 0 &\hspace{0.20cm} 0 &\hspace{0.20cm} 0 &\hspace{0.20cm} 0 &\hspace{0.20cm} 0 &\hspace{0.20cm} 0 &\hspace{0.20cm} 3 &\hspace{0.20cm} 0 &\hspace{0.20cm} 0 &\hspace{0.20cm} 0 &\hspace{0.20cm} 0 &\hspace{0.20cm} 0 &\hspace{0.20cm} 0 \\
		2 &\hspace{0.20cm} 0 &\hspace{0.20cm} 0 &\hspace{0.20cm} 0 &\hspace{0.20cm} 0 &\hspace{0.20cm} 0 &\hspace{0.20cm} 0 &\hspace{0.20cm} 0 &\hspace{0.20cm} 3 &\hspace{0.20cm} 0 &\hspace{0.20cm} 0 &\hspace{0.20cm} 0 &\hspace{0.20cm} 0 &\hspace{0.20cm} 0 &\hspace{0.20cm} 0 &\hspace{0.20cm} 0 \\
		2 &\hspace{0.20cm} 0 &\hspace{0.20cm} 0 &\hspace{0.20cm} 0 &\hspace{0.20cm} 0 &\hspace{0.20cm} 0 &\hspace{0.20cm} 0 &\hspace{0.20cm} 3 &\hspace{0.20cm} 0 &\hspace{0.20cm} 0 &\hspace{0.20cm} 0 &\hspace{0.20cm} 0 &\hspace{0.20cm} 0 &\hspace{0.20cm} 0 &\hspace{0.20cm} 0 &\hspace{0.20cm} 0 \\
		\hspace{-0.25cm}-3 &\hspace{0.20cm} 0 &\hspace{0.20cm} 0 &\hspace{0.20cm} 0 &\hspace{0.20cm} 0 &\hspace{0.20cm} 0 &\hspace{0.20cm} 3 &\hspace{0.20cm} 0 &\hspace{0.20cm} 0 &\hspace{0.20cm} 0 &\hspace{0.20cm} 0 &\hspace{0.20cm} 0 &\hspace{0.20cm} 0 &\hspace{0.20cm} 0 &\hspace{0.20cm} 0 &\hspace{0.20cm} 0 \\
		\hspace{-0.25cm}-3 &\hspace{0.20cm} 0 &\hspace{0.20cm} 0 &\hspace{0.20cm} 0 &\hspace{0.20cm} 0 &\hspace{0.20cm} 3 &\hspace{0.20cm} 0 &\hspace{0.20cm} 0 &\hspace{0.20cm} 0 &\hspace{0.20cm} 0 &\hspace{0.20cm} 0 &\hspace{0.20cm} 0 &\hspace{0.20cm} 0 &\hspace{0.20cm} 0 &\hspace{0.20cm} 0 &\hspace{0.20cm} 0 \\
		12 &\hspace{0.20cm} 0 &\hspace{0.20cm} 0 &\hspace{0.20cm} 0 &\hspace{0.20cm} 3 &\hspace{0.20cm} 0 &\hspace{0.20cm} 0 &\hspace{0.20cm} 0 &\hspace{0.20cm} 0 &\hspace{0.20cm} 0 &\hspace{0.20cm} 0 &\hspace{0.20cm} 0 &\hspace{0.20cm} 0 &\hspace{0.20cm} 0 &\hspace{0.20cm} 0 &\hspace{0.20cm} 0 \\
		\hspace{-0.25cm}-3 &\hspace{0.20cm} 0 &\hspace{0.20cm} 0 &\hspace{0.20cm} 3 &\hspace{0.20cm} 0 &\hspace{0.20cm} 0 &\hspace{0.20cm} 0 &\hspace{0.20cm} 0 &\hspace{0.20cm} 0 &\hspace{0.20cm} 0 &\hspace{0.20cm} 0 &\hspace{0.20cm} 0 &\hspace{0.20cm} 0 &\hspace{0.20cm} 0 &\hspace{0.20cm} 0 &\hspace{0.20cm} 0 \\
		2 &\hspace{0.20cm} 0 &\hspace{0.20cm} 3 &\hspace{0.20cm} 0 &\hspace{0.20cm} 0 &\hspace{0.20cm} 0 &\hspace{0.20cm} 0 &\hspace{0.20cm} 0 &\hspace{0.20cm} 0 &\hspace{0.20cm} 0 &\hspace{0.20cm} 0 &\hspace{0.20cm} 0 &\hspace{0.20cm} 0 &\hspace{0.20cm} 0 &\hspace{0.20cm} 0 &\hspace{0.20cm} 0 \\
		2 &\hspace{0.20cm} 3 &\hspace{0.20cm} 0 &\hspace{0.20cm} 0 &\hspace{0.20cm} 0 &\hspace{0.20cm} 0 &\hspace{0.20cm} 0 &\hspace{0.20cm} 0 &\hspace{0.20cm} 0 &\hspace{0.20cm} 0 &\hspace{0.20cm} 0 &\hspace{0.20cm} 0 &\hspace{0.20cm} 0 &\hspace{0.20cm} 0 &\hspace{0.20cm} 0 &\hspace{0.20cm} 0 \\
	\end{array}
\right) \, , \quad  D=(\mathbf{1}_1,0).
\end{equation} \\
This mixing matrix  follows all the known properties and the only exponentiated colour factor of this Cweb is given by, 
\begin{align}
	(YC)_1=& - f^{abr}f^{adp}f^{dcg}f^{ecm}\tp{1} \trr{2} \tg{3} \tb{4} \tm{5} \te{6}\, .
\end{align}
In table~\ref{tab:BASES}, we list down all the new basis Cwebs that appear at five loops and in the appendix~\ref{sec:appdxBasis}, we present the explicit calculation of these Cwebs. 
\subsection{Orphan Cwebs}
As noted above these new type of Cwebs  start appearing at five loops; we call these new type as Orphan Cwebs. \\

\noindent {\it Orphan Cwebs}:  The Orphan Cwebs at a given order are obtained by connecting the highest number of Wilson lines,  that are neither a basis Cweb nor a member of any family that appears at lower orders.\footnote{An orphan Cweb at a given loop order will form its own family with Cwebs at higher perturbative orders.}
\\

\begin{figure}[h]
	\centering
	\includegraphics[scale=0.09]{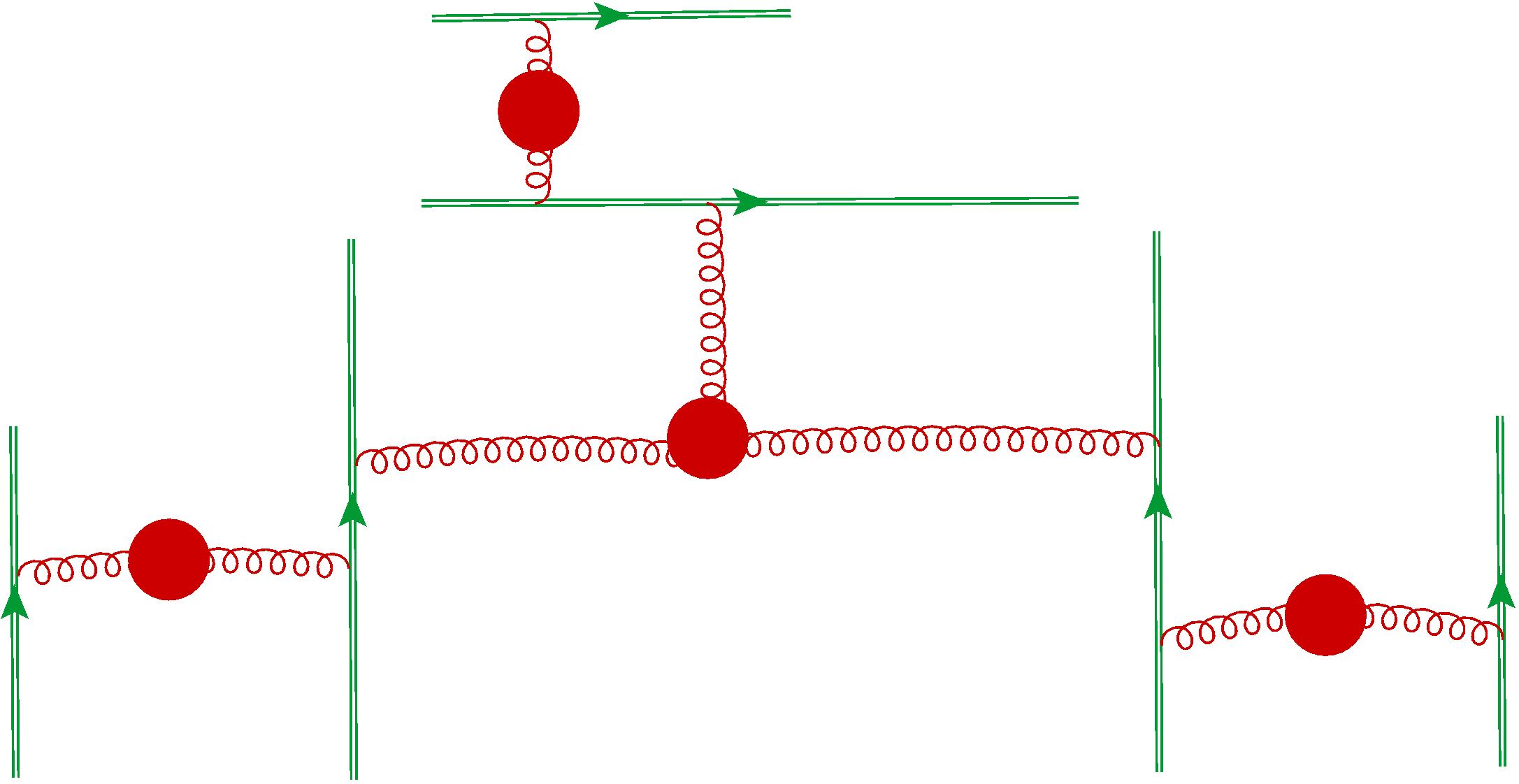}
	\caption{The Orphan Cweb $W^{(3,1)}_{6} (2, 2, 2, 1, 1, 1)$ present at five loops. It is evident from this representation that none of the gluon attachments can be removed without changing its shuffle and mixing matrix, thus this Cweb can not be traced to a basis Cweb.}
	\label{fig:orphan}
\end{figure}

\noindent Orphan Cwebs have all $ s(d_i)\neq0 $ but have at least one $ m $-point ($ m>2 $) correlator. 
The only Orphan Cweb  at five loops  is $W^{(3,1)}_{6} (2, 2, 2, 1, 1, 1)$ and is shown in fig.~\ref{fig:OrphanExmpl}. 
This Cweb is not a basis due to the presence of a three-point gluon correlator also it is not a member of a known family as its shuffle is not same as any of the known basis Cwebs
 which is evident from its {alternative} representation shown in fig.~\ref{fig:orphan}. 
This Cweb has eight diagrams, as there are two attachments on each of the lines 1, 2, and 3, one of the diagram is shown in fig.~\ref{fig:OrphanExmpl}. The sequences of diagrams and corresponding $s$-factors are provided in the table \ref{tab:orphanEXMPL}.
\vskip0.5cm
\begin{minipage}{0.45\textwidth}

\hspace{1.6cm}	\includegraphics[scale=0.5]{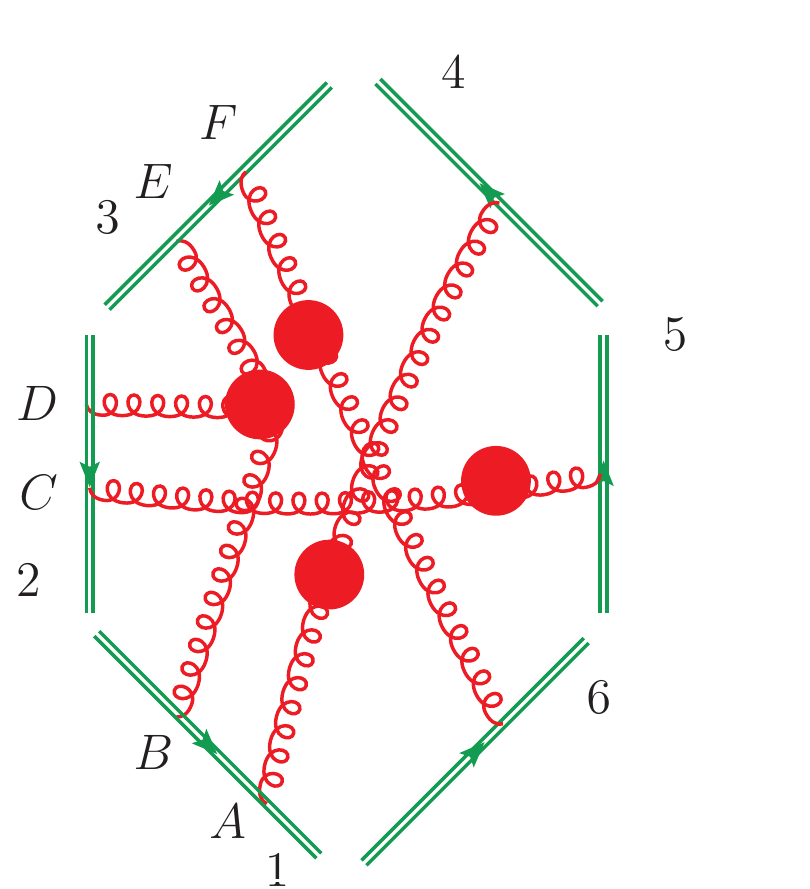}	
	\captionof{figure}{$W^{(3,1)}_{6} (2, 2, 2, 1, 1, 1)$}
	\label{fig:OrphanExmpl}
\end{minipage}
\begin{minipage}{0.45\textwidth}
	\footnotesize{
		\begin{tabular}{ | c | c | c |}
			\hline
			\textbf{Diagrams} & \textbf{Sequences} & \textbf{$s$-factors} \\ \hline
			$d_1$ & $\lbrace \lbrace AB \rbrace,  \lbrace CD \rbrace, \lbrace FE \rbrace\rbrace$ & 6 \\ 
			\hline
			$d_2$ & $\lbrace \lbrace AB \rbrace,  \lbrace CD \rbrace, \lbrace EF \rbrace\rbrace$ & 2 \\ 
			\hline	
			$d_3$ & $\lbrace \lbrace AB \rbrace,  \lbrace DC \rbrace, \lbrace EF \rbrace\rbrace$ & 2 \\ 
			\hline	
			$d_4$ & $\lbrace \lbrace AB \rbrace,  \lbrace DC \rbrace, \lbrace FE \rbrace\rbrace$ & 2 \\ 
			\hline
			$d_5$ & $\lbrace \lbrace BA \rbrace,  \lbrace CD \rbrace, \lbrace EF \rbrace\rbrace$ & 2 \\ 
			\hline
			$d_6$ & $\lbrace \lbrace BA \rbrace,  \lbrace CD \rbrace, \lbrace FE \rbrace\rbrace$ & 2 \\ 
			\hline
			$d_7$ & $\lbrace \lbrace BA \rbrace,  \lbrace DC \rbrace, \lbrace EF \rbrace\rbrace$ & 2 \\ 
			\hline
			$d_8$ & $\lbrace \lbrace BA \rbrace,  \lbrace DC \rbrace, \lbrace FE \rbrace\rbrace$ &6 \\ 
			\hline
		\end{tabular}
		\captionof{table}{Sequences and $s$-factors}
		\label{tab:orphanEXMPL}	}
\end{minipage} 
\vspace{0.5cm}

\noindent	The mixing matrix $R$, the diagonalizing matrix $Y$ and the diagonal matrix $D$ for this  Cweb  are
\begin{align}
	R=\frac{1}{6} \left( \,\,\,
	\begin{array}{cccccccc}
		0 &\hspace{0.20cm} 0 &\hspace{0.20cm} 0 &\hspace{0.20cm} 0 &\hspace{0.20cm} 0 &\hspace{0.20cm} 0 &\hspace{0.20cm} 0 &\hspace{0.20cm} 0 \\
		\hspace{-0.25cm}-1 &\hspace{0.20cm} 1 &\hspace{0.20cm} 1 &\hspace{0.20cm} \hspace{-0.25cm}-1 &\hspace{0.20cm} 1 &\hspace{0.20cm} \hspace{-0.25cm}-1 &\hspace{0.20cm} \hspace{-0.25cm}-1 &\hspace{0.20cm} 1 \\
		\hspace{-0.25cm}-1 &\hspace{0.20cm} 1 &\hspace{0.20cm} 1 &\hspace{0.20cm} \hspace{-0.25cm}-1 &\hspace{0.20cm} 1 &\hspace{0.20cm} \hspace{-0.25cm}-1 &\hspace{0.20cm} \hspace{-0.25cm}-1 &\hspace{0.20cm} 1 \\
		1 &\hspace{0.20cm} \hspace{-0.25cm}-1 &\hspace{0.20cm} \hspace{-0.25cm}-1 &\hspace{0.20cm} 1 &\hspace{0.20cm} \hspace{-0.25cm}-1 &\hspace{0.20cm} 1 &\hspace{0.20cm} 1 &\hspace{0.20cm} \hspace{-0.25cm}-1 \\
		\hspace{-0.25cm}-1 &\hspace{0.20cm} 1 &\hspace{0.20cm} 1 &\hspace{0.20cm} \hspace{-0.25cm}-1 &\hspace{0.20cm} 1 &\hspace{0.20cm} \hspace{-0.25cm}-1 &\hspace{0.20cm} \hspace{-0.25cm}-1 &\hspace{0.20cm} 1 \\
		1 &\hspace{0.20cm} \hspace{-0.25cm}-1 &\hspace{0.20cm} \hspace{-0.25cm}-1 &\hspace{0.20cm} 1 &\hspace{0.20cm} \hspace{-0.25cm}-1 &\hspace{0.20cm} 1 &\hspace{0.20cm} 1 &\hspace{0.20cm} \hspace{-0.25cm}-1 \\
		1 &\hspace{0.20cm} \hspace{-0.25cm}-1 &\hspace{0.20cm} \hspace{-0.25cm}-1 &\hspace{0.20cm} 1 &\hspace{0.20cm} \hspace{-0.25cm}-1 &\hspace{0.20cm} 1 &\hspace{0.20cm} 1 &\hspace{0.20cm} \hspace{-0.25cm}-1 \\
		0 &\hspace{0.20cm} 0 &\hspace{0.20cm} 0 &\hspace{0.20cm} 0 &\hspace{0.20cm} 0 &\hspace{0.20cm} 0 &\hspace{0.20cm} 0 &\hspace{0.20cm} 0 \\
	\end{array}
	 \,\,\, \right)\,,\quad  Y=\left(\,\,\,
	\begin{array}{cccccccc}
		\hspace{-0.25cm}-1 &\hspace{0.20cm} 1 &\hspace{0.20cm} 1 &\hspace{0.20cm} \hspace{-0.25cm}-1 &\hspace{0.20cm} 1 &\hspace{0.20cm} \hspace{-0.25cm}-1 &\hspace{0.20cm} \hspace{-0.25cm}-1 &\hspace{0.20cm} 1 \\
		0 &\hspace{0.20cm} 0 &\hspace{0.20cm} 0 &\hspace{0.20cm} 0 &\hspace{0.20cm} 0 &\hspace{0.20cm} 0 &\hspace{0.20cm} 0 &\hspace{0.20cm} 1 \\
		0 &\hspace{0.20cm} 1 &\hspace{0.20cm} 0 &\hspace{0.20cm} 0 &\hspace{0.20cm} 0 &\hspace{0.20cm} 0 &\hspace{0.20cm} 1 &\hspace{0.20cm} 0 \\
		0 &\hspace{0.20cm} 1 &\hspace{0.20cm} 0 &\hspace{0.20cm} 0 &\hspace{0.20cm} 0 &\hspace{0.20cm} 1 &\hspace{0.20cm} 0 &\hspace{0.20cm} 0 \\
		0 &\hspace{0.20cm} \hspace{-0.25cm}-1 &\hspace{0.20cm} 0 &\hspace{0.20cm} 0 &\hspace{0.20cm} 1 &\hspace{0.20cm} 0 &\hspace{0.20cm} 0 &\hspace{0.20cm} 0 \\
		0 &\hspace{0.20cm} 1 &\hspace{0.20cm} 0 &\hspace{0.20cm} 1 &\hspace{0.20cm} 0 &\hspace{0.20cm} 0 &\hspace{0.20cm} 0 &\hspace{0.20cm} 0 \\
		0 &\hspace{0.20cm} \hspace{-0.25cm}-1 &\hspace{0.20cm} 1 &\hspace{0.20cm} 0 &\hspace{0.20cm} 0 &\hspace{0.20cm} 0 &\hspace{0.20cm} 0 &\hspace{0.20cm} 0 \\
		1 &\hspace{0.20cm} 0 &\hspace{0.20cm} 0 &\hspace{0.20cm} 0 &\hspace{0.20cm} 0 &\hspace{0.20cm} 0 &\hspace{0.20cm} 0 &\hspace{0.20cm} 0 \\
	\end{array}
\right)\,,
	& \quad  D=(\mathbf{1}_1,0).
	\label{MM:orphan}
\end{align}
The mixing matrix in eq.~(\ref{MM:orphan}) satisfies all the known properties and the only exponentiated colour factor of this Cweb is
\begin{align}
	(YC)_1=&  f^{abc}f^{aep}f^{dcr}f^{gbq} \tp{1} \tq{2} \trr{3} \te{4} \tg{5} \td{6} \, .
\end{align}
\begin{table}[t] 
	\begin{center}
		\begin{tabular}{|c|c|c|c|c|}
			\hline
			$ {\cal O}(\alpha_s^n) $ & Total Cwebs & Cwebs from existing families & Basis Cwebs & Orphan Cwebs \\
			\hline
			1 & 1 & 0 & 1 & 0 \\
			\hline
			2 & 1 & 0 & 1 & 0 \\
			\hline	
			3 & 4 & 1 & 2 & 0 \\
			\hline
			4 & 8 & 5 & 3 & 0 \\
			\hline
			5 & 22 & 15 & 6 & 1 \\
			\hline
		\end{tabular}	
		\caption{Classification of Cwebs appearing at maximum number of lines at a given loop order. The Cwebs that contain only one correlator are not included here at $ {\cal O}(\alpha_s^2) $ and above.\label{tab:BasisandLoops}}
	\end{center}	
\end{table}

To summarize, in this section we have learnt that Cwebs that are formed by connecting the maximum number of Wilson lines at five loops and beyond can be categorized into three groups: Cwebs belonging to existing families, new basis Cwebs, and Orphan Cwebs.  

	In table~\ref{tab:BasisandLoops}, we present the number of Cwebs that belong to these categories up to five loops.
	From the table it is evident that the fraction of total number of Cwebs (at the largest number of Wilson lines) that belong to the existing families increases with the loop order. Out of the 22 Cwebs that we get, the mixing matrices of 15 of them can be read off from the known results from previous studies. Fresh computations need to be done only for the 6 new basis Cwebs, and 1 Orphan Cweb. The results for mixing matrices of all the Cwebs including six new basis and one Orphan Cweb are presented in the appendix \ref{sec:appendix}.

\begin{figure}[h]
	\centering
	\vspace{-3mm}
	\hspace{-0.2cm}
	\subfloat[]{\includegraphics[scale=0.45]{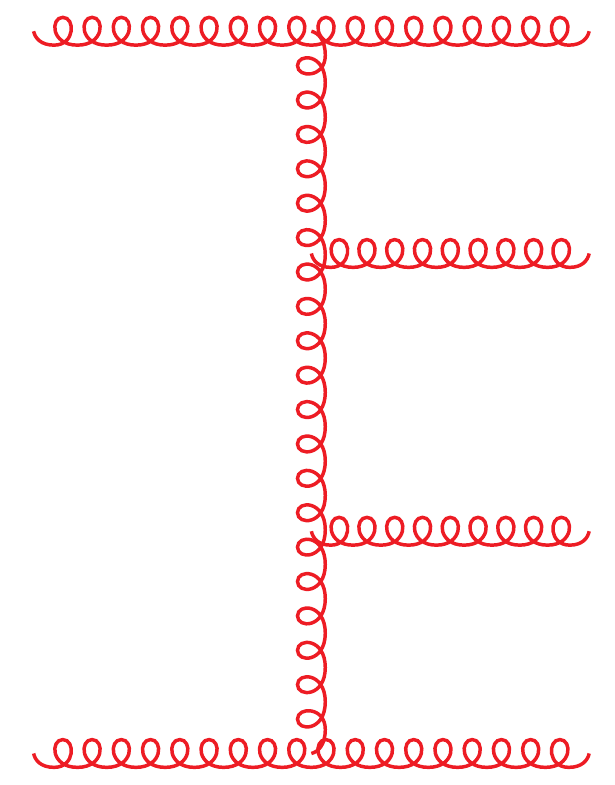} } 
	 \hspace{1.9cm}
	\subfloat[]{\includegraphics[scale=0.45]{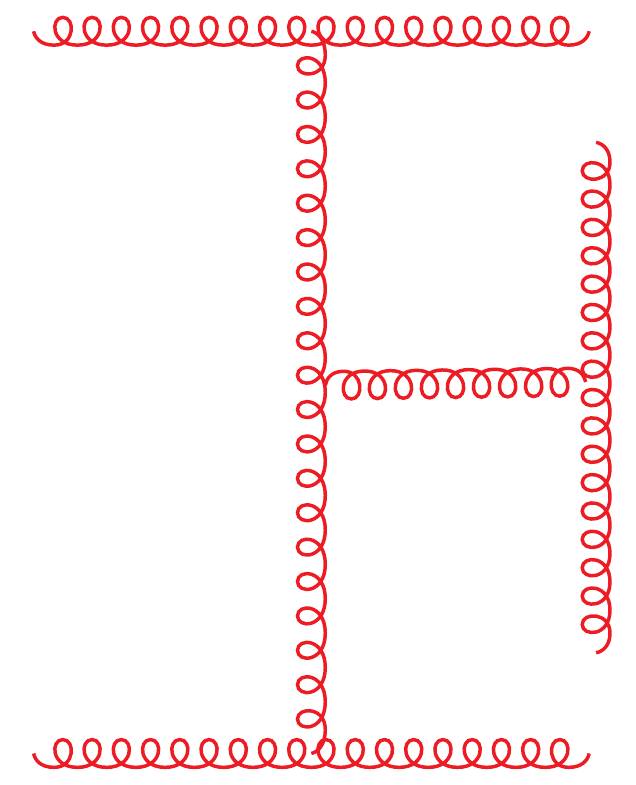} } 
	\caption{The two colour structures contributing to the soft anomalous dimension from five loops and six lines}
	\label{fig:final_structure}
\end{figure}

\section{Colour structure at five loops and six lines  }
\label{sec:colourstru}
The exponentiated colour factors obtained for all the Cwebs that connect six lines have the general structures shown in fig.~\ref{fig:final_structure}.
The explicit results shown in the previous sections and the appendix give 22 Cwebs connecting six Wilson lines at 5 loops. The total count of terms from the ECFs of all 22 Cwebs combined is 107. If ECFs of a Cweb $ k $ has $ n_k $ terms, then
	\begin{align}
		n_1 + n_2  + \cdots + n_{22}  =  107 \,.
	\end{align}
	Some of these terms are repeated in several of the ECFs and discarding the duplicates gives us 46 unique terms.
	Out of these, 36 belong to colour structure given in fig.~\ref{fig:final_structure}\textcolor{blue}{a} and the remaining 10 belong to the structure given in fig.~\ref{fig:final_structure}\textcolor{blue}{b}.
	We find that there are 17 unique Jacobi's relations among these 46 terms which finally give 29 independent colour factors.
	These Jacobi's relations involve terms from both the structures, and one can choose to eliminate all the terms of second structure completely in favour of the first structure. However, this leads to an increase in the number of basis colour factors. Either choice is equally good.
  The 29 basis colour factors $ {\cal B}_i $ are given by
\begin{align}
	{\cal B}_i \equiv F_{i,a b c d e g} \Ts \,.
\end{align}
where $ F_{i,a b c d e g} $ are given below (for the ease of notation we drop the colour indices on $F_{i,a b c d e g} $ and write it as 
$F_i$).

\begin{minipage}{0.45\textwidth}
	\begin{align}
		F_{1} &= f^{ayz}f^{bcx}f^{dey}f^{gxz}  , \nn \\
		F_{2} &= f^{ayz}f^{bcx}f^{dxz}f^{egy}  , \nn \\
		F_{3} &= f^{ayz}f^{bcx}f^{dxz}f^{egy}  , \nn \\
		F_{4} &= f^{axz}f^{bcx}f^{dyz}f^{egy}  , \nn \\
		F_{5} &= f^{ayz}f^{bdx}f^{cey}f^{gxz}  , \nn \\
		F_{6} &= f^{axz}f^{bdx}f^{cey}f^{gyz}  , \nn \\
		F_{7}&= f^{axz}f^{bdx}f^{cgy}f^{eyz}  , \nn \\
		F_{8}&= f^{axz}f^{bex}f^{cyz}f^{dgy}   , \nn\\
		F_{9}&= f^{aey}f^{bgx}f^{cyz}f^{dxz}  , \nn \\
		F_{10}&= f^{abx}f^{cgy}f^{dxz}f^{eyz}  , \nn \\ 
		F_{11}&= f^{agz}f^{bxy}f^{cdx}f^{eyz}  , \nn \\
		F_{12} &= f^{adz}f^{bxy}f^{cgx}f^{eyz}  , \nn \\
		F_{13} &= f^{aez}f^{bxy}f^{cgx}f^{dyz}  , \nn \\
		F_{14} &= f^{ady}f^{bxy}f^{cgz}f^{exz}  , \nn \\
		F_{15} &= f^{acy}f^{bxy}f^{dxz}f^{egz}  , \nn 
	\end{align}
\end{minipage} \hspace{0.3cm}
\begin{minipage}{0.45\textwidth}
	\begin{align}		
		F_{16} &= f^{acz}f^{bxy}f^{dgx}f^{eyz}  , \nn \\
		F_{17} &= f^{acz}f^{bxy}f^{dyz}f^{egx}  , \nn \\
		F_{18} &= f^{aey}f^{bxy}f^{cxz}f^{dgz}  , \nn \\
		F_{19} &= f^{adz}f^{bxy}f^{cxz}f^{egy}  , \nn \\ 
		F_{20} &= f^{abx}f^{cdy}f^{exz}f^{gyz}  , \nn \\
		F_{21} &= f^{xyz}f^{acy}f^{bgx}f^{dez}  , \nn \\
		F_{22} &= f^{ayz}f^{bgx}f^{cey}f^{dxz}  , \nn \\
		F_{23} &= f^{axz}f^{bcx}f^{dey}f^{gyz}  , \nn \\ 
		F_{24} &= f^{ady}f^{bgx}f^{cxz}f^{eyz}  , \nn \\
		F_{25} &= f^{xyz}f^{aez}f^{bcx}f^{dgy}  , \nn \\
		F_{26} &= f^{xyz}f^{agx}f^{bez}f^{cdy}  , \nn \\
		F_{27} &= f^{xyz}f^{acy}f^{bex}f^{dgz}  , \nn \\
		F_{28} &= f^{xyz}f^{adz}f^{bgx}f^{cey}  , \nn  \\
		F_{29} &= f^{xyz}f^{abx}f^{cdy}f^{egz}  . 
		\label{eq:BASIS}		
	\end{align}
\end{minipage}
\vspace{0.3cm}

\begin{figure}[h]
	\centering
	\vspace{-3mm}
	\subfloat[]{\hspace{0.6cm}\includegraphics[scale=0.45]{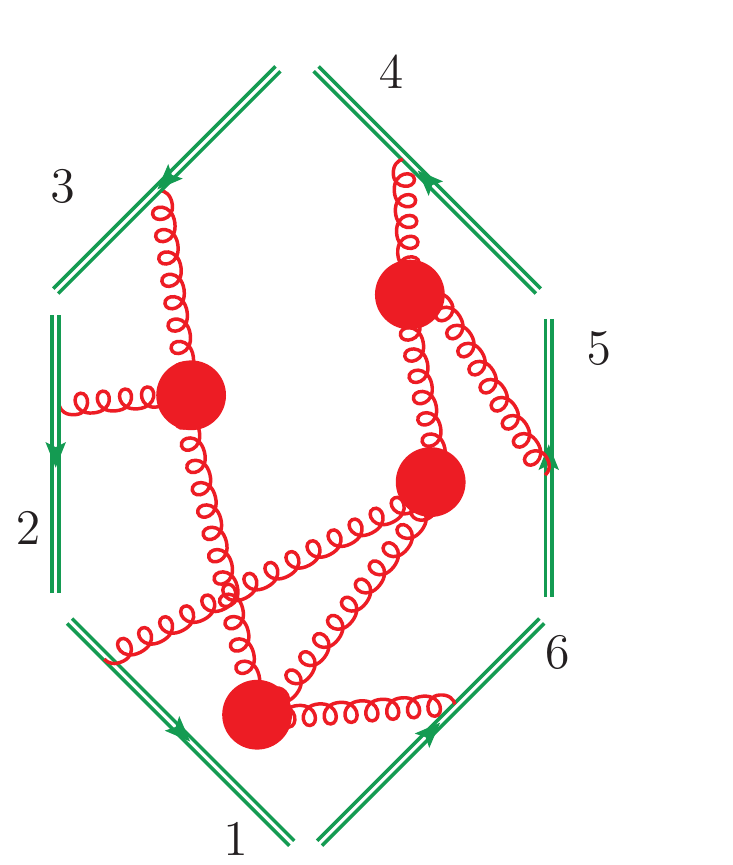} }  \qquad \qquad
	\subfloat[]{\hspace{0.6cm}\includegraphics[scale=0.45]{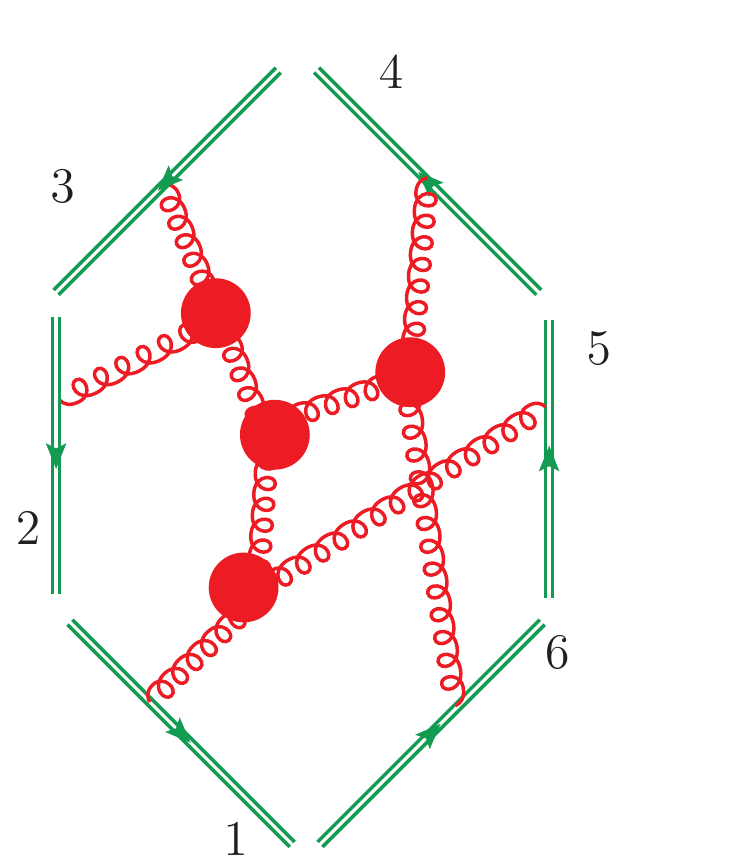} } 
	\caption{Diagrammatic representation of basis elements $ \mathcal{B}_1 $ and $ \mathcal{B}_{25} $ }
	\label{fig:Color_basis}
\end{figure}

\noindent Note that the 24 basis elements, $ \mathcal{B}_1-\mathcal{B}_{24} $,  and  5 basis elements, $ \mathcal{B}_{25}-\mathcal{B}_{29} $ correspond to the colour structure shown in  fig.~\ref{fig:final_structure}\textcolor{blue}{a} and fig.~\ref{fig:final_structure}\textcolor{blue}{b}, respectively.
Below, in fig.~\ref{fig:Color_basis}\textcolor{blue}{a} and fig.~\ref{fig:Color_basis}\textcolor{blue}{b} we diagrammatically show representative colour basis elements  $ \mathcal{B}_1 $ and $ \mathcal{B}_{25} $, which belong to the sets $ \mathcal{B}_1-\mathcal{B}_{24} $ and $ \mathcal{B}_{25}-\mathcal{B}_{29} $,  respectively. 
%
%
The ECFs of all the Cwebs, expressed in the colour basis, are presented in the  appendix~\ref{sec:appendix}.

\section{Summary and outlook}\label{sec:conclusion}

This era of precision calculation demands the knowledge of IR singularities at higher loop orders in the perturbation theory, which invites our study of five loop Cwebs connecting six Wilson lines. These IR singularities are controlled by the soft anomalous dimension. 
The calculation of full soft anomalous dimension at any given loop order is complicated and one needs to divide the calculation in two parts: $(i)$ the calculation of colour structures and $(ii)$ calculation of the corresponding kinematic contributions.
In this article, we focus on  the calculation of the colour structure contributing to the soft anomalous dimension at five loops.  
As the guiding symmetries of colour structures present in soft anomalous dimension constrain~\cite{Becher:2009cu,Becher:2019avh} the associated kinematic functions, the prior knowledge of colour structures simplify its explicit computation. Thus, the first step  towards the understanding of five loops structure is the computation of its colour building blocks. 

In this work, we study the Cwebs appearing at five loops and connecting six Wilson lines, and we start with their enumeration
using the algorithm introduced in~\cite{Agarwal:2020nyc}. {The results presented in this work apply both to massless and massive Wilson lines as self energy correlators do not appear when the largest number of Wilson lines are connected at a given order.} We have classified these Cwebs  into three categories: $ (i) $ Cwebs which are higher order members of the known families, $ (ii) $ new basis Cwebs, and $ (iii) $ Orphan Cwebs. 15 out of the 22 Cwebs that appear at five loops connecting six lines are members of the families that appear up to four loops, and thus their mixing matrices can be read off from the literature without explicit calculation once their column weight vectors are determined. The mixing matrices for remaining 7 Cwebs are obtained using  replica-trick. The computed mixing matrices obey all the known properties including the conjectured weighted column-sum rule. 
We observe that the exponentiated colour factors obtained for all the Cwebs in this work has the general structures shown in fig.~\ref{fig:final_structure}. 

{A complete study of soft anomalous dimension up to five loops will require the computation of all the  Cwebs connecting up to six Wilson lines -- the case of six lines considered here, and lower number of Wilson lines which will be more challenging to calculate. This requires understanding not only the colour structures but also the kinematic factors. Calculation of kinematic factors is extremely challenging and partial results at four loops are available in certain kinematical limits. Results presented in this article pave  the way for studying the five loop structure of the soft anomalous dimension. We believe that our results will be useful in constraining the kinematic factors associated with the five loop soft anomalous dimension.}

\section*{Acknowledgments}
SM would like to thank CSIR, Govt. of India, for an SRF fellowship
(09/1001(0052)/2019-EMR-I). SP and AT would like to thank Neelima Agarwal, and Lorenzo Magnea for collaboration on the related earlier projects. 
SP would like to thank Physical Research Laboratory (PRL), Department of Space, Govt.~of
India, for a PDF fellowship. AS would like to thank CSIR, Govt.~of India, for an SRF fellowship
(09/1001(0075)/2020-EMR-I).
\appendix
\section{Calculation of $ s $-factor in different representations} \label{sec:shrinking}
\begin{figure}[h]
	\centering
	\vspace{-3mm}
	\hspace{-0.2cm}
	\subfloat[]{\hspace{0.8cm}\includegraphics[scale=0.6]{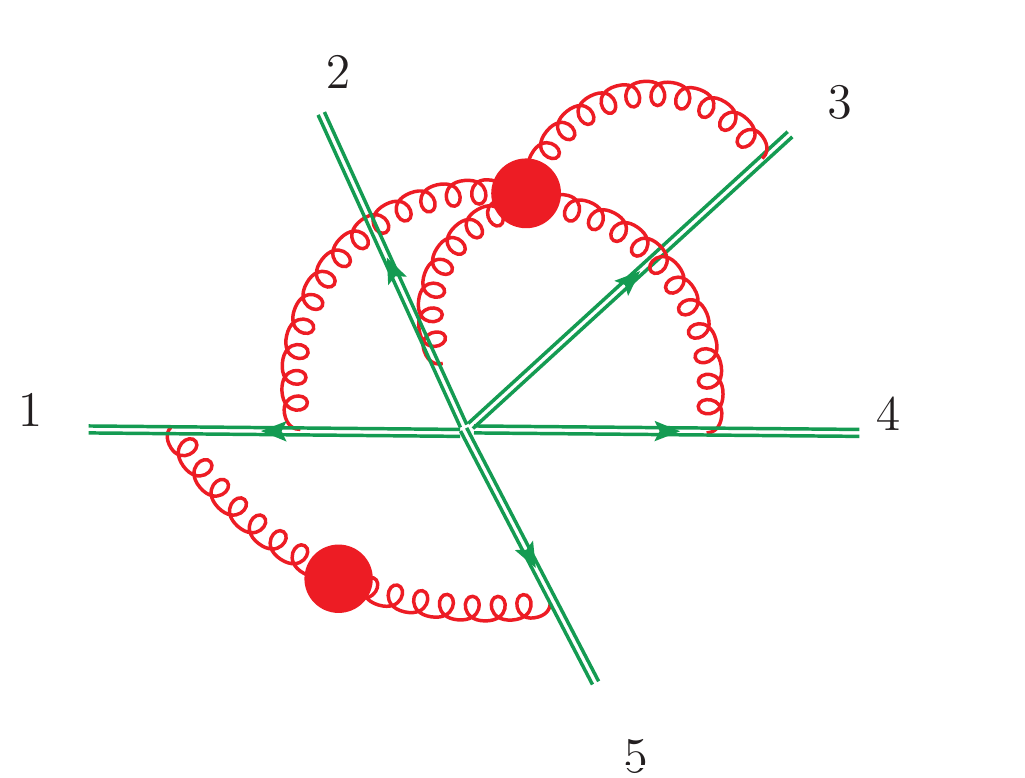} } 
	$ \; $
	\subfloat[]{\hspace{1.8cm}\includegraphics[scale=0.6]{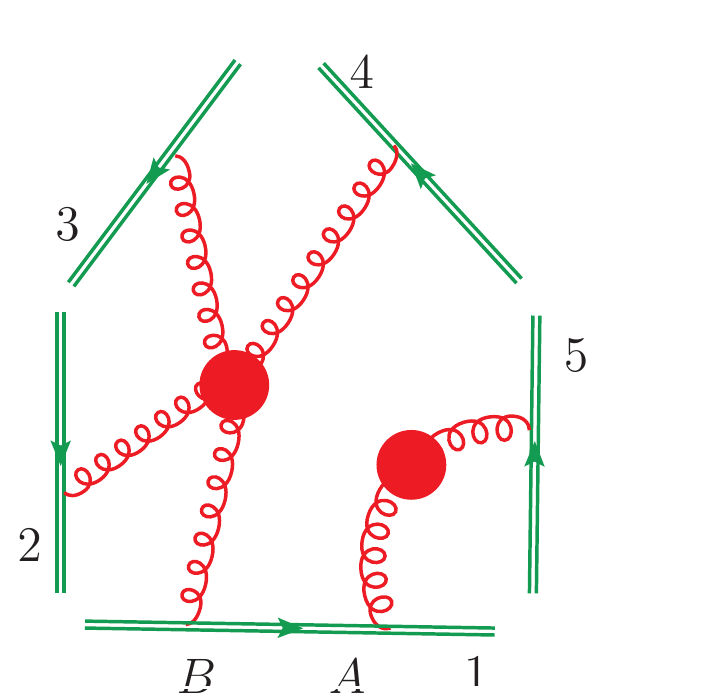} } 
	\caption{Two different representations of Cweb $W^{(1,0,1)}_5 (2,1, 1, 1, 1)$ }
	\label{fig:sfactorEx}
\end{figure}

In this section, we demonstrate the computation of $ s $-factors by shrinking correlators in two different representations. Let us consider the Cweb $W^{(1,0,1)}_5 (2, 1, 1, 1, 1)$. Here, in fig.~\ref{fig:sfactorEx}\textcolor{blue}{a} the hard interaction vertex is located at the point where tails of all the Wilson lines meet. The same diagram is drawn in fig.~\ref{fig:sfactorEx}\textcolor{blue}{b}  where the arrows on the Wilson lines are used to indicate the hard vertex: the tail ends of each of the Wilson lines meet at the hard vertex. 

In fig.~\ref{fig:sfactorEx}\textcolor{blue}{a} we first shrink the four gluon correlator to the hard interaction vertex, followed by the two gluon correlator this gives $ s=1$ for this diagram.
Now we describe the procedure for finding the $s$-factors when the diagrams are drawn as in 
 fig.~\ref{fig:sfactorEx}\textcolor{blue}{b} or fig.~\ref{fig:ElongBasis}.
Pull each of the gluon correlators to the tail ends of the associated Wilson lines without dragging the attachments from other correlators. 
For example in fig.~\ref{fig:sfactorEx}\textcolor{blue}{b}, we can pull the four gluon correlator to the tails of attached Wilson lines (1, 2, 3 and 4) followed by the pulling of two gluon correlator to the tails of Wilson lines 1 and 5. This is the only way we can sequentially pull these gluon correlators without dragging other correlators in the process, giving $s=1$. Following the same steps one can compute the $ s $-factors for any diagram in any of the representations.

\section{Results at five loops and six lines}\label{sec:appendix}
The Cwebs that appear at five loops and six lines are all listed in this appendix, along with their mixing matrix, diagonalizing matrix, $s$-factors, and ECFs. We note that all 22 computed matrices adhere to the row sum rule, conjectured column sum rule, and idempotence character as described in the literature. The computed ECFs are completely connected. However, the Uniqueness theorem can directly predict the mixing matrices of 15 of these Cwebs. We present the results below in three categories as discussed in section~\ref{sec:UNQ}. The dimension of mixing matrices for few Cwebs are quite huge and  presenting them here is not convenient. Thus we provide all the mixing matrices and the corresponding column weight vectors of the Cwebs present at five loops connecting six  Wilson lines in an ancillary Mathematica file named as 
\lq\,Cwebs5Loop6line.nb\,\rq. The order and name of Cwebs given in the Mathematica file are same as provided in this appendix. 

\subsection{Cwebs from existing families} \label{sec:appdxFamily}

In this section, we list down all the Cwebs at five loops and six lines that can be traced back to a basis appearing at lower perturbative order.
\subsection*{(1)
	\ $\mathbf{W^{(2,0,1)}_6 (3, 1, 1, 1, 1, 1)}$}
\label{C11}
This Cweb has three gluon attachments on the same line and one attachment to  the rest of the five lines, which results in six different shuffles and, thus six diagrams. We have shown one of the possible diagram in fig.~\ref{Diag:1}, and a table~\ref{Table:Diag1}, that includes all the diagrams with varying sequences of gluon attachments on the Wilson lines and their corresponding $s$-factors. 

\vskip0.5cm
\begin{minipage}{0.45\textwidth}
	\hspace{2cm}	\includegraphics[scale=0.5]{C11N}
	\captionof{figure}{$W^{(2,0,1)}_6 (3, 1, 1, 1, 1, 1)$ }
	\label{Diag:1}
\end{minipage}
\begin{minipage}{0.45\textwidth}
	
	\begin{tabular}{ | c | c | c |}
		\hline
		\textbf{Diagrams} & \textbf{Sequences} & \textbf{$s$-factors} \\ \hline
		$d_1$ & $\lbrace \lbrace ABC  \rbrace\rbrace$ & 1 \\ 
		\hline
		$d_2$ & $\lbrace \lbrace ACB \rbrace\rbrace$ & 1 \\ 
		\hline
		$d_3$ & $\lbrace \lbrace BAC\rbrace\rbrace$ & 1 \\ 
		\hline	
		$d_4$ & $\lbrace \lbrace BCA \rbrace\rbrace$ & 1 \\ 
		\hline	
		$d_5$ & $\lbrace \lbrace CAB \rbrace\rbrace$ & 1 \\ 
		\hline
		$d_6$ & $\lbrace \lbrace CBA \rbrace\rbrace$ & 1 \\ 
		\hline
	\end{tabular}	
	\captionof{table}{Sequences and $s$-factors}
	\label{Table:Diag1}
\end{minipage} 
\vskip0.2cm
\noindent 
The mixing matrix $R$, the diagonalizing matrix $Y$ and the diagonal matrix $D$ for the Cweb mentioned above are
\begin{align} 
	R = \frac{1}{6} \left( \,\,\,
	\begin{array}{cccccc}
		2 &\hspace{0.20cm} \hspace{-0.25cm}-1 &\hspace{0.20cm} \hspace{-0.25cm}-1 &\hspace{0.20cm} \hspace{-0.25cm}-1 &\hspace{0.20cm} \hspace{-0.25cm}-1 &\hspace{0.20cm} 2 \\
		\hspace{-0.25cm}-1 &\hspace{0.20cm} 2 &\hspace{0.20cm} \hspace{-0.25cm}-1 &\hspace{0.20cm} 2 &\hspace{0.20cm} \hspace{-0.25cm}-1 &\hspace{0.20cm} \hspace{-0.25cm}-1 \\
		\hspace{-0.25cm}-1 &\hspace{0.20cm} \hspace{-0.25cm}-1 &\hspace{0.20cm} 2 &\hspace{0.20cm} \hspace{-0.25cm}-1 &\hspace{0.20cm} 2 &\hspace{0.20cm} \hspace{-0.25cm}-1 \\
		\hspace{-0.25cm}-1 &\hspace{0.20cm} 2 &\hspace{0.20cm} \hspace{-0.25cm}-1 &\hspace{0.20cm} 2 &\hspace{0.20cm} \hspace{-0.25cm}-1 &\hspace{0.20cm} \hspace{-0.25cm}-1 \\
		\hspace{-0.25cm}-1 &\hspace{0.20cm} \hspace{-0.25cm}-1 &\hspace{0.20cm} 2 &\hspace{0.20cm} \hspace{-0.25cm}-1 &\hspace{0.20cm} 2 &\hspace{0.20cm} \hspace{-0.25cm}-1 \\
		2 &\hspace{0.20cm} \hspace{-0.25cm}-1 &\hspace{0.20cm} \hspace{-0.25cm}-1 &\hspace{0.20cm} \hspace{-0.25cm}-1 &\hspace{0.20cm} \hspace{-0.25cm}-1 &\hspace{0.20cm} 2 \\
	\end{array}
	\right)&\hspace{0.20cm} \quad , \quad  Y= \left( \,\,\,
	\begin{array}{cccccc}
		1 &\hspace{0.20cm} \hspace{-0.25cm}-1 &\hspace{0.20cm} 0 &\hspace{0.20cm} \hspace{-0.25cm}-1 &\hspace{0.20cm} 0 &\hspace{0.20cm} 1 \\
		0 &\hspace{0.20cm} \hspace{-0.25cm}-1 &\hspace{0.20cm} 1 &\hspace{0.20cm} \hspace{-0.25cm}-1 &\hspace{0.20cm} 1 &\hspace{0.20cm} 0 \\
		\hspace{-0.25cm}-1 &\hspace{0.20cm} 0 &\hspace{0.20cm} 0 &\hspace{0.20cm} 0 &\hspace{0.20cm} 0 &\hspace{0.20cm} 1 \\
		1 &\hspace{0.20cm} 1 &\hspace{0.20cm} 0 &\hspace{0.20cm} 0 &\hspace{0.20cm} 1 &\hspace{0.20cm} 0 \\
		0 &\hspace{0.20cm} \hspace{-0.25cm}-1 &\hspace{0.20cm} 0 &\hspace{0.20cm} 1 &\hspace{0.20cm} 0 &\hspace{0.20cm} 0 \\
		1 &\hspace{0.20cm} 1 &\hspace{0.20cm} 1 &\hspace{0.20cm} 0 &\hspace{0.20cm} 0 &\hspace{0.20cm} 0 \\
	\end{array}
	\right),\nonumber  \\ & D=(\mathbf{1}_2,0).
	\label{D11}
\end{align}
As the rank for the mixing matrix of this Cweb is two, thus there are two independent exponentiated colour factors of this Cweb and they are given below
\begin{align}
	(YC)_1=& - f^{abl}f^{cdl}f^{dgk}f^{kem} \tm{1} \ta{2} \tb{3} \tc{4} \te{5} \tg{6} + f^{abl}f^{cdl}f^{deh}f^{hgn} \tn{1} \ta{2} \tb{3} \tc{4} \te{5} \tg{6},\nonumber \\
	= & \; -{\cal{B}}_{2}-2{\cal{B}}_{3}\, . \nn \\
	(YC)_2=&+ f^{abl}f^{cdl}f^{dgk}f^{kem} \tm{1} \ta{2} \tb{3} \tc{4} \te{5} \tg{6}, \nn \\
	=  & \; {\cal{B}}_{3} \, .
\end{align}

\subsection*{(2)
	$\mathbf{W^{(2,0,1)}_{6} (2, 1, 1, 1, 2, 1)}$} \label{C12}
This Cweb has two gluon attachments on two different lines, which results in four different shuffles and, thus four diagrams. We have shown one of the possible diagram in fig.~\ref{Diag:2}, and a table~\ref{Table:Diag2}, that includes all the diagrams with varying sequences of gluon attachments on the Wilson lines and their corresponding $s$-factors.

\begin{minipage}{0.5\textwidth}
	\hspace{2cm}	\includegraphics[scale=0.5]{C12N}
	\captionof{figure}{$W^{(2,0,1)}_{6} (2, 1, 1, 1, 2, 1)$}
	\label{Diag:2}
\end{minipage}
\begin{minipage}{0.45\textwidth}

	\begin{tabular}{ | c | c | c |}
		\hline
		\textbf{Diagrams} & \textbf{Sequences} & \textbf{$s$-factors} \\ \hline
		$d_1$ & $\lbrace \lbrace AB \rbrace,\lbrace CD \rbrace\rbrace $ & 1 \\ 
		\hline
		$d_2$ & $\lbrace \lbrace AB \rbrace,\lbrace DC  \rbrace\rbrace$ & 2 \\ 
		\hline	
		$d_3$ & $\lbrace \lbrace BA \rbrace,\lbrace CD \rbrace\rbrace$ & 2 \\
		\hline
		$d_4$ & $\lbrace \lbrace BA \rbrace,\lbrace DC \rbrace\rbrace$ & 1 \\
		\hline	
		
	\end{tabular}	
	\captionof{table}{Sequences and $s$-factors}
	\label{Table:Diag2}
\end{minipage} 
\vspace{0.5cm}

\noindent         The mixing matrix $R$, the diagonalizing matrix $Y$ and the diagonal matrix $D$ for the Cweb mentioned above are

\begin{align}
	R =\frac{1}{6} \left( \,\,\,
	\begin{array}{cccc}
		2 &\hspace{0.20cm} \hspace{-0.25cm}-2 &\hspace{0.20cm} \hspace{-0.25cm}-2 &\hspace{0.20cm} 2 \\
		\hspace{-0.25cm}-1 &\hspace{0.20cm} 1 &\hspace{0.20cm} 1 &\hspace{0.20cm} \hspace{-0.25cm}-1 \\
		\hspace{-0.25cm}-1 &\hspace{0.20cm} 1 &\hspace{0.20cm} 1 &\hspace{0.20cm} \hspace{-0.25cm}-1 \\
		2 &\hspace{0.20cm} \hspace{-0.25cm}-2 &\hspace{0.20cm} \hspace{-0.25cm}-2 &\hspace{0.20cm} 2 \\
	\end{array}
	\right) \quad, \quad Y =\left( \,\,\,
	\begin{array}{cccc}
		1 &\hspace{0.20cm} \hspace{-0.25cm}-1 &\hspace{0.20cm} \hspace{-0.25cm}-1 &\hspace{0.20cm} 1 \\
		\hspace{-0.25cm}-1 &\hspace{0.20cm} 0 &\hspace{0.20cm} 0 &\hspace{0.20cm} 1 \\
		\frac{1}{2} &\hspace{0.20cm} 0 &\hspace{0.20cm} 1 &\hspace{0.20cm} 0 \\
		\frac{1}{2} &\hspace{0.20cm} 1 &\hspace{0.20cm} 0 &\hspace{0.20cm} 0 \\
	\end{array}
	\right)\quad , \quad  D=(\mathbf{1}_1,0).
	\label{D12}
\end{align}
As the rank for the mixing matrix of this Cweb is one, thus there is only one exponentiated colour factor of this Cweb which is given below
\begin{align}
	(YC)_1\;&=- f^{abl}f^{cdl}f^{deh}f^{egk} \thh{1} \ta{2} \tb{3} \tc{4} \tkk{5} \tg{6} \, , \nn \\
	&=\; {\cal{B}}_{2} \, .
\end{align}

\subsection*{(3) \  $\mathbf{W^{(2,0,1)}_{6} (2, 1, 1, 2, 1, 1)}$}  \label{C13}

This Cweb has four diagrams due to two shuffles each at lines 1 and 4. We have shown one of the possible diagram in fig.~\ref{Diag:3}, and a table~\ref{Table:Diag3}, that includes all the diagrams with varying sequences of gluon attachments on the Wilson lines and their corresponding $s$-factors.
\vskip0.5cm
\begin{minipage}{0.45\textwidth}
	\hspace{1.5cm}	\includegraphics[scale=0.5]{C13N}
	\captionof{figure}{$W^{(2,0,1)}_{6} (2, 1, 1, 2, 1, 1)$}
	\label{Diag:3}
\end{minipage}
\begin{minipage}{0.45\textwidth}
	\begin{tabular}{ | c | c | c |}
		\hline
		\textbf{Diagrams} & \textbf{Sequences} & \textbf{$s$-factors} \\ \hline
		$d_1$ & $\lbrace \lbrace AB \rbrace,\lbrace CD  \rbrace\rbrace$ & 2 \\  
		\hline
		$d_2$ & $\lbrace \lbrace AB \rbrace,\lbrace DC \rbrace\rbrace$ & 1 \\
		\hline	
		$d_3$ & $\lbrace \lbrace BA \rbrace,\lbrace CD \rbrace\rbrace$ & 1 \\  
		\hline	
		$d_4$ & $\lbrace \lbrace BA \rbrace,\lbrace DC \rbrace\rbrace$ & 2 \\ 
		\hline
	\end{tabular}		
	\captionof{table}{Sequences and $s$-factors}
	\label{Table:Diag3}
\end{minipage}

\vspace{0.5cm}
\noindent The mixing matrix $R$, the diagonalizing matrix $Y$ and the diagonal matrix $D$ for the Cweb mentioned above are
\begin{equation} 
	R=\frac{1}{6} \left(\,\,\,
	\begin{array}{cccc}
		1 &\hspace{0.20cm} \hspace{-0.25cm}-1 &\hspace{0.20cm} \hspace{-0.25cm}-1 &\hspace{0.20cm} 1 \\
		\hspace{-0.25cm}-2 &\hspace{0.20cm} 2 &\hspace{0.20cm} 2 &\hspace{0.20cm} \hspace{-0.25cm}-2 \\
		\hspace{-0.25cm}-2 &\hspace{0.20cm} 2 &\hspace{0.20cm} 2 &\hspace{0.20cm} \hspace{-0.25cm}-2 \\
		1 &\hspace{0.20cm} \hspace{-0.25cm}-1 &\hspace{0.20cm} \hspace{-0.25cm}-1 &\hspace{0.20cm} 1 \\
	\end{array}
	\right) \quad, \quad Y=\left(\,\,\,
	\begin{array}{cccc}
		1 &\hspace{0.20cm} \hspace{-0.25cm}-1 &\hspace{0.20cm} \hspace{-0.25cm}-1 &\hspace{0.20cm} 1 \\
		\hspace{-0.25cm}-1 &\hspace{0.20cm} 0 &\hspace{0.20cm} 0 &\hspace{0.20cm} 1 \\
		2 &\hspace{0.20cm} 0 &\hspace{0.20cm} 1 &\hspace{0.20cm} 0 \\
		2 &\hspace{0.20cm} 1 &\hspace{0.20cm} 0 &\hspace{0.20cm} 0 \\
	\end{array}
	\right) \quad , \quad  D=(\mathbf{1}_1,0).
	\label{D13}
\end{equation}
The above mixing matrix has rank one. The only exponentiated colour factor of this Cweb is
\begin{align}
	(YC)_1&= f^{abl}f^{cdl}f^{dgh}f^{eck}\thh{1} \ta{2} \tb{3} \tkk{4} \tg{5} \te{6}\, ,\nn \\
	&=\;{\cal{B}}_{25} \, .
\end{align}

\subsection*{(4) \  $\mathbf{W^{(0,1,1)}_6 (2, 1, 1, 1, 1, 1)}$ } 
\label{C16}
This Cweb has two diagrams due to two shuffles at line 1. We have shown one of the possible diagram in fig.~\ref{Diag:4}, and a table~\ref{Table:Diag4}, that includes all the diagrams with varying sequences of gluon attachments on the Wilson lines and their corresponding $s$-factors.

\vskip0.5cm

\begin{minipage}{0.45\textwidth}
	\hspace{1.5cm}	\includegraphics[scale=0.5]{C16N}
	\captionof{figure}{$W^{(0,1,1)}_6 (2, 1, 1, 1, 1, 1)$}
	\label{Diag:4}
\end{minipage}
\begin{minipage}{0.45\textwidth}
	
	\begin{tabular}{ | c | c | c |}
		\hline
		\textbf{Diagrams} & \textbf{Sequences} & \textbf{$s$-factors} \\ \hline
		$d_1$ & $\lbrace \lbrace AB \rbrace\rbrace$ & 1 \\ 
		\hline
		$d_2$ & $\lbrace \lbrace BA \rbrace\rbrace$ & 1 \\ 
		\hline
	\end{tabular}	
	\captionof{table}{Sequences and $s$-factors}
	\label{Table:Diag4}
\end{minipage} 

\vspace{0.5cm}

\noindent
The mixing matrix $R$, the diagonalizing matrix $Y$ and the diagonal matrix $D$ for the Cweb mentioned above are

\begin{equation}
	R =\frac{1}{2} \left(\,\,\,
	\begin{array}{cc}
		1 &\hspace{0.20cm} \hspace{-0.25cm}-1 \\
		\hspace{-0.25cm}-1 &\hspace{0.20cm} 1 \\
	\end{array}
	\right)\quad , \quad Y=\left(\,\,\,
	\begin{array}{cc}
		\hspace{-0.25cm}-1 &\hspace{0.20cm} 1 \\
		1 &\hspace{0.20cm} 1 \\
	\end{array}
	\right)\quad , \quad  D=(\mathbf{1}_1,0).
	\label{D16}
\end{equation}
The above mixing matrix has rank one. The only exponentiated colour factor of this Cweb is
\begin{align}
	(YC)_1&=- f^{abl}f^{agh}f^{cdl}f^{ekg} \thh{1} \tb{2} \tc{3} \td{4} \te{5} \tkk{6}\, , \nn\\
	&=\;-{\cal{B}}_{7}-{\cal{B}}_{10} \, .	
\end{align}

\subsection*{(5)  $\mathbf{W^{(1,0,0,1)}_6 (2, 1, 1, 1, 1, 1)}$ }
\label{C17}

This Cweb again has two diagrams. One of the diagrams is shown in fig.~\ref{Diag:1}. The sequences of diagrams and their corresponding $s$-factors are provided in table~\ref{Table:Diag5}.

\begin{minipage}{0.45\textwidth}
	\hspace{1.5cm}	\includegraphics[scale=0.5]{C17N}
	\captionof{figure}{$W^{(1,0,0,1)}_6 (2, 1, 1, 1, 1, 1)$}
	\label{Diag:5}
\end{minipage}
\begin{minipage}{0.45\textwidth}
	
	\begin{tabular}{ | c | c | c |}
		\hline
		\textbf{Diagrams} & \textbf{Sequences} & \textbf{$s$-factors} \\ \hline
		$d_1$ & $\lbrace \lbrace AB \rbrace\rbrace$ & 1 \\ 
		\hline
		$d_2$ & $\lbrace \lbrace BA \rbrace\rbrace$ & 1 \\ 
		\hline
	\end{tabular}	
	\captionof{table}{Sequences and $s$-factors}
	\label{Table:Diag5}
\end{minipage} 

\vspace{0.5cm}
\noindent  The mixing matrix $R$, the diagonalizing matrix $Y$ and the diagonal matrix $D$ for the Cweb mentioned above are

\begin{equation}
	R =\frac{1}{2} \left(\,\,\,
	\begin{array}{cc}
		1 &\hspace{0.20cm} \hspace{-0.25cm}-1 \\
		\hspace{-0.25cm}-1 &\hspace{0.20cm} 1 \\
	\end{array}
	\right)\quad , \quad Y=\left(\,\,\,
	\begin{array}{cc}
		\hspace{-0.25cm}-1 &\hspace{0.20cm} 1 \\
		1 &\hspace{0.20cm} 1 \\
	\end{array}
	\right) \quad , \quad  D=(\mathbf{1}_1,0).
	\label{D17}
\end{equation}
The only exponentiated colour factor of this Cweb is
\begin{align}
	(YC)_1&=- f^{adg}f^{ahm}f^{bcl}f^{egl} \tm{1} \tb{2} \tc{3} \td{4} \thh{5} \te{6}\, ,\nn\\
	&=\; -{\cal{B}}_{3}-{\cal{B}}_{25} \, .
\end{align}

\subsection*{(6) $\mathbf{W^{(1,2)}_6 (3, 1, 1, 1, 1, 1)}$ }  \label{C21}

As there are three attachments on line 1 resulting in six diagrams for this Cweb. One of the diagrams is shown in fig.~\ref{Diag:6}. The sequences of diagrams and their corresponding $s$-factors are provided in table~\ref{Table:Diag6}.
\vskip0.5cm
\begin{minipage}{0.45\textwidth}
	\hspace{2cm}	\includegraphics[scale=0.5]{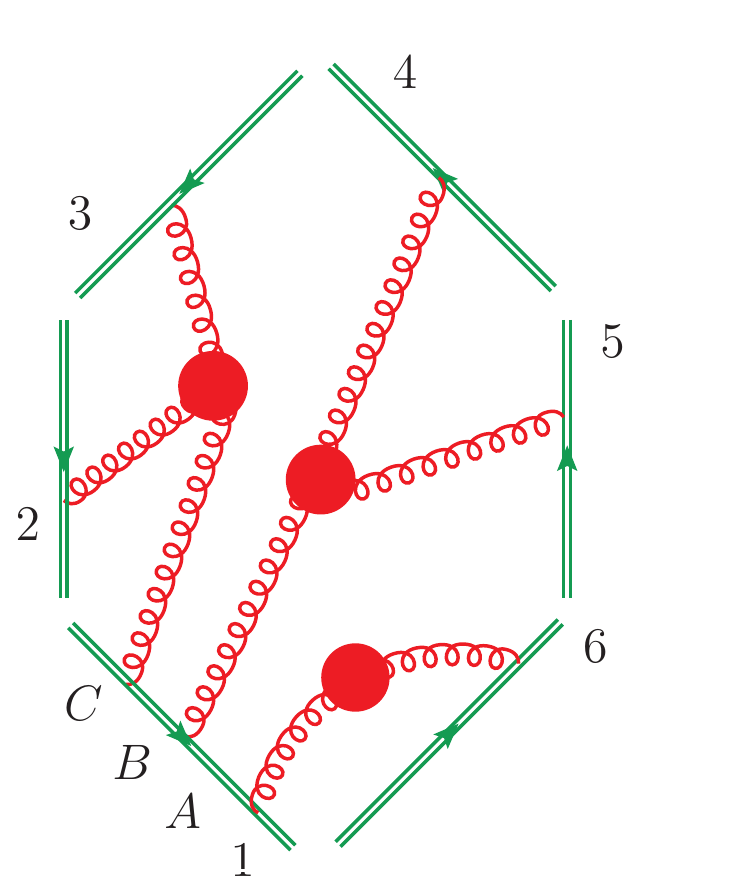}
	\captionof{figure}{$W^{(1,2)}_6 (3, 1, 1, 1, 1, 1)$}
	\label{Diag:6}
\end{minipage}
\begin{minipage}{0.45\textwidth}
	
	\begin{tabular}{ | c | c | c |}
		\hline
		\textbf{Diagrams} & \textbf{Sequences} & \textbf{$s$-factors} \\ \hline
		$d_1$ & $\lbrace \lbrace ABC \rbrace\rbrace$ & 1 \\ 
		\hline
		$d_2$ & $\lbrace \lbrace ACB \rbrace\rbrace$ & 1 \\ 
		\hline
		$d_3$ & $\lbrace \lbrace BAC \rbrace\rbrace$ & 1 \\ 
		\hline	
		$d_4$ & $\lbrace \lbrace BCA \rbrace\rbrace$ & 1 \\ 
		\hline	
		$d_5$ & $\lbrace \lbrace CAB \rbrace\rbrace$ & 1 \\ 
		\hline	
		$d_6$ & $\lbrace \lbrace CBA \rbrace\rbrace$ & 1 \\ 
		\hline	
	\end{tabular}	
	\captionof{table}{Sequences and $s$-factors}
	\label{Table:Diag6}
\end{minipage} 

\vspace{0.5cm}
\noindent   The mixing matrix $R$, the diagonalizing matrix $Y$ and the diagonal matrix $D$ for the Cweb mentioned above are

\begin{align} 
	R =\frac{1}{6} \left(\,\,\,
	\begin{array}{cccccc}
		2 &\hspace{0.20cm} \hspace{-0.25cm}-1 &\hspace{0.20cm} \hspace{-0.25cm}-1 &\hspace{0.20cm} \hspace{-0.25cm}-1 &\hspace{0.20cm} \hspace{-0.25cm}-1 &\hspace{0.20cm} 2 \\
		\hspace{-0.25cm}-1 &\hspace{0.20cm} 2 &\hspace{0.20cm} \hspace{-0.25cm}-1 &\hspace{0.20cm} 2 &\hspace{0.20cm} \hspace{-0.25cm}-1 &\hspace{0.20cm} \hspace{-0.25cm}-1 \\
		\hspace{-0.25cm}-1 &\hspace{0.20cm} \hspace{-0.25cm}-1 &\hspace{0.20cm} 2 &\hspace{0.20cm} \hspace{-0.25cm}-1 &\hspace{0.20cm} 2 &\hspace{0.20cm} \hspace{-0.25cm}-1 \\
		\hspace{-0.25cm}-1 &\hspace{0.20cm} 2 &\hspace{0.20cm} \hspace{-0.25cm}-1 &\hspace{0.20cm} 2 &\hspace{0.20cm} \hspace{-0.25cm}-1 &\hspace{0.20cm} \hspace{-0.25cm}-1 \\
		\hspace{-0.25cm}-1 &\hspace{0.20cm} \hspace{-0.25cm}-1 &\hspace{0.20cm} 2 &\hspace{0.20cm} \hspace{-0.25cm}-1 &\hspace{0.20cm} 2 &\hspace{0.20cm} \hspace{-0.25cm}-1 \\
		2 &\hspace{0.20cm} \hspace{-0.25cm}-1 &\hspace{0.20cm} \hspace{-0.25cm}-1 &\hspace{0.20cm} \hspace{-0.25cm}-1 &\hspace{0.20cm} \hspace{-0.25cm}-1 &\hspace{0.20cm} 2 \\
	\end{array}
	\right) &\hspace{0.20cm}
	\quad, \quad Y= \left( \,\,\,
	\begin{array}{cccccc}
		1 &\hspace{0.20cm} \hspace{-0.25cm}-1 &\hspace{0.20cm} 0 &\hspace{0.20cm} \hspace{-0.25cm}-1 &\hspace{0.20cm} 0 &\hspace{0.20cm} 1 \\
		0 &\hspace{0.20cm} \hspace{-0.25cm}-1 &\hspace{0.20cm} 1 &\hspace{0.20cm} \hspace{-0.25cm}-1 &\hspace{0.20cm} 1 &\hspace{0.20cm} 0 \\
		\hspace{-0.25cm}-1 &\hspace{0.20cm} 0 &\hspace{0.20cm} 0 &\hspace{0.20cm} 0 &\hspace{0.20cm} 0 &\hspace{0.20cm} 1 \\
		1 &\hspace{0.20cm} 1 &\hspace{0.20cm} 0 &\hspace{0.20cm} 0 &\hspace{0.20cm} 1 &\hspace{0.20cm} 0 \\
		0 &\hspace{0.20cm} \hspace{-0.25cm}-1 &\hspace{0.20cm} 0 &\hspace{0.20cm} 1 &\hspace{0.20cm} 0 &\hspace{0.20cm} 0 \\
		1 &\hspace{0.20cm} 1 &\hspace{0.20cm} 1 &\hspace{0.20cm} 0 &\hspace{0.20cm} 0 &\hspace{0.20cm} 0 \\
	\end{array}
	\right), \nonumber \\ & \hspace{-1cm} \quad  \quad  D=(\mathbf{1}_2,0).
	\label{D21}
\end{align}
The two exponentiated colour factors of this Cweb are
\begin{align}
	(YC)_1=& \,f^{abc}f^{ahm}f^{deg}f^{gmp}\tp{1} \tb{2} \tc{3} \td{4} \te{5} \thh{6}-f^{abc}f^{agn}f^{deg}f^{hnr} \trr{1} \tb{2} \tc{3} \td{4} \te{5} \thh{6} \,, \nonumber \\
	&=\;{\cal{B}}_{23} \, .\\
	(YC)_2=& \,f^{abc}f^{ahm}f^{deg}f^{gmp} \tp{1} \tb{2} \tc{3} \td{4} \te{5} \thh{6}\, ,\nn\\
	&=\;-{\cal{B}}_{1} \, .
\end{align}

\subsection*{(7)  \ $\mathbf{W^{(1,2)}_{6} (2, 1, 1, 1, 2, 1)}$} \label{C22}

This Cweb has four diagrams, one of the diagrams is shown in fig.~\ref{Diag:7}. The sequences and the corresponding $s$-factors of all the diagrams are provided in the table~\ref{Table:Diag7}.
\vskip0.5cm
\begin{minipage}{0.45\textwidth}
	\hspace{1.5cm}	\includegraphics[scale=0.5]{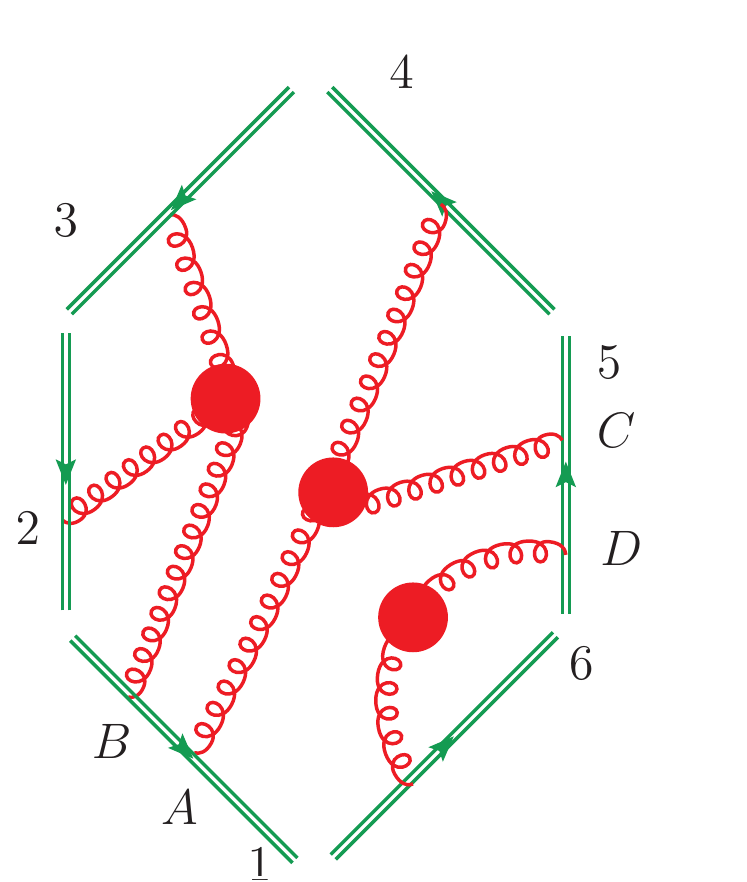}
	\captionof{figure}{$W^{(1,2)}_{6} (2, 1, 1, 1, 2, 1)$}
	\label{Diag:7}
\end{minipage}
\begin{minipage}{0.45\textwidth}
	
	\begin{tabular}{ | c | c | c |}
		\hline
		\textbf{Diagrams} & \textbf{Sequences} & \textbf{$s$-factors} \\ \hline
		$d_1$ & $\lbrace \lbrace  AB \rbrace,\lbrace CD \rbrace\rbrace$ & 1 \\ 
		\hline
		$d_2$ & $\lbrace \lbrace  AB \rbrace,\lbrace DC \rbrace\rbrace$ & 2 \\ 
		\hline
		$d_3$ & $\lbrace \lbrace  BA \rbrace,\lbrace CD \rbrace\rbrace$ & 2  \\ 
		\hline
		$d_4$ & $\lbrace \lbrace  BA \rbrace,\lbrace DC \rbrace\rbrace$ & 1 \\ 
		\hline	
		
	\end{tabular}	
	\captionof{table}{Sequences and $s$-factors}
	\label{Table:Diag7}
\end{minipage}

\vspace{0.5cm}
\noindent  The mixing matrix $R$, the diagonalizing matrix $Y$ and the diagonal matrix $D$ for the Cweb mentioned above are

\begin{equation}  
	R =\frac{1}{6} \left(\,\,\,
	\begin{array}{cccc}
		2 &\hspace{0.20cm} \hspace{-0.25cm}-2 &\hspace{0.20cm} \hspace{-0.25cm}-2 &\hspace{0.20cm} 2 \\
		\hspace{-0.25cm}-1 &\hspace{0.20cm} 1 &\hspace{0.20cm} 1 &\hspace{0.20cm} \hspace{-0.25cm}-1 \\
		\hspace{-0.25cm}-1 &\hspace{0.20cm} 1 &\hspace{0.20cm} 1 &\hspace{0.20cm} \hspace{-0.25cm}-1 \\
		2 &\hspace{0.20cm} \hspace{-0.25cm}-2 &\hspace{0.20cm} \hspace{-0.25cm}-2 &\hspace{0.20cm} 2 \\
	\end{array}
	\right)\quad, \quad Y=\left(\,\,\,
	\begin{array}{cccc}
		1 &\hspace{0.20cm} \hspace{-0.25cm}-1 &\hspace{0.20cm} \hspace{-0.25cm}-1 &\hspace{0.20cm} 1 \\
		\hspace{-0.25cm}-1 &\hspace{0.20cm} 0 &\hspace{0.20cm} 0 &\hspace{0.20cm} 1 \\
		\frac{1}{2} &\hspace{0.20cm} 0 &\hspace{0.20cm} 1 &\hspace{0.20cm} 0 \\
		\frac{1}{2} &\hspace{0.20cm} 1 &\hspace{0.20cm} 0 &\hspace{0.20cm} 0 \\
	\end{array}
	\right)\quad , \quad  D=(\mathbf{1}_1,0).
	\label{D22}
\end{equation}
The exponentiated colour factor of this Cweb is
\begin{align}
	(YC)_1&=- f^{abc}f^{agn}f^{deg}f^{her}\tn{1} \tb{2} \tc{3} \td{4} \trr{5} \thh{6}\, ,\nn\\
	&=\;-{\cal{B}}_{4} \, .
\end{align}

\subsection*{(8) \ $\mathbf{W^{(3,1)}_6 (4, 1, 1, 1, 1, 1)}$}  \label{C31}

This Cweb has twenty-four diagrams as there are four attachments on line 1, one of the diagram is shown in fig.~\ref{Diag:8}. The sequences of diagrams and their corresponding $s$-factors are provided in table~\ref{Table:Diag8}.
\vskip0.5cm
\begin{minipage}{0.45\textwidth}
	\hspace{2cm}	\includegraphics[scale=0.5]{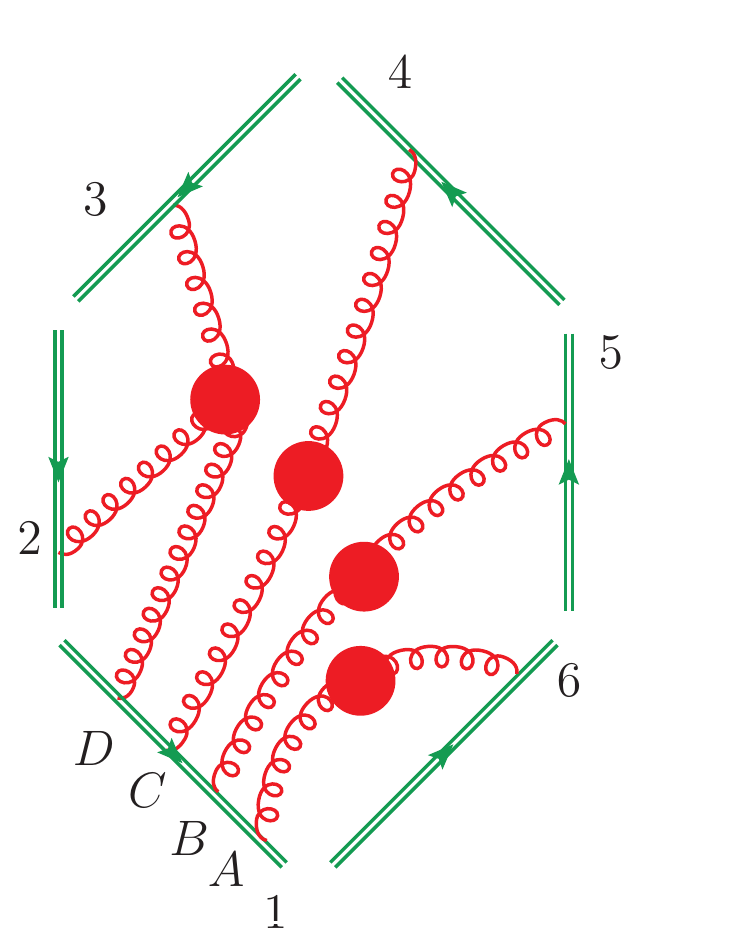}
	\captionof{figure}{$W^{(3,1)}_6 (4, 1, 1, 1, 1, 1)$}
	\label{Diag:8}	
\end{minipage}
\hspace{1cm}\begin{minipage}{0.45\textwidth}
	\footnotesize{
		\begin{tabular}{ | c | c | c |}
			\hline
			\textbf{Diagrams} & \textbf{Sequences} & \textbf{$s$-factors} \\ \hline
			$d_1$ & $\lbrace \lbrace ABCD \rbrace\rbrace$ &1 \\ 
			\hline
			$d_2$ & $\lbrace \lbrace ABDC \rbrace\rbrace$ & 1 \\ 
			\hline
			$d_3$ & $\lbrace \lbrace ACBD \rbrace\rbrace$ & 1 \\ 
			\hline	
			$d_4$ & $\lbrace \lbrace ACDB \rbrace\rbrace$ & 1 \\ 
			\hline	
			$d_5$ & $\lbrace \lbrace ADBC \rbrace\rbrace$ & 1 \\ 
			\hline
			$d_6$ & $\lbrace \lbrace ADCB \rbrace\rbrace$ & 1 \\ 
			\hline
			$d_7$ & $\lbrace \lbrace BACD \rbrace\rbrace$ & 1 \\ 
			\hline
			$d_8$ & $\lbrace \lbrace BADC \rbrace\rbrace$ & 1 \\ 
			\hline
			$d_9$ & $\lbrace \lbrace BCAD \rbrace\rbrace$ & 1 \\ 
			\hline
			$d_{10}$ & $\lbrace \lbrace BCDA \rbrace\rbrace$ & 1 \\ 
			\hline
			$d_{11}$ & $\lbrace \lbrace BDAC \rbrace\rbrace$ & 1 \\ 
			\hline
			$d_{12}$ & $\lbrace \lbrace BDCA \rbrace\rbrace$ & 1 \\ 
			\hline
			$d_{13}$ & $\lbrace \lbrace CABD \rbrace\rbrace$ & 1 \\ 
			\hline
			$d_{14}$ & $\lbrace \lbrace CADB \rbrace\rbrace$ & 1 \\ 
			\hline
			$d_{15}$ & $\lbrace \lbrace CBAD \rbrace\rbrace$ & 1 \\ 
			\hline
			$d_{16}$ & $\lbrace \lbrace CBDA \rbrace\rbrace$ & 1 \\ 
			\hline
			$d_{17}$ & $\lbrace \lbrace CDAB \rbrace\rbrace$ & 1 \\ 
			\hline
			$d_{18}$ & $\lbrace \lbrace CDBA \rbrace\rbrace$ & 1 \\ 
			\hline
			$d_{19}$ & $\lbrace \lbrace DABC \rbrace\rbrace$ & 1 \\ 
			\hline
			$d_{20}$ & $\lbrace \lbrace DACB \rbrace\rbrace$ & 1 \\ 
			\hline
			$d_{21}$ & $\lbrace \lbrace DBAC \rbrace\rbrace$ & 1 \\ 
			\hline
			$d_{22}$ & $\lbrace \lbrace DBCA \rbrace\rbrace$ & 1 \\ 
			\hline
			$d_{23}$ & $\lbrace \lbrace DCAB \rbrace\rbrace$ & 1 \\ 
			\hline
			$d_{24}$ & $\lbrace \lbrace DCBA \rbrace\rbrace$ & 1 \\ 
			\hline
		\end{tabular}
		\captionof{table}{Sequences and $s$-factors}
		\label{Table:Diag8}	}
\end{minipage} 
\vspace{0.4cm}

\vspace{0.5cm}
\noindent   The mixing matrix $R$, the diagonalizing matrix $Y$ and the diagonal matrix $D$ for the Cweb mentioned above are

\begin{equation}
	\scriptsize { R =\frac{1}{12} \left(\,\,\,
		\begin{array}{cccccccccccccccccccccccc}
			3 &\hspace{0.20cm} \hspace{-0.25cm}-1 &\hspace{0.20cm} \hspace{-0.25cm}-1 &\hspace{0.20cm} \hspace{-0.25cm}-1 &\hspace{0.20cm} \hspace{-0.25cm}-1 &\hspace{0.20cm} 1 &\hspace{0.20cm} \hspace{-0.25cm}-1 &\hspace{0.20cm} 1 &\hspace{0.20cm} \hspace{-0.25cm}-1 &\hspace{0.20cm} \hspace{-0.25cm}-1 &\hspace{0.20cm} \hspace{-0.25cm}-1 &\hspace{0.20cm} 1 &\hspace{0.20cm} \hspace{-0.25cm}-1 &\hspace{0.20cm} 1 &\hspace{0.20cm} 1 &\hspace{0.20cm} 1 &\hspace{0.20cm} \hspace{-0.25cm}-1 &\hspace{0.20cm} 1 &\hspace{0.20cm} \hspace{-0.25cm}-1 &\hspace{0.20cm} 1 &\hspace{0.20cm} 1 &\hspace{0.20cm} 1 &\hspace{0.20cm} 1 &\hspace{0.20cm} \hspace{-0.25cm}-3 \\
			\hspace{-0.25cm}-1 &\hspace{0.20cm} 3 &\hspace{0.20cm} \hspace{-0.25cm}-1 &\hspace{0.20cm} 1 &\hspace{0.20cm} \hspace{-0.25cm}-1 &\hspace{0.20cm} \hspace{-0.25cm}-1 &\hspace{0.20cm} 1 &\hspace{0.20cm} \hspace{-0.25cm}-1 &\hspace{0.20cm} \hspace{-0.25cm}-1 &\hspace{0.20cm} 1 &\hspace{0.20cm} \hspace{-0.25cm}-1 &\hspace{0.20cm} \hspace{-0.25cm}-1 &\hspace{0.20cm} \hspace{-0.25cm}-1 &\hspace{0.20cm} 1 &\hspace{0.20cm} 1 &\hspace{0.20cm} 1 &\hspace{0.20cm} 1 &\hspace{0.20cm} \hspace{-0.25cm}-3 &\hspace{0.20cm} \hspace{-0.25cm}-1 &\hspace{0.20cm} 1 &\hspace{0.20cm} 1 &\hspace{0.20cm} 1 &\hspace{0.20cm} \hspace{-0.25cm}-1 &\hspace{0.20cm} 1 \\
			\hspace{-0.25cm}-1 &\hspace{0.20cm} \hspace{-0.25cm}-1 &\hspace{0.20cm} 3 &\hspace{0.20cm} \hspace{-0.25cm}-1 &\hspace{0.20cm} 1 &\hspace{0.20cm} \hspace{-0.25cm}-1 &\hspace{0.20cm} \hspace{-0.25cm}-1 &\hspace{0.20cm} 1 &\hspace{0.20cm} 1 &\hspace{0.20cm} 1 &\hspace{0.20cm} \hspace{-0.25cm}-1 &\hspace{0.20cm} 1 &\hspace{0.20cm} \hspace{-0.25cm}-1 &\hspace{0.20cm} 1 &\hspace{0.20cm} \hspace{-0.25cm}-1 &\hspace{0.20cm} \hspace{-0.25cm}-1 &\hspace{0.20cm} \hspace{-0.25cm}-1 &\hspace{0.20cm} 1 &\hspace{0.20cm} 1 &\hspace{0.20cm} \hspace{-0.25cm}-1 &\hspace{0.20cm} 1 &\hspace{0.20cm} \hspace{-0.25cm}-3 &\hspace{0.20cm} 1 &\hspace{0.20cm} 1 \\
			\hspace{-0.25cm}-1 &\hspace{0.20cm} 1 &\hspace{0.20cm} \hspace{-0.25cm}-1 &\hspace{0.20cm} 3 &\hspace{0.20cm} \hspace{-0.25cm}-1 &\hspace{0.20cm} \hspace{-0.25cm}-1 &\hspace{0.20cm} \hspace{-0.25cm}-1 &\hspace{0.20cm} 1 &\hspace{0.20cm} 1 &\hspace{0.20cm} 1 &\hspace{0.20cm} 1 &\hspace{0.20cm} \hspace{-0.25cm}-3 &\hspace{0.20cm} 1 &\hspace{0.20cm} \hspace{-0.25cm}-1 &\hspace{0.20cm} \hspace{-0.25cm}-1 &\hspace{0.20cm} 1 &\hspace{0.20cm} \hspace{-0.25cm}-1 &\hspace{0.20cm} \hspace{-0.25cm}-1 &\hspace{0.20cm} 1 &\hspace{0.20cm} \hspace{-0.25cm}-1 &\hspace{0.20cm} \hspace{-0.25cm}-1 &\hspace{0.20cm} 1 &\hspace{0.20cm} 1 &\hspace{0.20cm} 1 \\
			\hspace{-0.25cm}-1 &\hspace{0.20cm} \hspace{-0.25cm}-1 &\hspace{0.20cm} 1 &\hspace{0.20cm} \hspace{-0.25cm}-1 &\hspace{0.20cm} 3 &\hspace{0.20cm} \hspace{-0.25cm}-1 &\hspace{0.20cm} 1 &\hspace{0.20cm} \hspace{-0.25cm}-1 &\hspace{0.20cm} \hspace{-0.25cm}-1 &\hspace{0.20cm} 1 &\hspace{0.20cm} 1 &\hspace{0.20cm} 1 &\hspace{0.20cm} 1 &\hspace{0.20cm} \hspace{-0.25cm}-1 &\hspace{0.20cm} 1 &\hspace{0.20cm} \hspace{-0.25cm}-3 &\hspace{0.20cm} 1 &\hspace{0.20cm} 1 &\hspace{0.20cm} \hspace{-0.25cm}-1 &\hspace{0.20cm} 1 &\hspace{0.20cm} \hspace{-0.25cm}-1 &\hspace{0.20cm} \hspace{-0.25cm}-1 &\hspace{0.20cm} \hspace{-0.25cm}-1 &\hspace{0.20cm} 1 \\
			1 &\hspace{0.20cm} \hspace{-0.25cm}-1 &\hspace{0.20cm} \hspace{-0.25cm}-1 &\hspace{0.20cm} \hspace{-0.25cm}-1 &\hspace{0.20cm} \hspace{-0.25cm}-1 &\hspace{0.20cm} 3 &\hspace{0.20cm} 1 &\hspace{0.20cm} \hspace{-0.25cm}-1 &\hspace{0.20cm} 1 &\hspace{0.20cm} \hspace{-0.25cm}-3 &\hspace{0.20cm} 1 &\hspace{0.20cm} 1 &\hspace{0.20cm} 1 &\hspace{0.20cm} \hspace{-0.25cm}-1 &\hspace{0.20cm} \hspace{-0.25cm}-1 &\hspace{0.20cm} 1 &\hspace{0.20cm} 1 &\hspace{0.20cm} 1 &\hspace{0.20cm} 1 &\hspace{0.20cm} \hspace{-0.25cm}-1 &\hspace{0.20cm} \hspace{-0.25cm}-1 &\hspace{0.20cm} 1 &\hspace{0.20cm} \hspace{-0.25cm}-1 &\hspace{0.20cm} \hspace{-0.25cm}-1 \\
			\hspace{-0.25cm}-1 &\hspace{0.20cm} 1 &\hspace{0.20cm} \hspace{-0.25cm}-1 &\hspace{0.20cm} \hspace{-0.25cm}-1 &\hspace{0.20cm} \hspace{-0.25cm}-1 &\hspace{0.20cm} 1 &\hspace{0.20cm} 3 &\hspace{0.20cm} \hspace{-0.25cm}-1 &\hspace{0.20cm} \hspace{-0.25cm}-1 &\hspace{0.20cm} \hspace{-0.25cm}-1 &\hspace{0.20cm} \hspace{-0.25cm}-1 &\hspace{0.20cm} 1 &\hspace{0.20cm} 1 &\hspace{0.20cm} 1 &\hspace{0.20cm} \hspace{-0.25cm}-1 &\hspace{0.20cm} 1 &\hspace{0.20cm} 1 &\hspace{0.20cm} \hspace{-0.25cm}-1 &\hspace{0.20cm} 1 &\hspace{0.20cm} 1 &\hspace{0.20cm} \hspace{-0.25cm}-1 &\hspace{0.20cm} 1 &\hspace{0.20cm} \hspace{-0.25cm}-3 &\hspace{0.20cm} 1 \\
			1 &\hspace{0.20cm} \hspace{-0.25cm}-1 &\hspace{0.20cm} \hspace{-0.25cm}-1 &\hspace{0.20cm} 1 &\hspace{0.20cm} \hspace{-0.25cm}-1 &\hspace{0.20cm} \hspace{-0.25cm}-1 &\hspace{0.20cm} \hspace{-0.25cm}-1 &\hspace{0.20cm} 3 &\hspace{0.20cm} \hspace{-0.25cm}-1 &\hspace{0.20cm} 1 &\hspace{0.20cm} \hspace{-0.25cm}-1 &\hspace{0.20cm} \hspace{-0.25cm}-1 &\hspace{0.20cm} 1 &\hspace{0.20cm} 1 &\hspace{0.20cm} \hspace{-0.25cm}-1 &\hspace{0.20cm} 1 &\hspace{0.20cm} \hspace{-0.25cm}-3 &\hspace{0.20cm} 1 &\hspace{0.20cm} 1 &\hspace{0.20cm} 1 &\hspace{0.20cm} \hspace{-0.25cm}-1 &\hspace{0.20cm} 1 &\hspace{0.20cm} 1 &\hspace{0.20cm} \hspace{-0.25cm}-1 \\
			\hspace{-0.25cm}-1 &\hspace{0.20cm} 1 &\hspace{0.20cm} 1 &\hspace{0.20cm} 1 &\hspace{0.20cm} \hspace{-0.25cm}-1 &\hspace{0.20cm} 1 &\hspace{0.20cm} \hspace{-0.25cm}-1 &\hspace{0.20cm} \hspace{-0.25cm}-1 &\hspace{0.20cm} 3 &\hspace{0.20cm} \hspace{-0.25cm}-1 &\hspace{0.20cm} 1 &\hspace{0.20cm} \hspace{-0.25cm}-1 &\hspace{0.20cm} \hspace{-0.25cm}-1 &\hspace{0.20cm} \hspace{-0.25cm}-1 &\hspace{0.20cm} \hspace{-0.25cm}-1 &\hspace{0.20cm} 1 &\hspace{0.20cm} 1 &\hspace{0.20cm} \hspace{-0.25cm}-1 &\hspace{0.20cm} 1 &\hspace{0.20cm} \hspace{-0.25cm}-3 &\hspace{0.20cm} 1 &\hspace{0.20cm} \hspace{-0.25cm}-1 &\hspace{0.20cm} 1 &\hspace{0.20cm} 1 \\
			\hspace{-0.25cm}-1 &\hspace{0.20cm} 1 &\hspace{0.20cm} 1 &\hspace{0.20cm} 1 &\hspace{0.20cm} 1 &\hspace{0.20cm} \hspace{-0.25cm}-3 &\hspace{0.20cm} \hspace{-0.25cm}-1 &\hspace{0.20cm} 1 &\hspace{0.20cm} \hspace{-0.25cm}-1 &\hspace{0.20cm} 3 &\hspace{0.20cm} \hspace{-0.25cm}-1 &\hspace{0.20cm} \hspace{-0.25cm}-1 &\hspace{0.20cm} \hspace{-0.25cm}-1 &\hspace{0.20cm} 1 &\hspace{0.20cm} 1 &\hspace{0.20cm} \hspace{-0.25cm}-1 &\hspace{0.20cm} \hspace{-0.25cm}-1 &\hspace{0.20cm} \hspace{-0.25cm}-1 &\hspace{0.20cm} \hspace{-0.25cm}-1 &\hspace{0.20cm} 1 &\hspace{0.20cm} 1 &\hspace{0.20cm} \hspace{-0.25cm}-1 &\hspace{0.20cm} 1 &\hspace{0.20cm} 1 \\
			1 &\hspace{0.20cm} \hspace{-0.25cm}-1 &\hspace{0.20cm} \hspace{-0.25cm}-1 &\hspace{0.20cm} 1 &\hspace{0.20cm} 1 &\hspace{0.20cm} 1 &\hspace{0.20cm} \hspace{-0.25cm}-1 &\hspace{0.20cm} \hspace{-0.25cm}-1 &\hspace{0.20cm} 1 &\hspace{0.20cm} \hspace{-0.25cm}-1 &\hspace{0.20cm} 3 &\hspace{0.20cm} \hspace{-0.25cm}-1 &\hspace{0.20cm} 1 &\hspace{0.20cm} \hspace{-0.25cm}-3 &\hspace{0.20cm} 1 &\hspace{0.20cm} \hspace{-0.25cm}-1 &\hspace{0.20cm} 1 &\hspace{0.20cm} 1 &\hspace{0.20cm} \hspace{-0.25cm}-1 &\hspace{0.20cm} \hspace{-0.25cm}-1 &\hspace{0.20cm} \hspace{-0.25cm}-1 &\hspace{0.20cm} 1 &\hspace{0.20cm} 1 &\hspace{0.20cm} \hspace{-0.25cm}-1 \\
			1 &\hspace{0.20cm} \hspace{-0.25cm}-1 &\hspace{0.20cm} 1 &\hspace{0.20cm} \hspace{-0.25cm}-3 &\hspace{0.20cm} 1 &\hspace{0.20cm} 1 &\hspace{0.20cm} 1 &\hspace{0.20cm} \hspace{-0.25cm}-1 &\hspace{0.20cm} \hspace{-0.25cm}-1 &\hspace{0.20cm} \hspace{-0.25cm}-1 &\hspace{0.20cm} \hspace{-0.25cm}-1 &\hspace{0.20cm} 3 &\hspace{0.20cm} \hspace{-0.25cm}-1 &\hspace{0.20cm} 1 &\hspace{0.20cm} 1 &\hspace{0.20cm} \hspace{-0.25cm}-1 &\hspace{0.20cm} 1 &\hspace{0.20cm} 1 &\hspace{0.20cm} \hspace{-0.25cm}-1 &\hspace{0.20cm} 1 &\hspace{0.20cm} 1 &\hspace{0.20cm} \hspace{-0.25cm}-1 &\hspace{0.20cm} \hspace{-0.25cm}-1 &\hspace{0.20cm} \hspace{-0.25cm}-1 \\
			\hspace{-0.25cm}-1 &\hspace{0.20cm} \hspace{-0.25cm}-1 &\hspace{0.20cm} \hspace{-0.25cm}-1 &\hspace{0.20cm} 1 &\hspace{0.20cm} 1 &\hspace{0.20cm} \hspace{-0.25cm}-1 &\hspace{0.20cm} 1 &\hspace{0.20cm} 1 &\hspace{0.20cm} \hspace{-0.25cm}-1 &\hspace{0.20cm} 1 &\hspace{0.20cm} 1 &\hspace{0.20cm} \hspace{-0.25cm}-1 &\hspace{0.20cm} 3 &\hspace{0.20cm} \hspace{-0.25cm}-1 &\hspace{0.20cm} \hspace{-0.25cm}-1 &\hspace{0.20cm} \hspace{-0.25cm}-1 &\hspace{0.20cm} \hspace{-0.25cm}-1 &\hspace{0.20cm} 1 &\hspace{0.20cm} 1 &\hspace{0.20cm} 1 &\hspace{0.20cm} \hspace{-0.25cm}-3 &\hspace{0.20cm} 1 &\hspace{0.20cm} \hspace{-0.25cm}-1 &\hspace{0.20cm} 1 \\
			\hspace{-0.25cm}-1 &\hspace{0.20cm} 1 &\hspace{0.20cm} 1 &\hspace{0.20cm} \hspace{-0.25cm}-1 &\hspace{0.20cm} \hspace{-0.25cm}-1 &\hspace{0.20cm} \hspace{-0.25cm}-1 &\hspace{0.20cm} 1 &\hspace{0.20cm} 1 &\hspace{0.20cm} \hspace{-0.25cm}-1 &\hspace{0.20cm} 1 &\hspace{0.20cm} \hspace{-0.25cm}-3 &\hspace{0.20cm} 1 &\hspace{0.20cm} \hspace{-0.25cm}-1 &\hspace{0.20cm} 3 &\hspace{0.20cm} \hspace{-0.25cm}-1 &\hspace{0.20cm} 1 &\hspace{0.20cm} \hspace{-0.25cm}-1 &\hspace{0.20cm} \hspace{-0.25cm}-1 &\hspace{0.20cm} 1 &\hspace{0.20cm} 1 &\hspace{0.20cm} 1 &\hspace{0.20cm} \hspace{-0.25cm}-1 &\hspace{0.20cm} \hspace{-0.25cm}-1 &\hspace{0.20cm} 1 \\
			1 &\hspace{0.20cm} 1 &\hspace{0.20cm} \hspace{-0.25cm}-1 &\hspace{0.20cm} 1 &\hspace{0.20cm} 1 &\hspace{0.20cm} \hspace{-0.25cm}-1 &\hspace{0.20cm} \hspace{-0.25cm}-1 &\hspace{0.20cm} \hspace{-0.25cm}-1 &\hspace{0.20cm} \hspace{-0.25cm}-1 &\hspace{0.20cm} 1 &\hspace{0.20cm} 1 &\hspace{0.20cm} \hspace{-0.25cm}-1 &\hspace{0.20cm} \hspace{-0.25cm}-1 &\hspace{0.20cm} \hspace{-0.25cm}-1 &\hspace{0.20cm} 3 &\hspace{0.20cm} \hspace{-0.25cm}-1 &\hspace{0.20cm} 1 &\hspace{0.20cm} \hspace{-0.25cm}-1 &\hspace{0.20cm} \hspace{-0.25cm}-3 &\hspace{0.20cm} 1 &\hspace{0.20cm} 1 &\hspace{0.20cm} 1 &\hspace{0.20cm} 1 &\hspace{0.20cm} \hspace{-0.25cm}-1 \\
			1 &\hspace{0.20cm} 1 &\hspace{0.20cm} \hspace{-0.25cm}-1 &\hspace{0.20cm} 1 &\hspace{0.20cm} \hspace{-0.25cm}-3 &\hspace{0.20cm} 1 &\hspace{0.20cm} \hspace{-0.25cm}-1 &\hspace{0.20cm} 1 &\hspace{0.20cm} 1 &\hspace{0.20cm} \hspace{-0.25cm}-1 &\hspace{0.20cm} \hspace{-0.25cm}-1 &\hspace{0.20cm} \hspace{-0.25cm}-1 &\hspace{0.20cm} \hspace{-0.25cm}-1 &\hspace{0.20cm} 1 &\hspace{0.20cm} \hspace{-0.25cm}-1 &\hspace{0.20cm} 3 &\hspace{0.20cm} \hspace{-0.25cm}-1 &\hspace{0.20cm} \hspace{-0.25cm}-1 &\hspace{0.20cm} 1 &\hspace{0.20cm} \hspace{-0.25cm}-1 &\hspace{0.20cm} 1 &\hspace{0.20cm} 1 &\hspace{0.20cm} 1 &\hspace{0.20cm} \hspace{-0.25cm}-1 \\
			\hspace{-0.25cm}-1 &\hspace{0.20cm} 1 &\hspace{0.20cm} 1 &\hspace{0.20cm} \hspace{-0.25cm}-1 &\hspace{0.20cm} 1 &\hspace{0.20cm} 1 &\hspace{0.20cm} 1 &\hspace{0.20cm} \hspace{-0.25cm}-3 &\hspace{0.20cm} 1 &\hspace{0.20cm} \hspace{-0.25cm}-1 &\hspace{0.20cm} 1 &\hspace{0.20cm} 1 &\hspace{0.20cm} \hspace{-0.25cm}-1 &\hspace{0.20cm} \hspace{-0.25cm}-1 &\hspace{0.20cm} 1 &\hspace{0.20cm} \hspace{-0.25cm}-1 &\hspace{0.20cm} 3 &\hspace{0.20cm} \hspace{-0.25cm}-1 &\hspace{0.20cm} \hspace{-0.25cm}-1 &\hspace{0.20cm} \hspace{-0.25cm}-1 &\hspace{0.20cm} 1 &\hspace{0.20cm} \hspace{-0.25cm}-1 &\hspace{0.20cm} \hspace{-0.25cm}-1 &\hspace{0.20cm} 1 \\
			1 &\hspace{0.20cm} \hspace{-0.25cm}-3 &\hspace{0.20cm} 1 &\hspace{0.20cm} \hspace{-0.25cm}-1 &\hspace{0.20cm} 1 &\hspace{0.20cm} 1 &\hspace{0.20cm} \hspace{-0.25cm}-1 &\hspace{0.20cm} 1 &\hspace{0.20cm} 1 &\hspace{0.20cm} \hspace{-0.25cm}-1 &\hspace{0.20cm} 1 &\hspace{0.20cm} 1 &\hspace{0.20cm} 1 &\hspace{0.20cm} \hspace{-0.25cm}-1 &\hspace{0.20cm} \hspace{-0.25cm}-1 &\hspace{0.20cm} \hspace{-0.25cm}-1 &\hspace{0.20cm} \hspace{-0.25cm}-1 &\hspace{0.20cm} 3 &\hspace{0.20cm} 1 &\hspace{0.20cm} \hspace{-0.25cm}-1 &\hspace{0.20cm} \hspace{-0.25cm}-1 &\hspace{0.20cm} \hspace{-0.25cm}-1 &\hspace{0.20cm} 1 &\hspace{0.20cm} \hspace{-0.25cm}-1 \\
			\hspace{-0.25cm}-1 &\hspace{0.20cm} \hspace{-0.25cm}-1 &\hspace{0.20cm} 1 &\hspace{0.20cm} \hspace{-0.25cm}-1 &\hspace{0.20cm} \hspace{-0.25cm}-1 &\hspace{0.20cm} 1 &\hspace{0.20cm} 1 &\hspace{0.20cm} 1 &\hspace{0.20cm} 1 &\hspace{0.20cm} \hspace{-0.25cm}-1 &\hspace{0.20cm} \hspace{-0.25cm}-1 &\hspace{0.20cm} 1 &\hspace{0.20cm} 1 &\hspace{0.20cm} 1 &\hspace{0.20cm} \hspace{-0.25cm}-3 &\hspace{0.20cm} 1 &\hspace{0.20cm} \hspace{-0.25cm}-1 &\hspace{0.20cm} 1 &\hspace{0.20cm} 3 &\hspace{0.20cm} \hspace{-0.25cm}-1 &\hspace{0.20cm} \hspace{-0.25cm}-1 &\hspace{0.20cm} \hspace{-0.25cm}-1 &\hspace{0.20cm} \hspace{-0.25cm}-1 &\hspace{0.20cm} 1 \\
			1 &\hspace{0.20cm} \hspace{-0.25cm}-1 &\hspace{0.20cm} \hspace{-0.25cm}-1 &\hspace{0.20cm} \hspace{-0.25cm}-1 &\hspace{0.20cm} 1 &\hspace{0.20cm} \hspace{-0.25cm}-1 &\hspace{0.20cm} 1 &\hspace{0.20cm} 1 &\hspace{0.20cm} \hspace{-0.25cm}-3 &\hspace{0.20cm} 1 &\hspace{0.20cm} \hspace{-0.25cm}-1 &\hspace{0.20cm} 1 &\hspace{0.20cm} 1 &\hspace{0.20cm} 1 &\hspace{0.20cm} 1 &\hspace{0.20cm} \hspace{-0.25cm}-1 &\hspace{0.20cm} \hspace{-0.25cm}-1 &\hspace{0.20cm} 1 &\hspace{0.20cm} \hspace{-0.25cm}-1 &\hspace{0.20cm} 3 &\hspace{0.20cm} \hspace{-0.25cm}-1 &\hspace{0.20cm} 1 &\hspace{0.20cm} \hspace{-0.25cm}-1 &\hspace{0.20cm} \hspace{-0.25cm}-1 \\
			1 &\hspace{0.20cm} 1 &\hspace{0.20cm} 1 &\hspace{0.20cm} \hspace{-0.25cm}-1 &\hspace{0.20cm} \hspace{-0.25cm}-1 &\hspace{0.20cm} 1 &\hspace{0.20cm} \hspace{-0.25cm}-1 &\hspace{0.20cm} \hspace{-0.25cm}-1 &\hspace{0.20cm} 1 &\hspace{0.20cm} \hspace{-0.25cm}-1 &\hspace{0.20cm} \hspace{-0.25cm}-1 &\hspace{0.20cm} 1 &\hspace{0.20cm} \hspace{-0.25cm}-3 &\hspace{0.20cm} 1 &\hspace{0.20cm} 1 &\hspace{0.20cm} 1 &\hspace{0.20cm} 1 &\hspace{0.20cm} \hspace{-0.25cm}-1 &\hspace{0.20cm} \hspace{-0.25cm}-1 &\hspace{0.20cm} \hspace{-0.25cm}-1 &\hspace{0.20cm} 3 &\hspace{0.20cm} \hspace{-0.25cm}-1 &\hspace{0.20cm} 1 &\hspace{0.20cm} \hspace{-0.25cm}-1 \\
			1 &\hspace{0.20cm} 1 &\hspace{0.20cm} \hspace{-0.25cm}-3 &\hspace{0.20cm} 1 &\hspace{0.20cm} \hspace{-0.25cm}-1 &\hspace{0.20cm} 1 &\hspace{0.20cm} 1 &\hspace{0.20cm} \hspace{-0.25cm}-1 &\hspace{0.20cm} \hspace{-0.25cm}-1 &\hspace{0.20cm} \hspace{-0.25cm}-1 &\hspace{0.20cm} 1 &\hspace{0.20cm} \hspace{-0.25cm}-1 &\hspace{0.20cm} 1 &\hspace{0.20cm} \hspace{-0.25cm}-1 &\hspace{0.20cm} 1 &\hspace{0.20cm} 1 &\hspace{0.20cm} 1 &\hspace{0.20cm} \hspace{-0.25cm}-1 &\hspace{0.20cm} \hspace{-0.25cm}-1 &\hspace{0.20cm} 1 &\hspace{0.20cm} \hspace{-0.25cm}-1 &\hspace{0.20cm} 3 &\hspace{0.20cm} \hspace{-0.25cm}-1 &\hspace{0.20cm} \hspace{-0.25cm}-1 \\
			1 &\hspace{0.20cm} \hspace{-0.25cm}-1 &\hspace{0.20cm} 1 &\hspace{0.20cm} 1 &\hspace{0.20cm} 1 &\hspace{0.20cm} \hspace{-0.25cm}-1 &\hspace{0.20cm} \hspace{-0.25cm}-3 &\hspace{0.20cm} 1 &\hspace{0.20cm} 1 &\hspace{0.20cm} 1 &\hspace{0.20cm} 1 &\hspace{0.20cm} \hspace{-0.25cm}-1 &\hspace{0.20cm} \hspace{-0.25cm}-1 &\hspace{0.20cm} \hspace{-0.25cm}-1 &\hspace{0.20cm} 1 &\hspace{0.20cm} \hspace{-0.25cm}-1 &\hspace{0.20cm} \hspace{-0.25cm}-1 &\hspace{0.20cm} 1 &\hspace{0.20cm} \hspace{-0.25cm}-1 &\hspace{0.20cm} \hspace{-0.25cm}-1 &\hspace{0.20cm} 1 &\hspace{0.20cm} \hspace{-0.25cm}-1 &\hspace{0.20cm} 3 &\hspace{0.20cm} \hspace{-0.25cm}-1 \\
			\hspace{-0.25cm}-3 &\hspace{0.20cm} 1 &\hspace{0.20cm} 1 &\hspace{0.20cm} 1 &\hspace{0.20cm} 1 &\hspace{0.20cm} \hspace{-0.25cm}-1 &\hspace{0.20cm} 1 &\hspace{0.20cm} \hspace{-0.25cm}-1 &\hspace{0.20cm} 1 &\hspace{0.20cm} 1 &\hspace{0.20cm} 1 &\hspace{0.20cm} \hspace{-0.25cm}-1 &\hspace{0.20cm} 1 &\hspace{0.20cm} \hspace{-0.25cm}-1 &\hspace{0.20cm} \hspace{-0.25cm}-1 &\hspace{0.20cm} \hspace{-0.25cm}-1 &\hspace{0.20cm} 1 &\hspace{0.20cm} \hspace{-0.25cm}-1 &\hspace{0.20cm} 1 &\hspace{0.20cm} \hspace{-0.25cm}-1 &\hspace{0.20cm} \hspace{-0.25cm}-1 &\hspace{0.20cm} \hspace{-0.25cm}-1 &\hspace{0.20cm} \hspace{-0.25cm}-1 &\hspace{0.20cm} 3 \\
		\end{array}
		\right)}, \nonumber
\end{equation}

\begin{equation}\scriptsize{
		Y=\left( \,\,\,
		\begin{array}{cccccccccccccccccccccccc}
			\hspace{-0.25cm}-1 &\hspace{0.20cm} 1 &\hspace{0.20cm} 0 &\hspace{0.20cm} 1 &\hspace{0.20cm} 0 &\hspace{0.20cm} \hspace{-0.25cm}-1 &\hspace{0.20cm} 0 &\hspace{0.20cm} 0 &\hspace{0.20cm} 0 &\hspace{0.20cm} 1 &\hspace{0.20cm} 0 &\hspace{0.20cm} \hspace{-0.25cm}-1 &\hspace{0.20cm} 0 &\hspace{0.20cm} 0 &\hspace{0.20cm} 0 &\hspace{0.20cm} 0 &\hspace{0.20cm} 0 &\hspace{0.20cm} \hspace{-0.25cm}-1 &\hspace{0.20cm} 0 &\hspace{0.20cm} 0 &\hspace{0.20cm} 0 &\hspace{0.20cm} 0 &\hspace{0.20cm} 0 &\hspace{0.20cm} 1 \\
			0 &\hspace{0.20cm} 0 &\hspace{0.20cm} 0 &\hspace{0.20cm} 1 &\hspace{0.20cm} 0 &\hspace{0.20cm} \hspace{-0.25cm}-1 &\hspace{0.20cm} \hspace{-0.25cm}-1 &\hspace{0.20cm} 1 &\hspace{0.20cm} 0 &\hspace{0.20cm} 1 &\hspace{0.20cm} 0 &\hspace{0.20cm} \hspace{-0.25cm}-1 &\hspace{0.20cm} 0 &\hspace{0.20cm} 0 &\hspace{0.20cm} 0 &\hspace{0.20cm} 0 &\hspace{0.20cm} \hspace{-0.25cm}-1 &\hspace{0.20cm} 0 &\hspace{0.20cm} 0 &\hspace{0.20cm} 0 &\hspace{0.20cm} 0 &\hspace{0.20cm} 0 &\hspace{0.20cm} 1 &\hspace{0.20cm} 0 \\
			0 &\hspace{0.20cm} 1 &\hspace{0.20cm} \hspace{-0.25cm}-1 &\hspace{0.20cm} 1 &\hspace{0.20cm} \hspace{-0.25cm}-1 &\hspace{0.20cm} 0 &\hspace{0.20cm} 0 &\hspace{0.20cm} 0 &\hspace{0.20cm} 0 &\hspace{0.20cm} 0 &\hspace{0.20cm} 0 &\hspace{0.20cm} \hspace{-0.25cm}-1 &\hspace{0.20cm} 0 &\hspace{0.20cm} 0 &\hspace{0.20cm} 0 &\hspace{0.20cm} 1 &\hspace{0.20cm} 0 &\hspace{0.20cm} \hspace{-0.25cm}-1 &\hspace{0.20cm} 0 &\hspace{0.20cm} 0 &\hspace{0.20cm} 0 &\hspace{0.20cm} 1 &\hspace{0.20cm} 0 &\hspace{0.20cm} 0 \\
			0 &\hspace{0.20cm} 1 &\hspace{0.20cm} 0 &\hspace{0.20cm} 0 &\hspace{0.20cm} \hspace{-0.25cm}-1 &\hspace{0.20cm} 0 &\hspace{0.20cm} 0 &\hspace{0.20cm} 0 &\hspace{0.20cm} 0 &\hspace{0.20cm} 0 &\hspace{0.20cm} \hspace{-0.25cm}-1 &\hspace{0.20cm} 0 &\hspace{0.20cm} \hspace{-0.25cm}-1 &\hspace{0.20cm} 1 &\hspace{0.20cm} 0 &\hspace{0.20cm} 1 &\hspace{0.20cm} 0 &\hspace{0.20cm} \hspace{-0.25cm}-1 &\hspace{0.20cm} 0 &\hspace{0.20cm} 0 &\hspace{0.20cm} 1 &\hspace{0.20cm} 0 &\hspace{0.20cm} 0 &\hspace{0.20cm} 0 \\
			0 &\hspace{0.20cm} 0 &\hspace{0.20cm} 0 &\hspace{0.20cm} 0 &\hspace{0.20cm} 0 &\hspace{0.20cm} \hspace{-0.25cm}-1 &\hspace{0.20cm} 0 &\hspace{0.20cm} 1 &\hspace{0.20cm} \hspace{-0.25cm}-1 &\hspace{0.20cm} 1 &\hspace{0.20cm} \hspace{-0.25cm}-1 &\hspace{0.20cm} 0 &\hspace{0.20cm} 0 &\hspace{0.20cm} 1 &\hspace{0.20cm} 0 &\hspace{0.20cm} 0 &\hspace{0.20cm} \hspace{-0.25cm}-1 &\hspace{0.20cm} 0 &\hspace{0.20cm} 0 &\hspace{0.20cm} 1 &\hspace{0.20cm} 0 &\hspace{0.20cm} 0 &\hspace{0.20cm} 0 &\hspace{0.20cm} 0 \\
			0 &\hspace{0.20cm} 0 &\hspace{0.20cm} 0 &\hspace{0.20cm} 0 &\hspace{0.20cm} \hspace{-0.25cm}-1 &\hspace{0.20cm} 0 &\hspace{0.20cm} 0 &\hspace{0.20cm} 1 &\hspace{0.20cm} 0 &\hspace{0.20cm} 0 &\hspace{0.20cm} \hspace{-0.25cm}-1 &\hspace{0.20cm} 0 &\hspace{0.20cm} 0 &\hspace{0.20cm} 1 &\hspace{0.20cm} \hspace{-0.25cm}-1 &\hspace{0.20cm} 1 &\hspace{0.20cm} \hspace{-0.25cm}-1 &\hspace{0.20cm} 0 &\hspace{0.20cm} 1 &\hspace{0.20cm} 0 &\hspace{0.20cm} 0 &\hspace{0.20cm} 0 &\hspace{0.20cm} 0 &\hspace{0.20cm} 0 \\
			1 &\hspace{0.20cm} 0 &\hspace{0.20cm} 0 &\hspace{0.20cm} 0 &\hspace{0.20cm} 0 &\hspace{0.20cm} 0 &\hspace{0.20cm} 0 &\hspace{0.20cm} 0 &\hspace{0.20cm} 0 &\hspace{0.20cm} 0 &\hspace{0.20cm} 0 &\hspace{0.20cm} 0 &\hspace{0.20cm} 0 &\hspace{0.20cm} 0 &\hspace{0.20cm} 0 &\hspace{0.20cm} 0 &\hspace{0.20cm} 0 &\hspace{0.20cm} 0 &\hspace{0.20cm} 0 &\hspace{0.20cm} 0 &\hspace{0.20cm} 0 &\hspace{0.20cm} 0 &\hspace{0.20cm} 0 &\hspace{0.20cm} 1 \\
			0 &\hspace{0.20cm} 0 &\hspace{0.20cm} 0 &\hspace{0.20cm} 0 &\hspace{0.20cm} 0 &\hspace{0.20cm} 0 &\hspace{0.20cm} 1 &\hspace{0.20cm} 0 &\hspace{0.20cm} 0 &\hspace{0.20cm} 0 &\hspace{0.20cm} 0 &\hspace{0.20cm} 0 &\hspace{0.20cm} 0 &\hspace{0.20cm} 0 &\hspace{0.20cm} 0 &\hspace{0.20cm} 0 &\hspace{0.20cm} 0 &\hspace{0.20cm} 0 &\hspace{0.20cm} 0 &\hspace{0.20cm} 0 &\hspace{0.20cm} 0 &\hspace{0.20cm} 0 &\hspace{0.20cm} 1 &\hspace{0.20cm} 0 \\
			0 &\hspace{0.20cm} 0 &\hspace{0.20cm} 1 &\hspace{0.20cm} 0 &\hspace{0.20cm} 0 &\hspace{0.20cm} 0 &\hspace{0.20cm} 0 &\hspace{0.20cm} 0 &\hspace{0.20cm} 0 &\hspace{0.20cm} 0 &\hspace{0.20cm} 0 &\hspace{0.20cm} 0 &\hspace{0.20cm} 0 &\hspace{0.20cm} 0 &\hspace{0.20cm} 0 &\hspace{0.20cm} 0 &\hspace{0.20cm} 0 &\hspace{0.20cm} 0 &\hspace{0.20cm} 0 &\hspace{0.20cm} 0 &\hspace{0.20cm} 0 &\hspace{0.20cm} 1 &\hspace{0.20cm} 0 &\hspace{0.20cm} 0 \\
			1 &\hspace{0.20cm} \hspace{-0.25cm}-\frac{1}{3} &\hspace{0.20cm} \frac{1}{3} &\hspace{0.20cm} \frac{5}{3} &\hspace{0.20cm} \frac{4}{3} &\hspace{0.20cm} 0 &\hspace{0.20cm} 1 &\hspace{0.20cm} 0 &\hspace{0.20cm} 0 &\hspace{0.20cm} 0 &\hspace{0.20cm} 0 &\hspace{0.20cm} 0 &\hspace{0.20cm} 0 &\hspace{0.20cm} 0 &\hspace{0.20cm} 0 &\hspace{0.20cm} 0 &\hspace{0.20cm} 0 &\hspace{0.20cm} 0 &\hspace{0.20cm} 0 &\hspace{0.20cm} 0 &\hspace{0.20cm} 1 &\hspace{0.20cm} 0 &\hspace{0.20cm} 0 &\hspace{0.20cm} 0 \\
			\hspace{-0.25cm}-2 &\hspace{0.20cm} \hspace{-0.25cm}-\frac{1}{3} &\hspace{0.20cm} \hspace{-0.25cm}-\frac{2}{3} &\hspace{0.20cm} \hspace{-0.25cm}-\frac{4}{3} &\hspace{0.20cm} \hspace{-0.25cm}-\frac{5}{3} &\hspace{0.20cm} 0 &\hspace{0.20cm} \hspace{-0.25cm}-1 &\hspace{0.20cm} 0 &\hspace{0.20cm} 0 &\hspace{0.20cm} 0 &\hspace{0.20cm} 0 &\hspace{0.20cm} 0 &\hspace{0.20cm} 0 &\hspace{0.20cm} 0 &\hspace{0.20cm} 0 &\hspace{0.20cm} 0 &\hspace{0.20cm} 0 &\hspace{0.20cm} 0 &\hspace{0.20cm} 0 &\hspace{0.20cm} 1 &\hspace{0.20cm} 0 &\hspace{0.20cm} 0 &\hspace{0.20cm} 0 &\hspace{0.20cm} 0 \\
			0 &\hspace{0.20cm} \frac{2}{3} &\hspace{0.20cm} \hspace{-0.25cm}-\frac{2}{3} &\hspace{0.20cm} \hspace{-0.25cm}-\frac{1}{3} &\hspace{0.20cm} \frac{1}{3} &\hspace{0.20cm} 0 &\hspace{0.20cm} \hspace{-0.25cm}-1 &\hspace{0.20cm} 0 &\hspace{0.20cm} 0 &\hspace{0.20cm} 0 &\hspace{0.20cm} 0 &\hspace{0.20cm} 0 &\hspace{0.20cm} 0 &\hspace{0.20cm} 0 &\hspace{0.20cm} 0 &\hspace{0.20cm} 0 &\hspace{0.20cm} 0 &\hspace{0.20cm} 0 &\hspace{0.20cm} 1 &\hspace{0.20cm} 0 &\hspace{0.20cm} 0 &\hspace{0.20cm} 0 &\hspace{0.20cm} 0 &\hspace{0.20cm} 0 \\
			0 &\hspace{0.20cm} 1 &\hspace{0.20cm} 0 &\hspace{0.20cm} 0 &\hspace{0.20cm} 0 &\hspace{0.20cm} 0 &\hspace{0.20cm} 0 &\hspace{0.20cm} 0 &\hspace{0.20cm} 0 &\hspace{0.20cm} 0 &\hspace{0.20cm} 0 &\hspace{0.20cm} 0 &\hspace{0.20cm} 0 &\hspace{0.20cm} 0 &\hspace{0.20cm} 0 &\hspace{0.20cm} 0 &\hspace{0.20cm} 0 &\hspace{0.20cm} 1 &\hspace{0.20cm} 0 &\hspace{0.20cm} 0 &\hspace{0.20cm} 0 &\hspace{0.20cm} 0 &\hspace{0.20cm} 0 &\hspace{0.20cm} 0 \\
			2 &\hspace{0.20cm} 0 &\hspace{0.20cm} 1 &\hspace{0.20cm} 2 &\hspace{0.20cm} 1 &\hspace{0.20cm} 0 &\hspace{0.20cm} 1 &\hspace{0.20cm} 0 &\hspace{0.20cm} 0 &\hspace{0.20cm} 0 &\hspace{0.20cm} 0 &\hspace{0.20cm} 0 &\hspace{0.20cm} 0 &\hspace{0.20cm} 0 &\hspace{0.20cm} 0 &\hspace{0.20cm} 0 &\hspace{0.20cm} 1 &\hspace{0.20cm} 0 &\hspace{0.20cm} 0 &\hspace{0.20cm} 0 &\hspace{0.20cm} 0 &\hspace{0.20cm} 0 &\hspace{0.20cm} 0 &\hspace{0.20cm} 0 \\
			0 &\hspace{0.20cm} 0 &\hspace{0.20cm} 0 &\hspace{0.20cm} 0 &\hspace{0.20cm} 1 &\hspace{0.20cm} 0 &\hspace{0.20cm} 0 &\hspace{0.20cm} 0 &\hspace{0.20cm} 0 &\hspace{0.20cm} 0 &\hspace{0.20cm} 0 &\hspace{0.20cm} 0 &\hspace{0.20cm} 0 &\hspace{0.20cm} 0 &\hspace{0.20cm} 0 &\hspace{0.20cm} 1 &\hspace{0.20cm} 0 &\hspace{0.20cm} 0 &\hspace{0.20cm} 0 &\hspace{0.20cm} 0 &\hspace{0.20cm} 0 &\hspace{0.20cm} 0 &\hspace{0.20cm} 0 &\hspace{0.20cm} 0 \\
			0 &\hspace{0.20cm} \hspace{-0.25cm}-\frac{2}{3} &\hspace{0.20cm} \frac{2}{3} &\hspace{0.20cm} \frac{1}{3} &\hspace{0.20cm} \hspace{-0.25cm}-\frac{1}{3} &\hspace{0.20cm} 0 &\hspace{0.20cm} 1 &\hspace{0.20cm} 0 &\hspace{0.20cm} 0 &\hspace{0.20cm} 0 &\hspace{0.20cm} 0 &\hspace{0.20cm} 0 &\hspace{0.20cm} 0 &\hspace{0.20cm} 0 &\hspace{0.20cm} 1 &\hspace{0.20cm} 0 &\hspace{0.20cm} 0 &\hspace{0.20cm} 0 &\hspace{0.20cm} 0 &\hspace{0.20cm} 0 &\hspace{0.20cm} 0 &\hspace{0.20cm} 0 &\hspace{0.20cm} 0 &\hspace{0.20cm} 0 \\
			\hspace{-0.25cm}-1 &\hspace{0.20cm} \hspace{-0.25cm}-\frac{2}{3} &\hspace{0.20cm} \hspace{-0.25cm}-\frac{4}{3} &\hspace{0.20cm} \hspace{-0.25cm}-\frac{2}{3} &\hspace{0.20cm} \hspace{-0.25cm}-\frac{1}{3} &\hspace{0.20cm} 0 &\hspace{0.20cm} \hspace{-0.25cm}-1 &\hspace{0.20cm} 0 &\hspace{0.20cm} 0 &\hspace{0.20cm} 0 &\hspace{0.20cm} 0 &\hspace{0.20cm} 0 &\hspace{0.20cm} 0 &\hspace{0.20cm} 1 &\hspace{0.20cm} 0 &\hspace{0.20cm} 0 &\hspace{0.20cm} 0 &\hspace{0.20cm} 0 &\hspace{0.20cm} 0 &\hspace{0.20cm} 0 &\hspace{0.20cm} 0 &\hspace{0.20cm} 0 &\hspace{0.20cm} 0 &\hspace{0.20cm} 0 \\
			\hspace{-0.25cm}-1 &\hspace{0.20cm} \frac{1}{3} &\hspace{0.20cm} \hspace{-0.25cm}-\frac{1}{3} &\hspace{0.20cm} \hspace{-0.25cm}-\frac{5}{3} &\hspace{0.20cm} \hspace{-0.25cm}-\frac{4}{3} &\hspace{0.20cm} 0 &\hspace{0.20cm} \hspace{-0.25cm}-1 &\hspace{0.20cm} 0 &\hspace{0.20cm} 0 &\hspace{0.20cm} 0 &\hspace{0.20cm} 0 &\hspace{0.20cm} 0 &\hspace{0.20cm} 1 &\hspace{0.20cm} 0 &\hspace{0.20cm} 0 &\hspace{0.20cm} 0 &\hspace{0.20cm} 0 &\hspace{0.20cm} 0 &\hspace{0.20cm} 0 &\hspace{0.20cm} 0 &\hspace{0.20cm} 0 &\hspace{0.20cm} 0 &\hspace{0.20cm} 0 &\hspace{0.20cm} 0 \\
			0 &\hspace{0.20cm} 0 &\hspace{0.20cm} 0 &\hspace{0.20cm} 1 &\hspace{0.20cm} 0 &\hspace{0.20cm} 0 &\hspace{0.20cm} 0 &\hspace{0.20cm} 0 &\hspace{0.20cm} 0 &\hspace{0.20cm} 0 &\hspace{0.20cm} 0 &\hspace{0.20cm} 1 &\hspace{0.20cm} 0 &\hspace{0.20cm} 0 &\hspace{0.20cm} 0 &\hspace{0.20cm} 0 &\hspace{0.20cm} 0 &\hspace{0.20cm} 0 &\hspace{0.20cm} 0 &\hspace{0.20cm} 0 &\hspace{0.20cm} 0 &\hspace{0.20cm} 0 &\hspace{0.20cm} 0 &\hspace{0.20cm} 0 \\
			1 &\hspace{0.20cm} \frac{2}{3} &\hspace{0.20cm} \frac{4}{3} &\hspace{0.20cm} \frac{2}{3} &\hspace{0.20cm} \frac{1}{3} &\hspace{0.20cm} 0 &\hspace{0.20cm} 1 &\hspace{0.20cm} 0 &\hspace{0.20cm} 0 &\hspace{0.20cm} 0 &\hspace{0.20cm} 1 &\hspace{0.20cm} 0 &\hspace{0.20cm} 0 &\hspace{0.20cm} 0 &\hspace{0.20cm} 0 &\hspace{0.20cm} 0 &\hspace{0.20cm} 0 &\hspace{0.20cm} 0 &\hspace{0.20cm} 0 &\hspace{0.20cm} 0 &\hspace{0.20cm} 0 &\hspace{0.20cm} 0 &\hspace{0.20cm} 0 &\hspace{0.20cm} 0 \\
			\hspace{-0.25cm}-1 &\hspace{0.20cm} \hspace{-0.25cm}-1 &\hspace{0.20cm} \hspace{-0.25cm}-1 &\hspace{0.20cm} \hspace{-0.25cm}-1 &\hspace{0.20cm} \hspace{-0.25cm}-1 &\hspace{0.20cm} 0 &\hspace{0.20cm} 0 &\hspace{0.20cm} 0 &\hspace{0.20cm} 0 &\hspace{0.20cm} 1 &\hspace{0.20cm} 0 &\hspace{0.20cm} 0 &\hspace{0.20cm} 0 &\hspace{0.20cm} 0 &\hspace{0.20cm} 0 &\hspace{0.20cm} 0 &\hspace{0.20cm} 0 &\hspace{0.20cm} 0 &\hspace{0.20cm} 0 &\hspace{0.20cm} 0 &\hspace{0.20cm} 0 &\hspace{0.20cm} 0 &\hspace{0.20cm} 0 &\hspace{0.20cm} 0 \\
			2 &\hspace{0.20cm} \frac{1}{3} &\hspace{0.20cm} \frac{2}{3} &\hspace{0.20cm} \frac{4}{3} &\hspace{0.20cm} \frac{5}{3} &\hspace{0.20cm} 0 &\hspace{0.20cm} 1 &\hspace{0.20cm} 0 &\hspace{0.20cm} 1 &\hspace{0.20cm} 0 &\hspace{0.20cm} 0 &\hspace{0.20cm} 0 &\hspace{0.20cm} 0 &\hspace{0.20cm} 0 &\hspace{0.20cm} 0 &\hspace{0.20cm} 0 &\hspace{0.20cm} 0 &\hspace{0.20cm} 0 &\hspace{0.20cm} 0 &\hspace{0.20cm} 0 &\hspace{0.20cm} 0 &\hspace{0.20cm} 0 &\hspace{0.20cm} 0 &\hspace{0.20cm} 0 \\
			\hspace{-0.25cm}-2 &\hspace{0.20cm} 0 &\hspace{0.20cm} \hspace{-0.25cm}-1 &\hspace{0.20cm} \hspace{-0.25cm}-2 &\hspace{0.20cm} \hspace{-0.25cm}-1 &\hspace{0.20cm} 0 &\hspace{0.20cm} \hspace{-0.25cm}-1 &\hspace{0.20cm} 1 &\hspace{0.20cm} 0 &\hspace{0.20cm} 0 &\hspace{0.20cm} 0 &\hspace{0.20cm} 0 &\hspace{0.20cm} 0 &\hspace{0.20cm} 0 &\hspace{0.20cm} 0 &\hspace{0.20cm} 0 &\hspace{0.20cm} 0 &\hspace{0.20cm} 0 &\hspace{0.20cm} 0 &\hspace{0.20cm} 0 &\hspace{0.20cm} 0 &\hspace{0.20cm} 0 &\hspace{0.20cm} 0 &\hspace{0.20cm} 0 \\
			1 &\hspace{0.20cm} 1 &\hspace{0.20cm} 1 &\hspace{0.20cm} 1 &\hspace{0.20cm} 1 &\hspace{0.20cm} 1 &\hspace{0.20cm} 0 &\hspace{0.20cm} 0 &\hspace{0.20cm} 0 &\hspace{0.20cm} 0 &\hspace{0.20cm} 0 &\hspace{0.20cm} 0 &\hspace{0.20cm} 0 &\hspace{0.20cm} 0 &\hspace{0.20cm} 0 &\hspace{0.20cm} 0 &\hspace{0.20cm} 0 &\hspace{0.20cm} 0 &\hspace{0.20cm} 0 &\hspace{0.20cm} 0 &\hspace{0.20cm} 0 &\hspace{0.20cm} 0 &\hspace{0.20cm} 0 &\hspace{0.20cm} 0 \\
		\end{array}
		\right),}
	\label{D31} \nonumber
\end{equation}
\begin{align}
	D=(\mathbf{1}_6,0).
\end{align}
The mixing matrix of this Cweb has rank six. Thus there are six exponentiated colour factors of this Cweb, which are given below
\begin{align}
	(YC)_1&=- f^{abc}f^{amn}f^{egp}f^{pdm} \tn{1} \tb{2} \tc{3} \td{4} \te{5} \tg{6}\,, \nonumber \\
	&=\; -{\cal{B}}_{2}-{\cal{B}}_{3} \, . \nn\\
	(YC)_2&=- f^{abc}f^{dmn}f^{egp}f^{pam} \tn{1} \tb{2} \tc{3} \td{4} \te{5} \tg{6}\,, \nonumber \\
	&=\;{\cal{B}}_{2}+{\cal{B}}_{4}	\, . \nn\\
	(YC)_3&=\,  f^{abc}f^{amn}f^{dgr}f^{erm} \tn{1} \tb{2} \tc{3} \td{4} \te{5} \tg{6}\,, \\
	&=\;-{\cal{B}}_{4}-{\cal{B}}_{23} \, .\nn\\
	(YC)_4&= \,  f^{abc}f^{arm}f^{dgr}f^{emn} \tn{1} \tb{2} \tc{3} \td{4} \te{5} \tg{6}\,, \nonumber \\
	&=\;{\cal{B}}_{25} \, . \nn\\
	(YC)_5&=\,  f^{abc}f^{agp}f^{dpm}f^{emn}\tn{1} \tb{2} \tc{3} \td{4} \te{5} \tg{6} 
	- f^{abc}f^{agp}f^{deq}f^{pqo} \too{1} \tb{2} \tc{3} \td{4} \te{5} \tg{6}\,, \nonumber \\
	&=\;-{\cal{B}}_{1}-{\cal{B}}_{3}-{\cal{B}}_{25} \, .	 \nn\\
	(YC)_6&=- f^{abc}f^{agp}f^{dpm}f^{men}\tn{1} \tb{2} \tc{3} \td{4} \te{5} \tg{6} \nonumber\, ,\\
	&=\;-{\cal{B}}_{3}-{\cal{B}}_{25}	 \, .
\end{align}

\subsection*{(9)  \ $\mathbf{W^{(3,1)}_{6} (3, 1, 1, 1, 2, 1)}$}  \label{C32}

This Cweb has twelve diagrams, as there are three attachments on line 1 and two on line 5. One of the diagrams is shown in fig.~\ref{Diag:9}. The sequences of diagrams and their corresponding $s$-factors are provided in table~\ref{Table:Diag9}.
\vskip0.5cm

\begin{minipage}{0.45\textwidth}
	\hspace{1.5cm}	\includegraphics[scale=0.5]{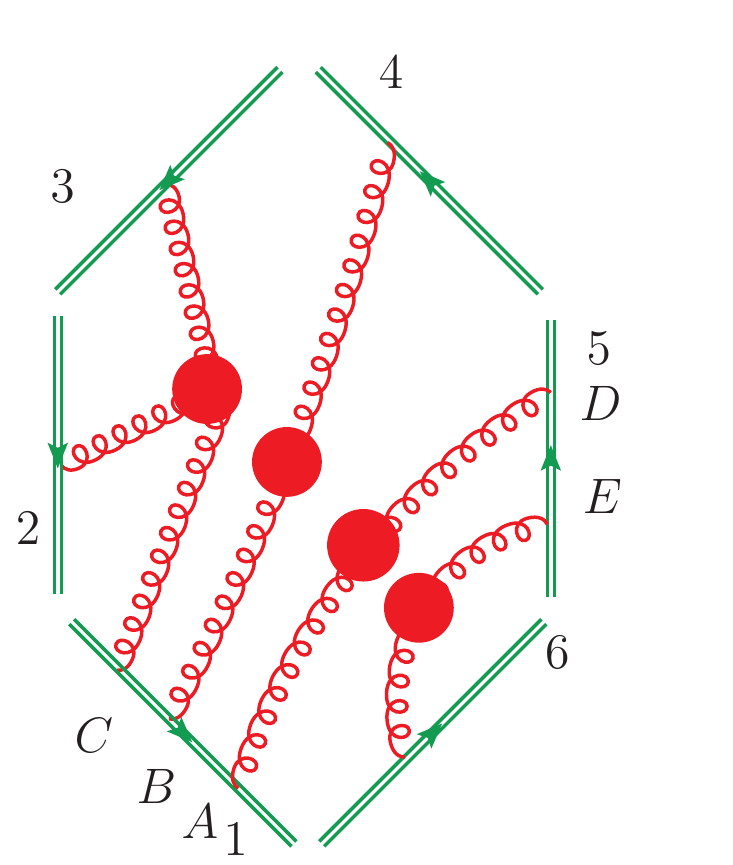}
	\captionof{figure}{$W^{(3,1)}_{6} (3, 1, 1, 1, 2, 1)$}	
	\label{Diag:9}
\end{minipage}
\begin{minipage}{0.45\textwidth}
	\footnotesize{
		\begin{tabular}{ | c | c | c |}
			\hline
			\textbf{Diagrams} & \textbf{Sequences} & \textbf{$s$-factors} \\ \hline
			$d_1$ & $\lbrace \lbrace ABC \rbrace,\lbrace DE  \rbrace\rbrace$ &1 \\ 
			\hline
			$d_2$ & $\lbrace \lbrace ABC \rbrace,\lbrace ED \rbrace\rbrace$ & 3  \\ 
			\hline
			$d_3$ & $\lbrace \lbrace ACB \rbrace,\lbrace DE \rbrace\rbrace$ & 2  \\ 
			\hline
			$d_4$ & $\lbrace \lbrace ACB \rbrace,\lbrace ED \rbrace\rbrace$ & 2  \\ 
			\hline
			$d_{5}$ & $\lbrace \lbrace BAC \rbrace,\lbrace DE \rbrace\rbrace$ & 1  \\ 
			\hline	
			$d_{6}$ & $\lbrace \lbrace BAC \rbrace,\lbrace ED \rbrace\rbrace$ & 3 \\ 
			\hline
			$d_7$ & $\lbrace \lbrace BCA \rbrace,\lbrace DE \rbrace\rbrace$ & 2  \\ 
			\hline
			$d_8$ & $\lbrace \lbrace BCA \rbrace,\lbrace ED \rbrace\rbrace$ & 2  \\ 
			\hline	
			$d_{9}$ & $\lbrace \lbrace CAB \rbrace,\lbrace DE \rbrace\rbrace$ & 3 \\ 
			\hline
			$d_{10}$ & $\lbrace \lbrace CAB \rbrace,\lbrace ED \rbrace\rbrace$ & 1 \\ 
			\hline
			$d_{11}$ & $\lbrace \lbrace CBA \rbrace,\lbrace DE \rbrace\rbrace$ & 3 \\ 
			\hline
			$d_{12}$ & $\lbrace \lbrace CBA \rbrace,\lbrace ED \rbrace\rbrace$ & 1 \\ 
			\hline
			
		\end{tabular}
		\captionof{table}{Sequences and $s$-factors}
		\label{Table:Diag9}	}
\end{minipage} 

\vspace{0.5cm}
\noindent   The mixing matrix $R$, the diagonalizing matrix $Y$ and the diagonal matrix $D$ for the Cweb mentioned above are

\begin{equation} 
	R=	\frac{1}{12} \left( \,\,\,
	\begin{array}{cccccccccccc}
		3 &\hspace{0.20cm} \hspace{-0.25cm}-3 &\hspace{0.20cm} \hspace{-0.25cm}-2 &\hspace{0.20cm} 2 &\hspace{0.20cm} \hspace{-0.25cm}-1 &\hspace{0.20cm} 1 &\hspace{0.20cm} \hspace{-0.25cm}-2 &\hspace{0.20cm} 2 &\hspace{0.20cm} \hspace{-0.25cm}-1 &\hspace{0.20cm} 1 &\hspace{0.20cm} 3 &\hspace{0.20cm} \hspace{-0.25cm}-3 \\
		\hspace{-0.25cm}-1 &\hspace{0.20cm} 1 &\hspace{0.20cm} 0 &\hspace{0.20cm} 0 &\hspace{0.20cm} 1 &\hspace{0.20cm} \hspace{-0.25cm}-1 &\hspace{0.20cm} 0 &\hspace{0.20cm} 0 &\hspace{0.20cm} 1 &\hspace{0.20cm} \hspace{-0.25cm}-1 &\hspace{0.20cm} \hspace{-0.25cm}-1 &\hspace{0.20cm} 1 \\
		\hspace{-0.25cm}-1 &\hspace{0.20cm} 1 &\hspace{0.20cm} 2 &\hspace{0.20cm} \hspace{-0.25cm}-2 &\hspace{0.20cm} \hspace{-0.25cm}-1 &\hspace{0.20cm} 1 &\hspace{0.20cm} 2 &\hspace{0.20cm} \hspace{-0.25cm}-2 &\hspace{0.20cm} \hspace{-0.25cm}-1 &\hspace{0.20cm} 1 &\hspace{0.20cm} \hspace{-0.25cm}-1 &\hspace{0.20cm} 1 \\
		1 &\hspace{0.20cm} \hspace{-0.25cm}-1 &\hspace{0.20cm} \hspace{-0.25cm}-2 &\hspace{0.20cm} 2 &\hspace{0.20cm} 1 &\hspace{0.20cm} \hspace{-0.25cm}-1 &\hspace{0.20cm} \hspace{-0.25cm}-2 &\hspace{0.20cm} 2 &\hspace{0.20cm} 1 &\hspace{0.20cm} \hspace{-0.25cm}-1 &\hspace{0.20cm} 1 &\hspace{0.20cm} \hspace{-0.25cm}-1 \\
		\hspace{-0.25cm}-1 &\hspace{0.20cm} 1 &\hspace{0.20cm} \hspace{-0.25cm}-2 &\hspace{0.20cm} 2 &\hspace{0.20cm} 3 &\hspace{0.20cm} \hspace{-0.25cm}-3 &\hspace{0.20cm} \hspace{-0.25cm}-2 &\hspace{0.20cm} 2 &\hspace{0.20cm} 3 &\hspace{0.20cm} \hspace{-0.25cm}-3 &\hspace{0.20cm} \hspace{-0.25cm}-1 &\hspace{0.20cm} 1 \\
		1 &\hspace{0.20cm} \hspace{-0.25cm}-1 &\hspace{0.20cm} 0 &\hspace{0.20cm} 0 &\hspace{0.20cm} \hspace{-0.25cm}-1 &\hspace{0.20cm} 1 &\hspace{0.20cm} 0 &\hspace{0.20cm} 0 &\hspace{0.20cm} \hspace{-0.25cm}-1 &\hspace{0.20cm} 1 &\hspace{0.20cm} 1 &\hspace{0.20cm} \hspace{-0.25cm}-1 \\
		\hspace{-0.25cm}-1 &\hspace{0.20cm} 1 &\hspace{0.20cm} 2 &\hspace{0.20cm} \hspace{-0.25cm}-2 &\hspace{0.20cm} \hspace{-0.25cm}-1 &\hspace{0.20cm} 1 &\hspace{0.20cm} 2 &\hspace{0.20cm} \hspace{-0.25cm}-2 &\hspace{0.20cm} \hspace{-0.25cm}-1 &\hspace{0.20cm} 1 &\hspace{0.20cm} \hspace{-0.25cm}-1 &\hspace{0.20cm} 1 \\
		1 &\hspace{0.20cm} \hspace{-0.25cm}-1 &\hspace{0.20cm} \hspace{-0.25cm}-2 &\hspace{0.20cm} 2 &\hspace{0.20cm} 1 &\hspace{0.20cm} \hspace{-0.25cm}-1 &\hspace{0.20cm} \hspace{-0.25cm}-2 &\hspace{0.20cm} 2 &\hspace{0.20cm} 1 &\hspace{0.20cm} \hspace{-0.25cm}-1 &\hspace{0.20cm} 1 &\hspace{0.20cm} \hspace{-0.25cm}-1 \\
		\hspace{-0.25cm}-1 &\hspace{0.20cm} 1 &\hspace{0.20cm} 0 &\hspace{0.20cm} 0 &\hspace{0.20cm} 1 &\hspace{0.20cm} \hspace{-0.25cm}-1 &\hspace{0.20cm} 0 &\hspace{0.20cm} 0 &\hspace{0.20cm} 1 &\hspace{0.20cm} \hspace{-0.25cm}-1 &\hspace{0.20cm} \hspace{-0.25cm}-1 &\hspace{0.20cm} 1 \\
		1 &\hspace{0.20cm} \hspace{-0.25cm}-1 &\hspace{0.20cm} 2 &\hspace{0.20cm} \hspace{-0.25cm}-2 &\hspace{0.20cm} \hspace{-0.25cm}-3 &\hspace{0.20cm} 3 &\hspace{0.20cm} 2 &\hspace{0.20cm} \hspace{-0.25cm}-2 &\hspace{0.20cm} \hspace{-0.25cm}-3 &\hspace{0.20cm} 3 &\hspace{0.20cm} 1 &\hspace{0.20cm} \hspace{-0.25cm}-1 \\
		1 &\hspace{0.20cm} \hspace{-0.25cm}-1 &\hspace{0.20cm} 0 &\hspace{0.20cm} 0 &\hspace{0.20cm} \hspace{-0.25cm}-1 &\hspace{0.20cm} 1 &\hspace{0.20cm} 0 &\hspace{0.20cm} 0 &\hspace{0.20cm} \hspace{-0.25cm}-1 &\hspace{0.20cm} 1 &\hspace{0.20cm} 1 &\hspace{0.20cm} \hspace{-0.25cm}-1 \\
		\hspace{-0.25cm}-3 &\hspace{0.20cm} 3 &\hspace{0.20cm} 2 &\hspace{0.20cm} \hspace{-0.25cm}-2 &\hspace{0.20cm} 1 &\hspace{0.20cm} \hspace{-0.25cm}-1 &\hspace{0.20cm} 2 &\hspace{0.20cm} \hspace{-0.25cm}-2 &\hspace{0.20cm} 1 &\hspace{0.20cm} \hspace{-0.25cm}-1 &\hspace{0.20cm} \hspace{-0.25cm}-3 &\hspace{0.20cm} 3 \\
	\end{array}
	\right),
\end{equation}

\begin{equation}	
	\hspace{0.6cm}	 Y=\left( \,\,\,
	\begin{array}{cccccccccccc}
		\hspace{-0.25cm}-1 &\hspace{0.20cm} 1 &\hspace{0.20cm} 1 &\hspace{0.20cm} \hspace{-0.25cm}-1 &\hspace{0.20cm} 0 &\hspace{0.20cm} 0 &\hspace{0.20cm} 1 &\hspace{0.20cm} \hspace{-0.25cm}-1 &\hspace{0.20cm} 0 &\hspace{0.20cm} 0 &\hspace{0.20cm} \hspace{-0.25cm}-1 &\hspace{0.20cm} 1 \\
		0 &\hspace{0.20cm} 0 &\hspace{0.20cm} 1 &\hspace{0.20cm} \hspace{-0.25cm}-1 &\hspace{0.20cm} \hspace{-0.25cm}-1 &\hspace{0.20cm} 1 &\hspace{0.20cm} 1 &\hspace{0.20cm} \hspace{-0.25cm}-1 &\hspace{0.20cm} \hspace{-0.25cm}-1 &\hspace{0.20cm} 1 &\hspace{0.20cm} 0 &\hspace{0.20cm} 0 \\
		1 &\hspace{0.20cm} 0 &\hspace{0.20cm} 0 &\hspace{0.20cm} 0 &\hspace{0.20cm} 0 &\hspace{0.20cm} 0 &\hspace{0.20cm} 0 &\hspace{0.20cm} 0 &\hspace{0.20cm} 0 &\hspace{0.20cm} 0 &\hspace{0.20cm} 0 &\hspace{0.20cm} 1 \\
		0 &\hspace{0.20cm} 1 &\hspace{0.20cm} 0 &\hspace{0.20cm} 0 &\hspace{0.20cm} 0 &\hspace{0.20cm} 0 &\hspace{0.20cm} 0 &\hspace{0.20cm} 0 &\hspace{0.20cm} 0 &\hspace{0.20cm} 0 &\hspace{0.20cm} 1 &\hspace{0.20cm} 0 \\
		1 &\hspace{0.20cm} 4 &\hspace{0.20cm} 0 &\hspace{0.20cm} 0 &\hspace{0.20cm} 0 &\hspace{0.20cm} 0 &\hspace{0.20cm} 0 &\hspace{0.20cm} 0 &\hspace{0.20cm} 0 &\hspace{0.20cm} 1 &\hspace{0.20cm} 0 &\hspace{0.20cm} 0 \\
		0 &\hspace{0.20cm} \hspace{-0.25cm}-1 &\hspace{0.20cm} 0 &\hspace{0.20cm} 0 &\hspace{0.20cm} 0 &\hspace{0.20cm} 0 &\hspace{0.20cm} 0 &\hspace{0.20cm} 0 &\hspace{0.20cm} 1 &\hspace{0.20cm} 0 &\hspace{0.20cm} 0 &\hspace{0.20cm} 0 \\
		\hspace{-0.25cm}-1 &\hspace{0.20cm} \hspace{-0.25cm}-2 &\hspace{0.20cm} 0 &\hspace{0.20cm} 0 &\hspace{0.20cm} 0 &\hspace{0.20cm} 0 &\hspace{0.20cm} 0 &\hspace{0.20cm} 1 &\hspace{0.20cm} 0 &\hspace{0.20cm} 0 &\hspace{0.20cm} 0 &\hspace{0.20cm} 0 \\
		1 &\hspace{0.20cm} 2 &\hspace{0.20cm} 0 &\hspace{0.20cm} 0 &\hspace{0.20cm} 0 &\hspace{0.20cm} 0 &\hspace{0.20cm} 1 &\hspace{0.20cm} 0 &\hspace{0.20cm} 0 &\hspace{0.20cm} 0 &\hspace{0.20cm} 0 &\hspace{0.20cm} 0 \\
		0 &\hspace{0.20cm} 1 &\hspace{0.20cm} 0 &\hspace{0.20cm} 0 &\hspace{0.20cm} 0 &\hspace{0.20cm} 1 &\hspace{0.20cm} 0 &\hspace{0.20cm} 0 &\hspace{0.20cm} 0 &\hspace{0.20cm} 0 &\hspace{0.20cm} 0 &\hspace{0.20cm} 0 \\
		\hspace{-0.25cm}-1 &\hspace{0.20cm} \hspace{-0.25cm}-4 &\hspace{0.20cm} 0 &\hspace{0.20cm} 0 &\hspace{0.20cm} 1 &\hspace{0.20cm} 0 &\hspace{0.20cm} 0 &\hspace{0.20cm} 0 &\hspace{0.20cm} 0 &\hspace{0.20cm} 0 &\hspace{0.20cm} 0 &\hspace{0.20cm} 0 \\
		\hspace{-0.25cm}-1 &\hspace{0.20cm} \hspace{-0.25cm}-2 &\hspace{0.20cm} 0 &\hspace{0.20cm} 1 &\hspace{0.20cm} 0 &\hspace{0.20cm} 0 &\hspace{0.20cm} 0 &\hspace{0.20cm} 0 &\hspace{0.20cm} 0 &\hspace{0.20cm} 0 &\hspace{0.20cm} 0 &\hspace{0.20cm} 0 \\
		1 &\hspace{0.20cm} 2 &\hspace{0.20cm} 1 &\hspace{0.20cm} 0 &\hspace{0.20cm} 0 &\hspace{0.20cm} 0 &\hspace{0.20cm} 0 &\hspace{0.20cm} 0 &\hspace{0.20cm} 0 &\hspace{0.20cm} 0 &\hspace{0.20cm} 0 &\hspace{0.20cm} 0 \\
	\end{array}
	\right),  \nonumber
\end{equation} 

\begin{align}
	D=(\mathbf{1}_2,0).
	\label{D32}
\end{align}
The rank of the above mixing matrix is two, and the two exponentiated colour factors of this Cweb are
\begin{align}
	(YC)_1=&  f^{abc}f^{adp}f^{egr}f^{pen}\tn{1} \tb{2} \tc{3} \td{4} \trr{5} \tg{6} +  f^{abc}f^{aem}f^{dmn}f^{egr}\tn{1} \tb{2} \tc{3} \td{4} \trr{5} \tg{6}, \nonumber \\
	&=\;2{\cal{B}}_{2}+{\cal{B}}_{4} \, . \nn\\
	(YC)_2=&   f^{abc}f^{aem}f^{dmn}f^{egr}\tn{1} \tb{2} \tc{3} \td{4} \trr{5} \tg{6}\, ,\nn\\
	&=\;-{\cal{B}}_{2}-{\cal{B}}_{4} \, .
\end{align}

\subsection*{(10)  \ $\mathbf{W^{(3,1)}_{6,\text{I}} (3, 1, 2, 1, 1, 1)}$}  \label{C34}

This Cweb has twelve diagrams, with three attachments on line 1 and two on line 3. One of the diagrams is shown in fig.~\ref{Diag:10}.
The sequences of diagrams and their corresponding $s$-factors are provided in table~\ref{Table:Diag10}.
\vskip0.5cm
\begin{minipage}{0.45\textwidth}
	\hspace{1.5cm}	\includegraphics[scale=0.5]{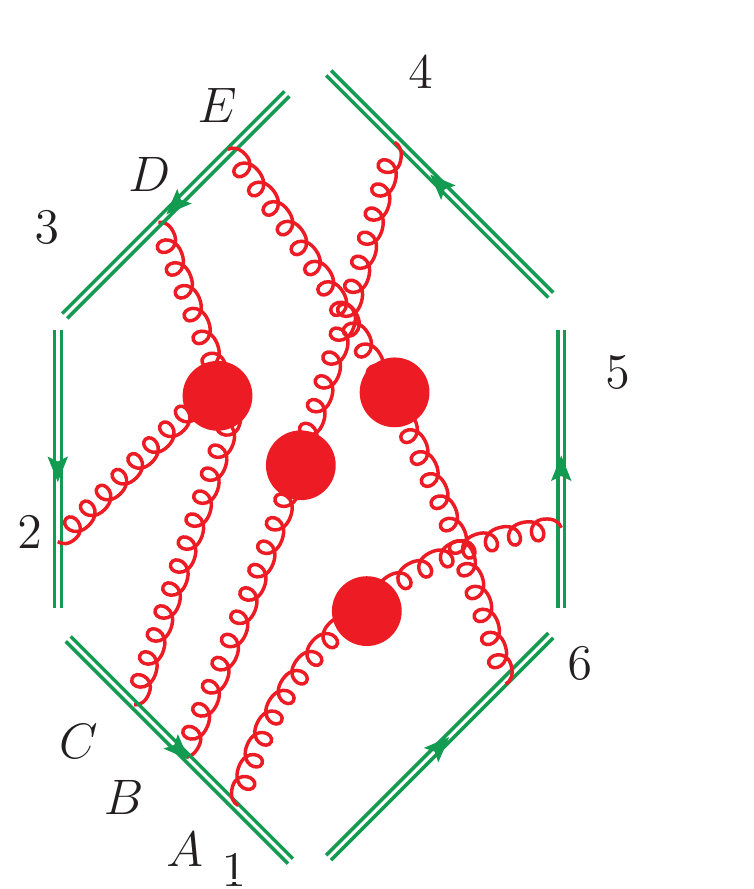}
	\captionof{figure}{$W^{(3,1)}_{6,\text{I}} (3, 1, 2, 1, 1, 1)$}	
	\label{Diag:10}
\end{minipage}
\begin{minipage}{0.45\textwidth}
	\footnotesize{
		\begin{tabular}{ | c | c | c |}
			\hline
			\textbf{Diagrams} & \textbf{Sequences} & \textbf{$s$-factors} \\ \hline
			$d_1$ & $\lbrace \lbrace  ABC \rbrace,\lbrace DE \rbrace\rbrace$ & 3 \\ 
			\hline
			$d_2$ & $\lbrace \lbrace  ABC \rbrace,\lbrace ED \rbrace\rbrace$ & 1\\ 
			\hline
			$d_{3}$ & $\lbrace \lbrace ACB \rbrace,\lbrace DE \rbrace\rbrace$ & 3  \\ 
			\hline
			$d_4$ & $\lbrace \lbrace ACB \rbrace,\lbrace ED \rbrace\rbrace$ & 1 \\ 
			\hline
			$d_5$ & $\lbrace \lbrace BAC \rbrace,\lbrace DE \rbrace\rbrace$ & 2 \\ 
			\hline
			$d_{6}$ & $\lbrace \lbrace BAC \rbrace,\lbrace ED \rbrace\rbrace$ & 2 \\ 
			\hline
			$d_{7}$ & $\lbrace \lbrace BCA \rbrace,\lbrace DE \rbrace\rbrace$ & 1 \\ 
			\hline
			$d_{8}$ & $\lbrace \lbrace BCA \rbrace,\lbrace ED \rbrace\rbrace$ & 3 \\ 
			\hline
			$d_9$ & $\lbrace \lbrace  CAB \rbrace,\lbrace DE \rbrace\rbrace$ & 2 \\ 
			\hline										
			$d_{10}$ & $\lbrace \lbrace  CAB \rbrace,\lbrace ED \rbrace\rbrace$ & 2  \\ 
			\hline
			$d_{11}$ & $\lbrace \lbrace  CBA \rbrace,\lbrace DE \rbrace\rbrace$ & 1 \\ 
			\hline
			$d_{12}$ & $\lbrace \lbrace  CBA \rbrace,\lbrace ED \rbrace\rbrace$ &3 \\ 
			\hline
		\end{tabular}
		\captionof{table}{Sequences and $s$-factors}
		\label{Table:Diag10}	}
\end{minipage} 

\vspace{0.4cm}

\noindent The mixing matrix $R$, the diagonalizing matrix $Y$ and the diagonal matrix $D$ for the Cweb mentioned above are

\begin{equation}
	R=  \frac{1}{12}\left(\,\,\,
	\begin{array}{cccccccccccc}
		1 &\hspace{0.20cm} \hspace{-0.25cm}-1 &\hspace{0.20cm} \hspace{-0.25cm}-1 &\hspace{0.20cm} 1 &\hspace{0.20cm} 0 &\hspace{0.20cm} 0 &\hspace{0.20cm} \hspace{-0.25cm}-1 &\hspace{0.20cm} 1 &\hspace{0.20cm} 0 &\hspace{0.20cm} 0 &\hspace{0.20cm} 1 &\hspace{0.20cm} \hspace{-0.25cm}-1 \\
		\hspace{-0.25cm}-3 &\hspace{0.20cm} 3 &\hspace{0.20cm} 1 &\hspace{0.20cm} \hspace{-0.25cm}-1 &\hspace{0.20cm} 2 &\hspace{0.20cm} \hspace{-0.25cm}-2 &\hspace{0.20cm} 1 &\hspace{0.20cm} \hspace{-0.25cm}-1 &\hspace{0.20cm} 2 &\hspace{0.20cm} \hspace{-0.25cm}-2 &\hspace{0.20cm} \hspace{-0.25cm}-3 &\hspace{0.20cm} 3 \\
		\hspace{-0.25cm}-1 &\hspace{0.20cm} 1 &\hspace{0.20cm} 1 &\hspace{0.20cm} \hspace{-0.25cm}-1 &\hspace{0.20cm} 0 &\hspace{0.20cm} 0 &\hspace{0.20cm} 1 &\hspace{0.20cm} \hspace{-0.25cm}-1 &\hspace{0.20cm} 0 &\hspace{0.20cm} 0 &\hspace{0.20cm} \hspace{-0.25cm}-1 &\hspace{0.20cm} 1 \\
		1 &\hspace{0.20cm} \hspace{-0.25cm}-1 &\hspace{0.20cm} \hspace{-0.25cm}-3 &\hspace{0.20cm} 3 &\hspace{0.20cm} 2 &\hspace{0.20cm} \hspace{-0.25cm}-2 &\hspace{0.20cm} \hspace{-0.25cm}-3 &\hspace{0.20cm} 3 &\hspace{0.20cm} 2 &\hspace{0.20cm} \hspace{-0.25cm}-2 &\hspace{0.20cm} 1 &\hspace{0.20cm} \hspace{-0.25cm}-1 \\
		\hspace{-0.25cm}-1 &\hspace{0.20cm} 1 &\hspace{0.20cm} \hspace{-0.25cm}-1 &\hspace{0.20cm} 1 &\hspace{0.20cm} 2 &\hspace{0.20cm} \hspace{-0.25cm}-2 &\hspace{0.20cm} \hspace{-0.25cm}-1 &\hspace{0.20cm} 1 &\hspace{0.20cm} 2 &\hspace{0.20cm} \hspace{-0.25cm}-2 &\hspace{0.20cm} \hspace{-0.25cm}-1 &\hspace{0.20cm} 1 \\
		1 &\hspace{0.20cm} \hspace{-0.25cm}-1 &\hspace{0.20cm} 1 &\hspace{0.20cm} \hspace{-0.25cm}-1 &\hspace{0.20cm} \hspace{-0.25cm}-2 &\hspace{0.20cm} 2 &\hspace{0.20cm} 1 &\hspace{0.20cm} \hspace{-0.25cm}-1 &\hspace{0.20cm} \hspace{-0.25cm}-2 &\hspace{0.20cm} 2 &\hspace{0.20cm} 1 &\hspace{0.20cm} \hspace{-0.25cm}-1 \\
		\hspace{-0.25cm}-1 &\hspace{0.20cm} 1 &\hspace{0.20cm} 3 &\hspace{0.20cm} \hspace{-0.25cm}-3 &\hspace{0.20cm} \hspace{-0.25cm}-2 &\hspace{0.20cm} 2 &\hspace{0.20cm} 3 &\hspace{0.20cm} \hspace{-0.25cm}-3 &\hspace{0.20cm} \hspace{-0.25cm}-2 &\hspace{0.20cm} 2 &\hspace{0.20cm} \hspace{-0.25cm}-1 &\hspace{0.20cm} 1 \\
		1 &\hspace{0.20cm} \hspace{-0.25cm}-1 &\hspace{0.20cm} \hspace{-0.25cm}-1 &\hspace{0.20cm} 1 &\hspace{0.20cm} 0 &\hspace{0.20cm} 0 &\hspace{0.20cm} \hspace{-0.25cm}-1 &\hspace{0.20cm} 1 &\hspace{0.20cm} 0 &\hspace{0.20cm} 0 &\hspace{0.20cm} 1 &\hspace{0.20cm} \hspace{-0.25cm}-1 \\
		\hspace{-0.25cm}-1 &\hspace{0.20cm} 1 &\hspace{0.20cm} \hspace{-0.25cm}-1 &\hspace{0.20cm} 1 &\hspace{0.20cm} 2 &\hspace{0.20cm} \hspace{-0.25cm}-2 &\hspace{0.20cm} \hspace{-0.25cm}-1 &\hspace{0.20cm} 1 &\hspace{0.20cm} 2 &\hspace{0.20cm} \hspace{-0.25cm}-2 &\hspace{0.20cm} \hspace{-0.25cm}-1 &\hspace{0.20cm} 1 \\
		1 &\hspace{0.20cm} \hspace{-0.25cm}-1 &\hspace{0.20cm} 1 &\hspace{0.20cm} \hspace{-0.25cm}-1 &\hspace{0.20cm} \hspace{-0.25cm}-2 &\hspace{0.20cm} 2 &\hspace{0.20cm} 1 &\hspace{0.20cm} \hspace{-0.25cm}-1 &\hspace{0.20cm} \hspace{-0.25cm}-2 &\hspace{0.20cm} 2 &\hspace{0.20cm} 1 &\hspace{0.20cm} \hspace{-0.25cm}-1 \\
		3 &\hspace{0.20cm} \hspace{-0.25cm}-3 &\hspace{0.20cm} \hspace{-0.25cm}-1 &\hspace{0.20cm} 1 &\hspace{0.20cm} \hspace{-0.25cm}-2 &\hspace{0.20cm} 2 &\hspace{0.20cm} \hspace{-0.25cm}-1 &\hspace{0.20cm} 1 &\hspace{0.20cm} \hspace{-0.25cm}-2 &\hspace{0.20cm} 2 &\hspace{0.20cm} 3 &\hspace{0.20cm} \hspace{-0.25cm}-3 \\
		\hspace{-0.25cm}-1 &\hspace{0.20cm} 1 &\hspace{0.20cm} 1 &\hspace{0.20cm} \hspace{-0.25cm}-1 &\hspace{0.20cm} 0 &\hspace{0.20cm} 0 &\hspace{0.20cm} 1 &\hspace{0.20cm} \hspace{-0.25cm}-1 &\hspace{0.20cm} 0 &\hspace{0.20cm} 0 &\hspace{0.20cm} \hspace{-0.25cm}-1 &\hspace{0.20cm} 1 \\
	\end{array}
	\right),
\end{equation}	
\begin{equation}
	\hspace{0.5cm}		  Y=\left( \,\,\,	
	\begin{array}{cccccccccccc}
		\hspace{-0.25cm}-1 &\hspace{0.20cm} 1 &\hspace{0.20cm} 1 &\hspace{0.20cm} \hspace{-0.25cm}-1 &\hspace{0.20cm} 0 &\hspace{0.20cm} 0 &\hspace{0.20cm} 1 &\hspace{0.20cm} \hspace{-0.25cm}-1 &\hspace{0.20cm} 0 &\hspace{0.20cm} 0 &\hspace{0.20cm} \hspace{-0.25cm}-1 &\hspace{0.20cm} 1 \\
		0 &\hspace{0.20cm} 0 &\hspace{0.20cm} 1 &\hspace{0.20cm} \hspace{-0.25cm}-1 &\hspace{0.20cm} \hspace{-0.25cm}-1 &\hspace{0.20cm} 1 &\hspace{0.20cm} 1 &\hspace{0.20cm} \hspace{-0.25cm}-1 &\hspace{0.20cm} \hspace{-0.25cm}-1 &\hspace{0.20cm} 1 &\hspace{0.20cm} 0 &\hspace{0.20cm} 0 \\
		1 &\hspace{0.20cm} 0 &\hspace{0.20cm} 0 &\hspace{0.20cm} 0 &\hspace{0.20cm} 0 &\hspace{0.20cm} 0 &\hspace{0.20cm} 0 &\hspace{0.20cm} 0 &\hspace{0.20cm} 0 &\hspace{0.20cm} 0 &\hspace{0.20cm} 0 &\hspace{0.20cm} 1 \\
		0 &\hspace{0.20cm} 1 &\hspace{0.20cm} 0 &\hspace{0.20cm} 0 &\hspace{0.20cm} 0 &\hspace{0.20cm} 0 &\hspace{0.20cm} 0 &\hspace{0.20cm} 0 &\hspace{0.20cm} 0 &\hspace{0.20cm} 0 &\hspace{0.20cm} 1 &\hspace{0.20cm} 0 \\
		2 &\hspace{0.20cm} 1 &\hspace{0.20cm} 0 &\hspace{0.20cm} 0 &\hspace{0.20cm} 0 &\hspace{0.20cm} 0 &\hspace{0.20cm} 0 &\hspace{0.20cm} 0 &\hspace{0.20cm} 0 &\hspace{0.20cm} 1 &\hspace{0.20cm} 0 &\hspace{0.20cm} 0 \\
		\hspace{-0.25cm}-2 &\hspace{0.20cm} \hspace{-0.25cm}-1 &\hspace{0.20cm} 0 &\hspace{0.20cm} 0 &\hspace{0.20cm} 0 &\hspace{0.20cm} 0 &\hspace{0.20cm} 0 &\hspace{0.20cm} 0 &\hspace{0.20cm} 1 &\hspace{0.20cm} 0 &\hspace{0.20cm} 0 &\hspace{0.20cm} 0 \\
		\hspace{-0.25cm}-1 &\hspace{0.20cm} 0 &\hspace{0.20cm} 0 &\hspace{0.20cm} 0 &\hspace{0.20cm} 0 &\hspace{0.20cm} 0 &\hspace{0.20cm} 0 &\hspace{0.20cm} 1 &\hspace{0.20cm} 0 &\hspace{0.20cm} 0 &\hspace{0.20cm} 0 &\hspace{0.20cm} 0 \\
		4 &\hspace{0.20cm} 1 &\hspace{0.20cm} 0 &\hspace{0.20cm} 0 &\hspace{0.20cm} 0 &\hspace{0.20cm} 0 &\hspace{0.20cm} 1 &\hspace{0.20cm} 0 &\hspace{0.20cm} 0 &\hspace{0.20cm} 0 &\hspace{0.20cm} 0 &\hspace{0.20cm} 0 \\
		2 &\hspace{0.20cm} 1 &\hspace{0.20cm} 0 &\hspace{0.20cm} 0 &\hspace{0.20cm} 0 &\hspace{0.20cm} 1 &\hspace{0.20cm} 0 &\hspace{0.20cm} 0 &\hspace{0.20cm} 0 &\hspace{0.20cm} 0 &\hspace{0.20cm} 0 &\hspace{0.20cm} 0 \\
		\hspace{-0.25cm}-2 &\hspace{0.20cm} \hspace{-0.25cm}-1 &\hspace{0.20cm} 0 &\hspace{0.20cm} 0 &\hspace{0.20cm} 1 &\hspace{0.20cm} 0 &\hspace{0.20cm} 0 &\hspace{0.20cm} 0 &\hspace{0.20cm} 0 &\hspace{0.20cm} 0 &\hspace{0.20cm} 0 &\hspace{0.20cm} 0 \\
		\hspace{-0.25cm}-4 &\hspace{0.20cm} \hspace{-0.25cm}-1 &\hspace{0.20cm} 0 &\hspace{0.20cm} 1 &\hspace{0.20cm} 0 &\hspace{0.20cm} 0 &\hspace{0.20cm} 0 &\hspace{0.20cm} 0 &\hspace{0.20cm} 0 &\hspace{0.20cm} 0 &\hspace{0.20cm} 0 &\hspace{0.20cm} 0 \\
		1 &\hspace{0.20cm} 0 &\hspace{0.20cm} 1 &\hspace{0.20cm} 0 &\hspace{0.20cm} 0 &\hspace{0.20cm} 0 &\hspace{0.20cm} 0 &\hspace{0.20cm} 0 &\hspace{0.20cm} 0 &\hspace{0.20cm} 0 &\hspace{0.20cm} 0 &\hspace{0.20cm} 0 \\
	\end{array}
	\,\,\,\right), \nonumber
\end{equation}
\begin{align} 
	D=(\mathbf{1}_2,0)\,.
	\label{D34}
\end{align}
The two exponentiated colour factors of this Cweb are
\begin{align}
	(YC)_1&=  f^{abc}f^{agq}f^{dcr}f^{eqm}\tm{1} \tb{2} \trr{3} \te{4} \tg{5} \td{6} +  f^{abc}f^{aep}f^{dcr}f^{pgn} \tn{1} \tb{2} \trr{3} \te{4} \tg{5} \td{6}, \nonumber \\
	&=\;{\cal{B}}_{12}-{\cal{B}}_{13} \, . \nn\\
	(YC)_2&=  f^{abc}f^{aep}f^{dcr}f^{pgn}\tn{1} \tb{2} \trr{3} \te{4} \tg{5} \td{6} \, , \nn\\
	&=\;-{\cal{B}}_{13} \, .
\end{align}

\subsection*{(11)  \ $\mathbf{W^{(3,1)}_{6} (2, 2, 1, 1, 2, 1)}$}  \label{C42}

This Cweb has eight diagrams as there are two attachments each on lines 1, 2 and 5, one of the diagrams is shown in fig.~\ref{Diag:11}. The sequences of diagrams and their corresponding $s$-factors are provided in table~\ref{Table:Diag11}.
\vskip0.5cm
\begin{minipage}{0.45\textwidth}
	\hspace{1.5cm}	\includegraphics[scale=0.5]{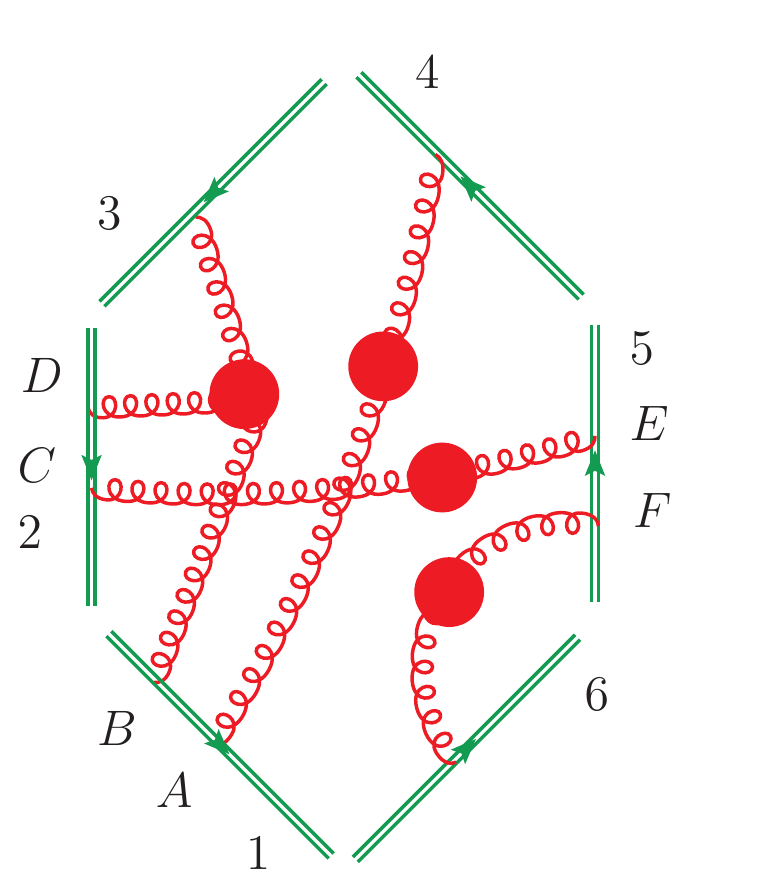}
	\captionof{figure}{$W^{(3,1)}_{6} (2, 2, 1, 1, 2, 1)$}	
	\label{Diag:11}
\end{minipage}
\begin{minipage}{0.45\textwidth}
	\footnotesize{
		\begin{tabular}{ | c | c | c |}
			\hline
			\textbf{Diagrams} & \textbf{Sequences} & \textbf{$s$-factors} \\ \hline
			$d_1$ & $\lbrace \lbrace AB \rbrace,  \lbrace CD \rbrace, \lbrace EF \rbrace \rbrace    $ &3  \\ 
			\hline	
			$d_2$ & $\lbrace \lbrace AB \rbrace,  \lbrace CD \rbrace, \lbrace FE \rbrace\rbrace$ & 5 \\ 
			\hline
			$d_3$ & $\lbrace \lbrace AB \rbrace,  \lbrace DC \rbrace, \lbrace EF \rbrace\rbrace$ & 3 \\ 
			\hline
			$d_4$ & $\lbrace \lbrace AB \rbrace,  \lbrace DC \rbrace, \lbrace FE \rbrace\rbrace$ & 1  \\ 
			\hline	
			$d_5$ & $\lbrace \lbrace BA \rbrace,  \lbrace CD \rbrace, \lbrace EF \rbrace\rbrace$ & 1 \\ 
			\hline	
			$d_6$ & $\lbrace \lbrace BA \rbrace,  \lbrace CD \rbrace, \lbrace FE \rbrace\rbrace$ & 3 \\ 
			\hline
			$d_7$ & $\lbrace \lbrace BA \rbrace,  \lbrace DC \rbrace, \lbrace EF \rbrace\rbrace$ & 5 \\ 
			\hline
			$d_8$ & $\lbrace \lbrace BA \rbrace,  \lbrace DC \rbrace, \lbrace FE \rbrace\rbrace$ & 3 \\ 
			\hline
		\end{tabular}
		\captionof{table}{Sequences and $s$-factors}
		\label{Table:Diag11}	}
\end{minipage} 

\vspace{0.5cm}
\noindent   The mixing matrix $R$, the diagonalizing matrix $Y$ and the diagonal matrix $D$ for the Cweb mentioned above are

\begin{equation} 
	R=\frac{1}{12} \left( \,\,\,
	\begin{array}{cccccccc}
		1 &\hspace{0.20cm} \hspace{-0.25cm}-1 &\hspace{0.20cm} \hspace{-0.25cm}-1 &\hspace{0.20cm} 1 &\hspace{0.20cm} \hspace{-0.25cm}-1 &\hspace{0.20cm} 1 &\hspace{0.20cm} 1 &\hspace{0.20cm} \hspace{-0.25cm}-1 \\
		\hspace{-0.25cm}-1 &\hspace{0.20cm} 1 &\hspace{0.20cm} 1 &\hspace{0.20cm} \hspace{-0.25cm}-1 &\hspace{0.20cm} 1 &\hspace{0.20cm} \hspace{-0.25cm}-1 &\hspace{0.20cm} \hspace{-0.25cm}-1 &\hspace{0.20cm} 1 \\
		\hspace{-0.25cm}-1 &\hspace{0.20cm} 1 &\hspace{0.20cm} 1 &\hspace{0.20cm} \hspace{-0.25cm}-1 &\hspace{0.20cm} 1 &\hspace{0.20cm} \hspace{-0.25cm}-1 &\hspace{0.20cm} \hspace{-0.25cm}-1 &\hspace{0.20cm} 1 \\
		3 &\hspace{0.20cm} \hspace{-0.25cm}-3 &\hspace{0.20cm} \hspace{-0.25cm}-3 &\hspace{0.20cm} 3 &\hspace{0.20cm} \hspace{-0.25cm}-3 &\hspace{0.20cm} 3 &\hspace{0.20cm} 3 &\hspace{0.20cm} \hspace{-0.25cm}-3 \\
		\hspace{-0.25cm}-3 &\hspace{0.20cm} 3 &\hspace{0.20cm} 3 &\hspace{0.20cm} \hspace{-0.25cm}-3 &\hspace{0.20cm} 3 &\hspace{0.20cm} \hspace{-0.25cm}-3 &\hspace{0.20cm} \hspace{-0.25cm}-3 &\hspace{0.20cm} 3 \\
		1 &\hspace{0.20cm} \hspace{-0.25cm}-1 &\hspace{0.20cm} \hspace{-0.25cm}-1 &\hspace{0.20cm} 1 &\hspace{0.20cm} \hspace{-0.25cm}-1 &\hspace{0.20cm} 1 &\hspace{0.20cm} 1 &\hspace{0.20cm} \hspace{-0.25cm}-1 \\
		1 &\hspace{0.20cm} \hspace{-0.25cm}-1 &\hspace{0.20cm} \hspace{-0.25cm}-1 &\hspace{0.20cm} 1 &\hspace{0.20cm} \hspace{-0.25cm}-1 &\hspace{0.20cm} 1 &\hspace{0.20cm} 1 &\hspace{0.20cm} \hspace{-0.25cm}-1 \\
		\hspace{-0.25cm}-1 &\hspace{0.20cm} 1 &\hspace{0.20cm} 1 &\hspace{0.20cm} \hspace{-0.25cm}-1 &\hspace{0.20cm} 1 &\hspace{0.20cm} \hspace{-0.25cm}-1 &\hspace{0.20cm} \hspace{-0.25cm}-1 &\hspace{0.20cm} 1 \\
	\end{array}
	\,\,\,\right), Y=\left(\,\,\,
	\begin{array}{cccccccc}
		\hspace{-0.25cm}-1 &\hspace{0.20cm} 1 &\hspace{0.20cm} 1 &\hspace{0.20cm} \hspace{-0.25cm}-1 &\hspace{0.20cm} 1 &\hspace{0.20cm} \hspace{-0.25cm}-1 &\hspace{0.20cm} \hspace{-0.25cm}-1 &\hspace{0.20cm} 1 \\
		1 &\hspace{0.20cm} 0 &\hspace{0.20cm} 0 &\hspace{0.20cm} 0 &\hspace{0.20cm} 0 &\hspace{0.20cm} 0 &\hspace{0.20cm} 0 &\hspace{0.20cm} 1 \\
		\hspace{-0.25cm}-1 &\hspace{0.20cm} 0 &\hspace{0.20cm} 0 &\hspace{0.20cm} 0 &\hspace{0.20cm} 0 &\hspace{0.20cm} 0 &\hspace{0.20cm} 1 &\hspace{0.20cm} 0 \\
		\hspace{-0.25cm}-1 &\hspace{0.20cm} 0 &\hspace{0.20cm} 0 &\hspace{0.20cm} 0 &\hspace{0.20cm} 0 &\hspace{0.20cm} 1 &\hspace{0.20cm} 0 &\hspace{0.20cm} 0 \\
		3 &\hspace{0.20cm} 0 &\hspace{0.20cm} 0 &\hspace{0.20cm} 0 &\hspace{0.20cm} 1 &\hspace{0.20cm} 0 &\hspace{0.20cm} 0 &\hspace{0.20cm} 0 \\
		\hspace{-0.25cm}-3 &\hspace{0.20cm} 0 &\hspace{0.20cm} 0 &\hspace{0.20cm} 1 &\hspace{0.20cm} 0 &\hspace{0.20cm} 0 &\hspace{0.20cm} 0 &\hspace{0.20cm} 0 \\
		1 &\hspace{0.20cm} 0 &\hspace{0.20cm} 1 &\hspace{0.20cm} 0 &\hspace{0.20cm} 0 &\hspace{0.20cm} 0 &\hspace{0.20cm} 0 &\hspace{0.20cm} 0 \\
		1 &\hspace{0.20cm} 1 &\hspace{0.20cm} 0 &\hspace{0.20cm} 0 &\hspace{0.20cm} 0 &\hspace{0.20cm} 0 &\hspace{0.20cm} 0 &\hspace{0.20cm} 0 \\
	\end{array}
	\,\,\,\right),  D=(\mathbf{1}_1,0).
	\label{D42}
\end{equation}
The rank of the mixing matrix is one and the only exponentiated colour factor of this Cweb is
\begin{align}
	(YC)_1=&  f^{abc}f^{adp}f^{ber}f^{egq}\tp{1} \trr{2} \tc{3} \td{4} \tq{5} \tg{6}\, ,\nn\\
	=&\;-{\cal{B}}_{19} \, .
\end{align}

\subsection*{(12)  \ $\mathbf{W^{(3,1)}_{6,\text{II}} (3, 1, 2, 1, 1, 1)}$ } 

This Cweb has twelve diagrams, as there are three attachments on lines 1 and two on line 3. One of the diagrams is shown in fig.~\ref{Diag:21}. The sequences of diagrams and their corresponding $s$-factors are provided in table~\ref{Table:Diag21}.
\vspace{0.7cm}

\begin{minipage}{0.5\textwidth}
	\hspace{1.5cm}	\includegraphics[scale=0.5]{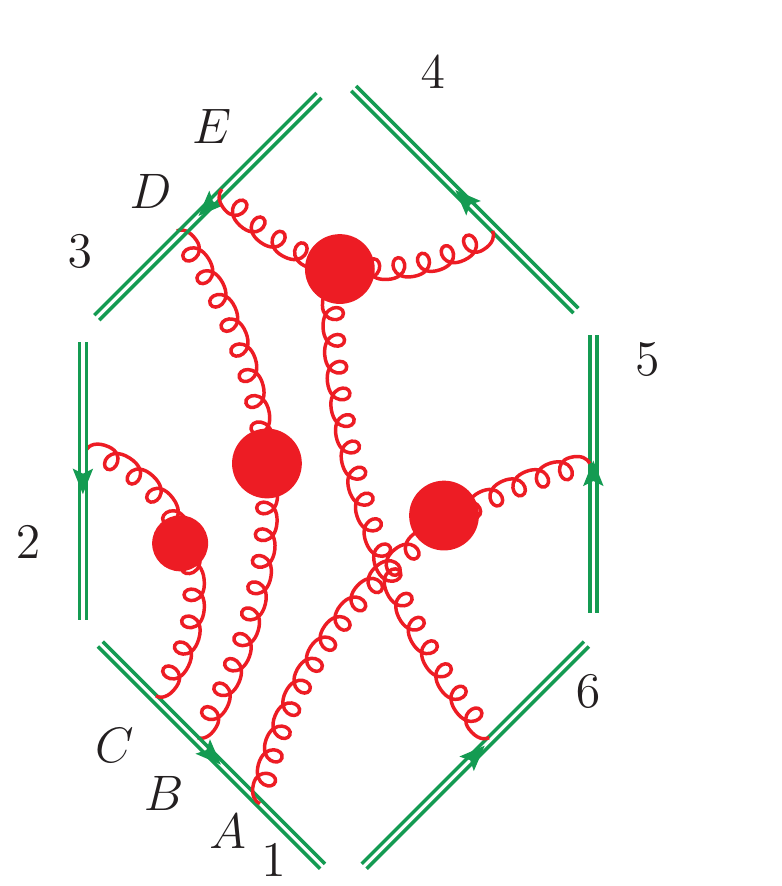}
	\captionof{figure}{$W^{(3,1)}_{6,\text{II}} (3, 1, 2, 1, 1, 1)$}
	\label{Diag:21}
\end{minipage}
\hspace{-1cm}\begin{minipage}{0.45\textwidth}
	\begin{tabular}{ | c | c | c |}
		\hline
		\textbf{Diagrams} & \textbf{Sequences} & \textbf{$s$-factors} \\ \hline
		$d_1$ & $\lbrace \lbrace ABC \rbrace,  \lbrace DE \rbrace\rbrace$ & 2 \\ 
		\hline
		$d_2$ & $\lbrace \lbrace ABC \rbrace,  \lbrace ED \rbrace\rbrace$ & 2 \\ 
		\hline
		$d_3$ & $\lbrace \lbrace ACB \rbrace,  \lbrace DE \rbrace\rbrace$ & 1 \\ 
		\hline
		$d_4$ & $\lbrace \lbrace ACB \rbrace,  \lbrace ED \rbrace\rbrace$ & 3 \\ 
		\hline	
		$d_{5}$ & $\lbrace \lbrace BCA \rbrace,  \lbrace DE \rbrace\rbrace$ & 3 \\ 
		\hline
		$d_6$ & $\lbrace \lbrace BAC \rbrace,  \lbrace ED \rbrace\rbrace$ & 1 \\ 
		\hline	
		$d_{7}$ & $\lbrace \lbrace BAC \rbrace,  \lbrace DE  \rbrace\rbrace$ & 3 \\ 
		\hline
		$d_8$ & $\lbrace \lbrace BCA \rbrace,  \lbrace ED \rbrace\rbrace$ & 1 \\ 
		\hline
		$d_9$ & $\lbrace \lbrace CAB \rbrace,  \lbrace DE \rbrace\rbrace$ & 1 \\ 
		\hline
		$d_{10}$ & $\lbrace \lbrace CAB \rbrace,  \lbrace ED \rbrace\rbrace$ & 3 \\ 
		\hline
		$d_{11}$ & $\lbrace \lbrace CBA \rbrace,  \lbrace DE \rbrace\rbrace$ & 2 \\ 
		\hline
		$d_{12}$ & $\lbrace \lbrace CBA \rbrace,  \lbrace ED \rbrace\rbrace$ & 2 \\ 
		\hline
	\end{tabular}	
	\captionof{table}{Sequences and $s$-factors}
	\label{Table:Diag21}
\end{minipage} 

\vspace{0.4cm}

\noindent   The mixing matrix $R$, the diagonalizing matrix $Y$ and the diagonal matrix $D$ for the Cweb mentioned above are

\begin{equation} 
	R=\frac{1}{12} \left(\,\,\,
	\begin{array}{cccccccccccc}
		2 &\hspace{0.20cm} \hspace{-0.25cm}-2 &\hspace{0.20cm} \hspace{-0.25cm}-1 &\hspace{0.20cm} 1 &\hspace{0.20cm} \hspace{-0.25cm}-1 &\hspace{0.20cm} 1 &\hspace{0.20cm} \hspace{-0.25cm}-1 &\hspace{0.20cm} 1 &\hspace{0.20cm} \hspace{-0.25cm}-1 &\hspace{0.20cm} 1 &\hspace{0.20cm} 2 &\hspace{0.20cm} \hspace{-0.25cm}-2 \\
		\hspace{-0.25cm}-2 &\hspace{0.20cm} 2 &\hspace{0.20cm} 1 &\hspace{0.20cm} \hspace{-0.25cm}-1 &\hspace{0.20cm} 1 &\hspace{0.20cm} \hspace{-0.25cm}-1 &\hspace{0.20cm} 1 &\hspace{0.20cm} \hspace{-0.25cm}-1 &\hspace{0.20cm} 1 &\hspace{0.20cm} \hspace{-0.25cm}-1 &\hspace{0.20cm} \hspace{-0.25cm}-2 &\hspace{0.20cm} 2 \\
		\hspace{-0.25cm}-2 &\hspace{0.20cm} 2 &\hspace{0.20cm} 3 &\hspace{0.20cm} \hspace{-0.25cm}-3 &\hspace{0.20cm} \hspace{-0.25cm}-1 &\hspace{0.20cm} 1 &\hspace{0.20cm} 3 &\hspace{0.20cm} \hspace{-0.25cm}-3 &\hspace{0.20cm} \hspace{-0.25cm}-1 &\hspace{0.20cm} 1 &\hspace{0.20cm} \hspace{-0.25cm}-2 &\hspace{0.20cm} 2 \\
		0 &\hspace{0.20cm} 0 &\hspace{0.20cm} \hspace{-0.25cm}-1 &\hspace{0.20cm} 1 &\hspace{0.20cm} 1 &\hspace{0.20cm} \hspace{-0.25cm}-1 &\hspace{0.20cm} \hspace{-0.25cm}-1 &\hspace{0.20cm} 1 &\hspace{0.20cm} 1 &\hspace{0.20cm} \hspace{-0.25cm}-1 &\hspace{0.20cm} 0 &\hspace{0.20cm} 0 \\
		0 &\hspace{0.20cm} 0 &\hspace{0.20cm} \hspace{-0.25cm}-1 &\hspace{0.20cm} 1 &\hspace{0.20cm} 1 &\hspace{0.20cm} \hspace{-0.25cm}-1 &\hspace{0.20cm} \hspace{-0.25cm}-1 &\hspace{0.20cm} 1 &\hspace{0.20cm} 1 &\hspace{0.20cm} \hspace{-0.25cm}-1 &\hspace{0.20cm} 0 &\hspace{0.20cm} 0 \\
		2 &\hspace{0.20cm} \hspace{-0.25cm}-2 &\hspace{0.20cm} 1 &\hspace{0.20cm} \hspace{-0.25cm}-1 &\hspace{0.20cm} \hspace{-0.25cm}-3 &\hspace{0.20cm} 3 &\hspace{0.20cm} 1 &\hspace{0.20cm} \hspace{-0.25cm}-1 &\hspace{0.20cm} \hspace{-0.25cm}-3 &\hspace{0.20cm} 3 &\hspace{0.20cm} 2 &\hspace{0.20cm} \hspace{-0.25cm}-2 \\
		0 &\hspace{0.20cm} 0 &\hspace{0.20cm} 1 &\hspace{0.20cm} \hspace{-0.25cm}-1 &\hspace{0.20cm} \hspace{-0.25cm}-1 &\hspace{0.20cm} 1 &\hspace{0.20cm} 1 &\hspace{0.20cm} \hspace{-0.25cm}-1 &\hspace{0.20cm} \hspace{-0.25cm}-1 &\hspace{0.20cm} 1 &\hspace{0.20cm} 0 &\hspace{0.20cm} 0 \\
		2 &\hspace{0.20cm} \hspace{-0.25cm}-2 &\hspace{0.20cm} \hspace{-0.25cm}-3 &\hspace{0.20cm} 3 &\hspace{0.20cm} 1 &\hspace{0.20cm} \hspace{-0.25cm}-1 &\hspace{0.20cm} \hspace{-0.25cm}-3 &\hspace{0.20cm} 3 &\hspace{0.20cm} 1 &\hspace{0.20cm} \hspace{-0.25cm}-1 &\hspace{0.20cm} 2 &\hspace{0.20cm} \hspace{-0.25cm}-2 \\
		\hspace{-0.25cm}-2 &\hspace{0.20cm} 2 &\hspace{0.20cm} \hspace{-0.25cm}-1 &\hspace{0.20cm} 1 &\hspace{0.20cm} 3 &\hspace{0.20cm} \hspace{-0.25cm}-3 &\hspace{0.20cm} \hspace{-0.25cm}-1 &\hspace{0.20cm} 1 &\hspace{0.20cm} 3 &\hspace{0.20cm} \hspace{-0.25cm}-3 &\hspace{0.20cm} \hspace{-0.25cm}-2 &\hspace{0.20cm} 2 \\
		0 &\hspace{0.20cm} 0 &\hspace{0.20cm} 1 &\hspace{0.20cm} \hspace{-0.25cm}-1 &\hspace{0.20cm} \hspace{-0.25cm}-1 &\hspace{0.20cm} 1 &\hspace{0.20cm} 1 &\hspace{0.20cm} \hspace{-0.25cm}-1 &\hspace{0.20cm} \hspace{-0.25cm}-1 &\hspace{0.20cm} 1 &\hspace{0.20cm} 0 &\hspace{0.20cm} 0 \\
		2 &\hspace{0.20cm} \hspace{-0.25cm}-2 &\hspace{0.20cm} \hspace{-0.25cm}-1 &\hspace{0.20cm} 1 &\hspace{0.20cm} \hspace{-0.25cm}-1 &\hspace{0.20cm} 1 &\hspace{0.20cm} \hspace{-0.25cm}-1 &\hspace{0.20cm} 1 &\hspace{0.20cm} \hspace{-0.25cm}-1 &\hspace{0.20cm} 1 &\hspace{0.20cm} 2 &\hspace{0.20cm} \hspace{-0.25cm}-2 \\
		\hspace{-0.25cm}-2 &\hspace{0.20cm} 2 &\hspace{0.20cm} 1 &\hspace{0.20cm} \hspace{-0.25cm}-1 &\hspace{0.20cm} 1 &\hspace{0.20cm} \hspace{-0.25cm}-1 &\hspace{0.20cm} 1 &\hspace{0.20cm} \hspace{-0.25cm}-1 &\hspace{0.20cm} 1 &\hspace{0.20cm} \hspace{-0.25cm}-1 &\hspace{0.20cm} \hspace{-0.25cm}-2 &\hspace{0.20cm} 2 \\
	\end{array}
	\right),
\end{equation}
\begin{equation}
	\hspace{0.5cm}	  Y=\left(\,\,\,
	\begin{array}{cccccccccccc}
		\hspace{-0.25cm}-1 &\hspace{0.20cm} 1 &\hspace{0.20cm} 1 &\hspace{0.20cm} \hspace{-0.25cm}-1 &\hspace{0.20cm} 0 &\hspace{0.20cm} 0 &\hspace{0.20cm} 1 &\hspace{0.20cm} \hspace{-0.25cm}-1 &\hspace{0.20cm} 0 &\hspace{0.20cm} 0 &\hspace{0.20cm} \hspace{-0.25cm}-1 &\hspace{0.20cm} 1 \\
		0 &\hspace{0.20cm} 0 &\hspace{0.20cm} 1 &\hspace{0.20cm} \hspace{-0.25cm}-1 &\hspace{0.20cm} \hspace{-0.25cm}-1 &\hspace{0.20cm} 1 &\hspace{0.20cm} 1 &\hspace{0.20cm} \hspace{-0.25cm}-1 &\hspace{0.20cm} \hspace{-0.25cm}-1 &\hspace{0.20cm} 1 &\hspace{0.20cm} 0 &\hspace{0.20cm} 0 \\
		1 &\hspace{0.20cm} 0 &\hspace{0.20cm} 0 &\hspace{0.20cm} 0 &\hspace{0.20cm} 0 &\hspace{0.20cm} 0 &\hspace{0.20cm} 0 &\hspace{0.20cm} 0 &\hspace{0.20cm} 0 &\hspace{0.20cm} 0 &\hspace{0.20cm} 0 &\hspace{0.20cm} 1 \\
		\hspace{-0.25cm}-1 &\hspace{0.20cm} 0 &\hspace{0.20cm} 0 &\hspace{0.20cm} 0 &\hspace{0.20cm} 0 &\hspace{0.20cm} 0 &\hspace{0.20cm} 0 &\hspace{0.20cm} 0 &\hspace{0.20cm} 0 &\hspace{0.20cm} 0 &\hspace{0.20cm} 1 &\hspace{0.20cm} 0 \\
		\hspace{-0.25cm}-\frac{1}{2} &\hspace{0.20cm} 0 &\hspace{0.20cm} \hspace{-0.25cm}-\frac{1}{2} &\hspace{0.20cm} 0 &\hspace{0.20cm} 0 &\hspace{0.20cm} 0 &\hspace{0.20cm} 0 &\hspace{0.20cm} 0 &\hspace{0.20cm} 0 &\hspace{0.20cm} 1 &\hspace{0.20cm} 0 &\hspace{0.20cm} 0 \\
		2 &\hspace{0.20cm} 0 &\hspace{0.20cm} 1 &\hspace{0.20cm} 0 &\hspace{0.20cm} 0 &\hspace{0.20cm} 0 &\hspace{0.20cm} 0 &\hspace{0.20cm} 0 &\hspace{0.20cm} 1 &\hspace{0.20cm} 0 &\hspace{0.20cm} 0 &\hspace{0.20cm} 0 \\
		0 &\hspace{0.20cm} 0 &\hspace{0.20cm} 1 &\hspace{0.20cm} 0 &\hspace{0.20cm} 0 &\hspace{0.20cm} 0 &\hspace{0.20cm} 0 &\hspace{0.20cm} 1 &\hspace{0.20cm} 0 &\hspace{0.20cm} 0 &\hspace{0.20cm} 0 &\hspace{0.20cm} 0 \\
		\hspace{-0.25cm}-\frac{1}{2} &\hspace{0.20cm} 0 &\hspace{0.20cm} \hspace{-0.25cm}-\frac{1}{2} &\hspace{0.20cm} 0 &\hspace{0.20cm} 0 &\hspace{0.20cm} 0 &\hspace{0.20cm} 1 &\hspace{0.20cm} 0 &\hspace{0.20cm} 0 &\hspace{0.20cm} 0 &\hspace{0.20cm} 0 &\hspace{0.20cm} 0 \\
		\hspace{-0.25cm}-2 &\hspace{0.20cm} 0 &\hspace{0.20cm} \hspace{-0.25cm}-1 &\hspace{0.20cm} 0 &\hspace{0.20cm} 0 &\hspace{0.20cm} 1 &\hspace{0.20cm} 0 &\hspace{0.20cm} 0 &\hspace{0.20cm} 0 &\hspace{0.20cm} 0 &\hspace{0.20cm} 0 &\hspace{0.20cm} 0 \\
		\frac{1}{2} &\hspace{0.20cm} 0 &\hspace{0.20cm} \frac{1}{2} &\hspace{0.20cm} 0 &\hspace{0.20cm} 1 &\hspace{0.20cm} 0 &\hspace{0.20cm} 0 &\hspace{0.20cm} 0 &\hspace{0.20cm} 0 &\hspace{0.20cm} 0 &\hspace{0.20cm} 0 &\hspace{0.20cm} 0 \\
		\frac{1}{2} &\hspace{0.20cm} 0 &\hspace{0.20cm} \frac{1}{2} &\hspace{0.20cm} 1 &\hspace{0.20cm} 0 &\hspace{0.20cm} 0 &\hspace{0.20cm} 0 &\hspace{0.20cm} 0 &\hspace{0.20cm} 0 &\hspace{0.20cm} 0 &\hspace{0.20cm} 0 &\hspace{0.20cm} 0 \\
		1 &\hspace{0.20cm} 1 &\hspace{0.20cm} 0 &\hspace{0.20cm} 0 &\hspace{0.20cm} 0 &\hspace{0.20cm} 0 &\hspace{0.20cm} 0 &\hspace{0.20cm} 0 &\hspace{0.20cm} 0 &\hspace{0.20cm} 0 &\hspace{0.20cm} 0 &\hspace{0.20cm} 0 \\
	\end{array}
	\right), 
	\label{RY78}
	\nonumber
\end{equation}
\begin{align}
	D=(\mathbf{1}_2,0).
	\label{D78}
\end{align}
The two exponentiated colour factors of this Cweb are given below
\begin{align}
	(YC)_1&= - f^{abq}f^{cbm}f^{cde}f^{qgo} \too{1} \ta{2} \tm{3} \tg{4} \td{5} \te{6}   + f^{abr}f^{aem}f^{dcq}f^{qeo} \trr{1} \ta{2} \tm{3} \te{4} \tg{5} \td{6} , \nonumber \\
	=&\;-{\cal{B}}_{8}+{\cal{B}}_{18} \, . \nn \\
	(YC)_2&= + f^{abr}f^{aem}f^{dcq}f^{qeo} \trr{1} \ta{2} \tm{3} \te{4} \tg{5} \td{6} \, , \nn\\
	=&\;-{\cal{B}}_{8} \, .
\end{align}

\subsection*{(13)  \ $\mathbf{W^{(3,1)}_{6} (2, 1, 1, 3, 1, 1)}$}  \label{C54}

This Cweb has twelve diagrams as there are two attachments on line 1 and three on line 4 giving twelve shuffles. One of the diagrams is shown in fig.~\ref{Diag:13}. The sequences of diagrams and their corresponding $s$-factors are provided in table~\ref{Table:Diag13}.

\begin{minipage}{0.45\textwidth}
	\hspace{1.5cm}	\includegraphics[scale=0.5]{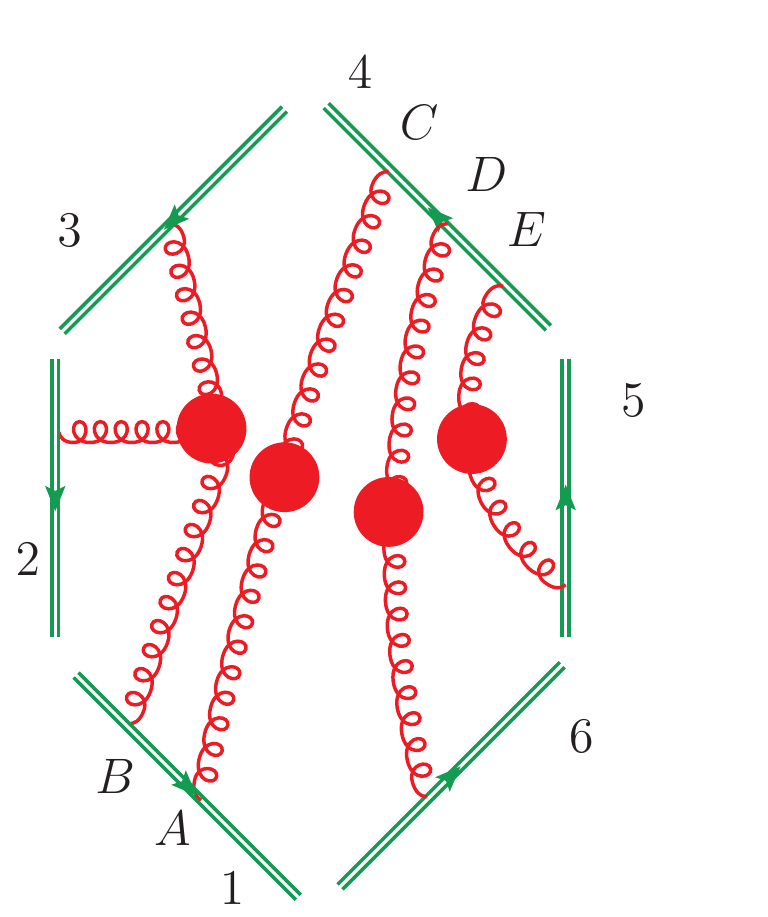}
	\captionof{figure}{$W^{(3,1)}_{6} (2, 1, 1, 3, 1, 1)$}
	\label{Diag:13}	
\end{minipage}
\begin{minipage}{0.45\textwidth}
	\footnotesize{
		\begin{tabular}{ | c | c | c |}
			\hline
			\textbf{Diagrams} & \textbf{Sequences} & \textbf{$s$-factors} \\ \hline
			$d_1$ & $\lbrace \lbrace AB \rbrace,\lbrace CDE  \rbrace\rbrace$ & 1 \\ 
			\hline
			$d_2$ & $\lbrace \lbrace  AB \rbrace,\lbrace CED \rbrace\rbrace$ & 1 \\ 
			\hline
			$d_3$ & $\lbrace \lbrace  AB \rbrace,\lbrace DCE \rbrace\rbrace$ & 2 \\ 
			\hline
			$d_4$ & $\lbrace \lbrace  AB \rbrace,\lbrace DEC \rbrace\rbrace$ & 3  \\ 
			\hline
			$d_5$ & $\lbrace \lbrace  AB \rbrace,\lbrace ECD \rbrace\rbrace$ & 2 \\ 
			\hline
			$d_{6}$ & $\lbrace \lbrace AB \rbrace,\lbrace EDC \rbrace\rbrace$ & 3  \\ 
			\hline	
			$d_{7}$ & $\lbrace \lbrace  BA \rbrace,\lbrace CDE   \rbrace\rbrace$ & 3 \\ 
			\hline
			$d_{8}$ & $\lbrace \lbrace BA \rbrace,\lbrace CED \rbrace\rbrace$ & 3 \\  
			\hline
			$d_{9}$ & $\lbrace \lbrace BA \rbrace,\lbrace DCE \rbrace\rbrace$ & 2 \\ 
			\hline		
			$d_{10}$ & $\lbrace \lbrace BA \rbrace,\lbrace DEC \rbrace\rbrace$ & 1 \\ 
			\hline
			$d_{11}$ & $\lbrace \lbrace BA \rbrace,\lbrace ECD \rbrace\rbrace$ & 2  \\ 
			\hline
			$d_{12}$ & $\lbrace \lbrace BA \rbrace,\lbrace EDC \rbrace\rbrace$ & 1 \\ 
			\hline

		\end{tabular}
		\captionof{table}{Sequences and $s$-factors}
		\label{Table:Diag13}	}
\end{minipage} 

\vspace{0.5cm}
\noindent  The mixing matrix $R$, the diagonalizing matrix $Y$ and the diagonal matrix $D$ for the Cweb mentioned above are

\begin{equation}
	R=\frac{1}{12}\left(\,\,\,
	\begin{array}{cccccccccccc}
		3 &\hspace{0.20cm} \hspace{-0.25cm}-1 &\hspace{0.20cm} \hspace{-0.25cm}-2 &\hspace{0.20cm} \hspace{-0.25cm}-1 &\hspace{0.20cm} \hspace{-0.25cm}-2 &\hspace{0.20cm} 3 &\hspace{0.20cm} \hspace{-0.25cm}-3 &\hspace{0.20cm} 1 &\hspace{0.20cm} 2 &\hspace{0.20cm} 1 &\hspace{0.20cm} 2 &\hspace{0.20cm} \hspace{-0.25cm}-3 \\
		\hspace{-0.25cm}-1 &\hspace{0.20cm} 3 &\hspace{0.20cm} \hspace{-0.25cm}-2 &\hspace{0.20cm} 3 &\hspace{0.20cm} \hspace{-0.25cm}-2 &\hspace{0.20cm} \hspace{-0.25cm}-1 &\hspace{0.20cm} 1 &\hspace{0.20cm} \hspace{-0.25cm}-3 &\hspace{0.20cm} 2 &\hspace{0.20cm} \hspace{-0.25cm}-3 &\hspace{0.20cm} 2 &\hspace{0.20cm} 1 \\
		\hspace{-0.25cm}-1 &\hspace{0.20cm} \hspace{-0.25cm}-1 &\hspace{0.20cm} 2 &\hspace{0.20cm} \hspace{-0.25cm}-1 &\hspace{0.20cm} 2 &\hspace{0.20cm} \hspace{-0.25cm}-1 &\hspace{0.20cm} 1 &\hspace{0.20cm} 1 &\hspace{0.20cm} \hspace{-0.25cm}-2 &\hspace{0.20cm} 1 &\hspace{0.20cm} \hspace{-0.25cm}-2 &\hspace{0.20cm} 1 \\
		\hspace{-0.25cm}-1 &\hspace{0.20cm} 1 &\hspace{0.20cm} 0 &\hspace{0.20cm} 1 &\hspace{0.20cm} 0 &\hspace{0.20cm} \hspace{-0.25cm}-1 &\hspace{0.20cm} 1 &\hspace{0.20cm} \hspace{-0.25cm}-1 &\hspace{0.20cm} 0 &\hspace{0.20cm} \hspace{-0.25cm}-1 &\hspace{0.20cm} 0 &\hspace{0.20cm} 1 \\
		\hspace{-0.25cm}-1 &\hspace{0.20cm} \hspace{-0.25cm}-1 &\hspace{0.20cm} 2 &\hspace{0.20cm} \hspace{-0.25cm}-1 &\hspace{0.20cm} 2 &\hspace{0.20cm} \hspace{-0.25cm}-1 &\hspace{0.20cm} 1 &\hspace{0.20cm} 1 &\hspace{0.20cm} \hspace{-0.25cm}-2 &\hspace{0.20cm} 1 &\hspace{0.20cm} \hspace{-0.25cm}-2 &\hspace{0.20cm} 1 \\
		1 &\hspace{0.20cm} \hspace{-0.25cm}-1 &\hspace{0.20cm} 0 &\hspace{0.20cm} \hspace{-0.25cm}-1 &\hspace{0.20cm} 0 &\hspace{0.20cm} 1 &\hspace{0.20cm} \hspace{-0.25cm}-1 &\hspace{0.20cm} 1 &\hspace{0.20cm} 0 &\hspace{0.20cm} 1 &\hspace{0.20cm} 0 &\hspace{0.20cm} \hspace{-0.25cm}-1 \\
		\hspace{-0.25cm}-1 &\hspace{0.20cm} 1 &\hspace{0.20cm} 0 &\hspace{0.20cm} 1 &\hspace{0.20cm} 0 &\hspace{0.20cm} \hspace{-0.25cm}-1 &\hspace{0.20cm} 1 &\hspace{0.20cm} \hspace{-0.25cm}-1 &\hspace{0.20cm} 0 &\hspace{0.20cm} \hspace{-0.25cm}-1 &\hspace{0.20cm} 0 &\hspace{0.20cm} 1 \\
		1 &\hspace{0.20cm} \hspace{-0.25cm}-1 &\hspace{0.20cm} 0 &\hspace{0.20cm} \hspace{-0.25cm}-1 &\hspace{0.20cm} 0 &\hspace{0.20cm} 1 &\hspace{0.20cm} \hspace{-0.25cm}-1 &\hspace{0.20cm} 1 &\hspace{0.20cm} 0 &\hspace{0.20cm} 1 &\hspace{0.20cm} 0 &\hspace{0.20cm} \hspace{-0.25cm}-1 \\
		1 &\hspace{0.20cm} 1 &\hspace{0.20cm} \hspace{-0.25cm}-2 &\hspace{0.20cm} 1 &\hspace{0.20cm} \hspace{-0.25cm}-2 &\hspace{0.20cm} 1 &\hspace{0.20cm} \hspace{-0.25cm}-1 &\hspace{0.20cm} \hspace{-0.25cm}-1 &\hspace{0.20cm} 2 &\hspace{0.20cm} \hspace{-0.25cm}-1 &\hspace{0.20cm} 2 &\hspace{0.20cm} \hspace{-0.25cm}-1 \\
		1 &\hspace{0.20cm} \hspace{-0.25cm}-3 &\hspace{0.20cm} 2 &\hspace{0.20cm} \hspace{-0.25cm}-3 &\hspace{0.20cm} 2 &\hspace{0.20cm} 1 &\hspace{0.20cm} \hspace{-0.25cm}-1 &\hspace{0.20cm} 3 &\hspace{0.20cm} \hspace{-0.25cm}-2 &\hspace{0.20cm} 3 &\hspace{0.20cm} \hspace{-0.25cm}-2 &\hspace{0.20cm} \hspace{-0.25cm}-1 \\
		1 &\hspace{0.20cm} 1 &\hspace{0.20cm} \hspace{-0.25cm}-2 &\hspace{0.20cm} 1 &\hspace{0.20cm} \hspace{-0.25cm}-2 &\hspace{0.20cm} 1 &\hspace{0.20cm} \hspace{-0.25cm}-1 &\hspace{0.20cm} \hspace{-0.25cm}-1 &\hspace{0.20cm} 2 &\hspace{0.20cm} \hspace{-0.25cm}-1 &\hspace{0.20cm} 2 &\hspace{0.20cm} \hspace{-0.25cm}-1 \\
		\hspace{-0.25cm}-3 &\hspace{0.20cm} 1 &\hspace{0.20cm} 2 &\hspace{0.20cm} 1 &\hspace{0.20cm} 2 &\hspace{0.20cm} \hspace{-0.25cm}-3 &\hspace{0.20cm} 3 &\hspace{0.20cm} \hspace{-0.25cm}-1 &\hspace{0.20cm} \hspace{-0.25cm}-2 &\hspace{0.20cm} \hspace{-0.25cm}-1 &\hspace{0.20cm} \hspace{-0.25cm}-2 &\hspace{0.20cm} 3 \\
	\end{array}
	\right),
\end{equation}	
\begin{equation}
	\hspace{0.2cm}	 Y=\left(\,\,\,
	\begin{array}{cccccccccccc}
		\hspace{-0.25cm}-4 &\hspace{0.20cm} 4 &\hspace{0.20cm} 0 &\hspace{0.20cm} 4 &\hspace{0.20cm} 0 &\hspace{0.20cm} \hspace{-0.25cm}-4 &\hspace{0.20cm} 4 &\hspace{0.20cm} \hspace{-0.25cm}-4 &\hspace{0.20cm} 0 &\hspace{0.20cm} \hspace{-0.25cm}-4 &\hspace{0.20cm} 0 &\hspace{0.20cm} 4 \\
		0 &\hspace{0.20cm} 4 &\hspace{0.20cm} \hspace{-0.25cm}-4 &\hspace{0.20cm} 4 &\hspace{0.20cm} \hspace{-0.25cm}-4 &\hspace{0.20cm} 0 &\hspace{0.20cm} 0 &\hspace{0.20cm} \hspace{-0.25cm}-4 &\hspace{0.20cm} 4 &\hspace{0.20cm} \hspace{-0.25cm}-4 &\hspace{0.20cm} 4 &\hspace{0.20cm} 0 \\
		4 &\hspace{0.20cm} 0 &\hspace{0.20cm} 0 &\hspace{0.20cm} 0 &\hspace{0.20cm} 0 &\hspace{0.20cm} 0 &\hspace{0.20cm} 0 &\hspace{0.20cm} 0 &\hspace{0.20cm} 0 &\hspace{0.20cm} 0 &\hspace{0.20cm} 0 &\hspace{0.20cm} 4 \\
		\hspace{-0.25cm}-2 &\hspace{0.20cm} \hspace{-0.25cm}-2 &\hspace{0.20cm} 0 &\hspace{0.20cm} 0 &\hspace{0.20cm} 0 &\hspace{0.20cm} 0 &\hspace{0.20cm} 0 &\hspace{0.20cm} 0 &\hspace{0.20cm} 0 &\hspace{0.20cm} 0 &\hspace{0.20cm} 4 &\hspace{0.20cm} 0 \\
		0 &\hspace{0.20cm} 4 &\hspace{0.20cm} 0 &\hspace{0.20cm} 0 &\hspace{0.20cm} 0 &\hspace{0.20cm} 0 &\hspace{0.20cm} 0 &\hspace{0.20cm} 0 &\hspace{0.20cm} 0 &\hspace{0.20cm} 4 &\hspace{0.20cm} 0 &\hspace{0.20cm} 0 \\
		\hspace{-0.25cm}-2 &\hspace{0.20cm} \hspace{-0.25cm}-2 &\hspace{0.20cm} 0 &\hspace{0.20cm} 0 &\hspace{0.20cm} 0 &\hspace{0.20cm} 0 &\hspace{0.20cm} 0 &\hspace{0.20cm} 0 &\hspace{0.20cm} 4 &\hspace{0.20cm} 0 &\hspace{0.20cm} 0 &\hspace{0.20cm} 0 \\
		\hspace{-0.25cm}-1 &\hspace{0.20cm} 1 &\hspace{0.20cm} 0 &\hspace{0.20cm} 0 &\hspace{0.20cm} 0 &\hspace{0.20cm} 0 &\hspace{0.20cm} 0 &\hspace{0.20cm} 4 &\hspace{0.20cm} 0 &\hspace{0.20cm} 0 &\hspace{0.20cm} 0 &\hspace{0.20cm} 0 \\
		1 &\hspace{0.20cm} \hspace{-0.25cm}-1 &\hspace{0.20cm} 0 &\hspace{0.20cm} 0 &\hspace{0.20cm} 0 &\hspace{0.20cm} 0 &\hspace{0.20cm} 4 &\hspace{0.20cm} 0 &\hspace{0.20cm} 0 &\hspace{0.20cm} 0 &\hspace{0.20cm} 0 &\hspace{0.20cm} 0 \\
		\hspace{-0.25cm}-1 &\hspace{0.20cm} 1 &\hspace{0.20cm} 0 &\hspace{0.20cm} 0 &\hspace{0.20cm} 0 &\hspace{0.20cm} 4 &\hspace{0.20cm} 0 &\hspace{0.20cm} 0 &\hspace{0.20cm} 0 &\hspace{0.20cm} 0 &\hspace{0.20cm} 0 &\hspace{0.20cm} 0 \\
		2 &\hspace{0.20cm} 2 &\hspace{0.20cm} 0 &\hspace{0.20cm} 0 &\hspace{0.20cm} 4 &\hspace{0.20cm} 0 &\hspace{0.20cm} 0 &\hspace{0.20cm} 0 &\hspace{0.20cm} 0 &\hspace{0.20cm} 0 &\hspace{0.20cm} 0 &\hspace{0.20cm} 0 \\
		1 &\hspace{0.20cm} \hspace{-0.25cm}-1 &\hspace{0.20cm} 0 &\hspace{0.20cm} 4 &\hspace{0.20cm} 0 &\hspace{0.20cm} 0 &\hspace{0.20cm} 0 &\hspace{0.20cm} 0 &\hspace{0.20cm} 0 &\hspace{0.20cm} 0 &\hspace{0.20cm} 0 &\hspace{0.20cm} 0 \\
		2 &\hspace{0.20cm} 2 &\hspace{0.20cm} 4 &\hspace{0.20cm} 0 &\hspace{0.20cm} 0 &\hspace{0.20cm} 0 &\hspace{0.20cm} 0 &\hspace{0.20cm} 0 &\hspace{0.20cm} 0 &\hspace{0.20cm} 0 &\hspace{0.20cm} 0 &\hspace{0.20cm} 0 \\
	\end{array}
	\,\,\,\right),
	\nonumber
\end{equation}
\begin{align}
	D=(\mathbf{1}_2, 0).
	\label{D54}
\end{align}

The two exponentiated colour factors of this Cweb are given below
\begin{align}
	(YC)_1=&  f^{abc}f^{agp}f^{dgm}f^{emn}\tp{1} \tb{2} \tc{3} \tn{4} \te{5} \td{6} +  f^{abc}f^{aep}f^{egq}f^{qdo} \tp{1} \tb{2} \tc{3} \too{4} \te{5} \td{6} , \nonumber \\
	=&\;-{\cal{B}}_{4}-2{\cal{B}}_{23} \, . \nn\\
	(YC)_2=&  f^{abc}f^{aep}f^{egq}f^{qdo}\tp{1} \tb{2} \tc{3} \too{4} \te{5} \td{6} \, , \nn\\ 
	=&\;-{\cal{B}}_{4}-{\cal{B}}_{23} \, .
\end{align}

\subsection*{(14)  \ $\mathbf{W^{(3,1)}_{6} (2, 1, 1, 2, 2, 1)}$}  \label{C55}
This Cweb has eight diagrams as there are two attachments each on lines 1, 4 and 5. One of the diagrams is shown in fig.~\ref{Diag:14}. The sequences of diagrams and their corresponding $s$-factors are provided in table~\ref{Table:Diag14}.

\begin{minipage}{0.45\textwidth}
	\hspace{1.5cm}	\includegraphics[scale=0.5]{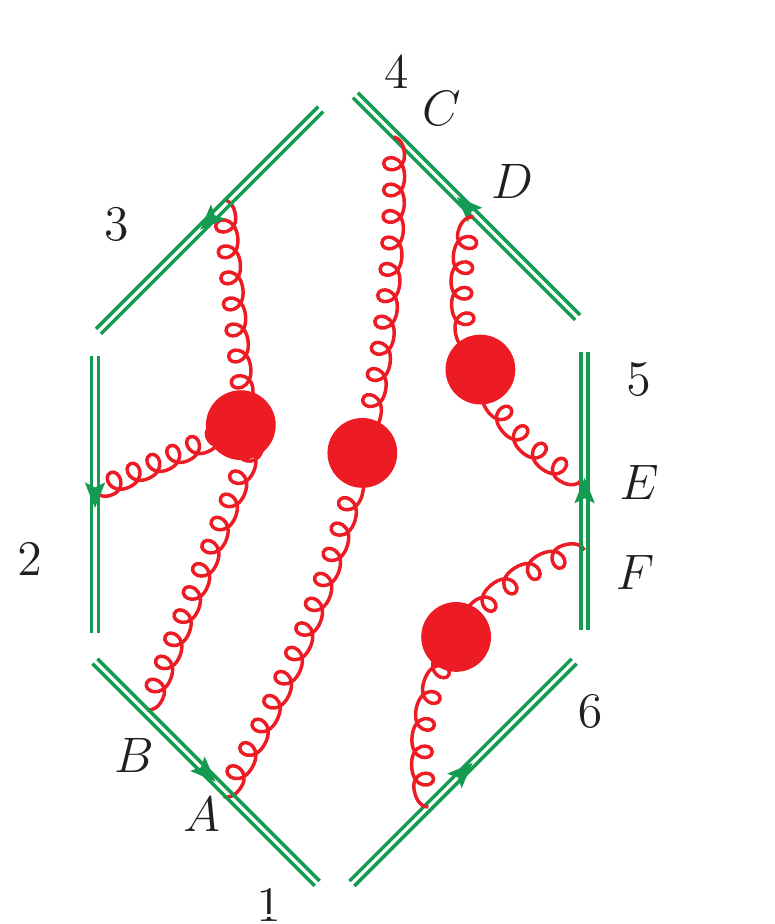}
	\captionof{figure}{$W^{(3,1)}_{6} (2, 1, 1, 2, 2, 1)$}
	\label{Diag:14}	
\end{minipage}
\begin{minipage}{0.45\textwidth}
	\footnotesize{
		\begin{tabular}{ | c | c | c |}
			\hline
			\textbf{Diagrams} & \textbf{Sequences} & \textbf{$s$-factors} \\ \hline
			$d_1$ & $\lbrace \lbrace AB \rbrace,  \lbrace CD \rbrace, \lbrace EF \rbrace\rbrace$ &1 \\ 
			\hline
			$d_2$ & $\lbrace \lbrace AB \rbrace,  \lbrace CD \rbrace, \lbrace FE \rbrace\rbrace$ & 3 \\ 
			\hline
			$d_3$ & $\lbrace \lbrace AB \rbrace,  \lbrace DC \rbrace, \lbrace EF \rbrace\rbrace$ & 5\\ 
			\hline
			$d_4$ & $\lbrace \lbrace AB \rbrace,  \lbrace DC \rbrace, \lbrace FE \rbrace\rbrace$ & 3 \\ 
			\hline		
			$d_5$ & $\lbrace \lbrace BA \rbrace,  \lbrace CD \rbrace, \lbrace EF \rbrace\rbrace$ & 3 \\ 
			\hline	
			$d_6$ & $\lbrace \lbrace BA \rbrace,  \lbrace CD \rbrace, \lbrace FE \rbrace\rbrace$ & 5 \\ 
			\hline
			$d_7$ & $\lbrace \lbrace BA \rbrace,  \lbrace DC \rbrace, \lbrace EF \rbrace\rbrace$ & 3  \\ 
			\hline
			$d_8$ & $\lbrace \lbrace BA \rbrace,  \lbrace DC \rbrace, \lbrace FE \rbrace\rbrace$ & 1 \\ 
			\hline

		\end{tabular}
		\captionof{table}{Sequences and $s$-factors}
		\label{Table:Diag14}	}
\end{minipage} 

\vspace{0.5cm}
\noindent   The mixing matrix $R$, the diagonalizing matrix $Y$ and the diagonal matrix $D$ for the Cweb mentioned above are
\vspace{0.5cm}

\begin{align} 
	R=\frac{1}{12}\left(\,\,\,
	\begin{array}{cccccccc}
		3 &\hspace{0.20cm} \hspace{-0.25cm}-3 &\hspace{0.20cm} \hspace{-0.25cm}-3 &\hspace{0.20cm} 3 &\hspace{0.20cm} \hspace{-0.25cm}-3 &\hspace{0.20cm} 3 &\hspace{0.20cm} 3 &\hspace{0.20cm} \hspace{-0.25cm}-3 \\
		\hspace{-0.25cm}-1 &\hspace{0.20cm} 1 &\hspace{0.20cm} 1 &\hspace{0.20cm} \hspace{-0.25cm}-1 &\hspace{0.20cm} 1 &\hspace{0.20cm} \hspace{-0.25cm}-1 &\hspace{0.20cm} \hspace{-0.25cm}-1 &\hspace{0.20cm} 1 \\
		\hspace{-0.25cm}-1 &\hspace{0.20cm} 1 &\hspace{0.20cm} 1 &\hspace{0.20cm} \hspace{-0.25cm}-1 &\hspace{0.20cm} 1 &\hspace{0.20cm} \hspace{-0.25cm}-1 &\hspace{0.20cm} \hspace{-0.25cm}-1 &\hspace{0.20cm} 1 \\
		1 &\hspace{0.20cm} \hspace{-0.25cm}-1 &\hspace{0.20cm} \hspace{-0.25cm}-1 &\hspace{0.20cm} 1 &\hspace{0.20cm} \hspace{-0.25cm}-1 &\hspace{0.20cm} 1 &\hspace{0.20cm} 1 &\hspace{0.20cm} \hspace{-0.25cm}-1 \\
		\hspace{-0.25cm}-1 &\hspace{0.20cm} 1 &\hspace{0.20cm} 1 &\hspace{0.20cm} \hspace{-0.25cm}-1 &\hspace{0.20cm} 1 &\hspace{0.20cm} \hspace{-0.25cm}-1 &\hspace{0.20cm} \hspace{-0.25cm}-1 &\hspace{0.20cm} 1 \\
		1 &\hspace{0.20cm} \hspace{-0.25cm}-1 &\hspace{0.20cm} \hspace{-0.25cm}-1 &\hspace{0.20cm} 1 &\hspace{0.20cm} \hspace{-0.25cm}-1 &\hspace{0.20cm} 1 &\hspace{0.20cm} 1 &\hspace{0.20cm} \hspace{-0.25cm}-1 \\
		1 &\hspace{0.20cm} \hspace{-0.25cm}-1 &\hspace{0.20cm} \hspace{-0.25cm}-1 &\hspace{0.20cm} 1 &\hspace{0.20cm} \hspace{-0.25cm}-1 &\hspace{0.20cm} 1 &\hspace{0.20cm} 1 &\hspace{0.20cm} \hspace{-0.25cm}-1 \\
		\hspace{-0.25cm}-3 &\hspace{0.20cm} 3 &\hspace{0.20cm} 3 &\hspace{0.20cm} \hspace{-0.25cm}-3 &\hspace{0.20cm} 3 &\hspace{0.20cm} \hspace{-0.25cm}-3 &\hspace{0.20cm} \hspace{-0.25cm}-3 &\hspace{0.20cm} 3 \\
	\end{array}
	\right)&\hspace{0.20cm}, \quad Y=\left(\,\,\,
	\begin{array}{cccccccc}
		\hspace{-0.25cm}-3 &\hspace{0.20cm} 3 &\hspace{0.20cm} 3 &\hspace{0.20cm} \hspace{-0.25cm}-3 &\hspace{0.20cm} 3 &\hspace{0.20cm} \hspace{-0.25cm}-3 &\hspace{0.20cm} \hspace{-0.25cm}-3 &\hspace{0.20cm} 3 \\
		3 &\hspace{0.20cm} 0 &\hspace{0.20cm} 0 &\hspace{0.20cm} 0 &\hspace{0.20cm} 0 &\hspace{0.20cm} 0 &\hspace{0.20cm} 0 &\hspace{0.20cm} 3 \\
		\hspace{-0.25cm}-1 &\hspace{0.20cm} 0 &\hspace{0.20cm} 0 &\hspace{0.20cm} 0 &\hspace{0.20cm} 0 &\hspace{0.20cm} 0 &\hspace{0.20cm} 3 &\hspace{0.20cm} 0 \\
		\hspace{-0.25cm}-1 &\hspace{0.20cm} 0 &\hspace{0.20cm} 0 &\hspace{0.20cm} 0 &\hspace{0.20cm} 0 &\hspace{0.20cm} 3 &\hspace{0.20cm} 0 &\hspace{0.20cm} 0 \\
		1 &\hspace{0.20cm} 0 &\hspace{0.20cm} 0 &\hspace{0.20cm} 0 &\hspace{0.20cm} 3 &\hspace{0.20cm} 0 &\hspace{0.20cm} 0 &\hspace{0.20cm} 0 \\
		\hspace{-0.25cm}-1 &\hspace{0.20cm} 0 &\hspace{0.20cm} 0 &\hspace{0.20cm} 3 &\hspace{0.20cm} 0 &\hspace{0.20cm} 0 &\hspace{0.20cm} 0 &\hspace{0.20cm} 0 \\
		1 &\hspace{0.20cm} 0 &\hspace{0.20cm} 3 &\hspace{0.20cm} 0 &\hspace{0.20cm} 0 &\hspace{0.20cm} 0 &\hspace{0.20cm} 0 &\hspace{0.20cm} 0 \\
		1 &\hspace{0.20cm} 3 &\hspace{0.20cm} 0 &\hspace{0.20cm} 0 &\hspace{0.20cm} 0 &\hspace{0.20cm} 0 &\hspace{0.20cm} 0 &\hspace{0.20cm} 0 \\
	\end{array}
	\right), \nonumber  \\
	& D=(\mathbf{1}_1,0).   
	\label{D55}
\end{align}

The only exponentiated colour factor of this Cweb is
\begin{align}
	(YC)_1=& - f^{abc}f^{aep}f^{der}f^{dgm}\tp{1} \tb{2} \tc{3} \trr{4} \tm{5} \tg{6} \, , \nn\\
	=&\;{\cal{B}}_{4} \, .
\end{align}

\subsubsection*{(15)  \ $\mathbf{W^{(1,2)}_{6} (2, 1, 1, 2, 1, 1)}$} \label{C56}
This Cweb has four diagrams, one of the diagram is shown in fig.~\ref{Diag:15}. The sequences and their corresponding $s$-factors of all the diagrams are provided in table \ref{Table:Diag15}.

\begin{minipage}{0.45\textwidth}
	\hspace{1.5cm}	\includegraphics[scale=0.5]{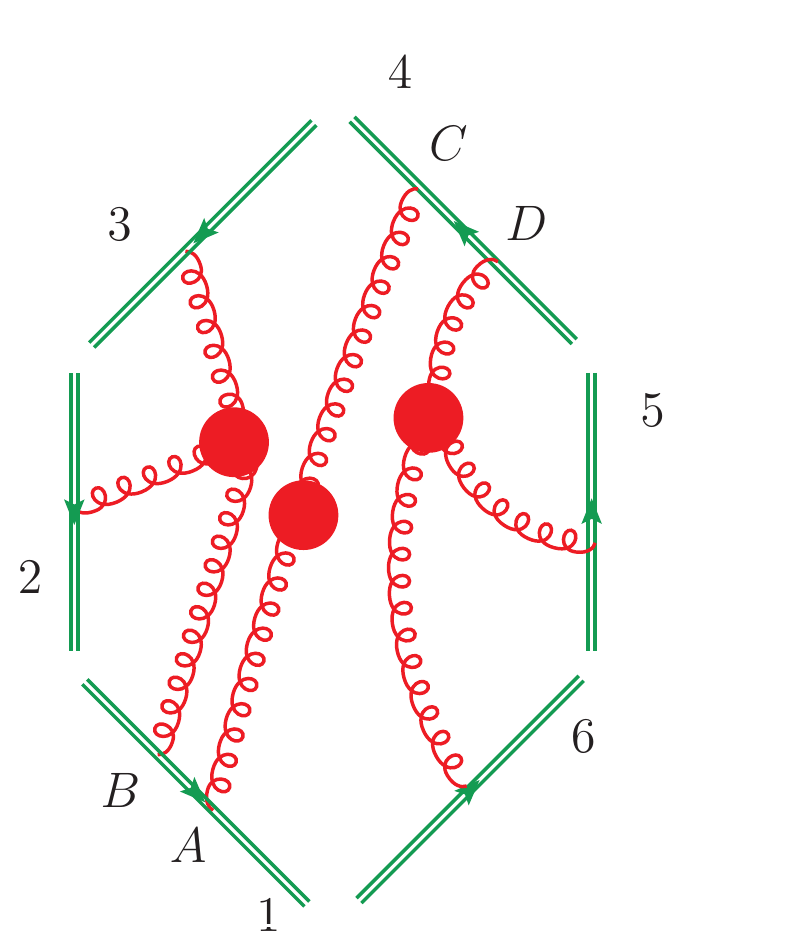}	
	\captionof{figure}{$W^{(1,2)}_{6} (2, 1, 1, 2, 1, 1)$}
	\label{Diag:15}
\end{minipage}
\begin{minipage}{0.45\textwidth}
	\footnotesize{
		\begin{tabular}{ | c | c | c |}
			\hline
			\textbf{Diagrams} & \textbf{Sequences} & \textbf{$s$-factors} \\ \hline
			$d_1$ & $\lbrace \lbrace AB \rbrace,  \lbrace CD \rbrace\rbrace$ &1 \\ 
			\hline
			$d_2$ & $\lbrace \lbrace AB \rbrace,  \lbrace DC \rbrace\rbrace$ & 2 \\ 
			\hline
			$d_3$ & $\lbrace \lbrace BA \rbrace,  \lbrace CD \rbrace\rbrace$ & 2  \\ 
			\hline
			$d_4$ & $\lbrace \lbrace BA \rbrace,  \lbrace DC \rbrace\rbrace$ & 1 \\ 
			\hline	
			
		\end{tabular}	
		\captionof{table}{Sequences and $s$-factors}
		\label{Table:Diag15}	}
\end{minipage} 

\vspace{0.5cm}
\noindent  The mixing matrix $R$, the diagonalizing matrix $Y$ and the diagonal matrix $D$ for the Cweb mentioned above are
\begin{equation} 
	R=\frac{1}{6} \left(\,\,\,
	\begin{array}{cccc}
		2 &\hspace{0.20cm} \hspace{-0.25cm}-2 &\hspace{0.20cm} \hspace{-0.25cm}-2 &\hspace{0.20cm} 2 \\
		\hspace{-0.25cm}-1 &\hspace{0.20cm} 1 &\hspace{0.20cm} 1 &\hspace{0.20cm} \hspace{-0.25cm}-1 \\
		\hspace{-0.25cm}-1 &\hspace{0.20cm} 1 &\hspace{0.20cm} 1 &\hspace{0.20cm} \hspace{-0.25cm}-1 \\
		2 &\hspace{0.20cm} \hspace{-0.25cm}-2 &\hspace{0.20cm} \hspace{-0.25cm}-2 &\hspace{0.20cm} 2 \\
	\end{array}
	\right),  Y=\left(\,\,\,
	\begin{array}{cccc}
		1 &\hspace{0.20cm} \hspace{-0.25cm}-1 &\hspace{0.20cm} \hspace{-0.25cm}-1 &\hspace{0.20cm} 1 \\
		\hspace{-0.25cm}-1 &\hspace{0.20cm} 0 &\hspace{0.20cm} 0 &\hspace{0.20cm} 1 \\
		\frac{1}{2} &\hspace{0.20cm} 0 &\hspace{0.20cm} 1 &\hspace{0.20cm} 0 \\
		\frac{1}{2} &\hspace{0.20cm} 1 &\hspace{0.20cm} 0 &\hspace{0.20cm} 0 \\
	\end{array}
	\right)   ,   D=(\mathbf{1}_1,0) .
	\label{D56}
\end{equation}

The exponentiated colour factor of this Cweb is
\begin{align}
	(YC)_1=& - f^{abc}f^{aep}f^{der}f^{dgh}\tp{1} \tb{2} \tc{3} \trr{4} \tg{5} \thh{6}\, ,\nn\\
	=&\;-{\cal{B}}_{4} \, .
\end{align}


\subsection{New basis Cwebs} \label{sec:appdxBasis}

In this section, we list down all the new basis Cwebs that are appearing at five loops.

\subsection*{(16)  \ $\mathbf{W^{(5)}_{6} (2, 2, 2, 1, 2, 1)}$}  \label{C61}

This Cweb consists of 5 two-point correlators and six Wilson lines thus, hence it is  a basis Cweb. This Cweb has sixteen diagrams, one of the diagrams is shown in fig.~\ref{Diag:16}. The sequences of diagrams and their corresponding $s$-factors are provided in table~\ref{Table:Diag16}.

\vskip1cm

\begin{minipage}{0.45\textwidth}
	\hspace{1.5cm}	\includegraphics[scale=0.5]{C61N}
	\captionof{figure}{$W^{(5)}_{6} (2, 2, 2, 1, 2, 1)$}
	\label{Diag:16}	
\end{minipage}
\begin{minipage}{0.35\textwidth}
	\footnotesize{
		\begin{tabular}{ | c | c | c |}
			\hline
			\textbf{Diagrams} & \textbf{Sequences} & \textbf{$s$-factors} \\ \hline
			$d_{1}$ & $\{\{AB\},\{CD\},\{EF\},\{GH\}\}$  &  4\\ \hline
			$d_{2}$ & $\{\{AB\},\{CD\},\{EF\},\{HG\}\}$  &  11\\ \hline
			$d_{3}$ & $\{\{AB\},\{CD\},\{FE\},\{GH\}\}$  &  16\\ \hline
			$d_{4}$ & $\{\{AB\},\{CD\},\{FE\},\{HG\}\}$  &  9\\ \hline
			$d_{5}$ & $\{\{  AB\},\{DC\},\{EF\},\{GH\}\}$  & 1 \\ \hline
			$d_{6}$ & $\{\{  AB\},\{DC\},\{EF\},\{HG\}\}$  & 4 \\ \hline
			$d_{7}$ & $\{\{  AB\},\{DC\},\{FE\},\{GH\}\}$  &9  \\ \hline
			$d_{8}$ & $\{\{  AB\},\{DC\},\{FE\},\{HG\}\}$  & 6 \\ \hline
			$d_{9}$ & $\{\{  BA\},\{CD\},\{EF\},\{GH\}\}$  &  6\\ \hline
			$d_{10}$ & $\{\{  BA\},\{CD\},\{EF\},\{HG\}\}$  &  9\\ \hline
			$d_{11}$ & $\{\{  BA\},\{CD\},\{FE\},\{GH\}\}$  &  4\\ \hline
			$d_{12}$ & $\{\{  BA\},\{CD\},\{FE\},\{HG\}\}$  & 1 \\ \hline
			$d_{13}$ & $\{\{  BA\},\{DC\},\{EF\},\{GH\}\}$  & 9 \\ \hline
			$d_{14}$ & $\{\{  BA\},\{DC\},\{EF\},\{HG\}\}$  & 16 \\ \hline
			$d_{15}$ & $\{\{  BA\},\{DC\},\{FE\},\{GH\}\}$  &  11\\ \hline
			$d_{16}$ & $\{\{  BA\},\{DC\},\{FE\},\{HG\}\}$  &  4\\ \hline
		\end{tabular}
		\captionof{table}{Sequences and $s$-factors}
		\label{Table:Diag16}	}
\end{minipage}

\vspace{0.5cm}

\noindent   The mixing matrix $R$, the diagonalizing matrix $Y$ and the diagonal matrix $D$ for the Cweb mentioned above are

\begin{equation*} 
	R=\frac{1}{60} \left(\,\,\,
	\begin{array}{cccccccccccccccc}
		3 &\hspace{0.20cm} \hspace{-0.25cm}-3 &\hspace{0.20cm} \hspace{-0.25cm}-3 &\hspace{0.20cm} 3 &\hspace{0.20cm} \hspace{-0.25cm}-3 &\hspace{0.20cm} 3 &\hspace{0.20cm} 3 &\hspace{0.20cm} \hspace{-0.25cm}-3 &\hspace{0.20cm} \hspace{-0.25cm}-3 &\hspace{0.20cm} 3 &\hspace{0.20cm} 3 &\hspace{0.20cm} \hspace{-0.25cm}-3 &\hspace{0.20cm} 3 &\hspace{0.20cm} \hspace{-0.25cm}-3 &\hspace{0.20cm} \hspace{-0.25cm}-3 &\hspace{0.20cm} 3 \\
		\hspace{-0.25cm}-2 &\hspace{0.20cm} 2 &\hspace{0.20cm} 2 &\hspace{0.20cm} \hspace{-0.25cm}-2 &\hspace{0.20cm} 2 &\hspace{0.20cm} \hspace{-0.25cm}-2 &\hspace{0.20cm} \hspace{-0.25cm}-2 &\hspace{0.20cm} 2 &\hspace{0.20cm} 2 &\hspace{0.20cm} \hspace{-0.25cm}-2 &\hspace{0.20cm} \hspace{-0.25cm}-2 &\hspace{0.20cm} 2 &\hspace{0.20cm} \hspace{-0.25cm}-2 &\hspace{0.20cm} 2 &\hspace{0.20cm} 2 &\hspace{0.20cm} \hspace{-0.25cm}-2 \\
		\hspace{-0.25cm}-2 &\hspace{0.20cm} 2 &\hspace{0.20cm} 2 &\hspace{0.20cm} \hspace{-0.25cm}-2 &\hspace{0.20cm} 2 &\hspace{0.20cm} \hspace{-0.25cm}-2 &\hspace{0.20cm} \hspace{-0.25cm}-2 &\hspace{0.20cm} 2 &\hspace{0.20cm} 2 &\hspace{0.20cm} \hspace{-0.25cm}-2 &\hspace{0.20cm} \hspace{-0.25cm}-2 &\hspace{0.20cm} 2 &\hspace{0.20cm} \hspace{-0.25cm}-2 &\hspace{0.20cm} 2 &\hspace{0.20cm} 2 &\hspace{0.20cm} \hspace{-0.25cm}-2 \\
		3 &\hspace{0.20cm} \hspace{-0.25cm}-3 &\hspace{0.20cm} \hspace{-0.25cm}-3 &\hspace{0.20cm} 3 &\hspace{0.20cm} \hspace{-0.25cm}-3 &\hspace{0.20cm} 3 &\hspace{0.20cm} 3 &\hspace{0.20cm} \hspace{-0.25cm}-3 &\hspace{0.20cm} \hspace{-0.25cm}-3 &\hspace{0.20cm} 3 &\hspace{0.20cm} 3 &\hspace{0.20cm} \hspace{-0.25cm}-3 &\hspace{0.20cm} 3 &\hspace{0.20cm} \hspace{-0.25cm}-3 &\hspace{0.20cm} \hspace{-0.25cm}-3 &\hspace{0.20cm} 3 \\
		\hspace{-0.25cm}-12 &\hspace{0.20cm} 12 &\hspace{0.20cm} 12 &\hspace{0.20cm} \hspace{-0.25cm}-12 &\hspace{0.20cm} 12 &\hspace{0.20cm} \hspace{-0.25cm}-12 &\hspace{0.20cm} \hspace{-0.25cm}-12 &\hspace{0.20cm} 12 &\hspace{0.20cm} 12 &\hspace{0.20cm} \hspace{-0.25cm}-12 &\hspace{0.20cm} \hspace{-0.25cm}-12 &\hspace{0.20cm} 12 &\hspace{0.20cm} \hspace{-0.25cm}-12 &\hspace{0.20cm} 12 &\hspace{0.20cm} 12 &\hspace{0.20cm} \hspace{-0.25cm}-12 \\
		3 &\hspace{0.20cm} \hspace{-0.25cm}-3 &\hspace{0.20cm} \hspace{-0.25cm}-3 &\hspace{0.20cm} 3 &\hspace{0.20cm} \hspace{-0.25cm}-3 &\hspace{0.20cm} 3 &\hspace{0.20cm} 3 &\hspace{0.20cm} \hspace{-0.25cm}-3 &\hspace{0.20cm} \hspace{-0.25cm}-3 &\hspace{0.20cm} 3 &\hspace{0.20cm} 3 &\hspace{0.20cm} \hspace{-0.25cm}-3 &\hspace{0.20cm} 3 &\hspace{0.20cm} \hspace{-0.25cm}-3 &\hspace{0.20cm} \hspace{-0.25cm}-3 &\hspace{0.20cm} 3 \\
		3 &\hspace{0.20cm} \hspace{-0.25cm}-3 &\hspace{0.20cm} \hspace{-0.25cm}-3 &\hspace{0.20cm} 3 &\hspace{0.20cm} \hspace{-0.25cm}-3 &\hspace{0.20cm} 3 &\hspace{0.20cm} 3 &\hspace{0.20cm} \hspace{-0.25cm}-3 &\hspace{0.20cm} \hspace{-0.25cm}-3 &\hspace{0.20cm} 3 &\hspace{0.20cm} 3 &\hspace{0.20cm} \hspace{-0.25cm}-3 &\hspace{0.20cm} 3 &\hspace{0.20cm} \hspace{-0.25cm}-3 &\hspace{0.20cm} \hspace{-0.25cm}-3 &\hspace{0.20cm} 3 \\
		\hspace{-0.25cm}-2 &\hspace{0.20cm} 2 &\hspace{0.20cm} 2 &\hspace{0.20cm} \hspace{-0.25cm}-2 &\hspace{0.20cm} 2 &\hspace{0.20cm} \hspace{-0.25cm}-2 &\hspace{0.20cm} \hspace{-0.25cm}-2 &\hspace{0.20cm} 2 &\hspace{0.20cm} 2 &\hspace{0.20cm} \hspace{-0.25cm}-2 &\hspace{0.20cm} \hspace{-0.25cm}-2 &\hspace{0.20cm} 2 &\hspace{0.20cm} \hspace{-0.25cm}-2 &\hspace{0.20cm} 2 &\hspace{0.20cm} 2 &\hspace{0.20cm} \hspace{-0.25cm}-2 \\
		\hspace{-0.25cm}-2 &\hspace{0.20cm} 2 &\hspace{0.20cm} 2 &\hspace{0.20cm} \hspace{-0.25cm}-2 &\hspace{0.20cm} 2 &\hspace{0.20cm} \hspace{-0.25cm}-2 &\hspace{0.20cm} \hspace{-0.25cm}-2 &\hspace{0.20cm} 2 &\hspace{0.20cm} 2 &\hspace{0.20cm} \hspace{-0.25cm}-2 &\hspace{0.20cm} \hspace{-0.25cm}-2 &\hspace{0.20cm} 2 &\hspace{0.20cm} \hspace{-0.25cm}-2 &\hspace{0.20cm} 2 &\hspace{0.20cm} 2 &\hspace{0.20cm} \hspace{-0.25cm}-2 \\
		3 &\hspace{0.20cm} \hspace{-0.25cm}-3 &\hspace{0.20cm} \hspace{-0.25cm}-3 &\hspace{0.20cm} 3 &\hspace{0.20cm} \hspace{-0.25cm}-3 &\hspace{0.20cm} 3 &\hspace{0.20cm} 3 &\hspace{0.20cm} \hspace{-0.25cm}-3 &\hspace{0.20cm} \hspace{-0.25cm}-3 &\hspace{0.20cm} 3 &\hspace{0.20cm} 3 &\hspace{0.20cm} \hspace{-0.25cm}-3 &\hspace{0.20cm} 3 &\hspace{0.20cm} \hspace{-0.25cm}-3 &\hspace{0.20cm} \hspace{-0.25cm}-3 &\hspace{0.20cm} 3 \\
		3 &\hspace{0.20cm} \hspace{-0.25cm}-3 &\hspace{0.20cm} \hspace{-0.25cm}-3 &\hspace{0.20cm} 3 &\hspace{0.20cm} \hspace{-0.25cm}-3 &\hspace{0.20cm} 3 &\hspace{0.20cm} 3 &\hspace{0.20cm} \hspace{-0.25cm}-3 &\hspace{0.20cm} \hspace{-0.25cm}-3 &\hspace{0.20cm} 3 &\hspace{0.20cm} 3 &\hspace{0.20cm} \hspace{-0.25cm}-3 &\hspace{0.20cm} 3 &\hspace{0.20cm} \hspace{-0.25cm}-3 &\hspace{0.20cm} \hspace{-0.25cm}-3 &\hspace{0.20cm} 3 \\
		\hspace{-0.25cm}-12 &\hspace{0.20cm} 12 &\hspace{0.20cm} 12 &\hspace{0.20cm} \hspace{-0.25cm}-12 &\hspace{0.20cm} 12 &\hspace{0.20cm} \hspace{-0.25cm}-12 &\hspace{0.20cm} \hspace{-0.25cm}-12 &\hspace{0.20cm} 12 &\hspace{0.20cm} 12 &\hspace{0.20cm} \hspace{-0.25cm}-12 &\hspace{0.20cm} \hspace{-0.25cm}-12 &\hspace{0.20cm} 12 &\hspace{0.20cm} \hspace{-0.25cm}-12 &\hspace{0.20cm} 12 &\hspace{0.20cm} 12 &\hspace{0.20cm} \hspace{-0.25cm}-12 \\
		3 &\hspace{0.20cm} \hspace{-0.25cm}-3 &\hspace{0.20cm} \hspace{-0.25cm}-3 &\hspace{0.20cm} 3 &\hspace{0.20cm} \hspace{-0.25cm}-3 &\hspace{0.20cm} 3 &\hspace{0.20cm} 3 &\hspace{0.20cm} \hspace{-0.25cm}-3 &\hspace{0.20cm} \hspace{-0.25cm}-3 &\hspace{0.20cm} 3 &\hspace{0.20cm} 3 &\hspace{0.20cm} \hspace{-0.25cm}-3 &\hspace{0.20cm} 3 &\hspace{0.20cm} \hspace{-0.25cm}-3 &\hspace{0.20cm} \hspace{-0.25cm}-3 &\hspace{0.20cm} 3 \\
		\hspace{-0.25cm}-2 &\hspace{0.20cm} 2 &\hspace{0.20cm} 2 &\hspace{0.20cm} \hspace{-0.25cm}-2 &\hspace{0.20cm} 2 &\hspace{0.20cm} \hspace{-0.25cm}-2 &\hspace{0.20cm} \hspace{-0.25cm}-2 &\hspace{0.20cm} 2 &\hspace{0.20cm} 2 &\hspace{0.20cm} \hspace{-0.25cm}-2 &\hspace{0.20cm} \hspace{-0.25cm}-2 &\hspace{0.20cm} 2 &\hspace{0.20cm} \hspace{-0.25cm}-2 &\hspace{0.20cm} 2 &\hspace{0.20cm} 2 &\hspace{0.20cm} \hspace{-0.25cm}-2 \\
		\hspace{-0.25cm}-2 &\hspace{0.20cm} 2 &\hspace{0.20cm} 2 &\hspace{0.20cm} \hspace{-0.25cm}-2 &\hspace{0.20cm} 2 &\hspace{0.20cm} \hspace{-0.25cm}-2 &\hspace{0.20cm} \hspace{-0.25cm}-2 &\hspace{0.20cm} 2 &\hspace{0.20cm} 2 &\hspace{0.20cm} \hspace{-0.25cm}-2 &\hspace{0.20cm} \hspace{-0.25cm}-2 &\hspace{0.20cm} 2 &\hspace{0.20cm} \hspace{-0.25cm}-2 &\hspace{0.20cm} 2 &\hspace{0.20cm} 2 &\hspace{0.20cm} \hspace{-0.25cm}-2 \\
		3 &\hspace{0.20cm} \hspace{-0.25cm}-3 &\hspace{0.20cm} \hspace{-0.25cm}-3 &\hspace{0.20cm} 3 &\hspace{0.20cm} \hspace{-0.25cm}-3 &\hspace{0.20cm} 3 &\hspace{0.20cm} 3 &\hspace{0.20cm} \hspace{-0.25cm}-3 &\hspace{0.20cm} \hspace{-0.25cm}-3 &\hspace{0.20cm} 3 &\hspace{0.20cm} 3 &\hspace{0.20cm} \hspace{-0.25cm}-3 &\hspace{0.20cm} 3 &\hspace{0.20cm} \hspace{-0.25cm}-3 &\hspace{0.20cm} \hspace{-0.25cm}-3 &\hspace{0.20cm} 3. \\
	\end{array}
	\right),
\end{equation*}
\begin{equation}
	\hspace{-0.25cm} Y=\frac{1}{3}\left(\,\,\,
	\begin{array}{cccccccccccccccc}
		3 &\hspace{0.20cm} \hspace{-0.25cm}-3 &\hspace{0.20cm} \hspace{-0.25cm}-3 &\hspace{0.20cm} 3 &\hspace{0.20cm} \hspace{-0.25cm}-3 &\hspace{0.20cm} 3 &\hspace{0.20cm} 3 &\hspace{0.20cm} \hspace{-0.25cm}-3 &\hspace{0.20cm} \hspace{-0.25cm}-3 &\hspace{0.20cm} 3 &\hspace{0.20cm} 3 &\hspace{0.20cm} \hspace{-0.25cm}-3 &\hspace{0.20cm} 3 &\hspace{0.20cm} \hspace{-0.25cm}-3 &\hspace{0.20cm} \hspace{-0.25cm}-3 &\hspace{0.20cm} 3 \\
		\hspace{-0.25cm}-3 &\hspace{0.20cm} 0 &\hspace{0.20cm} 0 &\hspace{0.20cm} 0 &\hspace{0.20cm} 0 &\hspace{0.20cm} 0 &\hspace{0.20cm} 0 &\hspace{0.20cm} 0 &\hspace{0.20cm} 0 &\hspace{0.20cm} 0 &\hspace{0.20cm} 0 &\hspace{0.20cm} 0 &\hspace{0.20cm} 0 &\hspace{0.20cm} 0 &\hspace{0.20cm} 0 &\hspace{0.20cm} 3 \\
		2 &\hspace{0.20cm} 0 &\hspace{0.20cm} 0 &\hspace{0.20cm} 0 &\hspace{0.20cm} 0 &\hspace{0.20cm} 0 &\hspace{0.20cm} 0 &\hspace{0.20cm} 0 &\hspace{0.20cm} 0 &\hspace{0.20cm} 0 &\hspace{0.20cm} 0 &\hspace{0.20cm} 0 &\hspace{0.20cm} 0 &\hspace{0.20cm} 0 &\hspace{0.20cm} 3 &\hspace{0.20cm} 0 \\
		2 &\hspace{0.20cm} 0 &\hspace{0.20cm} 0 &\hspace{0.20cm} 0 &\hspace{0.20cm} 0 &\hspace{0.20cm} 0 &\hspace{0.20cm} 0 &\hspace{0.20cm} 0 &\hspace{0.20cm} 0 &\hspace{0.20cm} 0 &\hspace{0.20cm} 0 &\hspace{0.20cm} 0 &\hspace{0.20cm} 0 &\hspace{0.20cm} 3 &\hspace{0.20cm} 0 &\hspace{0.20cm} 0 \\
		\hspace{-0.25cm}-3 &\hspace{0.20cm} 0 &\hspace{0.20cm} 0 &\hspace{0.20cm} 0 &\hspace{0.20cm} 0 &\hspace{0.20cm} 0 &\hspace{0.20cm} 0 &\hspace{0.20cm} 0 &\hspace{0.20cm} 0 &\hspace{0.20cm} 0 &\hspace{0.20cm} 0 &\hspace{0.20cm} 0 &\hspace{0.20cm} 3 &\hspace{0.20cm} 0 &\hspace{0.20cm} 0 &\hspace{0.20cm} 0 \\
		12 &\hspace{0.20cm} 0 &\hspace{0.20cm} 0 &\hspace{0.20cm} 0 &\hspace{0.20cm} 0 &\hspace{0.20cm} 0 &\hspace{0.20cm} 0 &\hspace{0.20cm} 0 &\hspace{0.20cm} 0 &\hspace{0.20cm} 0 &\hspace{0.20cm} 0 &\hspace{0.20cm} 3 &\hspace{0.20cm} 0 &\hspace{0.20cm} 0 &\hspace{0.20cm} 0 &\hspace{0.20cm} 0 \\
		\hspace{-0.25cm}-3 &\hspace{0.20cm} 0 &\hspace{0.20cm} 0 &\hspace{0.20cm} 0 &\hspace{0.20cm} 0 &\hspace{0.20cm} 0 &\hspace{0.20cm} 0 &\hspace{0.20cm} 0 &\hspace{0.20cm} 0 &\hspace{0.20cm} 0 &\hspace{0.20cm} 3 &\hspace{0.20cm} 0 &\hspace{0.20cm} 0 &\hspace{0.20cm} 0 &\hspace{0.20cm} 0 &\hspace{0.20cm} 0 \\
		\hspace{-0.25cm}-3 &\hspace{0.20cm} 0 &\hspace{0.20cm} 0 &\hspace{0.20cm} 0 &\hspace{0.20cm} 0 &\hspace{0.20cm} 0 &\hspace{0.20cm} 0 &\hspace{0.20cm} 0 &\hspace{0.20cm} 0 &\hspace{0.20cm} 3 &\hspace{0.20cm} 0 &\hspace{0.20cm} 0 &\hspace{0.20cm} 0 &\hspace{0.20cm} 0 &\hspace{0.20cm} 0 &\hspace{0.20cm} 0 \\
		2 &\hspace{0.20cm} 0 &\hspace{0.20cm} 0 &\hspace{0.20cm} 0 &\hspace{0.20cm} 0 &\hspace{0.20cm} 0 &\hspace{0.20cm} 0 &\hspace{0.20cm} 0 &\hspace{0.20cm} 3 &\hspace{0.20cm} 0 &\hspace{0.20cm} 0 &\hspace{0.20cm} 0 &\hspace{0.20cm} 0 &\hspace{0.20cm} 0 &\hspace{0.20cm} 0 &\hspace{0.20cm} 0 \\
		2 &\hspace{0.20cm} 0 &\hspace{0.20cm} 0 &\hspace{0.20cm} 0 &\hspace{0.20cm} 0 &\hspace{0.20cm} 0 &\hspace{0.20cm} 0 &\hspace{0.20cm} 3 &\hspace{0.20cm} 0 &\hspace{0.20cm} 0 &\hspace{0.20cm} 0 &\hspace{0.20cm} 0 &\hspace{0.20cm} 0 &\hspace{0.20cm} 0 &\hspace{0.20cm} 0 &\hspace{0.20cm} 0 \\
		\hspace{-0.25cm}-3 &\hspace{0.20cm} 0 &\hspace{0.20cm} 0 &\hspace{0.20cm} 0 &\hspace{0.20cm} 0 &\hspace{0.20cm} 0 &\hspace{0.20cm} 3 &\hspace{0.20cm} 0 &\hspace{0.20cm} 0 &\hspace{0.20cm} 0 &\hspace{0.20cm} 0 &\hspace{0.20cm} 0 &\hspace{0.20cm} 0 &\hspace{0.20cm} 0 &\hspace{0.20cm} 0 &\hspace{0.20cm} 0 \\
		\hspace{-0.25cm}-3 &\hspace{0.20cm} 0 &\hspace{0.20cm} 0 &\hspace{0.20cm} 0 &\hspace{0.20cm} 0 &\hspace{0.20cm} 3 &\hspace{0.20cm} 0 &\hspace{0.20cm} 0 &\hspace{0.20cm} 0 &\hspace{0.20cm} 0 &\hspace{0.20cm} 0 &\hspace{0.20cm} 0 &\hspace{0.20cm} 0 &\hspace{0.20cm} 0 &\hspace{0.20cm} 0 &\hspace{0.20cm} 0 \\
		12 &\hspace{0.20cm} 0 &\hspace{0.20cm} 0 &\hspace{0.20cm} 0 &\hspace{0.20cm} 3 &\hspace{0.20cm} 0 &\hspace{0.20cm} 0 &\hspace{0.20cm} 0 &\hspace{0.20cm} 0 &\hspace{0.20cm} 0 &\hspace{0.20cm} 0 &\hspace{0.20cm} 0 &\hspace{0.20cm} 0 &\hspace{0.20cm} 0 &\hspace{0.20cm} 0 &\hspace{0.20cm} 0 \\
		\hspace{-0.25cm}-3 &\hspace{0.20cm} 0 &\hspace{0.20cm} 0 &\hspace{0.20cm} 3 &\hspace{0.20cm} 0 &\hspace{0.20cm} 0 &\hspace{0.20cm} 0 &\hspace{0.20cm} 0 &\hspace{0.20cm} 0 &\hspace{0.20cm} 0 &\hspace{0.20cm} 0 &\hspace{0.20cm} 0 &\hspace{0.20cm} 0 &\hspace{0.20cm} 0 &\hspace{0.20cm} 0 &\hspace{0.20cm} 0 \\
		2 &\hspace{0.20cm} 0 &\hspace{0.20cm} 3 &\hspace{0.20cm} 0 &\hspace{0.20cm} 0 &\hspace{0.20cm} 0 &\hspace{0.20cm} 0 &\hspace{0.20cm} 0 &\hspace{0.20cm} 0 &\hspace{0.20cm} 0 &\hspace{0.20cm} 0 &\hspace{0.20cm} 0 &\hspace{0.20cm} 0 &\hspace{0.20cm} 0 &\hspace{0.20cm} 0 &\hspace{0.20cm} 0 \\
		2 &\hspace{0.20cm} 3 &\hspace{0.20cm} 0 &\hspace{0.20cm} 0 &\hspace{0.20cm} 0 &\hspace{0.20cm} 0 &\hspace{0.20cm} 0 &\hspace{0.20cm} 0 &\hspace{0.20cm} 0 &\hspace{0.20cm} 0 &\hspace{0.20cm} 0 &\hspace{0.20cm} 0 &\hspace{0.20cm} 0 &\hspace{0.20cm} 0 &\hspace{0.20cm} 0 &\hspace{0.20cm} 0 \\
	\end{array}
	\right) , \quad  D=(\mathbf{1}_1,0).
	\label{D61}
\end{equation}

The only exponentiated colour factor of this Cweb is
\begin{align}
	(YC)_1=& - f^{abr}f^{adp}f^{dcq}f^{ecm}\tp{1} \trr{2} \tq{3} \tb{4} \tm{5} \te{6} \, ,\nn\\
	=&\;-{\cal{B}}_{6}-{\cal{B}}_{7} \,.
\end{align}

\subsection*{(17)  \ $\mathbf{W^{(5)}_{6} (3, 2, 2, 1, 1, 1)}$}  \label{C62}

This Cweb is again a basis Cweb, and it has twenty-four diagrams as there are three attachments on line 1, two on line 2, and two on line 3. One of the diagrams is shown in fig.~\ref{Diag:17}. The sequences of diagrams and their corresponding $s$-factors are provided in table~\ref{Table:Diag17}.
\vspace{0.5cm}

\begin{minipage}{0.45\textwidth}
	\hspace{1.5cm}	\includegraphics[scale=0.5]{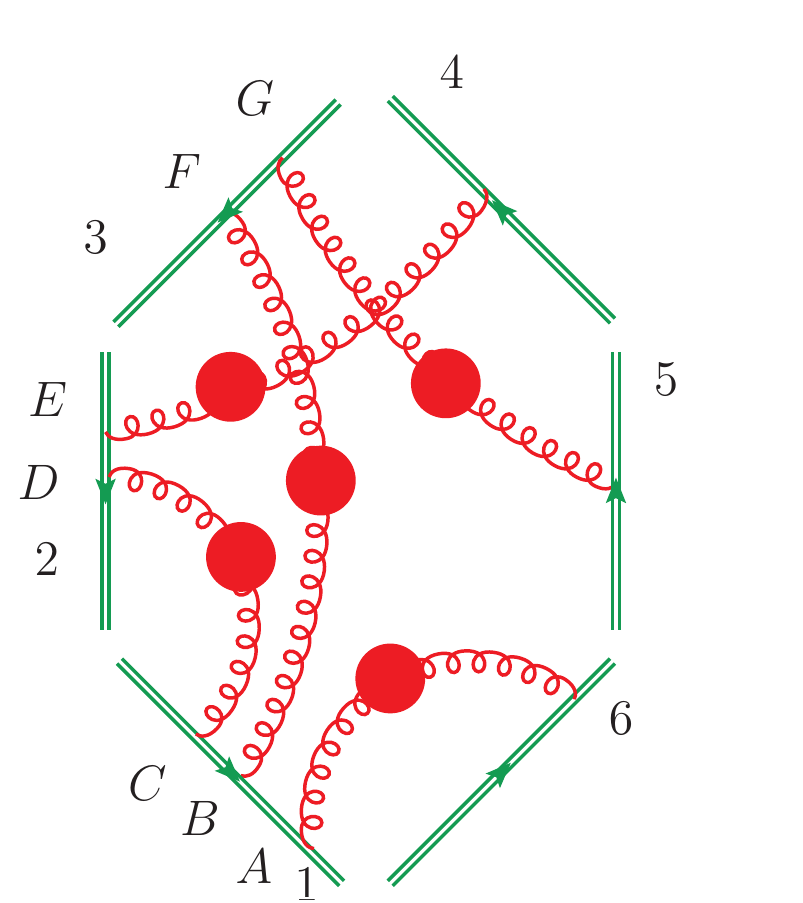}
	\captionof{figure}{$W^{(5)}_{6} (3, 2, 2, 1, 1, 1)$}	
	\label{Diag:17}
\end{minipage}
\begin{minipage}{0.45\textwidth}
	\footnotesize{
		\begin{tabular}{ | c | c | c |}
			\hline
			\textbf{Diagrams} & \textbf{Sequences} & \textbf{$s$-factors} \\ \hline
			$d_{1}$ & $\lbrace \lbrace ABC \rbrace,  \lbrace DE \rbrace, \lbrace FG \rbrace\rbrace$ &8 \\ 
			\hline
			$d_{2}$ & $\lbrace \lbrace ABC \rbrace,  \lbrace DE \rbrace, \lbrace GF \rbrace\rbrace$ & 7 \\ 
			\hline
			$d_{3}$ & $\lbrace \lbrace ABC \rbrace,  \lbrace ED \rbrace, \lbrace FG \rbrace\rbrace$ & 2 \\ 
			\hline
			$d_4$ & $\lbrace \lbrace ABC \rbrace,  \lbrace ED \rbrace, \lbrace GF \rbrace\rbrace$ & 3 \\ 
			\hline
			$d_{5}$ & $\lbrace \lbrace ACB \rbrace,  \lbrace DE \rbrace, \lbrace FG \rbrace\rbrace$ & 4 \\ 
			\hline
			$d_{6}$ & $\lbrace \lbrace ACB \rbrace,  \lbrace DE \rbrace, \lbrace GF \rbrace\rbrace$ & 11 \\ 
			\hline
			$d_7$ & $\lbrace \lbrace ACB \rbrace,  \lbrace ED \rbrace, \lbrace FG \rbrace\rbrace$ & 1 \\ 
			\hline
			$d_{8}$ & $\lbrace \lbrace ACB \rbrace,  \lbrace DE \rbrace, \lbrace GF \rbrace\rbrace$ & 4 \\ 
			\hline
			$d_{9}$ & $\lbrace \lbrace BAC \rbrace,  \lbrace DE \rbrace, \lbrace FG \rbrace\rbrace$ & 8 \\ 
			\hline
			$d_{10}$ & $\lbrace \lbrace BAC \rbrace,  \lbrace DE \rbrace, \lbrace GF \rbrace\rbrace$ & 2 \\ 
			\hline
			$d_{11}$ & $\lbrace \lbrace BAC \rbrace,  \lbrace ED \rbrace, \lbrace FG \rbrace\rbrace$ & 7 \\ 
			\hline
			$d_{12}$ & $\lbrace \lbrace BAC \rbrace,  \lbrace ED \rbrace, \lbrace GF \rbrace\rbrace$ & 3 \\ 
			\hline
			$d_{13}$ & $\lbrace \lbrace BCA \rbrace,  \lbrace DE \rbrace, \lbrace FG \rbrace\rbrace$ & 4 \\ 
			\hline	
			$d_{14}$ & $\lbrace \lbrace BCA \rbrace,  \lbrace DE \rbrace, \lbrace GF \rbrace\rbrace$ & 1 \\ 
			\hline
			$d_{15}$ & $\lbrace \lbrace BCA \rbrace,  \lbrace ED \rbrace, \lbrace FG \rbrace\rbrace$ & 11 \\ 
			\hline	
			$d_{16}$ & $\lbrace \lbrace BCA \rbrace,  \lbrace ED \rbrace, \lbrace GF \rbrace\rbrace$ & 4 \\ 
			\hline
			$d_{17}$ & $\lbrace \lbrace CAB \rbrace,  \lbrace DE \rbrace, \lbrace FG \rbrace\rbrace$ & 3 \\ 
			\hline
			$d_{18}$ & $\lbrace \lbrace CAB \rbrace,  \lbrace DE \rbrace, \lbrace GF \rbrace\rbrace$ & 7 \\ 
			\hline
			$d_{19}$ & $\lbrace \lbrace CAB \rbrace,  \lbrace ED \rbrace, \lbrace FG \rbrace\rbrace$ & 2 \\ 
			\hline		
			$d_{20}$ & $\lbrace \lbrace CAB \rbrace,  \lbrace ED \rbrace, \lbrace GF \rbrace\rbrace$ & 8 \\ 
			\hline		
			$d_{21}$ & $\lbrace \lbrace CBA \rbrace,  \lbrace DE \rbrace, \lbrace FG \rbrace\rbrace$ & 3 \\ 
			\hline
			$d_{22}$ & $\lbrace \lbrace CBA \rbrace,  \lbrace DE \rbrace, \lbrace GF \rbrace\rbrace$ & 2 \\ 
			\hline						
			$d_{23}$ & $\lbrace \lbrace CBA \rbrace,  \lbrace ED \rbrace, \lbrace FG \rbrace\rbrace$ & 7 \\ 
			\hline					
			$d_{24}$ & $\lbrace \lbrace CBA \rbrace,  \lbrace ED \rbrace, \lbrace GF \rbrace\rbrace$ & 8 \\ 
			\hline
			
		\end{tabular}
		\captionof{table}{Sequences and $s$-factors}
		\label{Table:Diag17}	}
\end{minipage} 
\vspace{0.7cm}

\vspace{0.5cm}
\noindent   The mixing matrix $R$, the diagonalizing matrix $Y$ and the diagonal matrix $D$ for the Cweb mentioned above are

\begin{equation}
	\hspace{-1cm} R= \frac{1}{60} \left( \,\,\,
	\begin{array}{cccccccccccccccccccccccc}
		1 &\hspace{0.20cm} \hspace{-0.25cm}-1 &\hspace{0.20cm} \hspace{-0.25cm}-1 &\hspace{0.20cm} 1 &\hspace{0.20cm} \hspace{-0.25cm}-2 &\hspace{0.20cm} 2 &\hspace{0.20cm} 2 &\hspace{0.20cm} \hspace{-0.25cm}-2 &\hspace{0.20cm} 1 &\hspace{0.20cm} \hspace{-0.25cm}-1 &\hspace{0.20cm} \hspace{-0.25cm}-1 &\hspace{0.20cm} 1 &\hspace{0.20cm} \hspace{-0.25cm}-2 &\hspace{0.20cm} 2 &\hspace{0.20cm} 2 &\hspace{0.20cm} \hspace{-0.25cm}-2 &\hspace{0.20cm} 1 &\hspace{0.20cm} \hspace{-0.25cm}-1 &\hspace{0.20cm} \hspace{-0.25cm}-1 &\hspace{0.20cm} 1 &\hspace{0.20cm} 1 &\hspace{0.20cm} \hspace{-0.25cm}-1 &\hspace{0.20cm} \hspace{-0.25cm}-1 &\hspace{0.20cm} 1 \\
		\hspace{-0.25cm}-4 &\hspace{0.20cm} 4 &\hspace{0.20cm} 4 &\hspace{0.20cm} \hspace{-0.25cm}-4 &\hspace{0.20cm} 3 &\hspace{0.20cm} \hspace{-0.25cm}-3 &\hspace{0.20cm} \hspace{-0.25cm}-3 &\hspace{0.20cm} 3 &\hspace{0.20cm} 1 &\hspace{0.20cm} \hspace{-0.25cm}-1 &\hspace{0.20cm} \hspace{-0.25cm}-1 &\hspace{0.20cm} 1 &\hspace{0.20cm} 3 &\hspace{0.20cm} \hspace{-0.25cm}-3 &\hspace{0.20cm} \hspace{-0.25cm}-3 &\hspace{0.20cm} 3 &\hspace{0.20cm} 1 &\hspace{0.20cm} \hspace{-0.25cm}-1 &\hspace{0.20cm} \hspace{-0.25cm}-1 &\hspace{0.20cm} 1 &\hspace{0.20cm} \hspace{-0.25cm}-4 &\hspace{0.20cm} 4 &\hspace{0.20cm} 4 &\hspace{0.20cm} \hspace{-0.25cm}-4 \\
		\hspace{-0.25cm}-9 &\hspace{0.20cm} 9 &\hspace{0.20cm} 9 &\hspace{0.20cm} \hspace{-0.25cm}-9 &\hspace{0.20cm} 3 &\hspace{0.20cm} \hspace{-0.25cm}-3 &\hspace{0.20cm} \hspace{-0.25cm}-3 &\hspace{0.20cm} 3 &\hspace{0.20cm} 6 &\hspace{0.20cm} \hspace{-0.25cm}-6 &\hspace{0.20cm} \hspace{-0.25cm}-6 &\hspace{0.20cm} 6 &\hspace{0.20cm} 3 &\hspace{0.20cm} \hspace{-0.25cm}-3 &\hspace{0.20cm} \hspace{-0.25cm}-3 &\hspace{0.20cm} 3 &\hspace{0.20cm} 6 &\hspace{0.20cm} \hspace{-0.25cm}-6 &\hspace{0.20cm} \hspace{-0.25cm}-6 &\hspace{0.20cm} 6 &\hspace{0.20cm} \hspace{-0.25cm}-9 &\hspace{0.20cm} 9 &\hspace{0.20cm} 9 &\hspace{0.20cm} \hspace{-0.25cm}-9 \\
		6 &\hspace{0.20cm} \hspace{-0.25cm}-6 &\hspace{0.20cm} \hspace{-0.25cm}-6 &\hspace{0.20cm} 6 &\hspace{0.20cm} \hspace{-0.25cm}-2 &\hspace{0.20cm} 2 &\hspace{0.20cm} 2 &\hspace{0.20cm} \hspace{-0.25cm}-2 &\hspace{0.20cm} \hspace{-0.25cm}-4 &\hspace{0.20cm} 4 &\hspace{0.20cm} 4 &\hspace{0.20cm} \hspace{-0.25cm}-4 &\hspace{0.20cm} \hspace{-0.25cm}-2 &\hspace{0.20cm} 2 &\hspace{0.20cm} 2 &\hspace{0.20cm} \hspace{-0.25cm}-2 &\hspace{0.20cm} \hspace{-0.25cm}-4 &\hspace{0.20cm} 4 &\hspace{0.20cm} 4 &\hspace{0.20cm} \hspace{-0.25cm}-4 &\hspace{0.20cm} 6 &\hspace{0.20cm} \hspace{-0.25cm}-6 &\hspace{0.20cm} \hspace{-0.25cm}-6 &\hspace{0.20cm} 6 \\
		\hspace{-0.25cm}-4 &\hspace{0.20cm} 4 &\hspace{0.20cm} 4 &\hspace{0.20cm} \hspace{-0.25cm}-4 &\hspace{0.20cm} 3 &\hspace{0.20cm} \hspace{-0.25cm}-3 &\hspace{0.20cm} \hspace{-0.25cm}-3 &\hspace{0.20cm} 3 &\hspace{0.20cm} 1 &\hspace{0.20cm} \hspace{-0.25cm}-1 &\hspace{0.20cm} \hspace{-0.25cm}-1 &\hspace{0.20cm} 1 &\hspace{0.20cm} 3 &\hspace{0.20cm} \hspace{-0.25cm}-3 &\hspace{0.20cm} \hspace{-0.25cm}-3 &\hspace{0.20cm} 3 &\hspace{0.20cm} 1 &\hspace{0.20cm} \hspace{-0.25cm}-1 &\hspace{0.20cm} \hspace{-0.25cm}-1 &\hspace{0.20cm} 1 &\hspace{0.20cm} \hspace{-0.25cm}-4 &\hspace{0.20cm} 4 &\hspace{0.20cm} 4 &\hspace{0.20cm} \hspace{-0.25cm}-4 \\
		1 &\hspace{0.20cm} \hspace{-0.25cm}-1 &\hspace{0.20cm} \hspace{-0.25cm}-1 &\hspace{0.20cm} 1 &\hspace{0.20cm} \hspace{-0.25cm}-2 &\hspace{0.20cm} 2 &\hspace{0.20cm} 2 &\hspace{0.20cm} \hspace{-0.25cm}-2 &\hspace{0.20cm} 1 &\hspace{0.20cm} \hspace{-0.25cm}-1 &\hspace{0.20cm} \hspace{-0.25cm}-1 &\hspace{0.20cm} 1 &\hspace{0.20cm} \hspace{-0.25cm}-2 &\hspace{0.20cm} 2 &\hspace{0.20cm} 2 &\hspace{0.20cm} \hspace{-0.25cm}-2 &\hspace{0.20cm} 1 &\hspace{0.20cm} \hspace{-0.25cm}-1 &\hspace{0.20cm} \hspace{-0.25cm}-1 &\hspace{0.20cm} 1 &\hspace{0.20cm} 1 &\hspace{0.20cm} \hspace{-0.25cm}-1 &\hspace{0.20cm} \hspace{-0.25cm}-1 &\hspace{0.20cm} 1 \\
		6 &\hspace{0.20cm} \hspace{-0.25cm}-6 &\hspace{0.20cm} \hspace{-0.25cm}-6 &\hspace{0.20cm} 6 &\hspace{0.20cm} \hspace{-0.25cm}-12 &\hspace{0.20cm} 12 &\hspace{0.20cm} 12 &\hspace{0.20cm} \hspace{-0.25cm}-12 &\hspace{0.20cm} 6 &\hspace{0.20cm} \hspace{-0.25cm}-6 &\hspace{0.20cm} \hspace{-0.25cm}-6 &\hspace{0.20cm} 6 &\hspace{0.20cm} \hspace{-0.25cm}-12 &\hspace{0.20cm} 12 &\hspace{0.20cm} 12 &\hspace{0.20cm} \hspace{-0.25cm}-12 &\hspace{0.20cm} 6 &\hspace{0.20cm} \hspace{-0.25cm}-6 &\hspace{0.20cm} \hspace{-0.25cm}-6 &\hspace{0.20cm} 6 &\hspace{0.20cm} 6 &\hspace{0.20cm} \hspace{-0.25cm}-6 &\hspace{0.20cm} \hspace{-0.25cm}-6 &\hspace{0.20cm} 6 \\
		1 &\hspace{0.20cm} \hspace{-0.25cm}-1 &\hspace{0.20cm} \hspace{-0.25cm}-1 &\hspace{0.20cm} 1 &\hspace{0.20cm} 3 &\hspace{0.20cm} \hspace{-0.25cm}-3 &\hspace{0.20cm} \hspace{-0.25cm}-3 &\hspace{0.20cm} 3 &\hspace{0.20cm} \hspace{-0.25cm}-4 &\hspace{0.20cm} 4 &\hspace{0.20cm} 4 &\hspace{0.20cm} \hspace{-0.25cm}-4 &\hspace{0.20cm} 3 &\hspace{0.20cm} \hspace{-0.25cm}-3 &\hspace{0.20cm} \hspace{-0.25cm}-3 &\hspace{0.20cm} 3 &\hspace{0.20cm} \hspace{-0.25cm}-4 &\hspace{0.20cm} 4 &\hspace{0.20cm} 4 &\hspace{0.20cm} \hspace{-0.25cm}-4 &\hspace{0.20cm} 1 &\hspace{0.20cm} \hspace{-0.25cm}-1 &\hspace{0.20cm} \hspace{-0.25cm}-1 &\hspace{0.20cm} 1 \\
		1 &\hspace{0.20cm} \hspace{-0.25cm}-1 &\hspace{0.20cm} \hspace{-0.25cm}-1 &\hspace{0.20cm} 1 &\hspace{0.20cm} \hspace{-0.25cm}-2 &\hspace{0.20cm} 2 &\hspace{0.20cm} 2 &\hspace{0.20cm} \hspace{-0.25cm}-2 &\hspace{0.20cm} 1 &\hspace{0.20cm} \hspace{-0.25cm}-1 &\hspace{0.20cm} \hspace{-0.25cm}-1 &\hspace{0.20cm} 1 &\hspace{0.20cm} \hspace{-0.25cm}-2 &\hspace{0.20cm} 2 &\hspace{0.20cm} 2 &\hspace{0.20cm} \hspace{-0.25cm}-2 &\hspace{0.20cm} 1 &\hspace{0.20cm} \hspace{-0.25cm}-1 &\hspace{0.20cm} \hspace{-0.25cm}-1 &\hspace{0.20cm} 1 &\hspace{0.20cm} 1 &\hspace{0.20cm} \hspace{-0.25cm}-1 &\hspace{0.20cm} \hspace{-0.25cm}-1 &\hspace{0.20cm} 1 \\
		6 &\hspace{0.20cm} \hspace{-0.25cm}-6 &\hspace{0.20cm} \hspace{-0.25cm}-6 &\hspace{0.20cm} 6 &\hspace{0.20cm} 3 &\hspace{0.20cm} \hspace{-0.25cm}-3 &\hspace{0.20cm} \hspace{-0.25cm}-3 &\hspace{0.20cm} 3 &\hspace{0.20cm} \hspace{-0.25cm}-9 &\hspace{0.20cm} 9 &\hspace{0.20cm} 9 &\hspace{0.20cm} \hspace{-0.25cm}-9 &\hspace{0.20cm} 3 &\hspace{0.20cm} \hspace{-0.25cm}-3 &\hspace{0.20cm} \hspace{-0.25cm}-3 &\hspace{0.20cm} 3 &\hspace{0.20cm} \hspace{-0.25cm}-9 &\hspace{0.20cm} 9 &\hspace{0.20cm} 9 &\hspace{0.20cm} \hspace{-0.25cm}-9 &\hspace{0.20cm} 6 &\hspace{0.20cm} \hspace{-0.25cm}-6 &\hspace{0.20cm} \hspace{-0.25cm}-6 &\hspace{0.20cm} 6 \\
		1 &\hspace{0.20cm} \hspace{-0.25cm}-1 &\hspace{0.20cm} \hspace{-0.25cm}-1 &\hspace{0.20cm} 1 &\hspace{0.20cm} 3 &\hspace{0.20cm} \hspace{-0.25cm}-3 &\hspace{0.20cm} \hspace{-0.25cm}-3 &\hspace{0.20cm} 3 &\hspace{0.20cm} \hspace{-0.25cm}-4 &\hspace{0.20cm} 4 &\hspace{0.20cm} 4 &\hspace{0.20cm} \hspace{-0.25cm}-4 &\hspace{0.20cm} 3 &\hspace{0.20cm} \hspace{-0.25cm}-3 &\hspace{0.20cm} \hspace{-0.25cm}-3 &\hspace{0.20cm} 3 &\hspace{0.20cm} \hspace{-0.25cm}-4 &\hspace{0.20cm} 4 &\hspace{0.20cm} 4 &\hspace{0.20cm} \hspace{-0.25cm}-4 &\hspace{0.20cm} 1 &\hspace{0.20cm} \hspace{-0.25cm}-1 &\hspace{0.20cm} \hspace{-0.25cm}-1 &\hspace{0.20cm} 1 \\
		\hspace{-0.25cm}-4 &\hspace{0.20cm} 4 &\hspace{0.20cm} 4 &\hspace{0.20cm} \hspace{-0.25cm}-4 &\hspace{0.20cm} \hspace{-0.25cm}-2 &\hspace{0.20cm} 2 &\hspace{0.20cm} 2 &\hspace{0.20cm} \hspace{-0.25cm}-2 &\hspace{0.20cm} 6 &\hspace{0.20cm} \hspace{-0.25cm}-6 &\hspace{0.20cm} \hspace{-0.25cm}-6 &\hspace{0.20cm} 6 &\hspace{0.20cm} \hspace{-0.25cm}-2 &\hspace{0.20cm} 2 &\hspace{0.20cm} 2 &\hspace{0.20cm} \hspace{-0.25cm}-2 &\hspace{0.20cm} 6 &\hspace{0.20cm} \hspace{-0.25cm}-6 &\hspace{0.20cm} \hspace{-0.25cm}-6 &\hspace{0.20cm} 6 &\hspace{0.20cm} \hspace{-0.25cm}-4 &\hspace{0.20cm} 4 &\hspace{0.20cm} 4 &\hspace{0.20cm} \hspace{-0.25cm}-4 \\
		1 &\hspace{0.20cm} \hspace{-0.25cm}-1 &\hspace{0.20cm} \hspace{-0.25cm}-1 &\hspace{0.20cm} 1 &\hspace{0.20cm} 3 &\hspace{0.20cm} \hspace{-0.25cm}-3 &\hspace{0.20cm} \hspace{-0.25cm}-3 &\hspace{0.20cm} 3 &\hspace{0.20cm} \hspace{-0.25cm}-4 &\hspace{0.20cm} 4 &\hspace{0.20cm} 4 &\hspace{0.20cm} \hspace{-0.25cm}-4 &\hspace{0.20cm} 3 &\hspace{0.20cm} \hspace{-0.25cm}-3 &\hspace{0.20cm} \hspace{-0.25cm}-3 &\hspace{0.20cm} 3 &\hspace{0.20cm} \hspace{-0.25cm}-4 &\hspace{0.20cm} 4 &\hspace{0.20cm} 4 &\hspace{0.20cm} \hspace{-0.25cm}-4 &\hspace{0.20cm} 1 &\hspace{0.20cm} \hspace{-0.25cm}-1 &\hspace{0.20cm} \hspace{-0.25cm}-1 &\hspace{0.20cm} 1 \\
		6 &\hspace{0.20cm} \hspace{-0.25cm}-6 &\hspace{0.20cm} \hspace{-0.25cm}-6 &\hspace{0.20cm} 6 &\hspace{0.20cm} \hspace{-0.25cm}-12 &\hspace{0.20cm} 12 &\hspace{0.20cm} 12 &\hspace{0.20cm} \hspace{-0.25cm}-12 &\hspace{0.20cm} 6 &\hspace{0.20cm} \hspace{-0.25cm}-6 &\hspace{0.20cm} \hspace{-0.25cm}-6 &\hspace{0.20cm} 6 &\hspace{0.20cm} \hspace{-0.25cm}-12 &\hspace{0.20cm} 12 &\hspace{0.20cm} 12 &\hspace{0.20cm} \hspace{-0.25cm}-12 &\hspace{0.20cm} 6 &\hspace{0.20cm} \hspace{-0.25cm}-6 &\hspace{0.20cm} \hspace{-0.25cm}-6 &\hspace{0.20cm} 6 &\hspace{0.20cm} 6 &\hspace{0.20cm} \hspace{-0.25cm}-6 &\hspace{0.20cm} \hspace{-0.25cm}-6 &\hspace{0.20cm} 6 \\
		1 &\hspace{0.20cm} \hspace{-0.25cm}-1 &\hspace{0.20cm} \hspace{-0.25cm}-1 &\hspace{0.20cm} 1 &\hspace{0.20cm} \hspace{-0.25cm}-2 &\hspace{0.20cm} 2 &\hspace{0.20cm} 2 &\hspace{0.20cm} \hspace{-0.25cm}-2 &\hspace{0.20cm} 1 &\hspace{0.20cm} \hspace{-0.25cm}-1 &\hspace{0.20cm} \hspace{-0.25cm}-1 &\hspace{0.20cm} 1 &\hspace{0.20cm} \hspace{-0.25cm}-2 &\hspace{0.20cm} 2 &\hspace{0.20cm} 2 &\hspace{0.20cm} \hspace{-0.25cm}-2 &\hspace{0.20cm} 1 &\hspace{0.20cm} \hspace{-0.25cm}-1 &\hspace{0.20cm} \hspace{-0.25cm}-1 &\hspace{0.20cm} 1 &\hspace{0.20cm} 1 &\hspace{0.20cm} \hspace{-0.25cm}-1 &\hspace{0.20cm} \hspace{-0.25cm}-1 &\hspace{0.20cm} 1 \\
		\hspace{-0.25cm}-4 &\hspace{0.20cm} 4 &\hspace{0.20cm} 4 &\hspace{0.20cm} \hspace{-0.25cm}-4 &\hspace{0.20cm} 3 &\hspace{0.20cm} \hspace{-0.25cm}-3 &\hspace{0.20cm} \hspace{-0.25cm}-3 &\hspace{0.20cm} 3 &\hspace{0.20cm} 1 &\hspace{0.20cm} \hspace{-0.25cm}-1 &\hspace{0.20cm} \hspace{-0.25cm}-1 &\hspace{0.20cm} 1 &\hspace{0.20cm} 3 &\hspace{0.20cm} \hspace{-0.25cm}-3 &\hspace{0.20cm} \hspace{-0.25cm}-3 &\hspace{0.20cm} 3 &\hspace{0.20cm} 1 &\hspace{0.20cm} \hspace{-0.25cm}-1 &\hspace{0.20cm} \hspace{-0.25cm}-1 &\hspace{0.20cm} 1 &\hspace{0.20cm} \hspace{-0.25cm}-4 &\hspace{0.20cm} 4 &\hspace{0.20cm} 4 &\hspace{0.20cm} \hspace{-0.25cm}-4 \\
		\hspace{-0.25cm}-4 &\hspace{0.20cm} 4 &\hspace{0.20cm} 4 &\hspace{0.20cm} \hspace{-0.25cm}-4 &\hspace{0.20cm} \hspace{-0.25cm}-2 &\hspace{0.20cm} 2 &\hspace{0.20cm} 2 &\hspace{0.20cm} \hspace{-0.25cm}-2 &\hspace{0.20cm} 6 &\hspace{0.20cm} \hspace{-0.25cm}-6 &\hspace{0.20cm} \hspace{-0.25cm}-6 &\hspace{0.20cm} 6 &\hspace{0.20cm} \hspace{-0.25cm}-2 &\hspace{0.20cm} 2 &\hspace{0.20cm} 2 &\hspace{0.20cm} \hspace{-0.25cm}-2 &\hspace{0.20cm} 6 &\hspace{0.20cm} \hspace{-0.25cm}-6 &\hspace{0.20cm} \hspace{-0.25cm}-6 &\hspace{0.20cm} 6 &\hspace{0.20cm} \hspace{-0.25cm}-4 &\hspace{0.20cm} 4 &\hspace{0.20cm} 4 &\hspace{0.20cm} \hspace{-0.25cm}-4 \\
		1 &\hspace{0.20cm} \hspace{-0.25cm}-1 &\hspace{0.20cm} \hspace{-0.25cm}-1 &\hspace{0.20cm} 1 &\hspace{0.20cm} 3 &\hspace{0.20cm} \hspace{-0.25cm}-3 &\hspace{0.20cm} \hspace{-0.25cm}-3 &\hspace{0.20cm} 3 &\hspace{0.20cm} \hspace{-0.25cm}-4 &\hspace{0.20cm} 4 &\hspace{0.20cm} 4 &\hspace{0.20cm} \hspace{-0.25cm}-4 &\hspace{0.20cm} 3 &\hspace{0.20cm} \hspace{-0.25cm}-3 &\hspace{0.20cm} \hspace{-0.25cm}-3 &\hspace{0.20cm} 3 &\hspace{0.20cm} \hspace{-0.25cm}-4 &\hspace{0.20cm} 4 &\hspace{0.20cm} 4 &\hspace{0.20cm} \hspace{-0.25cm}-4 &\hspace{0.20cm} 1 &\hspace{0.20cm} \hspace{-0.25cm}-1 &\hspace{0.20cm} \hspace{-0.25cm}-1 &\hspace{0.20cm} 1 \\
		6 &\hspace{0.20cm} \hspace{-0.25cm}-6 &\hspace{0.20cm} \hspace{-0.25cm}-6 &\hspace{0.20cm} 6 &\hspace{0.20cm} 3 &\hspace{0.20cm} \hspace{-0.25cm}-3 &\hspace{0.20cm} \hspace{-0.25cm}-3 &\hspace{0.20cm} 3 &\hspace{0.20cm} \hspace{-0.25cm}-9 &\hspace{0.20cm} 9 &\hspace{0.20cm} 9 &\hspace{0.20cm} \hspace{-0.25cm}-9 &\hspace{0.20cm} 3 &\hspace{0.20cm} \hspace{-0.25cm}-3 &\hspace{0.20cm} \hspace{-0.25cm}-3 &\hspace{0.20cm} 3 &\hspace{0.20cm} \hspace{-0.25cm}-9 &\hspace{0.20cm} 9 &\hspace{0.20cm} 9 &\hspace{0.20cm} \hspace{-0.25cm}-9 &\hspace{0.20cm} 6 &\hspace{0.20cm} \hspace{-0.25cm}-6 &\hspace{0.20cm} \hspace{-0.25cm}-6 &\hspace{0.20cm} 6 \\
		1 &\hspace{0.20cm} \hspace{-0.25cm}-1 &\hspace{0.20cm} \hspace{-0.25cm}-1 &\hspace{0.20cm} 1 &\hspace{0.20cm} \hspace{-0.25cm}-2 &\hspace{0.20cm} 2 &\hspace{0.20cm} 2 &\hspace{0.20cm} \hspace{-0.25cm}-2 &\hspace{0.20cm} 1 &\hspace{0.20cm} \hspace{-0.25cm}-1 &\hspace{0.20cm} \hspace{-0.25cm}-1 &\hspace{0.20cm} 1 &\hspace{0.20cm} \hspace{-0.25cm}-2 &\hspace{0.20cm} 2 &\hspace{0.20cm} 2 &\hspace{0.20cm} \hspace{-0.25cm}-2 &\hspace{0.20cm} 1 &\hspace{0.20cm} \hspace{-0.25cm}-1 &\hspace{0.20cm} \hspace{-0.25cm}-1 &\hspace{0.20cm} 1 &\hspace{0.20cm} 1 &\hspace{0.20cm} \hspace{-0.25cm}-1 &\hspace{0.20cm} \hspace{-0.25cm}-1 &\hspace{0.20cm} 1 \\
		6 &\hspace{0.20cm} \hspace{-0.25cm}-6 &\hspace{0.20cm} \hspace{-0.25cm}-6 &\hspace{0.20cm} 6 &\hspace{0.20cm} \hspace{-0.25cm}-2 &\hspace{0.20cm} 2 &\hspace{0.20cm} 2 &\hspace{0.20cm} \hspace{-0.25cm}-2 &\hspace{0.20cm} \hspace{-0.25cm}-4 &\hspace{0.20cm} 4 &\hspace{0.20cm} 4 &\hspace{0.20cm} \hspace{-0.25cm}-4 &\hspace{0.20cm} \hspace{-0.25cm}-2 &\hspace{0.20cm} 2 &\hspace{0.20cm} 2 &\hspace{0.20cm} \hspace{-0.25cm}-2 &\hspace{0.20cm} \hspace{-0.25cm}-4 &\hspace{0.20cm} 4 &\hspace{0.20cm} 4 &\hspace{0.20cm} \hspace{-0.25cm}-4 &\hspace{0.20cm} 6 &\hspace{0.20cm} \hspace{-0.25cm}-6 &\hspace{0.20cm} \hspace{-0.25cm}-6 &\hspace{0.20cm} 6 \\
		\hspace{-0.25cm}-9 &\hspace{0.20cm} 9 &\hspace{0.20cm} 9 &\hspace{0.20cm} \hspace{-0.25cm}-9 &\hspace{0.20cm} 3 &\hspace{0.20cm} \hspace{-0.25cm}-3 &\hspace{0.20cm} \hspace{-0.25cm}-3 &\hspace{0.20cm} 3 &\hspace{0.20cm} 6 &\hspace{0.20cm} \hspace{-0.25cm}-6 &\hspace{0.20cm} \hspace{-0.25cm}-6 &\hspace{0.20cm} 6 &\hspace{0.20cm} 3 &\hspace{0.20cm} \hspace{-0.25cm}-3 &\hspace{0.20cm} \hspace{-0.25cm}-3 &\hspace{0.20cm} 3 &\hspace{0.20cm} 6 &\hspace{0.20cm} \hspace{-0.25cm}-6 &\hspace{0.20cm} \hspace{-0.25cm}-6 &\hspace{0.20cm} 6 &\hspace{0.20cm} \hspace{-0.25cm}-9 &\hspace{0.20cm} 9 &\hspace{0.20cm} 9 &\hspace{0.20cm} \hspace{-0.25cm}-9 \\
		\hspace{-0.25cm}-4 &\hspace{0.20cm} 4 &\hspace{0.20cm} 4 &\hspace{0.20cm} \hspace{-0.25cm}-4 &\hspace{0.20cm} 3 &\hspace{0.20cm} \hspace{-0.25cm}-3 &\hspace{0.20cm} \hspace{-0.25cm}-3 &\hspace{0.20cm} 3 &\hspace{0.20cm} 1 &\hspace{0.20cm} \hspace{-0.25cm}-1 &\hspace{0.20cm} \hspace{-0.25cm}-1 &\hspace{0.20cm} 1 &\hspace{0.20cm} 3 &\hspace{0.20cm} \hspace{-0.25cm}-3 &\hspace{0.20cm} \hspace{-0.25cm}-3 &\hspace{0.20cm} 3 &\hspace{0.20cm} 1 &\hspace{0.20cm} \hspace{-0.25cm}-1 &\hspace{0.20cm} \hspace{-0.25cm}-1 &\hspace{0.20cm} 1 &\hspace{0.20cm} \hspace{-0.25cm}-4 &\hspace{0.20cm} 4 &\hspace{0.20cm} 4 &\hspace{0.20cm} \hspace{-0.25cm}-4 \\
		1 &\hspace{0.20cm} \hspace{-0.25cm}-1 &\hspace{0.20cm} \hspace{-0.25cm}-1 &\hspace{0.20cm} 1 &\hspace{0.20cm} \hspace{-0.25cm}-2 &\hspace{0.20cm} 2 &\hspace{0.20cm} 2 &\hspace{0.20cm} \hspace{-0.25cm}-2 &\hspace{0.20cm} 1 &\hspace{0.20cm} \hspace{-0.25cm}-1 &\hspace{0.20cm} \hspace{-0.25cm}-1 &\hspace{0.20cm} 1 &\hspace{0.20cm} \hspace{-0.25cm}-2 &\hspace{0.20cm} 2 &\hspace{0.20cm} 2 &\hspace{0.20cm} \hspace{-0.25cm}-2 &\hspace{0.20cm} 1 &\hspace{0.20cm} \hspace{-0.25cm}-1 &\hspace{0.20cm} \hspace{-0.25cm}-1 &\hspace{0.20cm} 1 &\hspace{0.20cm} 1 &\hspace{0.20cm} \hspace{-0.25cm}-1 &\hspace{0.20cm} \hspace{-0.25cm}-1 &\hspace{0.20cm} 1 \\
	\end{array}
	\right), \nonumber
\end{equation}
\begin{equation}
	Y=\left( \,\,\,
	\begin{array}{cccccccccccccccccccccccc}
		1 &\hspace{0.20cm} \hspace{-0.25cm}-1 &\hspace{0.20cm} \hspace{-0.25cm}-1 &\hspace{0.20cm} 1 &\hspace{0.20cm} \hspace{-0.25cm}-1 &\hspace{0.20cm} 1 &\hspace{0.20cm} 1 &\hspace{0.20cm} \hspace{-0.25cm}-1 &\hspace{0.20cm} 0 &\hspace{0.20cm} 0 &\hspace{0.20cm} 0 &\hspace{0.20cm} 0 &\hspace{0.20cm} \hspace{-0.25cm}-1 &\hspace{0.20cm} 1 &\hspace{0.20cm} 1 &\hspace{0.20cm} \hspace{-0.25cm}-1 &\hspace{0.20cm} 0 &\hspace{0.20cm} 0 &\hspace{0.20cm} 0 &\hspace{0.20cm} 0 &\hspace{0.20cm} 1 &\hspace{0.20cm} \hspace{-0.25cm}-1 &\hspace{0.20cm} \hspace{-0.25cm}-1 &\hspace{0.20cm} 1 \\
		0 &\hspace{0.20cm} 0 &\hspace{0.20cm} 0 &\hspace{0.20cm} 0 &\hspace{0.20cm} \hspace{-0.25cm}-1 &\hspace{0.20cm} 1 &\hspace{0.20cm} 1 &\hspace{0.20cm} \hspace{-0.25cm}-1 &\hspace{0.20cm} 1 &\hspace{0.20cm} \hspace{-0.25cm}-1 &\hspace{0.20cm} \hspace{-0.25cm}-1 &\hspace{0.20cm} 1 &\hspace{0.20cm} \hspace{-0.25cm}-1 &\hspace{0.20cm} 1 &\hspace{0.20cm} 1 &\hspace{0.20cm} \hspace{-0.25cm}-1 &\hspace{0.20cm} 1 &\hspace{0.20cm} \hspace{-0.25cm}-1 &\hspace{0.20cm} \hspace{-0.25cm}-1 &\hspace{0.20cm} 1 &\hspace{0.20cm} 0 &\hspace{0.20cm} 0 &\hspace{0.20cm} 0 &\hspace{0.20cm} 0 \\
		\hspace{-0.25cm}-1 &\hspace{0.20cm} 0 &\hspace{0.20cm} 0 &\hspace{0.20cm} 0 &\hspace{0.20cm} 0 &\hspace{0.20cm} 0 &\hspace{0.20cm} 0 &\hspace{0.20cm} 0 &\hspace{0.20cm} 0 &\hspace{0.20cm} 0 &\hspace{0.20cm} 0 &\hspace{0.20cm} 0 &\hspace{0.20cm} 0 &\hspace{0.20cm} 0 &\hspace{0.20cm} 0 &\hspace{0.20cm} 0 &\hspace{0.20cm} 0 &\hspace{0.20cm} 0 &\hspace{0.20cm} 0 &\hspace{0.20cm} 0 &\hspace{0.20cm} 0 &\hspace{0.20cm} 0 &\hspace{0.20cm} 0 &\hspace{0.20cm} 1 \\
		0 &\hspace{0.20cm} \hspace{-0.25cm}-1 &\hspace{0.20cm} 0 &\hspace{0.20cm} 0 &\hspace{0.20cm} 0 &\hspace{0.20cm} 0 &\hspace{0.20cm} 0 &\hspace{0.20cm} 0 &\hspace{0.20cm} 0 &\hspace{0.20cm} 0 &\hspace{0.20cm} 0 &\hspace{0.20cm} 0 &\hspace{0.20cm} 0 &\hspace{0.20cm} 0 &\hspace{0.20cm} 0 &\hspace{0.20cm} 0 &\hspace{0.20cm} 0 &\hspace{0.20cm} 0 &\hspace{0.20cm} 0 &\hspace{0.20cm} 0 &\hspace{0.20cm} 0 &\hspace{0.20cm} 0 &\hspace{0.20cm} 1 &\hspace{0.20cm} 0 \\
		\hspace{-0.25cm}-3 &\hspace{0.20cm} \hspace{-0.25cm}-3 &\hspace{0.20cm} 0 &\hspace{0.20cm} 0 &\hspace{0.20cm} 0 &\hspace{0.20cm} 0 &\hspace{0.20cm} 0 &\hspace{0.20cm} 0 &\hspace{0.20cm} 0 &\hspace{0.20cm} 0 &\hspace{0.20cm} 0 &\hspace{0.20cm} 0 &\hspace{0.20cm} 0 &\hspace{0.20cm} 0 &\hspace{0.20cm} 0 &\hspace{0.20cm} 0 &\hspace{0.20cm} 0 &\hspace{0.20cm} 0 &\hspace{0.20cm} 0 &\hspace{0.20cm} 0 &\hspace{0.20cm} 0 &\hspace{0.20cm} 1 &\hspace{0.20cm} 0 &\hspace{0.20cm} 0 \\
		2 &\hspace{0.20cm} 2 &\hspace{0.20cm} 0 &\hspace{0.20cm} 0 &\hspace{0.20cm} 0 &\hspace{0.20cm} 0 &\hspace{0.20cm} 0 &\hspace{0.20cm} 0 &\hspace{0.20cm} 0 &\hspace{0.20cm} 0 &\hspace{0.20cm} 0 &\hspace{0.20cm} 0 &\hspace{0.20cm} 0 &\hspace{0.20cm} 0 &\hspace{0.20cm} 0 &\hspace{0.20cm} 0 &\hspace{0.20cm} 0 &\hspace{0.20cm} 0 &\hspace{0.20cm} 0 &\hspace{0.20cm} 0 &\hspace{0.20cm} 1 &\hspace{0.20cm} 0 &\hspace{0.20cm} 0 &\hspace{0.20cm} 0 \\
		\hspace{-0.25cm}-1 &\hspace{0.20cm} 0 &\hspace{0.20cm} 0 &\hspace{0.20cm} 0 &\hspace{0.20cm} 0 &\hspace{0.20cm} 0 &\hspace{0.20cm} 0 &\hspace{0.20cm} 0 &\hspace{0.20cm} 0 &\hspace{0.20cm} 0 &\hspace{0.20cm} 0 &\hspace{0.20cm} 0 &\hspace{0.20cm} 0 &\hspace{0.20cm} 0 &\hspace{0.20cm} 0 &\hspace{0.20cm} 0 &\hspace{0.20cm} 0 &\hspace{0.20cm} 0 &\hspace{0.20cm} 0 &\hspace{0.20cm} 1 &\hspace{0.20cm} 0 &\hspace{0.20cm} 0 &\hspace{0.20cm} 0 &\hspace{0.20cm} 0 \\
		6 &\hspace{0.20cm} 3 &\hspace{0.20cm} 0 &\hspace{0.20cm} 0 &\hspace{0.20cm} 0 &\hspace{0.20cm} 0 &\hspace{0.20cm} 0 &\hspace{0.20cm} 0 &\hspace{0.20cm} 0 &\hspace{0.20cm} 0 &\hspace{0.20cm} 0 &\hspace{0.20cm} 0 &\hspace{0.20cm} 0 &\hspace{0.20cm} 0 &\hspace{0.20cm} 0 &\hspace{0.20cm} 0 &\hspace{0.20cm} 0 &\hspace{0.20cm} 0 &\hspace{0.20cm} 1 &\hspace{0.20cm} 0 &\hspace{0.20cm} 0 &\hspace{0.20cm} 0 &\hspace{0.20cm} 0 &\hspace{0.20cm} 0 \\
		3 &\hspace{0.20cm} 1 &\hspace{0.20cm} 0 &\hspace{0.20cm} 0 &\hspace{0.20cm} 0 &\hspace{0.20cm} 0 &\hspace{0.20cm} 0 &\hspace{0.20cm} 0 &\hspace{0.20cm} 0 &\hspace{0.20cm} 0 &\hspace{0.20cm} 0 &\hspace{0.20cm} 0 &\hspace{0.20cm} 0 &\hspace{0.20cm} 0 &\hspace{0.20cm} 0 &\hspace{0.20cm} 0 &\hspace{0.20cm} 0 &\hspace{0.20cm} 1 &\hspace{0.20cm} 0 &\hspace{0.20cm} 0 &\hspace{0.20cm} 0 &\hspace{0.20cm} 0 &\hspace{0.20cm} 0 &\hspace{0.20cm} 0 \\
		\hspace{-0.25cm}-4 &\hspace{0.20cm} \hspace{-0.25cm}-2 &\hspace{0.20cm} 0 &\hspace{0.20cm} 0 &\hspace{0.20cm} 0 &\hspace{0.20cm} 0 &\hspace{0.20cm} 0 &\hspace{0.20cm} 0 &\hspace{0.20cm} 0 &\hspace{0.20cm} 0 &\hspace{0.20cm} 0 &\hspace{0.20cm} 0 &\hspace{0.20cm} 0 &\hspace{0.20cm} 0 &\hspace{0.20cm} 0 &\hspace{0.20cm} 0 &\hspace{0.20cm} 1 &\hspace{0.20cm} 0 &\hspace{0.20cm} 0 &\hspace{0.20cm} 0 &\hspace{0.20cm} 0 &\hspace{0.20cm} 0 &\hspace{0.20cm} 0 &\hspace{0.20cm} 0 \\
		0 &\hspace{0.20cm} \hspace{-0.25cm}-1 &\hspace{0.20cm} 0 &\hspace{0.20cm} 0 &\hspace{0.20cm} 0 &\hspace{0.20cm} 0 &\hspace{0.20cm} 0 &\hspace{0.20cm} 0 &\hspace{0.20cm} 0 &\hspace{0.20cm} 0 &\hspace{0.20cm} 0 &\hspace{0.20cm} 0 &\hspace{0.20cm} 0 &\hspace{0.20cm} 0 &\hspace{0.20cm} 0 &\hspace{0.20cm} 1 &\hspace{0.20cm} 0 &\hspace{0.20cm} 0 &\hspace{0.20cm} 0 &\hspace{0.20cm} 0 &\hspace{0.20cm} 0 &\hspace{0.20cm} 0 &\hspace{0.20cm} 0 &\hspace{0.20cm} 0 \\
		\hspace{-0.25cm}-1 &\hspace{0.20cm} 0 &\hspace{0.20cm} 0 &\hspace{0.20cm} 0 &\hspace{0.20cm} 0 &\hspace{0.20cm} 0 &\hspace{0.20cm} 0 &\hspace{0.20cm} 0 &\hspace{0.20cm} 0 &\hspace{0.20cm} 0 &\hspace{0.20cm} 0 &\hspace{0.20cm} 0 &\hspace{0.20cm} 0 &\hspace{0.20cm} 0 &\hspace{0.20cm} 1 &\hspace{0.20cm} 0 &\hspace{0.20cm} 0 &\hspace{0.20cm} 0 &\hspace{0.20cm} 0 &\hspace{0.20cm} 0 &\hspace{0.20cm} 0 &\hspace{0.20cm} 0 &\hspace{0.20cm} 0 &\hspace{0.20cm} 0 \\
		\hspace{-0.25cm}-6 &\hspace{0.20cm} 0 &\hspace{0.20cm} 0 &\hspace{0.20cm} 0 &\hspace{0.20cm} 0 &\hspace{0.20cm} 0 &\hspace{0.20cm} 0 &\hspace{0.20cm} 0 &\hspace{0.20cm} 0 &\hspace{0.20cm} 0 &\hspace{0.20cm} 0 &\hspace{0.20cm} 0 &\hspace{0.20cm} 0 &\hspace{0.20cm} 1 &\hspace{0.20cm} 0 &\hspace{0.20cm} 0 &\hspace{0.20cm} 0 &\hspace{0.20cm} 0 &\hspace{0.20cm} 0 &\hspace{0.20cm} 0 &\hspace{0.20cm} 0 &\hspace{0.20cm} 0 &\hspace{0.20cm} 0 &\hspace{0.20cm} 0 \\
		3 &\hspace{0.20cm} 1 &\hspace{0.20cm} 0 &\hspace{0.20cm} 0 &\hspace{0.20cm} 0 &\hspace{0.20cm} 0 &\hspace{0.20cm} 0 &\hspace{0.20cm} 0 &\hspace{0.20cm} 0 &\hspace{0.20cm} 0 &\hspace{0.20cm} 0 &\hspace{0.20cm} 0 &\hspace{0.20cm} 1 &\hspace{0.20cm} 0 &\hspace{0.20cm} 0 &\hspace{0.20cm} 0 &\hspace{0.20cm} 0 &\hspace{0.20cm} 0 &\hspace{0.20cm} 0 &\hspace{0.20cm} 0 &\hspace{0.20cm} 0 &\hspace{0.20cm} 0 &\hspace{0.20cm} 0 &\hspace{0.20cm} 0 \\
		\hspace{-0.25cm}-4 &\hspace{0.20cm} \hspace{-0.25cm}-2 &\hspace{0.20cm} 0 &\hspace{0.20cm} 0 &\hspace{0.20cm} 0 &\hspace{0.20cm} 0 &\hspace{0.20cm} 0 &\hspace{0.20cm} 0 &\hspace{0.20cm} 0 &\hspace{0.20cm} 0 &\hspace{0.20cm} 0 &\hspace{0.20cm} 1 &\hspace{0.20cm} 0 &\hspace{0.20cm} 0 &\hspace{0.20cm} 0 &\hspace{0.20cm} 0 &\hspace{0.20cm} 0 &\hspace{0.20cm} 0 &\hspace{0.20cm} 0 &\hspace{0.20cm} 0 &\hspace{0.20cm} 0 &\hspace{0.20cm} 0 &\hspace{0.20cm} 0 &\hspace{0.20cm} 0 \\
		3 &\hspace{0.20cm} 1 &\hspace{0.20cm} 0 &\hspace{0.20cm} 0 &\hspace{0.20cm} 0 &\hspace{0.20cm} 0 &\hspace{0.20cm} 0 &\hspace{0.20cm} 0 &\hspace{0.20cm} 0 &\hspace{0.20cm} 0 &\hspace{0.20cm} 1 &\hspace{0.20cm} 0 &\hspace{0.20cm} 0 &\hspace{0.20cm} 0 &\hspace{0.20cm} 0 &\hspace{0.20cm} 0 &\hspace{0.20cm} 0 &\hspace{0.20cm} 0 &\hspace{0.20cm} 0 &\hspace{0.20cm} 0 &\hspace{0.20cm} 0 &\hspace{0.20cm} 0 &\hspace{0.20cm} 0 &\hspace{0.20cm} 0 \\
		6 &\hspace{0.20cm} 3 &\hspace{0.20cm} 0 &\hspace{0.20cm} 0 &\hspace{0.20cm} 0 &\hspace{0.20cm} 0 &\hspace{0.20cm} 0 &\hspace{0.20cm} 0 &\hspace{0.20cm} 0 &\hspace{0.20cm} 1 &\hspace{0.20cm} 0 &\hspace{0.20cm} 0 &\hspace{0.20cm} 0 &\hspace{0.20cm} 0 &\hspace{0.20cm} 0 &\hspace{0.20cm} 0 &\hspace{0.20cm} 0 &\hspace{0.20cm} 0 &\hspace{0.20cm} 0 &\hspace{0.20cm} 0 &\hspace{0.20cm} 0 &\hspace{0.20cm} 0 &\hspace{0.20cm} 0 &\hspace{0.20cm} 0 \\
		\hspace{-0.25cm}-1 &\hspace{0.20cm} 0 &\hspace{0.20cm} 0 &\hspace{0.20cm} 0 &\hspace{0.20cm} 0 &\hspace{0.20cm} 0 &\hspace{0.20cm} 0 &\hspace{0.20cm} 0 &\hspace{0.20cm} 1 &\hspace{0.20cm} 0 &\hspace{0.20cm} 0 &\hspace{0.20cm} 0 &\hspace{0.20cm} 0 &\hspace{0.20cm} 0 &\hspace{0.20cm} 0 &\hspace{0.20cm} 0 &\hspace{0.20cm} 0 &\hspace{0.20cm} 0 &\hspace{0.20cm} 0 &\hspace{0.20cm} 0 &\hspace{0.20cm} 0 &\hspace{0.20cm} 0 &\hspace{0.20cm} 0 &\hspace{0.20cm} 0 \\
		3 &\hspace{0.20cm} 1 &\hspace{0.20cm} 0 &\hspace{0.20cm} 0 &\hspace{0.20cm} 0 &\hspace{0.20cm} 0 &\hspace{0.20cm} 0 &\hspace{0.20cm} 1 &\hspace{0.20cm} 0 &\hspace{0.20cm} 0 &\hspace{0.20cm} 0 &\hspace{0.20cm} 0 &\hspace{0.20cm} 0 &\hspace{0.20cm} 0 &\hspace{0.20cm} 0 &\hspace{0.20cm} 0 &\hspace{0.20cm} 0 &\hspace{0.20cm} 0 &\hspace{0.20cm} 0 &\hspace{0.20cm} 0 &\hspace{0.20cm} 0 &\hspace{0.20cm} 0 &\hspace{0.20cm} 0 &\hspace{0.20cm} 0 \\
		\hspace{-0.25cm}-6 &\hspace{0.20cm} 0 &\hspace{0.20cm} 0 &\hspace{0.20cm} 0 &\hspace{0.20cm} 0 &\hspace{0.20cm} 0 &\hspace{0.20cm} 1 &\hspace{0.20cm} 0 &\hspace{0.20cm} 0 &\hspace{0.20cm} 0 &\hspace{0.20cm} 0 &\hspace{0.20cm} 0 &\hspace{0.20cm} 0 &\hspace{0.20cm} 0 &\hspace{0.20cm} 0 &\hspace{0.20cm} 0 &\hspace{0.20cm} 0 &\hspace{0.20cm} 0 &\hspace{0.20cm} 0 &\hspace{0.20cm} 0 &\hspace{0.20cm} 0 &\hspace{0.20cm} 0 &\hspace{0.20cm} 0 &\hspace{0.20cm} 0 \\
		\hspace{-0.25cm}-1 &\hspace{0.20cm} 0 &\hspace{0.20cm} 0 &\hspace{0.20cm} 0 &\hspace{0.20cm} 0 &\hspace{0.20cm} 1 &\hspace{0.20cm} 0 &\hspace{0.20cm} 0 &\hspace{0.20cm} 0 &\hspace{0.20cm} 0 &\hspace{0.20cm} 0 &\hspace{0.20cm} 0 &\hspace{0.20cm} 0 &\hspace{0.20cm} 0 &\hspace{0.20cm} 0 &\hspace{0.20cm} 0 &\hspace{0.20cm} 0 &\hspace{0.20cm} 0 &\hspace{0.20cm} 0 &\hspace{0.20cm} 0 &\hspace{0.20cm} 0 &\hspace{0.20cm} 0 &\hspace{0.20cm} 0 &\hspace{0.20cm} 0 \\
		0 &\hspace{0.20cm} \hspace{-0.25cm}-1 &\hspace{0.20cm} 0 &\hspace{0.20cm} 0 &\hspace{0.20cm} 1 &\hspace{0.20cm} 0 &\hspace{0.20cm} 0 &\hspace{0.20cm} 0 &\hspace{0.20cm} 0 &\hspace{0.20cm} 0 &\hspace{0.20cm} 0 &\hspace{0.20cm} 0 &\hspace{0.20cm} 0 &\hspace{0.20cm} 0 &\hspace{0.20cm} 0 &\hspace{0.20cm} 0 &\hspace{0.20cm} 0 &\hspace{0.20cm} 0 &\hspace{0.20cm} 0 &\hspace{0.20cm} 0 &\hspace{0.20cm} 0 &\hspace{0.20cm} 0 &\hspace{0.20cm} 0 &\hspace{0.20cm} 0 \\
		2 &\hspace{0.20cm} 2 &\hspace{0.20cm} 0 &\hspace{0.20cm} 1 &\hspace{0.20cm} 0 &\hspace{0.20cm} 0 &\hspace{0.20cm} 0 &\hspace{0.20cm} 0 &\hspace{0.20cm} 0 &\hspace{0.20cm} 0 &\hspace{0.20cm} 0 &\hspace{0.20cm} 0 &\hspace{0.20cm} 0 &\hspace{0.20cm} 0 &\hspace{0.20cm} 0 &\hspace{0.20cm} 0 &\hspace{0.20cm} 0 &\hspace{0.20cm} 0 &\hspace{0.20cm} 0 &\hspace{0.20cm} 0 &\hspace{0.20cm} 0 &\hspace{0.20cm} 0 &\hspace{0.20cm} 0 &\hspace{0.20cm} 0 \\
		\hspace{-0.25cm}-3 &\hspace{0.20cm} \hspace{-0.25cm}-3 &\hspace{0.20cm} 1 &\hspace{0.20cm} 0 &\hspace{0.20cm} 0 &\hspace{0.20cm} 0 &\hspace{0.20cm} 0 &\hspace{0.20cm} 0 &\hspace{0.20cm} 0 &\hspace{0.20cm} 0 &\hspace{0.20cm} 0 &\hspace{0.20cm} 0 &\hspace{0.20cm} 0 &\hspace{0.20cm} 0 &\hspace{0.20cm} 0 &\hspace{0.20cm} 0 &\hspace{0.20cm} 0 &\hspace{0.20cm} 0 &\hspace{0.20cm} 0 &\hspace{0.20cm} 0 &\hspace{0.20cm} 0 &\hspace{0.20cm} 0 &\hspace{0.20cm} 0 &\hspace{0.20cm} 0 \\
	\end{array}
	\right), \nn
\end{equation}

\begin{align}
	D=(\mathbf{1}_2,0).
\end{align}

The two exponentiated colour factors of this Cweb are
\begin{align}
	(YC)_1=&  f^{abr}f^{adp}f^{dcq}f^{pej} \tj{1} \trr{2} \tq{3} \tb{4} \tc{5} \te{6} +  f^{abr}f^{aem}f^{dcq}f^{dmo} \too{1} \trr{2} \tq{3} \tb{4} \tc{5} \te{6} \, , \nonumber \\
	&=\;{\cal{B}}_{6} \, . \nn\\
	(YC)_2=& f^{abr}f^{aem}f^{dcq}f^{dmo} \too{1} \trr{2} \tq{3} \tb{4} \tc{5} \te{6} \, , \nn\\
	&=\;-{\cal{B}}_{5} \, .
\end{align}

\subsection*{(18)  \ $\mathbf{W^{(5)}_{6} (2, 2, 3, 1, 1, 1)}$}  \label{C64}

This Cweb is the third basis Cweb appearing at five loops and six lines. This Cweb again has twenty-four diagrams. One of the diagrams is shown in fig.~\ref{Diag:18}. The sequences of diagrams and their corresponding $s$-factors are provided in table~\ref{Table:Diag18}.
\vspace{1cm}

\hspace{1cm} \begin{minipage}{0.45\textwidth}
	\hspace{1.5cm}	\includegraphics[scale=0.5]{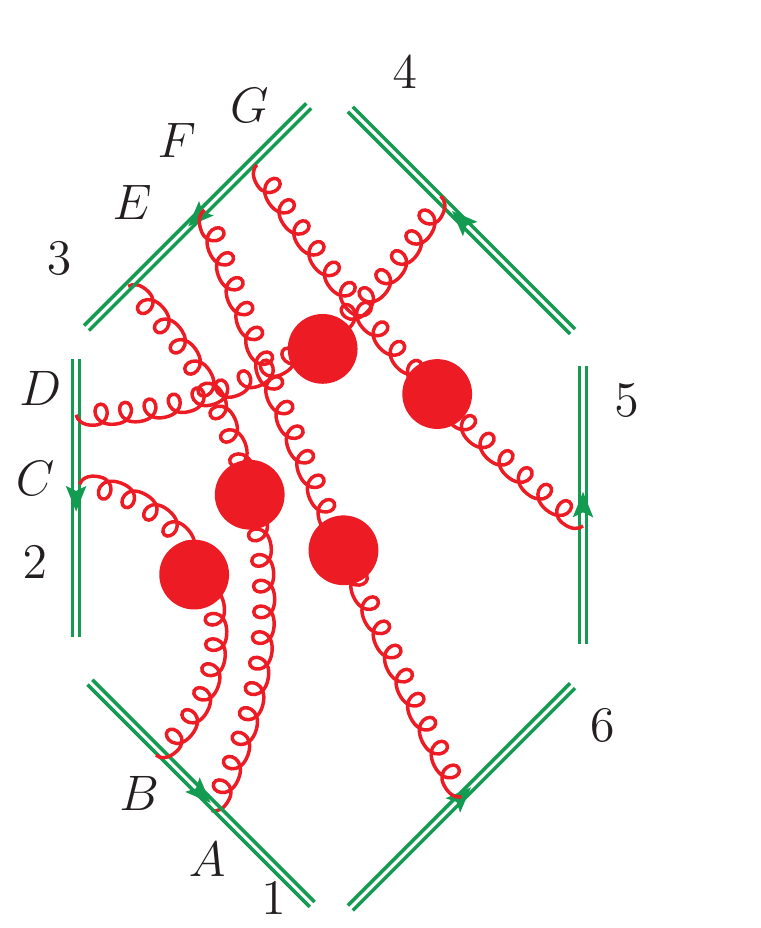}
	\captionof{figure}{$W^{(5)}_{6} (2, 2, 3, 1, 1, 1)$}
	\label{Diag:18}
\end{minipage}
\begin{minipage}{0.45\textwidth}
	\footnotesize{
		\begin{tabular}{ | c | c | c |}
			\hline
			\textbf{Diagrams} & \textbf{Sequences} & \textbf{$s$-factors} \\ \hline
			$d_1$ & $\lbrace \lbrace AB \rbrace,  \lbrace CD \rbrace, \lbrace EFG  \rbrace\rbrace$ & 4 \\ 
			\hline
			$d_{2}$ & $\lbrace \lbrace AB \rbrace,  \lbrace CD \rbrace, \lbrace EGF \rbrace\rbrace$ & 4 \\
			\hline
			$d_{3}$ & $\lbrace \lbrace AB \rbrace,  \lbrace CD \rbrace, \lbrace FEG \rbrace\rbrace$ & 7 \\ 
			\hline
			$d_{4}$ & $\lbrace \lbrace AB \rbrace,  \lbrace CD \rbrace, \lbrace FGE \rbrace\rbrace$ & 9  \\ 
			\hline
			$d_{5}$ & $\lbrace \lbrace AB \rbrace,  \lbrace CD \rbrace, \lbrace GEF \rbrace\rbrace$ & 7\\ 
			\hline
			$d_{6}$ & $\lbrace \lbrace AB \rbrace,  \lbrace CD \rbrace, \lbrace GFE \rbrace\rbrace$ & 9 \\ 
			\hline
			$d_7$ & $\lbrace \lbrace AB \rbrace,  \lbrace DC \rbrace, \lbrace EFG \rbrace\rbrace$ & 1 \\ 
			\hline
			$d_{8}$ & $\lbrace \lbrace AB \rbrace,  \lbrace DC \rbrace, \lbrace EGF \rbrace\rbrace$ & 1 \\ 
			\hline
			$d_9$ & $\lbrace \lbrace AB \rbrace,  \lbrace DC \rbrace, \lbrace FEG \rbrace\rbrace$ & 3  \\ 
			\hline
			$d_{10}$ & $\lbrace \lbrace AB \rbrace,  \lbrace DC \rbrace, \lbrace FGE \rbrace\rbrace$ & 6 \\ 
			\hline
			$d_{11}$ & $\lbrace \lbrace AB \rbrace,  \lbrace DC \rbrace, \lbrace GEF \rbrace\rbrace$ & 3 \\ 
			\hline
			$d_{12}$ & $\lbrace \lbrace AB \rbrace,  \lbrace DC \rbrace, \lbrace GFE \rbrace\rbrace$ & 6 \\ 
			\hline
			$d_{13}$ & $\lbrace \lbrace BA \rbrace,  \lbrace CD \rbrace, \lbrace EFG \rbrace\rbrace$ & 6 \\
			\hline
			$d_{14}$ & $\lbrace \lbrace BA \rbrace,  \lbrace CD \rbrace, \lbrace EGF \rbrace\rbrace$ & 6  \\ 
			\hline
			$d_{15}$ & $\lbrace \lbrace BA \rbrace,  \lbrace CD \rbrace, \lbrace FEG \rbrace\rbrace$ & 3 \\ 
			\hline
			$d_{16}$ & $\lbrace \lbrace BA \rbrace,  \lbrace CD \rbrace, \lbrace FGE \rbrace\rbrace$ & 1 \\ 
			\hline
			$d_{17}$ & $\lbrace \lbrace BA \rbrace,  \lbrace CD \rbrace, \lbrace GEF \rbrace\rbrace$ & 3 \\ 
			\hline	
			$d_{18}$ & $\lbrace \lbrace BA \rbrace,  \lbrace CD \rbrace, \lbrace GFE \rbrace\rbrace$ & 1 \\ 
			\hline
			$d_{19}$ & $\lbrace \lbrace BA \rbrace,  \lbrace DC \rbrace, \lbrace EFG \rbrace\rbrace$ & 9 \\ 
			\hline		
			$d_{20}$ & $\lbrace \lbrace BA \rbrace,  \lbrace DC \rbrace, \lbrace EGF \rbrace\rbrace$ & 9 \\ 
			\hline
			$d_{21}$ & $\lbrace \lbrace BA \rbrace,  \lbrace DC \rbrace, \lbrace FEG \rbrace\rbrace$ & 7\\ 
			\hline	
			$d_{22}$ & $\lbrace \lbrace BA \rbrace,  \lbrace DC \rbrace, \lbrace FGE \rbrace\rbrace$ & 4  \\ 
			\hline	
			$d_{23}$ & $\lbrace \lbrace BA \rbrace,  \lbrace DC \rbrace, \lbrace GEF \rbrace\rbrace$ & 7  \\ 
			\hline
			$d_{24}$ & $\lbrace \lbrace BA \rbrace,  \lbrace DC \rbrace, \lbrace FGE \rbrace\rbrace$ & 4 \\ 
			\hline		
		\end{tabular}
		\captionof{table}{Sequences and $s$-factors}
		\label{Table:Diag18}	}
\end{minipage}

\vspace{0.4cm}
\noindent   The mixing matrix $R$, the diagonalizing matrix $Y$ and the diagonal matrix $D$ for the Cweb mentioned above are 

\begin{equation}
	\hspace{-1cm}	R= \frac{1}{60} \left(\,\,\,
	\begin{array}{cccccccccccccccccccccccc}
		3 &\hspace{0.20cm} \hspace{-0.25cm}-2 &\hspace{0.20cm} \hspace{-0.25cm}-1 &\hspace{0.20cm} \hspace{-0.25cm}-2 &\hspace{0.20cm} \hspace{-0.25cm}-1 &\hspace{0.20cm} 3 &\hspace{0.20cm} \hspace{-0.25cm}-3 &\hspace{0.20cm} 2 &\hspace{0.20cm} 1 &\hspace{0.20cm} 2 &\hspace{0.20cm} 1 &\hspace{0.20cm} \hspace{-0.25cm}-3 &\hspace{0.20cm} \hspace{-0.25cm}-3 &\hspace{0.20cm} 2 &\hspace{0.20cm} 1 &\hspace{0.20cm} 2 &\hspace{0.20cm} 1 &\hspace{0.20cm} \hspace{-0.25cm}-3 &\hspace{0.20cm} 3 &\hspace{0.20cm} \hspace{-0.25cm}-2 &\hspace{0.20cm} \hspace{-0.25cm}-1 &\hspace{0.20cm} \hspace{-0.25cm}-2 &\hspace{0.20cm} \hspace{-0.25cm}-1 &\hspace{0.20cm} 3 \\
		\hspace{-0.25cm}-2 &\hspace{0.20cm} 3 &\hspace{0.20cm} \hspace{-0.25cm}-1 &\hspace{0.20cm} 3 &\hspace{0.20cm} \hspace{-0.25cm}-1 &\hspace{0.20cm} \hspace{-0.25cm}-2 &\hspace{0.20cm} 2 &\hspace{0.20cm} \hspace{-0.25cm}-3 &\hspace{0.20cm} 1 &\hspace{0.20cm} \hspace{-0.25cm}-3 &\hspace{0.20cm} 1 &\hspace{0.20cm} 2 &\hspace{0.20cm} 2 &\hspace{0.20cm} \hspace{-0.25cm}-3 &\hspace{0.20cm} 1 &\hspace{0.20cm} \hspace{-0.25cm}-3 &\hspace{0.20cm} 1 &\hspace{0.20cm} 2 &\hspace{0.20cm} \hspace{-0.25cm}-2 &\hspace{0.20cm} 3 &\hspace{0.20cm} \hspace{-0.25cm}-1 &\hspace{0.20cm} 3 &\hspace{0.20cm} \hspace{-0.25cm}-1 &\hspace{0.20cm} \hspace{-0.25cm}-2 \\
		\hspace{-0.25cm}-2 &\hspace{0.20cm} \hspace{-0.25cm}-2 &\hspace{0.20cm} 4 &\hspace{0.20cm} \hspace{-0.25cm}-2 &\hspace{0.20cm} 4 &\hspace{0.20cm} \hspace{-0.25cm}-2 &\hspace{0.20cm} 2 &\hspace{0.20cm} 2 &\hspace{0.20cm} \hspace{-0.25cm}-4 &\hspace{0.20cm} 2 &\hspace{0.20cm} \hspace{-0.25cm}-4 &\hspace{0.20cm} 2 &\hspace{0.20cm} 2 &\hspace{0.20cm} 2 &\hspace{0.20cm} \hspace{-0.25cm}-4 &\hspace{0.20cm} 2 &\hspace{0.20cm} \hspace{-0.25cm}-4 &\hspace{0.20cm} 2 &\hspace{0.20cm} \hspace{-0.25cm}-2 &\hspace{0.20cm} \hspace{-0.25cm}-2 &\hspace{0.20cm} 4 &\hspace{0.20cm} \hspace{-0.25cm}-2 &\hspace{0.20cm} 4 &\hspace{0.20cm} \hspace{-0.25cm}-2 \\
		\hspace{-0.25cm}-2 &\hspace{0.20cm} 3 &\hspace{0.20cm} \hspace{-0.25cm}-1 &\hspace{0.20cm} 3 &\hspace{0.20cm} \hspace{-0.25cm}-1 &\hspace{0.20cm} \hspace{-0.25cm}-2 &\hspace{0.20cm} 2 &\hspace{0.20cm} \hspace{-0.25cm}-3 &\hspace{0.20cm} 1 &\hspace{0.20cm} \hspace{-0.25cm}-3 &\hspace{0.20cm} 1 &\hspace{0.20cm} 2 &\hspace{0.20cm} 2 &\hspace{0.20cm} \hspace{-0.25cm}-3 &\hspace{0.20cm} 1 &\hspace{0.20cm} \hspace{-0.25cm}-3 &\hspace{0.20cm} 1 &\hspace{0.20cm} 2 &\hspace{0.20cm} \hspace{-0.25cm}-2 &\hspace{0.20cm} 3 &\hspace{0.20cm} \hspace{-0.25cm}-1 &\hspace{0.20cm} 3 &\hspace{0.20cm} \hspace{-0.25cm}-1 &\hspace{0.20cm} \hspace{-0.25cm}-2 \\
		\hspace{-0.25cm}-2 &\hspace{0.20cm} \hspace{-0.25cm}-2 &\hspace{0.20cm} 4 &\hspace{0.20cm} \hspace{-0.25cm}-2 &\hspace{0.20cm} 4 &\hspace{0.20cm} \hspace{-0.25cm}-2 &\hspace{0.20cm} 2 &\hspace{0.20cm} 2 &\hspace{0.20cm} \hspace{-0.25cm}-4 &\hspace{0.20cm} 2 &\hspace{0.20cm} \hspace{-0.25cm}-4 &\hspace{0.20cm} 2 &\hspace{0.20cm} 2 &\hspace{0.20cm} 2 &\hspace{0.20cm} \hspace{-0.25cm}-4 &\hspace{0.20cm} 2 &\hspace{0.20cm} \hspace{-0.25cm}-4 &\hspace{0.20cm} 2 &\hspace{0.20cm} \hspace{-0.25cm}-2 &\hspace{0.20cm} \hspace{-0.25cm}-2 &\hspace{0.20cm} 4 &\hspace{0.20cm} \hspace{-0.25cm}-2 &\hspace{0.20cm} 4 &\hspace{0.20cm} \hspace{-0.25cm}-2 \\
		3 &\hspace{0.20cm} \hspace{-0.25cm}-2 &\hspace{0.20cm} \hspace{-0.25cm}-1 &\hspace{0.20cm} \hspace{-0.25cm}-2 &\hspace{0.20cm} \hspace{-0.25cm}-1 &\hspace{0.20cm} 3 &\hspace{0.20cm} \hspace{-0.25cm}-3 &\hspace{0.20cm} 2 &\hspace{0.20cm} 1 &\hspace{0.20cm} 2 &\hspace{0.20cm} 1 &\hspace{0.20cm} \hspace{-0.25cm}-3 &\hspace{0.20cm} \hspace{-0.25cm}-3 &\hspace{0.20cm} 2 &\hspace{0.20cm} 1 &\hspace{0.20cm} 2 &\hspace{0.20cm} 1 &\hspace{0.20cm} \hspace{-0.25cm}-3 &\hspace{0.20cm} 3 &\hspace{0.20cm} \hspace{-0.25cm}-2 &\hspace{0.20cm} \hspace{-0.25cm}-1 &\hspace{0.20cm} \hspace{-0.25cm}-2 &\hspace{0.20cm} \hspace{-0.25cm}-1 &\hspace{0.20cm} 3 \\
		\hspace{-0.25cm}-12 &\hspace{0.20cm} 3 &\hspace{0.20cm} 9 &\hspace{0.20cm} 3 &\hspace{0.20cm} 9 &\hspace{0.20cm} \hspace{-0.25cm}-12 &\hspace{0.20cm} 12 &\hspace{0.20cm} \hspace{-0.25cm}-3 &\hspace{0.20cm} \hspace{-0.25cm}-9 &\hspace{0.20cm} \hspace{-0.25cm}-3 &\hspace{0.20cm} \hspace{-0.25cm}-9 &\hspace{0.20cm} 12 &\hspace{0.20cm} 12 &\hspace{0.20cm} \hspace{-0.25cm}-3 &\hspace{0.20cm} \hspace{-0.25cm}-9 &\hspace{0.20cm} \hspace{-0.25cm}-3 &\hspace{0.20cm} \hspace{-0.25cm}-9 &\hspace{0.20cm} 12 &\hspace{0.20cm} \hspace{-0.25cm}-12 &\hspace{0.20cm} 3 &\hspace{0.20cm} 9 &\hspace{0.20cm} 3 &\hspace{0.20cm} 9 &\hspace{0.20cm} \hspace{-0.25cm}-12 \\
		3 &\hspace{0.20cm} \hspace{-0.25cm}-12 &\hspace{0.20cm} 9 &\hspace{0.20cm} \hspace{-0.25cm}-12 &\hspace{0.20cm} 9 &\hspace{0.20cm} 3 &\hspace{0.20cm} \hspace{-0.25cm}-3 &\hspace{0.20cm} 12 &\hspace{0.20cm} \hspace{-0.25cm}-9 &\hspace{0.20cm} 12 &\hspace{0.20cm} \hspace{-0.25cm}-9 &\hspace{0.20cm} \hspace{-0.25cm}-3 &\hspace{0.20cm} \hspace{-0.25cm}-3 &\hspace{0.20cm} 12 &\hspace{0.20cm} \hspace{-0.25cm}-9 &\hspace{0.20cm} 12 &\hspace{0.20cm} \hspace{-0.25cm}-9 &\hspace{0.20cm} \hspace{-0.25cm}-3 &\hspace{0.20cm} 3 &\hspace{0.20cm} \hspace{-0.25cm}-12 &\hspace{0.20cm} 9 &\hspace{0.20cm} \hspace{-0.25cm}-12 &\hspace{0.20cm} 9 &\hspace{0.20cm} 3 \\
		3 &\hspace{0.20cm} 3 &\hspace{0.20cm} \hspace{-0.25cm}-6 &\hspace{0.20cm} 3 &\hspace{0.20cm} \hspace{-0.25cm}-6 &\hspace{0.20cm} 3 &\hspace{0.20cm} \hspace{-0.25cm}-3 &\hspace{0.20cm} \hspace{-0.25cm}-3 &\hspace{0.20cm} 6 &\hspace{0.20cm} \hspace{-0.25cm}-3 &\hspace{0.20cm} 6 &\hspace{0.20cm} \hspace{-0.25cm}-3 &\hspace{0.20cm} \hspace{-0.25cm}-3 &\hspace{0.20cm} \hspace{-0.25cm}-3 &\hspace{0.20cm} 6 &\hspace{0.20cm} \hspace{-0.25cm}-3 &\hspace{0.20cm} 6 &\hspace{0.20cm} \hspace{-0.25cm}-3 &\hspace{0.20cm} 3 &\hspace{0.20cm} 3 &\hspace{0.20cm} \hspace{-0.25cm}-6 &\hspace{0.20cm} 3 &\hspace{0.20cm} \hspace{-0.25cm}-6 &\hspace{0.20cm} 3 \\
		3 &\hspace{0.20cm} \hspace{-0.25cm}-2 &\hspace{0.20cm} \hspace{-0.25cm}-1 &\hspace{0.20cm} \hspace{-0.25cm}-2 &\hspace{0.20cm} \hspace{-0.25cm}-1 &\hspace{0.20cm} 3 &\hspace{0.20cm} \hspace{-0.25cm}-3 &\hspace{0.20cm} 2 &\hspace{0.20cm} 1 &\hspace{0.20cm} 2 &\hspace{0.20cm} 1 &\hspace{0.20cm} \hspace{-0.25cm}-3 &\hspace{0.20cm} \hspace{-0.25cm}-3 &\hspace{0.20cm} 2 &\hspace{0.20cm} 1 &\hspace{0.20cm} 2 &\hspace{0.20cm} 1 &\hspace{0.20cm} \hspace{-0.25cm}-3 &\hspace{0.20cm} 3 &\hspace{0.20cm} \hspace{-0.25cm}-2 &\hspace{0.20cm} \hspace{-0.25cm}-1 &\hspace{0.20cm} \hspace{-0.25cm}-2 &\hspace{0.20cm} \hspace{-0.25cm}-1 &\hspace{0.20cm} 3 \\
		3 &\hspace{0.20cm} 3 &\hspace{0.20cm} \hspace{-0.25cm}-6 &\hspace{0.20cm} 3 &\hspace{0.20cm} \hspace{-0.25cm}-6 &\hspace{0.20cm} 3 &\hspace{0.20cm} \hspace{-0.25cm}-3 &\hspace{0.20cm} \hspace{-0.25cm}-3 &\hspace{0.20cm} 6 &\hspace{0.20cm} \hspace{-0.25cm}-3 &\hspace{0.20cm} 6 &\hspace{0.20cm} \hspace{-0.25cm}-3 &\hspace{0.20cm} \hspace{-0.25cm}-3 &\hspace{0.20cm} \hspace{-0.25cm}-3 &\hspace{0.20cm} 6 &\hspace{0.20cm} \hspace{-0.25cm}-3 &\hspace{0.20cm} 6 &\hspace{0.20cm} \hspace{-0.25cm}-3 &\hspace{0.20cm} 3 &\hspace{0.20cm} 3 &\hspace{0.20cm} \hspace{-0.25cm}-6 &\hspace{0.20cm} 3 &\hspace{0.20cm} \hspace{-0.25cm}-6 &\hspace{0.20cm} 3 \\
		\hspace{-0.25cm}-2 &\hspace{0.20cm} 3 &\hspace{0.20cm} \hspace{-0.25cm}-1 &\hspace{0.20cm} 3 &\hspace{0.20cm} \hspace{-0.25cm}-1 &\hspace{0.20cm} \hspace{-0.25cm}-2 &\hspace{0.20cm} 2 &\hspace{0.20cm} \hspace{-0.25cm}-3 &\hspace{0.20cm} 1 &\hspace{0.20cm} \hspace{-0.25cm}-3 &\hspace{0.20cm} 1 &\hspace{0.20cm} 2 &\hspace{0.20cm} 2 &\hspace{0.20cm} \hspace{-0.25cm}-3 &\hspace{0.20cm} 1 &\hspace{0.20cm} \hspace{-0.25cm}-3 &\hspace{0.20cm} 1 &\hspace{0.20cm} 2 &\hspace{0.20cm} \hspace{-0.25cm}-2 &\hspace{0.20cm} 3 &\hspace{0.20cm} \hspace{-0.25cm}-1 &\hspace{0.20cm} 3 &\hspace{0.20cm} \hspace{-0.25cm}-1 &\hspace{0.20cm} \hspace{-0.25cm}-2 \\
		\hspace{-0.25cm}-2 &\hspace{0.20cm} 3 &\hspace{0.20cm} \hspace{-0.25cm}-1 &\hspace{0.20cm} 3 &\hspace{0.20cm} \hspace{-0.25cm}-1 &\hspace{0.20cm} \hspace{-0.25cm}-2 &\hspace{0.20cm} 2 &\hspace{0.20cm} \hspace{-0.25cm}-3 &\hspace{0.20cm} 1 &\hspace{0.20cm} \hspace{-0.25cm}-3 &\hspace{0.20cm} 1 &\hspace{0.20cm} 2 &\hspace{0.20cm} 2 &\hspace{0.20cm} \hspace{-0.25cm}-3 &\hspace{0.20cm} 1 &\hspace{0.20cm} \hspace{-0.25cm}-3 &\hspace{0.20cm} 1 &\hspace{0.20cm} 2 &\hspace{0.20cm} \hspace{-0.25cm}-2 &\hspace{0.20cm} 3 &\hspace{0.20cm} \hspace{-0.25cm}-1 &\hspace{0.20cm} 3 &\hspace{0.20cm} \hspace{-0.25cm}-1 &\hspace{0.20cm} \hspace{-0.25cm}-2 \\
		3 &\hspace{0.20cm} \hspace{-0.25cm}-2 &\hspace{0.20cm} \hspace{-0.25cm}-1 &\hspace{0.20cm} \hspace{-0.25cm}-2 &\hspace{0.20cm} \hspace{-0.25cm}-1 &\hspace{0.20cm} 3 &\hspace{0.20cm} \hspace{-0.25cm}-3 &\hspace{0.20cm} 2 &\hspace{0.20cm} 1 &\hspace{0.20cm} 2 &\hspace{0.20cm} 1 &\hspace{0.20cm} \hspace{-0.25cm}-3 &\hspace{0.20cm} \hspace{-0.25cm}-3 &\hspace{0.20cm} 2 &\hspace{0.20cm} 1 &\hspace{0.20cm} 2 &\hspace{0.20cm} 1 &\hspace{0.20cm} \hspace{-0.25cm}-3 &\hspace{0.20cm} 3 &\hspace{0.20cm} \hspace{-0.25cm}-2 &\hspace{0.20cm} \hspace{-0.25cm}-1 &\hspace{0.20cm} \hspace{-0.25cm}-2 &\hspace{0.20cm} \hspace{-0.25cm}-1 &\hspace{0.20cm} 3 \\
		3 &\hspace{0.20cm} 3 &\hspace{0.20cm} \hspace{-0.25cm}-6 &\hspace{0.20cm} 3 &\hspace{0.20cm} \hspace{-0.25cm}-6 &\hspace{0.20cm} 3 &\hspace{0.20cm} \hspace{-0.25cm}-3 &\hspace{0.20cm} \hspace{-0.25cm}-3 &\hspace{0.20cm} 6 &\hspace{0.20cm} \hspace{-0.25cm}-3 &\hspace{0.20cm} 6 &\hspace{0.20cm} \hspace{-0.25cm}-3 &\hspace{0.20cm} \hspace{-0.25cm}-3 &\hspace{0.20cm} \hspace{-0.25cm}-3 &\hspace{0.20cm} 6 &\hspace{0.20cm} \hspace{-0.25cm}-3 &\hspace{0.20cm} 6 &\hspace{0.20cm} \hspace{-0.25cm}-3 &\hspace{0.20cm} 3 &\hspace{0.20cm} 3 &\hspace{0.20cm} \hspace{-0.25cm}-6 &\hspace{0.20cm} 3 &\hspace{0.20cm} \hspace{-0.25cm}-6 &\hspace{0.20cm} 3 \\
		3 &\hspace{0.20cm} \hspace{-0.25cm}-12 &\hspace{0.20cm} 9 &\hspace{0.20cm} \hspace{-0.25cm}-12 &\hspace{0.20cm} 9 &\hspace{0.20cm} 3 &\hspace{0.20cm} \hspace{-0.25cm}-3 &\hspace{0.20cm} 12 &\hspace{0.20cm} \hspace{-0.25cm}-9 &\hspace{0.20cm} 12 &\hspace{0.20cm} \hspace{-0.25cm}-9 &\hspace{0.20cm} \hspace{-0.25cm}-3 &\hspace{0.20cm} \hspace{-0.25cm}-3 &\hspace{0.20cm} 12 &\hspace{0.20cm} \hspace{-0.25cm}-9 &\hspace{0.20cm} 12 &\hspace{0.20cm} \hspace{-0.25cm}-9 &\hspace{0.20cm} \hspace{-0.25cm}-3 &\hspace{0.20cm} 3 &\hspace{0.20cm} \hspace{-0.25cm}-12 &\hspace{0.20cm} 9 &\hspace{0.20cm} \hspace{-0.25cm}-12 &\hspace{0.20cm} 9 &\hspace{0.20cm} 3 \\
		3 &\hspace{0.20cm} 3 &\hspace{0.20cm} \hspace{-0.25cm}-6 &\hspace{0.20cm} 3 &\hspace{0.20cm} \hspace{-0.25cm}-6 &\hspace{0.20cm} 3 &\hspace{0.20cm} \hspace{-0.25cm}-3 &\hspace{0.20cm} \hspace{-0.25cm}-3 &\hspace{0.20cm} 6 &\hspace{0.20cm} \hspace{-0.25cm}-3 &\hspace{0.20cm} 6 &\hspace{0.20cm} \hspace{-0.25cm}-3 &\hspace{0.20cm} \hspace{-0.25cm}-3 &\hspace{0.20cm} \hspace{-0.25cm}-3 &\hspace{0.20cm} 6 &\hspace{0.20cm} \hspace{-0.25cm}-3 &\hspace{0.20cm} 6 &\hspace{0.20cm} \hspace{-0.25cm}-3 &\hspace{0.20cm} 3 &\hspace{0.20cm} 3 &\hspace{0.20cm} \hspace{-0.25cm}-6 &\hspace{0.20cm} 3 &\hspace{0.20cm} \hspace{-0.25cm}-6 &\hspace{0.20cm} 3 \\
		\hspace{-0.25cm}-12 &\hspace{0.20cm} 3 &\hspace{0.20cm} 9 &\hspace{0.20cm} 3 &\hspace{0.20cm} 9 &\hspace{0.20cm} \hspace{-0.25cm}-12 &\hspace{0.20cm} 12 &\hspace{0.20cm} \hspace{-0.25cm}-3 &\hspace{0.20cm} \hspace{-0.25cm}-9 &\hspace{0.20cm} \hspace{-0.25cm}-3 &\hspace{0.20cm} \hspace{-0.25cm}-9 &\hspace{0.20cm} 12 &\hspace{0.20cm} 12 &\hspace{0.20cm} \hspace{-0.25cm}-3 &\hspace{0.20cm} \hspace{-0.25cm}-9 &\hspace{0.20cm} \hspace{-0.25cm}-3 &\hspace{0.20cm} \hspace{-0.25cm}-9 &\hspace{0.20cm} 12 &\hspace{0.20cm} \hspace{-0.25cm}-12 &\hspace{0.20cm} 3 &\hspace{0.20cm} 9 &\hspace{0.20cm} 3 &\hspace{0.20cm} 9 &\hspace{0.20cm} \hspace{-0.25cm}-12 \\
		3 &\hspace{0.20cm} \hspace{-0.25cm}-2 &\hspace{0.20cm} \hspace{-0.25cm}-1 &\hspace{0.20cm} \hspace{-0.25cm}-2 &\hspace{0.20cm} \hspace{-0.25cm}-1 &\hspace{0.20cm} 3 &\hspace{0.20cm} \hspace{-0.25cm}-3 &\hspace{0.20cm} 2 &\hspace{0.20cm} 1 &\hspace{0.20cm} 2 &\hspace{0.20cm} 1 &\hspace{0.20cm} \hspace{-0.25cm}-3 &\hspace{0.20cm} \hspace{-0.25cm}-3 &\hspace{0.20cm} 2 &\hspace{0.20cm} 1 &\hspace{0.20cm} 2 &\hspace{0.20cm} 1 &\hspace{0.20cm} \hspace{-0.25cm}-3 &\hspace{0.20cm} 3 &\hspace{0.20cm} \hspace{-0.25cm}-2 &\hspace{0.20cm} \hspace{-0.25cm}-1 &\hspace{0.20cm} \hspace{-0.25cm}-2 &\hspace{0.20cm} \hspace{-0.25cm}-1 &\hspace{0.20cm} 3 \\
		\hspace{-0.25cm}-2 &\hspace{0.20cm} 3 &\hspace{0.20cm} \hspace{-0.25cm}-1 &\hspace{0.20cm} 3 &\hspace{0.20cm} \hspace{-0.25cm}-1 &\hspace{0.20cm} \hspace{-0.25cm}-2 &\hspace{0.20cm} 2 &\hspace{0.20cm} \hspace{-0.25cm}-3 &\hspace{0.20cm} 1 &\hspace{0.20cm} \hspace{-0.25cm}-3 &\hspace{0.20cm} 1 &\hspace{0.20cm} 2 &\hspace{0.20cm} 2 &\hspace{0.20cm} \hspace{-0.25cm}-3 &\hspace{0.20cm} 1 &\hspace{0.20cm} \hspace{-0.25cm}-3 &\hspace{0.20cm} 1 &\hspace{0.20cm} 2 &\hspace{0.20cm} \hspace{-0.25cm}-2 &\hspace{0.20cm} 3 &\hspace{0.20cm} \hspace{-0.25cm}-1 &\hspace{0.20cm} 3 &\hspace{0.20cm} \hspace{-0.25cm}-1 &\hspace{0.20cm} \hspace{-0.25cm}-2 \\
		\hspace{-0.25cm}-2 &\hspace{0.20cm} \hspace{-0.25cm}-2 &\hspace{0.20cm} 4 &\hspace{0.20cm} \hspace{-0.25cm}-2 &\hspace{0.20cm} 4 &\hspace{0.20cm} \hspace{-0.25cm}-2 &\hspace{0.20cm} 2 &\hspace{0.20cm} 2 &\hspace{0.20cm} \hspace{-0.25cm}-4 &\hspace{0.20cm} 2 &\hspace{0.20cm} \hspace{-0.25cm}-4 &\hspace{0.20cm} 2 &\hspace{0.20cm} 2 &\hspace{0.20cm} 2 &\hspace{0.20cm} \hspace{-0.25cm}-4 &\hspace{0.20cm} 2 &\hspace{0.20cm} \hspace{-0.25cm}-4 &\hspace{0.20cm} 2 &\hspace{0.20cm} \hspace{-0.25cm}-2 &\hspace{0.20cm} \hspace{-0.25cm}-2 &\hspace{0.20cm} 4 &\hspace{0.20cm} \hspace{-0.25cm}-2 &\hspace{0.20cm} 4 &\hspace{0.20cm} \hspace{-0.25cm}-2 \\
		\hspace{-0.25cm}-2 &\hspace{0.20cm} 3 &\hspace{0.20cm} \hspace{-0.25cm}-1 &\hspace{0.20cm} 3 &\hspace{0.20cm} \hspace{-0.25cm}-1 &\hspace{0.20cm} \hspace{-0.25cm}-2 &\hspace{0.20cm} 2 &\hspace{0.20cm} \hspace{-0.25cm}-3 &\hspace{0.20cm} 1 &\hspace{0.20cm} \hspace{-0.25cm}-3 &\hspace{0.20cm} 1 &\hspace{0.20cm} 2 &\hspace{0.20cm} 2 &\hspace{0.20cm} \hspace{-0.25cm}-3 &\hspace{0.20cm} 1 &\hspace{0.20cm} \hspace{-0.25cm}-3 &\hspace{0.20cm} 1 &\hspace{0.20cm} 2 &\hspace{0.20cm} \hspace{-0.25cm}-2 &\hspace{0.20cm} 3 &\hspace{0.20cm} \hspace{-0.25cm}-1 &\hspace{0.20cm} 3 &\hspace{0.20cm} \hspace{-0.25cm}-1 &\hspace{0.20cm} \hspace{-0.25cm}-2 \\
		\hspace{-0.25cm}-2 &\hspace{0.20cm} \hspace{-0.25cm}-2 &\hspace{0.20cm} 4 &\hspace{0.20cm} \hspace{-0.25cm}-2 &\hspace{0.20cm} 4 &\hspace{0.20cm} \hspace{-0.25cm}-2 &\hspace{0.20cm} 2 &\hspace{0.20cm} 2 &\hspace{0.20cm} \hspace{-0.25cm}-4 &\hspace{0.20cm} 2 &\hspace{0.20cm} \hspace{-0.25cm}-4 &\hspace{0.20cm} 2 &\hspace{0.20cm} 2 &\hspace{0.20cm} 2 &\hspace{0.20cm} \hspace{-0.25cm}-4 &\hspace{0.20cm} 2 &\hspace{0.20cm} \hspace{-0.25cm}-4 &\hspace{0.20cm} 2 &\hspace{0.20cm} \hspace{-0.25cm}-2 &\hspace{0.20cm} \hspace{-0.25cm}-2 &\hspace{0.20cm} 4 &\hspace{0.20cm} \hspace{-0.25cm}-2 &\hspace{0.20cm} 4 &\hspace{0.20cm} \hspace{-0.25cm}-2 \\
		3 &\hspace{0.20cm} \hspace{-0.25cm}-2 &\hspace{0.20cm} \hspace{-0.25cm}-1 &\hspace{0.20cm} \hspace{-0.25cm}-2 &\hspace{0.20cm} \hspace{-0.25cm}-1 &\hspace{0.20cm} 3 &\hspace{0.20cm} \hspace{-0.25cm}-3 &\hspace{0.20cm} 2 &\hspace{0.20cm} 1 &\hspace{0.20cm} 2 &\hspace{0.20cm} 1 &\hspace{0.20cm} \hspace{-0.25cm}-3 &\hspace{0.20cm} \hspace{-0.25cm}-3 &\hspace{0.20cm} 2 &\hspace{0.20cm} 1 &\hspace{0.20cm} 2 &\hspace{0.20cm} 1 &\hspace{0.20cm} \hspace{-0.25cm}-3 &\hspace{0.20cm} 3 &\hspace{0.20cm} \hspace{-0.25cm}-2 &\hspace{0.20cm} \hspace{-0.25cm}-1 &\hspace{0.20cm} \hspace{-0.25cm}-2 &\hspace{0.20cm} \hspace{-0.25cm}-1 &\hspace{0.20cm} 3 \\
	\end{array}
	\right), \nonumber 
\end{equation}

\begin{equation}
	Y=\left( \,\,\,
	\begin{array}{cccccccccccccccccccccccc}
		1 &\hspace{0.20cm} \hspace{-0.25cm}-1 &\hspace{0.20cm} 0 &\hspace{0.20cm} \hspace{-0.25cm}-1 &\hspace{0.20cm} 0 &\hspace{0.20cm} 1 &\hspace{0.20cm} \hspace{-0.25cm}-1 &\hspace{0.20cm} 1 &\hspace{0.20cm} 0 &\hspace{0.20cm} 1 &\hspace{0.20cm} 0 &\hspace{0.20cm} \hspace{-0.25cm}-1 &\hspace{0.20cm} \hspace{-0.25cm}-1 &\hspace{0.20cm} 1 &\hspace{0.20cm} 0 &\hspace{0.20cm} 1 &\hspace{0.20cm} 0 &\hspace{0.20cm} \hspace{-0.25cm}-1 &\hspace{0.20cm} 1 &\hspace{0.20cm} \hspace{-0.25cm}-1 &\hspace{0.20cm} 0 &\hspace{0.20cm} \hspace{-0.25cm}-1 &\hspace{0.20cm} 0 &\hspace{0.20cm} 1 \\
		0 &\hspace{0.20cm} \hspace{-0.25cm}-1 &\hspace{0.20cm} 1 &\hspace{0.20cm} \hspace{-0.25cm}-1 &\hspace{0.20cm} 1 &\hspace{0.20cm} 0 &\hspace{0.20cm} 0 &\hspace{0.20cm} 1 &\hspace{0.20cm} \hspace{-0.25cm}-1 &\hspace{0.20cm} 1 &\hspace{0.20cm} \hspace{-0.25cm}-1 &\hspace{0.20cm} 0 &\hspace{0.20cm} 0 &\hspace{0.20cm} 1 &\hspace{0.20cm} \hspace{-0.25cm}-1 &\hspace{0.20cm} 1 &\hspace{0.20cm} \hspace{-0.25cm}-1 &\hspace{0.20cm} 0 &\hspace{0.20cm} 0 &\hspace{0.20cm} \hspace{-0.25cm}-1 &\hspace{0.20cm} 1 &\hspace{0.20cm} \hspace{-0.25cm}-1 &\hspace{0.20cm} 1 &\hspace{0.20cm} 0 \\
		\hspace{-0.25cm}-1 &\hspace{0.20cm} 0 &\hspace{0.20cm} 0 &\hspace{0.20cm} 0 &\hspace{0.20cm} 0 &\hspace{0.20cm} 0 &\hspace{0.20cm} 0 &\hspace{0.20cm} 0 &\hspace{0.20cm} 0 &\hspace{0.20cm} 0 &\hspace{0.20cm} 0 &\hspace{0.20cm} 0 &\hspace{0.20cm} 0 &\hspace{0.20cm} 0 &\hspace{0.20cm} 0 &\hspace{0.20cm} 0 &\hspace{0.20cm} 0 &\hspace{0.20cm} 0 &\hspace{0.20cm} 0 &\hspace{0.20cm} 0 &\hspace{0.20cm} 0 &\hspace{0.20cm} 0 &\hspace{0.20cm} 0 &\hspace{0.20cm} 1 \\
		2 &\hspace{0.20cm} 2 &\hspace{0.20cm} 0 &\hspace{0.20cm} 0 &\hspace{0.20cm} 0 &\hspace{0.20cm} 0 &\hspace{0.20cm} 0 &\hspace{0.20cm} 0 &\hspace{0.20cm} 0 &\hspace{0.20cm} 0 &\hspace{0.20cm} 0 &\hspace{0.20cm} 0 &\hspace{0.20cm} 0 &\hspace{0.20cm} 0 &\hspace{0.20cm} 0 &\hspace{0.20cm} 0 &\hspace{0.20cm} 0 &\hspace{0.20cm} 0 &\hspace{0.20cm} 0 &\hspace{0.20cm} 0 &\hspace{0.20cm} 0 &\hspace{0.20cm} 0 &\hspace{0.20cm} 1 &\hspace{0.20cm} 0 \\
		0 &\hspace{0.20cm} \hspace{-0.25cm}-1 &\hspace{0.20cm} 0 &\hspace{0.20cm} 0 &\hspace{0.20cm} 0 &\hspace{0.20cm} 0 &\hspace{0.20cm} 0 &\hspace{0.20cm} 0 &\hspace{0.20cm} 0 &\hspace{0.20cm} 0 &\hspace{0.20cm} 0 &\hspace{0.20cm} 0 &\hspace{0.20cm} 0 &\hspace{0.20cm} 0 &\hspace{0.20cm} 0 &\hspace{0.20cm} 0 &\hspace{0.20cm} 0 &\hspace{0.20cm} 0 &\hspace{0.20cm} 0 &\hspace{0.20cm} 0 &\hspace{0.20cm} 0 &\hspace{0.20cm} 1 &\hspace{0.20cm} 0 &\hspace{0.20cm} 0 \\
		2 &\hspace{0.20cm} 2 &\hspace{0.20cm} 0 &\hspace{0.20cm} 0 &\hspace{0.20cm} 0 &\hspace{0.20cm} 0 &\hspace{0.20cm} 0 &\hspace{0.20cm} 0 &\hspace{0.20cm} 0 &\hspace{0.20cm} 0 &\hspace{0.20cm} 0 &\hspace{0.20cm} 0 &\hspace{0.20cm} 0 &\hspace{0.20cm} 0 &\hspace{0.20cm} 0 &\hspace{0.20cm} 0 &\hspace{0.20cm} 0 &\hspace{0.20cm} 0 &\hspace{0.20cm} 0 &\hspace{0.20cm} 0 &\hspace{0.20cm} 1 &\hspace{0.20cm} 0 &\hspace{0.20cm} 0 &\hspace{0.20cm} 0 \\
		0 &\hspace{0.20cm} \hspace{-0.25cm}-1 &\hspace{0.20cm} 0 &\hspace{0.20cm} 0 &\hspace{0.20cm} 0 &\hspace{0.20cm} 0 &\hspace{0.20cm} 0 &\hspace{0.20cm} 0 &\hspace{0.20cm} 0 &\hspace{0.20cm} 0 &\hspace{0.20cm} 0 &\hspace{0.20cm} 0 &\hspace{0.20cm} 0 &\hspace{0.20cm} 0 &\hspace{0.20cm} 0 &\hspace{0.20cm} 0 &\hspace{0.20cm} 0 &\hspace{0.20cm} 0 &\hspace{0.20cm} 0 &\hspace{0.20cm} 1 &\hspace{0.20cm} 0 &\hspace{0.20cm} 0 &\hspace{0.20cm} 0 &\hspace{0.20cm} 0 \\
		\hspace{-0.25cm}-1 &\hspace{0.20cm} 0 &\hspace{0.20cm} 0 &\hspace{0.20cm} 0 &\hspace{0.20cm} 0 &\hspace{0.20cm} 0 &\hspace{0.20cm} 0 &\hspace{0.20cm} 0 &\hspace{0.20cm} 0 &\hspace{0.20cm} 0 &\hspace{0.20cm} 0 &\hspace{0.20cm} 0 &\hspace{0.20cm} 0 &\hspace{0.20cm} 0 &\hspace{0.20cm} 0 &\hspace{0.20cm} 0 &\hspace{0.20cm} 0 &\hspace{0.20cm} 0 &\hspace{0.20cm} 1 &\hspace{0.20cm} 0 &\hspace{0.20cm} 0 &\hspace{0.20cm} 0 &\hspace{0.20cm} 0 &\hspace{0.20cm} 0 \\
		6 &\hspace{0.20cm} 3 &\hspace{0.20cm} 0 &\hspace{0.20cm} 0 &\hspace{0.20cm} 0 &\hspace{0.20cm} 0 &\hspace{0.20cm} 0 &\hspace{0.20cm} 0 &\hspace{0.20cm} 0 &\hspace{0.20cm} 0 &\hspace{0.20cm} 0 &\hspace{0.20cm} 0 &\hspace{0.20cm} 0 &\hspace{0.20cm} 0 &\hspace{0.20cm} 0 &\hspace{0.20cm} 0 &\hspace{0.20cm} 0 &\hspace{0.20cm} 1 &\hspace{0.20cm} 0 &\hspace{0.20cm} 0 &\hspace{0.20cm} 0 &\hspace{0.20cm} 0 &\hspace{0.20cm} 0 &\hspace{0.20cm} 0 \\
		\hspace{-0.25cm}-3 &\hspace{0.20cm} \hspace{-0.25cm}-3 &\hspace{0.20cm} 0 &\hspace{0.20cm} 0 &\hspace{0.20cm} 0 &\hspace{0.20cm} 0 &\hspace{0.20cm} 0 &\hspace{0.20cm} 0 &\hspace{0.20cm} 0 &\hspace{0.20cm} 0 &\hspace{0.20cm} 0 &\hspace{0.20cm} 0 &\hspace{0.20cm} 0 &\hspace{0.20cm} 0 &\hspace{0.20cm} 0 &\hspace{0.20cm} 0 &\hspace{0.20cm} 1 &\hspace{0.20cm} 0 &\hspace{0.20cm} 0 &\hspace{0.20cm} 0 &\hspace{0.20cm} 0 &\hspace{0.20cm} 0 &\hspace{0.20cm} 0 &\hspace{0.20cm} 0 \\
		3 &\hspace{0.20cm} 6 &\hspace{0.20cm} 0 &\hspace{0.20cm} 0 &\hspace{0.20cm} 0 &\hspace{0.20cm} 0 &\hspace{0.20cm} 0 &\hspace{0.20cm} 0 &\hspace{0.20cm} 0 &\hspace{0.20cm} 0 &\hspace{0.20cm} 0 &\hspace{0.20cm} 0 &\hspace{0.20cm} 0 &\hspace{0.20cm} 0 &\hspace{0.20cm} 0 &\hspace{0.20cm} 1 &\hspace{0.20cm} 0 &\hspace{0.20cm} 0 &\hspace{0.20cm} 0 &\hspace{0.20cm} 0 &\hspace{0.20cm} 0 &\hspace{0.20cm} 0 &\hspace{0.20cm} 0 &\hspace{0.20cm} 0 \\
		\hspace{-0.25cm}-3 &\hspace{0.20cm} \hspace{-0.25cm}-3 &\hspace{0.20cm} 0 &\hspace{0.20cm} 0 &\hspace{0.20cm} 0 &\hspace{0.20cm} 0 &\hspace{0.20cm} 0 &\hspace{0.20cm} 0 &\hspace{0.20cm} 0 &\hspace{0.20cm} 0 &\hspace{0.20cm} 0 &\hspace{0.20cm} 0 &\hspace{0.20cm} 0 &\hspace{0.20cm} 0 &\hspace{0.20cm} 1 &\hspace{0.20cm} 0 &\hspace{0.20cm} 0 &\hspace{0.20cm} 0 &\hspace{0.20cm} 0 &\hspace{0.20cm} 0 &\hspace{0.20cm} 0 &\hspace{0.20cm} 0 &\hspace{0.20cm} 0 &\hspace{0.20cm} 0 \\
		\hspace{-0.25cm}-1 &\hspace{0.20cm} 0 &\hspace{0.20cm} 0 &\hspace{0.20cm} 0 &\hspace{0.20cm} 0 &\hspace{0.20cm} 0 &\hspace{0.20cm} 0 &\hspace{0.20cm} 0 &\hspace{0.20cm} 0 &\hspace{0.20cm} 0 &\hspace{0.20cm} 0 &\hspace{0.20cm} 0 &\hspace{0.20cm} 0 &\hspace{0.20cm} 1 &\hspace{0.20cm} 0 &\hspace{0.20cm} 0 &\hspace{0.20cm} 0 &\hspace{0.20cm} 0 &\hspace{0.20cm} 0 &\hspace{0.20cm} 0 &\hspace{0.20cm} 0 &\hspace{0.20cm} 0 &\hspace{0.20cm} 0 &\hspace{0.20cm} 0 \\
		0 &\hspace{0.20cm} \hspace{-0.25cm}-1 &\hspace{0.20cm} 0 &\hspace{0.20cm} 0 &\hspace{0.20cm} 0 &\hspace{0.20cm} 0 &\hspace{0.20cm} 0 &\hspace{0.20cm} 0 &\hspace{0.20cm} 0 &\hspace{0.20cm} 0 &\hspace{0.20cm} 0 &\hspace{0.20cm} 0 &\hspace{0.20cm} 1 &\hspace{0.20cm} 0 &\hspace{0.20cm} 0 &\hspace{0.20cm} 0 &\hspace{0.20cm} 0 &\hspace{0.20cm} 0 &\hspace{0.20cm} 0 &\hspace{0.20cm} 0 &\hspace{0.20cm} 0 &\hspace{0.20cm} 0 &\hspace{0.20cm} 0 &\hspace{0.20cm} 0 \\
		0 &\hspace{0.20cm} \hspace{-0.25cm}-1 &\hspace{0.20cm} 0 &\hspace{0.20cm} 0 &\hspace{0.20cm} 0 &\hspace{0.20cm} 0 &\hspace{0.20cm} 0 &\hspace{0.20cm} 0 &\hspace{0.20cm} 0 &\hspace{0.20cm} 0 &\hspace{0.20cm} 0 &\hspace{0.20cm} 1 &\hspace{0.20cm} 0 &\hspace{0.20cm} 0 &\hspace{0.20cm} 0 &\hspace{0.20cm} 0 &\hspace{0.20cm} 0 &\hspace{0.20cm} 0 &\hspace{0.20cm} 0 &\hspace{0.20cm} 0 &\hspace{0.20cm} 0 &\hspace{0.20cm} 0 &\hspace{0.20cm} 0 &\hspace{0.20cm} 0 \\
		\hspace{-0.25cm}-3 &\hspace{0.20cm} \hspace{-0.25cm}-3 &\hspace{0.20cm} 0 &\hspace{0.20cm} 0 &\hspace{0.20cm} 0 &\hspace{0.20cm} 0 &\hspace{0.20cm} 0 &\hspace{0.20cm} 0 &\hspace{0.20cm} 0 &\hspace{0.20cm} 0 &\hspace{0.20cm} 1 &\hspace{0.20cm} 0 &\hspace{0.20cm} 0 &\hspace{0.20cm} 0 &\hspace{0.20cm} 0 &\hspace{0.20cm} 0 &\hspace{0.20cm} 0 &\hspace{0.20cm} 0 &\hspace{0.20cm} 0 &\hspace{0.20cm} 0 &\hspace{0.20cm} 0 &\hspace{0.20cm} 0 &\hspace{0.20cm} 0 &\hspace{0.20cm} 0 \\
		\hspace{-0.25cm}-1 &\hspace{0.20cm} 0 &\hspace{0.20cm} 0 &\hspace{0.20cm} 0 &\hspace{0.20cm} 0 &\hspace{0.20cm} 0 &\hspace{0.20cm} 0 &\hspace{0.20cm} 0 &\hspace{0.20cm} 0 &\hspace{0.20cm} 1 &\hspace{0.20cm} 0 &\hspace{0.20cm} 0 &\hspace{0.20cm} 0 &\hspace{0.20cm} 0 &\hspace{0.20cm} 0 &\hspace{0.20cm} 0 &\hspace{0.20cm} 0 &\hspace{0.20cm} 0 &\hspace{0.20cm} 0 &\hspace{0.20cm} 0 &\hspace{0.20cm} 0 &\hspace{0.20cm} 0 &\hspace{0.20cm} 0 &\hspace{0.20cm} 0 \\
		\hspace{-0.25cm}-3 &\hspace{0.20cm} \hspace{-0.25cm}-3 &\hspace{0.20cm} 0 &\hspace{0.20cm} 0 &\hspace{0.20cm} 0 &\hspace{0.20cm} 0 &\hspace{0.20cm} 0 &\hspace{0.20cm} 0 &\hspace{0.20cm} 1 &\hspace{0.20cm} 0 &\hspace{0.20cm} 0 &\hspace{0.20cm} 0 &\hspace{0.20cm} 0 &\hspace{0.20cm} 0 &\hspace{0.20cm} 0 &\hspace{0.20cm} 0 &\hspace{0.20cm} 0 &\hspace{0.20cm} 0 &\hspace{0.20cm} 0 &\hspace{0.20cm} 0 &\hspace{0.20cm} 0 &\hspace{0.20cm} 0 &\hspace{0.20cm} 0 &\hspace{0.20cm} 0 \\
		3 &\hspace{0.20cm} 6 &\hspace{0.20cm} 0 &\hspace{0.20cm} 0 &\hspace{0.20cm} 0 &\hspace{0.20cm} 0 &\hspace{0.20cm} 0 &\hspace{0.20cm} 1 &\hspace{0.20cm} 0 &\hspace{0.20cm} 0 &\hspace{0.20cm} 0 &\hspace{0.20cm} 0 &\hspace{0.20cm} 0 &\hspace{0.20cm} 0 &\hspace{0.20cm} 0 &\hspace{0.20cm} 0 &\hspace{0.20cm} 0 &\hspace{0.20cm} 0 &\hspace{0.20cm} 0 &\hspace{0.20cm} 0 &\hspace{0.20cm} 0 &\hspace{0.20cm} 0 &\hspace{0.20cm} 0 &\hspace{0.20cm} 0 \\
		6 &\hspace{0.20cm} 3 &\hspace{0.20cm} 0 &\hspace{0.20cm} 0 &\hspace{0.20cm} 0 &\hspace{0.20cm} 0 &\hspace{0.20cm} 1 &\hspace{0.20cm} 0 &\hspace{0.20cm} 0 &\hspace{0.20cm} 0 &\hspace{0.20cm} 0 &\hspace{0.20cm} 0 &\hspace{0.20cm} 0 &\hspace{0.20cm} 0 &\hspace{0.20cm} 0 &\hspace{0.20cm} 0 &\hspace{0.20cm} 0 &\hspace{0.20cm} 0 &\hspace{0.20cm} 0 &\hspace{0.20cm} 0 &\hspace{0.20cm} 0 &\hspace{0.20cm} 0 &\hspace{0.20cm} 0 &\hspace{0.20cm} 0 \\
		\hspace{-0.25cm}-1 &\hspace{0.20cm} 0 &\hspace{0.20cm} 0 &\hspace{0.20cm} 0 &\hspace{0.20cm} 0 &\hspace{0.20cm} 1 &\hspace{0.20cm} 0 &\hspace{0.20cm} 0 &\hspace{0.20cm} 0 &\hspace{0.20cm} 0 &\hspace{0.20cm} 0 &\hspace{0.20cm} 0 &\hspace{0.20cm} 0 &\hspace{0.20cm} 0 &\hspace{0.20cm} 0 &\hspace{0.20cm} 0 &\hspace{0.20cm} 0 &\hspace{0.20cm} 0 &\hspace{0.20cm} 0 &\hspace{0.20cm} 0 &\hspace{0.20cm} 0 &\hspace{0.20cm} 0 &\hspace{0.20cm} 0 &\hspace{0.20cm} 0 \\
		2 &\hspace{0.20cm} 2 &\hspace{0.20cm} 0 &\hspace{0.20cm} 0 &\hspace{0.20cm} 1 &\hspace{0.20cm} 0 &\hspace{0.20cm} 0 &\hspace{0.20cm} 0 &\hspace{0.20cm} 0 &\hspace{0.20cm} 0 &\hspace{0.20cm} 0 &\hspace{0.20cm} 0 &\hspace{0.20cm} 0 &\hspace{0.20cm} 0 &\hspace{0.20cm} 0 &\hspace{0.20cm} 0 &\hspace{0.20cm} 0 &\hspace{0.20cm} 0 &\hspace{0.20cm} 0 &\hspace{0.20cm} 0 &\hspace{0.20cm} 0 &\hspace{0.20cm} 0 &\hspace{0.20cm} 0 &\hspace{0.20cm} 0 \\
		0 &\hspace{0.20cm} \hspace{-0.25cm}-1 &\hspace{0.20cm} 0 &\hspace{0.20cm} 1 &\hspace{0.20cm} 0 &\hspace{0.20cm} 0 &\hspace{0.20cm} 0 &\hspace{0.20cm} 0 &\hspace{0.20cm} 0 &\hspace{0.20cm} 0 &\hspace{0.20cm} 0 &\hspace{0.20cm} 0 &\hspace{0.20cm} 0 &\hspace{0.20cm} 0 &\hspace{0.20cm} 0 &\hspace{0.20cm} 0 &\hspace{0.20cm} 0 &\hspace{0.20cm} 0 &\hspace{0.20cm} 0 &\hspace{0.20cm} 0 &\hspace{0.20cm} 0 &\hspace{0.20cm} 0 &\hspace{0.20cm} 0 &\hspace{0.20cm} 0 \\
		2 &\hspace{0.20cm} 2 &\hspace{0.20cm} 1 &\hspace{0.20cm} 0 &\hspace{0.20cm} 0 &\hspace{0.20cm} 0 &\hspace{0.20cm} 0 &\hspace{0.20cm} 0 &\hspace{0.20cm} 0 &\hspace{0.20cm} 0 &\hspace{0.20cm} 0 &\hspace{0.20cm} 0 &\hspace{0.20cm} 0 &\hspace{0.20cm} 0 &\hspace{0.20cm} 0 &\hspace{0.20cm} 0 &\hspace{0.20cm} 0 &\hspace{0.20cm} 0 &\hspace{0.20cm} 0 &\hspace{0.20cm} 0 &\hspace{0.20cm} 0 &\hspace{0.20cm} 0 &\hspace{0.20cm} 0 &\hspace{0.20cm} 0 \\
	\end{array}
	\right), \nn
	\label{D64}
\end{equation}
\begin{align}
	D=(\mathbf{1}_2,0).
\end{align}
The two exponentiated colour factors of this Cweb are
\begin{align}
	(YC)_1=&  f^{abr}f^{adp}f^{dem}f^{mcn} \tp{1} \trr{2} \tn{3} \tb{4} \tc{5} \te{6} +  f^{abr}f^{aem}f^{dcq}f^{qeo} \too{1} \trr{2} \tq{3} \tb{4} \tc{5} \te{6} \, , \nonumber \\
	&=\;{\cal{B}}_{6}+{\cal{B}}_{7} \, . \nn\\
	(YC)_2=& f^{abr}f^{aem}f^{dcq}f^{qeo} \too{1} \trr{2} \tq{3} \tb{4} \tc{5} \te{6} \, , \nn\\
	&=\;{\cal{B}}_{7} \, .
\end{align}

\subsection*{(19)  \ $\mathbf{W^{(5)}_{6} (4, 1, 2, 1, 1, 1)}$}  \label{C71}
This Cweb is the fourth basis Cweb appearing at five loops and six lines it has forty-eight diagrams as there are four attachments on line 1 and two on line 3. One of the diagrams is shown in fig.~\ref{Diag:19}. The sequences of diagrams and their corresponding $s$-factors are provided in table~\ref{Table:Diag19}.

\begin{minipage}{0.35\textwidth}
	\hspace{1cm}	\includegraphics[scale=0.5]{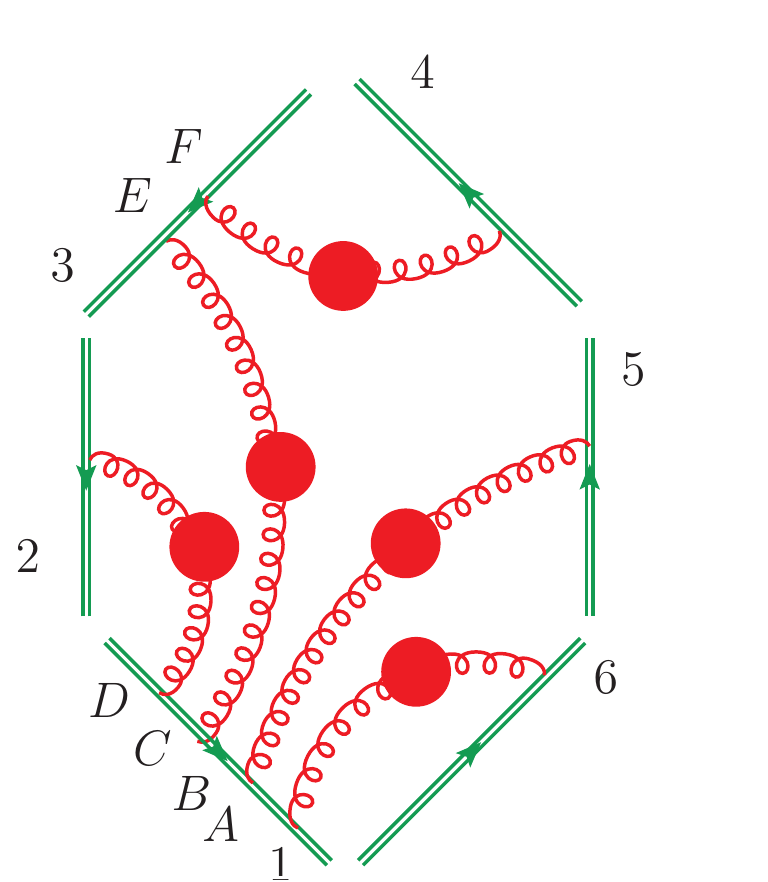}
	\captionof{figure}{$W^{(5)}_{6} (4, 1, 2, 1, 1, 1)$}
	\label{Diag:19}
\end{minipage}
\hspace{1.5cm}\begin{minipage}{0.35\textwidth}
	\footnotesize{
		\begin{tabular}{ | c | c | c |}
			\hline
			\textbf{Diagrams} & \textbf{Sequences} & \textbf{$s$-factors} \\ \hline
			$d_{1}$ & $\lbrace \lbrace ABCD \rbrace,  \lbrace EF \rbrace\rbrace$ & 3\\ 
			\hline
			$d_{2}$ & $\lbrace \lbrace ABCD \rbrace,  \lbrace FE \rbrace\rbrace$ & 2 \\ 
			\hline
			$d_{3}$ & $\lbrace \lbrace ABDC \rbrace,  \lbrace EF \rbrace\rbrace$ & 3 \\ 
			\hline	
			$d_{4}$ & $\lbrace \lbrace ABDC \rbrace,  \lbrace FE \rbrace\rbrace$ & 2 \\ 
			\hline
			$d_{5}$ & $\lbrace \lbrace ACBD \rbrace,  \lbrace EF \rbrace\rbrace$ & 2 \\ 
			\hline
			$d_{6}$ & $\lbrace \lbrace ACBD \rbrace,  \lbrace FE   \rbrace\rbrace$ & 3 \\ 
			\hline
			$d_7$ & $\lbrace \lbrace ACDB \rbrace,  \lbrace EF \rbrace\rbrace$ & 1 \\ 
			\hline
			$d_{8}$ & $\lbrace \lbrace ACDB \rbrace,  \lbrace FE \rbrace\rbrace$ & 4\\ 
			\hline
			$d_{9}$ & $\lbrace \lbrace ADBC \rbrace,  \lbrace EF \rbrace\rbrace$ & 2 \\ 
			\hline
			$d_{10}$ & $\lbrace \lbrace ADBC \rbrace,  \lbrace FE  \rbrace\rbrace$ & 3 \\ 
			\hline
			$d_{11}$ & $\lbrace \lbrace ADCB \rbrace,  \lbrace EF \rbrace\rbrace$ & 1 \\ 
			\hline
			$d_{12}$ & $\lbrace \lbrace ADCB \rbrace,  \lbrace FE \rbrace\rbrace$ & 4\\ 
			\hline	
			$d_{13}$ & $\lbrace \lbrace BACD \rbrace,  \lbrace EF \rbrace\rbrace$ & 4 \\ 
			\hline
			$d_{14}$ & $\lbrace \lbrace BACD \rbrace,  \lbrace FE \rbrace\rbrace$ & 1 \\ 
			\hline
			$d_{15}$ & $\lbrace \lbrace  BADC \rbrace,  \lbrace EF \rbrace\rbrace$ & 4\\ 
			\hline		
			$d_{16}$ & $\lbrace \lbrace BADC \rbrace,  \lbrace FE \rbrace\rbrace$ & 1\\ 
			\hline
			$d_{17}$ & $\lbrace \lbrace  BCAD \rbrace,  \lbrace EF \rbrace\rbrace$ & 4 \\ 
			\hline
			$d_{18}$ & $\lbrace \lbrace  BCAD \rbrace,  \lbrace FE \rbrace\rbrace$ & 1\\ 
			\hline
			$d_{19}$ & $\lbrace \lbrace BCDA \rbrace,  \lbrace EF \rbrace\rbrace$ & 4 \\ 
			\hline
			$d_{20}$ & $\lbrace \lbrace  BCDA \rbrace,  \lbrace FE \rbrace\rbrace$ & 1  \\ 
			\hline
			$d_{21}$ & $\lbrace \lbrace BDAC \rbrace,  \lbrace EF \rbrace\rbrace$ & 4\\ 
			\hline
			$d_{22}$ & $\lbrace \lbrace BDAC \rbrace,  \lbrace FE \rbrace\rbrace$ & 1 \\ 
			\hline
			$d_{23}$ & $\lbrace \lbrace BDCA \rbrace,  \lbrace EF \rbrace\rbrace$ & 4 \\ 
			\hline
			$d_{24}$ & $\lbrace \lbrace  BDCA \rbrace,  \lbrace FE \rbrace\rbrace$ & 1 \\ 
			\hline
			$d_{25}$ & $\lbrace \lbrace CABD \rbrace,  \lbrace EF \rbrace\rbrace$ & 2 \\ 
			\hline
			$d_{26}$ & $\lbrace \lbrace CABD \rbrace,  \lbrace FE \rbrace\rbrace$ & 3 \\ 
			\hline
			$d_{27}$ & $\lbrace \lbrace CADB \rbrace,  \lbrace EF \rbrace\rbrace$ & 1 \\ 
			\hline
			$d_{28}$ & $\lbrace \lbrace CADB \rbrace,  \lbrace FE \rbrace\rbrace$ & 4  \\ 
			\hline
			$d_{29}$ & $\lbrace \lbrace CBAD \rbrace,  \lbrace EF \rbrace\rbrace$ & 3 \\ 
			\hline
			$d_{30}$ & $\lbrace \lbrace CBAD \rbrace,  \lbrace FE \rbrace\rbrace$ & 2 \\ 
			\hline
			$d_{31}$ & $\lbrace \lbrace CBDA \rbrace,  \lbrace EF \rbrace\rbrace$ & 3 \\ 
			\hline
			$d_{32}$ & $\lbrace \lbrace CBDA \rbrace,  \lbrace FE \rbrace\rbrace$ & 2 \\ 
			\hline
			$d_{33}$ & $\lbrace \lbrace CDAB \rbrace,  \lbrace EF \rbrace\rbrace$ & 1  \\ 
			\hline
			$d_{34}$ & $\lbrace \lbrace CDAB \rbrace,  \lbrace FE \rbrace\rbrace$ & 4 \\ 
			\hline
			$d_{35}$ & $\lbrace \lbrace CDBA \rbrace,  \lbrace EF \rbrace\rbrace$ & 2 \\ 
			\hline
			$d_{36}$ & $\lbrace \lbrace  CDBA \rbrace,  \lbrace FE \rbrace\rbrace$ & 3 \\ 
			\hline	
			$d_{37}$ & $\lbrace \lbrace  DABC \rbrace,  \lbrace EF \rbrace\rbrace$ & 2 \\ 
			\hline
			$d_{38}$ & $\lbrace \lbrace DABC \rbrace,  \lbrace FE \rbrace\rbrace$ & 3 \\ 
			\hline
			$d_{39}$ & $\lbrace \lbrace DACB \rbrace,  \lbrace EF  \rbrace\rbrace$ & 1\\ 
			\hline
			$d_{40}$ & $\lbrace \lbrace DACB \rbrace,  \lbrace FE   \rbrace\rbrace$ & 4 \\ 
			\hline
			$d_{41}$ & $\lbrace \lbrace DBAC \rbrace,  \lbrace EF \rbrace\rbrace$ & 3  \\ 
			\hline
			$d_{42}$ & $\lbrace \lbrace DBAC \rbrace,  \lbrace FE \rbrace\rbrace$ & 2  \\ 
			\hline
			$d_{43}$ & $\lbrace \lbrace DBCA \rbrace,  \lbrace EF \rbrace\rbrace$ & 3 \\ 
			\hline
			$d_{44}$ & $\lbrace \lbrace DBCA \rbrace,  \lbrace FE \rbrace\rbrace$ & 2\\ 
			\hline
			$d_{45}$ & $\lbrace \lbrace DCAB \rbrace,  \lbrace EF \rbrace\rbrace$ & 1 \\ 
			\hline
			$d_{46}$ & $\lbrace \lbrace  DCAB \rbrace,  \lbrace FE  \rbrace\rbrace$ & 4 \\ 
			\hline
			$d_{47}$ & $\lbrace \lbrace DCBA \rbrace,  \lbrace FE \rbrace\rbrace$ & 2 \\ 
			\hline
			$d_{48}$ & $\lbrace \lbrace DCBA \rbrace,  \lbrace EF \rbrace\rbrace$ & 3 \\ 
			\hline
		\end{tabular}
		\captionof{table}{Sequences and $s$-factors}
		\label{Table:Diag19}	}
\end{minipage}

\vspace{0.5cm}
\noindent  The dimension of mixing matrix $R$ is quite large and is not presented here, instead as mentioned in the beginning of appendix, it is provided in a Mathematica file along with the matrices of other Cweb. The diagonal matrix $D$ for the Cweb mentioned above is
\begin{align}
	D=(\mathbf{1}_6,0).
	\label{D71}
\end{align}

The six exponentiated colour factors of this Cweb are listed below
\begin{align}
	(YC)_1=&- f^{aoj}f^{cbm}f^{edn}f^{nbo} \tj{1} \ta{2} \tm{3} \tc{4} \td{5} \te{6}, \nonumber \\
	&=\;{\cal{B}}_{29} \, .\nn\\
	(YC)_2=& +f^{abq}f^{cbm}f^{edn}f^{nqk}\tkk{1} \ta{2} \tm{3} \tc{4} \td{5} \te{6} - f^{aoj}f^{cbm}f^{edn}f^{nbo}\tj{1} \ta{2} \tm{3} \tc{4} \td{5} \te{6} , \nonumber \\
	&=\;-{\cal{B}}_{7}-{\cal{B}}_{10}+{\cal{B}}_{29} \, . \nn\\	
	(YC)_3=&- f^{aoj}f^{cbm}f^{ebr}f^{rdo}\tj{1} \ta{2} \tm{3} \tc{4} \td{5} \te{6}, \nonumber  \\
	&=\;{\cal{B}}_{20} \, . \nn\\
	(YC)_4=& +f^{adp}f^{cbm}f^{ebr}f^{rpk}\tkk{1} \ta{2} \tm{3} \tc{4} \td{5} \te{6}-      f^{aoj}f^{cbm}f^{ebr}f^{rdo}\tj{1} \ta{2} \tm{3} \tc{4} \td{5} \te{6}, \nonumber \\
	&=\;   -{\cal{B}}_{1}-{\cal{B}}_{8}+ {\cal{B}}_{20} \, . \nn\\
	(YC)_5=& -f^{adp}f^{cbm}f^{ebr}f^{prk}\tkk{1} \ta{2} \tm{3} \tc{4} \td{5} \te{6} 
	+ f^{abq}f^{cbm}f^{edn}f^{nql}\tl{1} \ta{2} \tm{3} \tc{4} \td{5} \te{6} \nonumber \\
	& + f^{adp}f^{cbm}f^{eod}f^{pbo}\tl{1} \ta{2} \tm{3} \tc{4} \td{5} \te{6} \nonumber + f^{abq}f^{cbm}f^{edn}f^{nql}\tl{1} \ta{2} \tm{3} \tc{4} \td{5} \te{6} ,\\
	&=\;{\cal{B}}_{1}+{\cal{B}}_{7}+{\cal{B}}_{8}+{\cal{B}}_{10}+{\cal{B}}_{26}+{\cal{B}}_{29} \, . \nn\\
	(YC)_6=& + f^{arj}f^{cbm}f^{djl}f^{ebr}\tl{1} \ta{2} \tm{3} \tc{4} \td{5} \te{6} + f^{abq}f^{cbm}f^{eos}f^{qdo}\ts{1} \ta{2} \tm{3} \tc{4} \td{5} \te{6} \nonumber \\
	& + f^{abq}f^{cbm}f^{edn}f^{qnt}\ttt{1} \ta{2} \tm{3} \tc{4} \td{5} \te{6} \, ,\nn\\
	&=\; -{\cal{B}}_{1}-{\cal{B}}_{7}-{\cal{B}}_{8}-{\cal{B}}_{10}+{\cal{B}}_{11}+{\cal{B}}_{20} \, .
\end{align}

\subsection*{(20)  \ $\mathbf{W^{(5)}_{6} (3, 1, 3, 1, 1, 1)}$}  \label{C74}
This Cweb is the fifth basis Cweb appearing at five loops and six lines it has thirty-six diagrams as there are three attachments on each lines 1 and 3. One of the diagrams is shown in fig.~\ref{Diag:20}. The sequences of diagrams and their corresponding $s$-factors are provided in table~\ref{Table:Diag20}.

\begin{minipage}{0.5\textwidth}
	\hspace{1.5cm}	\includegraphics[scale=0.5]{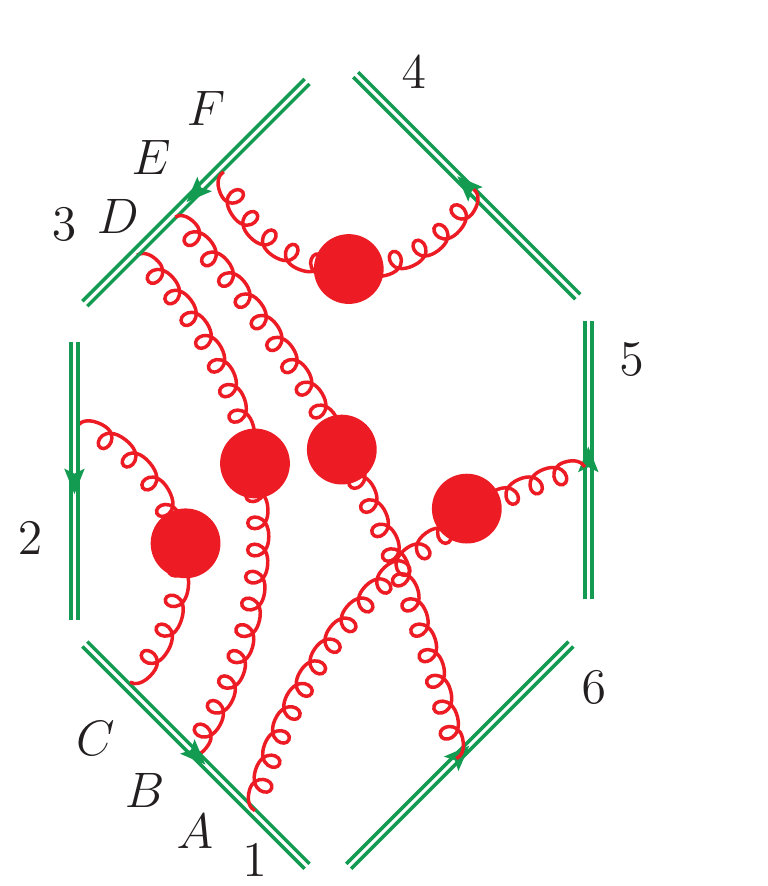}
	\captionof{figure}{$W^{(5)}_{6} (3, 1, 3, 1, 1, 1)$}
	\label{Diag:20}
	\label{d7-4}
\end{minipage}
\begin{minipage}{0.45\textwidth}
	\begin{tabular}{ | c | c | c |}
		\hline
		\textbf{Diagrams} & \textbf{Sequences} & \textbf{$s$-factors} \\ \hline
		$d_1$ & $\lbrace \lbrace ABC \rbrace,  \lbrace DEF \rbrace\rbrace$ & 3 \\ 
		\hline
		$d_{2}$ & $\lbrace \lbrace ABC \rbrace,  \lbrace DFE \rbrace\rbrace$ & 3 \\ 
		\hline
		$d_{3}$ & $\lbrace \lbrace ABC \rbrace,  \lbrace EDF \rbrace\rbrace$ & 4  \\ 
		\hline
		$d_{4}$ & $\lbrace \lbrace ABC \rbrace,  \lbrace EFD \rbrace\rbrace$ & 3  \\ 
		\hline
		$d_{5}$ & $\lbrace \lbrace ABC \rbrace,  \lbrace FDE \rbrace\rbrace$ & 4 \\ 
		\hline
		$d_{6}$ & $\lbrace \lbrace ABC \rbrace,  \lbrace FED \rbrace\rbrace$ & 3 \\ 
		\hline
		$d_{7}$ & $\lbrace \lbrace ACB \rbrace,  \lbrace DEF \rbrace\rbrace$ & 1 \\ 
		\hline
		$d_{8}$ & $\lbrace \lbrace ACB \rbrace,  \lbrace DFE \rbrace\rbrace$ & 1 \\ 
		\hline
		$d_{9}$ & $\lbrace \lbrace ACB \rbrace,  \lbrace EDF \rbrace\rbrace$ & 3 \\ 
		\hline
		$d_{10}$ & $\lbrace \lbrace ACB \rbrace,  \lbrace EFD \rbrace\rbrace$ & 6  \\ 
		\hline
		$d_{11}$ & $\lbrace \lbrace ACB \rbrace,  \lbrace FDE \rbrace\rbrace$ & 3 \\ 
		\hline
		$d_{12}$ & $\lbrace \lbrace ACB \rbrace,  \lbrace FED \rbrace\rbrace$ & 6 \\ 
		\hline
		$d_{13}$ & $\lbrace \lbrace BAC \rbrace,  \lbrace DEF \rbrace\rbrace$ & 6 \\ 
		\hline
		$d_{14}$ & $\lbrace \lbrace BAC \rbrace,  \lbrace DFE \rbrace\rbrace$ & 6 \\ 
		\hline
		$d_{15}$ & $\lbrace \lbrace BAC \rbrace,  \lbrace EDF \rbrace\rbrace$ & 3 \\ 
		\hline
		$d_{16}$ & $\lbrace \lbrace BAC \rbrace,  \lbrace EFD \rbrace\rbrace$ & 1  \\ 
		\hline
		$d_{17}$ & $\lbrace \lbrace BAC \rbrace,  \lbrace FDE \rbrace\rbrace$ & 3 \\ 
		\hline
		$d_{18}$ & $\lbrace \lbrace BAC \rbrace,  \lbrace FED \rbrace\rbrace$ & 1 \\ 
		\hline
		$d_{19}$ & $\lbrace \lbrace BCA \rbrace,  \lbrace DEF \rbrace\rbrace$ & 6 \\ 
		\hline
		$d_{20}$ & $\lbrace \lbrace BCA \rbrace,  \lbrace DFE \rbrace\rbrace$ & 6  \\ 
		\hline
		$d_{21}$ & $\lbrace \lbrace BCA \rbrace,  \lbrace EDF \rbrace\rbrace$ & 3 \\ 
		\hline
		$d_{22}$ & $\lbrace \lbrace BCA \rbrace,  \lbrace EFD \rbrace\rbrace$ & 1 \\ 
		\hline
		$d_{23}$ & $\lbrace \lbrace BCA \rbrace,  \lbrace FDE \rbrace\rbrace$ & 3 \\ 
		\hline
		$d_{24}$ & $\lbrace \lbrace BCA \rbrace,  \lbrace FED \rbrace\rbrace$ & 1\\ 
		\hline
		$d_{25}$ & $\lbrace \lbrace CAB \rbrace,  \lbrace DEF \rbrace\rbrace$ & 1 \\ 
		\hline
		$d_{26}$ & $\lbrace \lbrace CAB \rbrace,  \lbrace DFE \rbrace\rbrace$ & 1 \\ 
		\hline
		$d_{27}$ & $\lbrace \lbrace CAB \rbrace,  \lbrace EDF \rbrace\rbrace$ & 3 \\ 
		\hline		
		$d_{28}$ & $\lbrace \lbrace CAB \rbrace,  \lbrace EFD \rbrace\rbrace$ & 6 \\ 
		\hline
		$d_{29}$ & $\lbrace \lbrace CAB \rbrace,  \lbrace FDE \rbrace\rbrace$ & 3 \\ 
		\hline
		$d_{30}$ & $\lbrace \lbrace CAB \rbrace,  \lbrace FED \rbrace\rbrace$ & 6 \\ 
		\hline
		$d_{31}$ & $\lbrace \lbrace CBA \rbrace,  \lbrace DFE \rbrace\rbrace$ & 3 \\ 
		\hline
		$d_{32}$ & $\lbrace \lbrace CBA \rbrace,  \lbrace DEF \rbrace\rbrace$ & 3 \\ 
		\hline
		$d_{33}$ & $\lbrace \lbrace CBA \rbrace,  \lbrace EDF \rbrace\rbrace$ & 4  \\ 
		\hline
		$d_{34}$ & $\lbrace \lbrace CBA \rbrace,  \lbrace EFD \rbrace\rbrace$ & 3 \\ 
		\hline
		$d_{35}$ & $\lbrace \lbrace CBA \rbrace,  \lbrace FDE \rbrace\rbrace$ & 4  \\ 
		\hline
		$d_{36}$ & $\lbrace \lbrace CBA \rbrace,  \lbrace FED \rbrace\rbrace$ & 3  \\ 
		\hline
	\end{tabular}
	\captionof{table}{Sequences and $s$-factors}
	\label{Table:Diag20}
\end{minipage} 

\vspace{0.5cm}
\noindent  The dimension of mixing matrix $R$ is quite large and is not presented here, instead as mentioned in the beginning of appendix, it is provided in a Mathematica file along with the matrices of other Cweb. The diagonal matrix $ D $ for this Cweb is given as
\begin{align}
	D=(\mathbf{1}_4,0).
	\label{D74}
\end{align}
The four exponentiated colour factors of this Cweb are
\begin{align}
	(YC)_1=& + f^{arn}f^{cdp}f^{ebr}f^{pbm}\tn{1} \ta{2} \tm{3} \td{4} \te{5} \tc{6}, \nonumber \\
	&=\;-{\cal{B}}_{8}-{\cal{B}}_{18} \, . \nn\\
	(YC)_2=& + f^{arn}f^{cdp}f^{ebr}f^{pbm}\tn{1} \ta{2} \tm{3} \td{4} \te{5} \tc{6} - f^{arn}f^{cbj}f^{ebr}f^{jdo} \tn{1} \ta{2} \too{3} \td{4} \te{5} \tc{6} , \nonumber \\
	&=\;-{\cal{B}}_{8}-{\cal{B}}_{18}+{\cal{B}}_{20} \, . \nn\\
	(YC)_3=&- f^{abq}f^{cdp}f^{eqk}f^{pbm}\tkk{1} \ta{2} \tm{3} \td{4} \te{5} \tc{6}+ f^{arn}f^{cdp}f^{ebr}f^{pbm}\tn{1} \ta{2} \tm{3} \td{4} \te{5} \tc{6}, \nonumber  \\
	&=\;-{\cal{B}}_{8}-2{\cal{B}}_{18} \, . \nn\\
	(YC)_4=& -f^{abq}f^{cdp}f^{eqk}f^{pbm}\tkk{1} \ta{2} \tm{3} \td{4} \te{5} \tc{6}
	+ f^{abq}f^{cbj}f^{eqk}f^{jdo} \tkk{1} \ta{2} \too{3} \td{4} \te{5} \tc{6}\nonumber \\
	& + f^{arn}f^{cdp}f^{ebr}f^{pbm}\tn{1} \ta{2} \tm{3} \td{4} \te{5} \tc{6} \nonumber + f^{arn}f^{cbj}f^{ebr}f^{jdo}\tn{1} \ta{2} \too{3} \td{4} \te{5} \tc{6} ,\\
	&=\;{\cal{B}}_{1}+2{\cal{B}}_{8}-{\cal{B}}_{20}\, .\nn\\
\end{align}

\subsection*{(21)  \ $\mathbf{W^{(5)}_{6} (5, 1, 1, 1, 1, 1)}$}  \label{C81}

\begin{figure}[hbtp]
	\centering
	\includegraphics[scale=0.5]{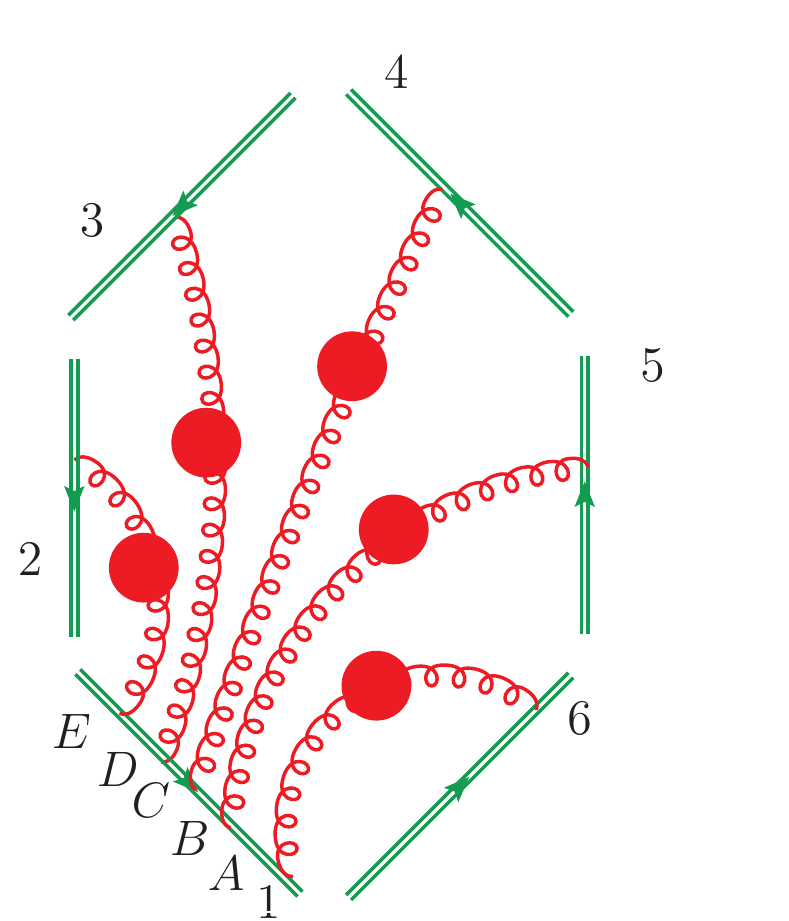}
	\caption{ The sixth basis Cweb $W^{(5)}_{6} (5, 1, 1, 1, 1, 1)$ at five loops and six lines} 
	\label{Diag:22}
\end{figure}

This Cweb is the sixth and last basis Cweb appearing at five loops and six lines it has one hundred-twenty diagrams as there are 5 attachments on line 1. One of the diagrams is shown in fig.~\ref{Diag:22}. This Cweb has the largest mixing matrix computed at five loops and six lines. The dimension of this mixing matrix is $120 \times 120$, and there are 24 ECFs associated with this Cweb. The $s$-factors of all $120$ diagrams is 1.

The mixing matrix and the diagonalizing matrix of this Cweb is $120 \times 120$ each. 
This mixing matrix is too large to report, and we present it in the attached ancillary file. The twenty-four ECFs and sequences for the $120$ diagrams with their corresponding $s$-factors are given in the  tables~\ref{Table:Diag22-1} and \ref{Table:Diag22-2}.

\hspace{2cm} \begin{minipage}{0.35\textwidth}
	\footnotesize{
		\begin{tabular}{ | c | c | c || c | c | c|}
			\hline
			\textbf{Diagrams} & \textbf{Sequences} & \textbf{$s$-factors} & \textbf{Diagrams} & \textbf{Sequences} & \textbf{$s$-factors} \\ \hline
			$d_1$ & $\lbrace \lbrace ABCDE \rbrace \rbrace$ & 1 & $d_{31}$ & $\lbrace \lbrace BCADE \rbrace\rbrace$ & 1 \\
			\hline
			$d_2$ & $\lbrace \lbrace ABCED \rbrace \rbrace$ & 1 & $d_{32}$ & $\lbrace \lbrace BCAED \rbrace \rbrace$ & 1 \\
			\hline
			$d_3$ & $\lbrace \lbrace ABDCE \rbrace \rbrace$ & 1 & $d_{33}$ & $\lbrace \lbrace BCDAE \rbrace \rbrace$ & 1 \\
			\hline
			$d_4$ & $\lbrace \lbrace ABDEC \rbrace\rbrace$ & 1 & $d_{34}$ & $\lbrace \lbrace BCDEA \rbrace \rbrace$ & 1 \\
			\hline
			$d_5$ & $\lbrace \lbrace ABECD \rbrace\rbrace$ & 1 & $d_{35}$ & $\lbrace \lbrace BCEAD \rbrace \rbrace$ & 1 \\
			\hline
			$d_6$ & $\lbrace \lbrace ABEDC \rbrace\rbrace$ & 1 & $d_{36}$ & $\lbrace \lbrace BCEDA \rbrace \rbrace$ & 1 \\
			\hline
			$d_7$ & $\lbrace \lbrace ACDBE \rbrace \rbrace$ & 1 & $d_{37}$ & $\lbrace \lbrace BDACE \rbrace \rbrace$ & 1 \\
			\hline
			$d_8$ & $\lbrace \lbrace ACBED \rbrace \rbrace$ & 1 & $d_{38}$ & $\lbrace \lbrace BDAEC \rbrace \rbrace$ & 1 \\
			\hline
			$d_9$ & $\lbrace \lbrace ACDBE \rbrace \rbrace$ & 1 & $d_{39}$ & $\lbrace \lbrace BDCAE \rbrace \rbrace$ & 1 \\
			\hline
			$d_{10}$ & $\lbrace \lbrace ACDEB \rbrace \rbrace$ & 1 & $d_{40}$ & $\lbrace \lbrace BDCEA \rbrace \rbrace$ & 1 \\
			\hline
			$d_{11}$ & $\lbrace \lbrace ACEBD \rbrace \rbrace$ & 1 & $d_{41}$ & $\lbrace \lbrace BDEAC \rbrace \rbrace$ & 1 \\
			\hline
			$d_{12}$ & $\lbrace \lbrace ACEDB \rbrace \rbrace$ & 1 & $d_{42}$ & $\lbrace \lbrace BDECA \rbrace \rbrace$ & 1 \\
			\hline
			$d_{13}$ & $\lbrace \lbrace ADBCE \rbrace \rbrace$ & 1 & $d_{43}$ & $\lbrace \lbrace BEACD \rbrace \rbrace$ & 1 \\
			\hline
			$d_{14}$ & $\lbrace \lbrace ADBEC \rbrace \rbrace$ & 1 & $d_{44}$ & $\lbrace \lbrace BEADC \rbrace \rbrace$ & 1 \\
			\hline
			$d_{15}$ & $\lbrace \lbrace ADCBE \rbrace \rbrace$ & 1 & $d_{45}$ & $\lbrace \lbrace BECAD \rbrace \rbrace$ & 1 \\
			\hline
			$d_{16}$ & $\lbrace \lbrace ADCEB \rbrace \rbrace$ & 1 & $d_{46}$ & $\lbrace \lbrace BECDA \rbrace \rbrace$ & 1 \\
			\hline
			$d_{17}$ & $\lbrace \lbrace ADEBC \rbrace \rbrace$ & 1 & $d_{47}$ & $\lbrace \lbrace BEDAC \rbrace \rbrace$ & 1 \\
			\hline
			$d_{18}$ & $\lbrace \lbrace ADECB\rbrace \rbrace$ & 1 & $d_{48}$ & $\lbrace \lbrace BEDCA \rbrace \rbrace$ & 1 \\
			\hline
			$d_{19}$ & $\lbrace \lbrace AEBCD \rbrace \rbrace$ & 1 & $d_{49}$ & $\lbrace \lbrace CABDE \rbrace \rbrace$ & 1 \\
			\hline
			$d_{20}$ & $\lbrace \lbrace AEBDC \rbrace \rbrace$ & 1 & $d_{50}$ & $\lbrace \lbrace CABED \rbrace \rbrace$ & 1 \\
			\hline
			$d_{21}$ & $\lbrace \lbrace AECBD \rbrace \rbrace$ & 1 & $d_{51}$ & $\lbrace \lbrace CADBE \rbrace \rbrace$ & 1 \\
			\hline
			$d_{22}$ & $\lbrace \lbrace AECDB \rbrace \rbrace$ & 1 & $d_{52}$ & $\lbrace \lbrace CADEB \rbrace \rbrace$ & 1 \\
			\hline
			$d_{23}$ & $\lbrace \lbrace AEDBC \rbrace \rbrace$ & 1 & $d_{53}$ & $\lbrace \lbrace CAEBD \rbrace \rbrace$ & 1 \\
			\hline
			$d_{24}$ & $\lbrace \lbrace AEDCB \rbrace \rbrace$ & 1 & $d_{54}$ & $\lbrace \lbrace CAEDB \rbrace \rbrace$ & 1 \\
			\hline
			$d_{25}$ & $\lbrace \lbrace BACDE \rbrace \rbrace$ & 1 & $d_{55}$ & $\lbrace \lbrace CBADE \rbrace \rbrace$ & 1 \\
			\hline
			$d_{26}$ & $\lbrace \lbrace BACED \rbrace \rbrace$ & 1 & $d_{56}$ & $\lbrace \lbrace CBAED  \rbrace \rbrace$ & 1 \\
			\hline
			$d_{27}$ & $\lbrace \lbrace BADCE \rbrace \rbrace$ & 1 & $d_{57}$ & $\lbrace \lbrace CBDAE \rbrace \rbrace$ & 1 \\
			\hline
			$d_{28}$ & $\lbrace \lbrace BADEC \rbrace \rbrace$ & 1 & $d_{58}$ & $\lbrace \lbrace CBDEA \rbrace \rbrace$ & 1 \\
			\hline
			$d_{29}$ & $\lbrace \lbrace BAECD  \rbrace \rbrace$ & 1 & $d_{59}$ & $\lbrace \lbrace CBEAD \rbrace \rbrace$ & 1 \\
			\hline
			$d_{30}$ & $\lbrace \lbrace BAEDC  \rbrace \rbrace$ & 1 & $d_{60}$ & $\lbrace \lbrace CBEDA \rbrace \rbrace$ & 1 \\
			\hline
		\end{tabular}
	}
\end{minipage} 
\captionof{table}{ The sequences and $s$-factors of first sixty diagrams for the Cweb $W^{(5)}_{6} (1, 1, 1, 1, 1, 5)$ as shown in fig.~\ref{Diag:22}. }
\label{Table:Diag22-1}
\hspace{2cm} \begin{minipage}{0.35\textwidth}
	\footnotesize{
		\begin{tabular}{ | c | c | c || c | c | c|}
			\hline
			\textbf{Diagrams} & \textbf{Sequences} & \textbf{$s$-factors} & \textbf{Diagrams} & \textbf{Sequences} & \textbf{$s$-factors} \\ \hline
			$d_{61}$ & $\lbrace \lbrace CDABE \rbrace \rbrace $ & 1 & $d_{91}$ & $\lbrace \lbrace DEABC \rbrace \rbrace$ & 1 \\
			\hline
			$d_{62}$ & $\lbrace \lbrace CDAEB  \rbrace\rbrace$ & 1 & $d_{92}$ & $\lbrace \lbrace DEACB  \rbrace\rbrace$ & 1 \\
			\hline
			$d_{63}$ & $\lbrace \lbrace CDBAE  \rbrace\rbrace$ & 1 & $d_{93}$ & $\lbrace \lbrace DEBAC  \rbrace\rbrace$ & 1 \\
			\hline
			$d_{64}$ & $\lbrace \lbrace CDBEA  \rbrace\rbrace$ & 1 & $d_{94}$ & $\lbrace \lbrace DEBCA \rbrace\rbrace$ & 1 \\
			\hline
			$d_{65}$ & $\lbrace \lbrace CDEAB \rbrace\rbrace$ & 1 & $d_{95}$ & $\lbrace \lbrace DECAB  \rbrace\rbrace$ & 1 \\
			\hline
			$d_{66}$ & $\lbrace \lbrace CDEBA  \rbrace\rbrace$ & 1 & $d_{96}$ & $\lbrace \lbrace DECBA \rbrace\rbrace$ & 1 \\
			\hline
			$d_{67}$ & $\lbrace \lbrace CEABD  \rbrace\rbrace$ & 1 & $d_{97}$ & $\lbrace \lbrace EABCD  \rbrace\rbrace$ & 1 \\
			\hline
			$d_{68}$ & $\lbrace \lbrace CEADB  \rbrace\rbrace$ & 1 & $d_{98}$ & $\lbrace \lbrace EABDC  \rbrace\rbrace$ & 1 \\
			\hline
			$d_{69}$ & $\lbrace \lbrace CEBAD  \rbrace\rbrace$ & 1 & $d_{99}$ & $\lbrace \lbrace EACBD  \rbrace\rbrace$ & 1 \\
			\hline
			$d_{70}$ & $\lbrace \lbrace CEBDA \rbrace\rbrace$ & 1 & $d_{100}$ & $\lbrace \lbrace EACDB  \rbrace\rbrace$ & 1 \\
			\hline
			$d_{71}$ & $\lbrace \lbrace CEDAB  \rbrace\rbrace$ & 1 & $d_{101}$ & $\lbrace \lbrace EADBC  \rbrace\rbrace$ & 1 \\
			\hline
			$d_{72}$ & $\lbrace \lbrace CEDBA  \rbrace\rbrace$ & 1 & $d_{102}$ & $\lbrace \lbrace EADCB  \rbrace\rbrace$ & 1 \\
			\hline
			$d_{73}$ & $\lbrace \lbrace DABCE  \rbrace\rbrace$ & 1 & $d_{103}$ & $\lbrace \lbrace EBACD \rbrace\rbrace$ & 1 \\
			\hline
			$d_{74}$ & $\lbrace \lbrace DABEC  \rbrace\rbrace$ & 1 & $d_{104}$ & $\lbrace \lbrace EBADC  \rbrace\rbrace$ & 1 \\
			\hline
			$d_{75}$ & $\lbrace \lbrace  DACBE  \rbrace\rbrace$ & 1 & $d_{105}$ & $\lbrace \lbrace EBCAD  \rbrace\rbrace$ & 1 \\
			\hline
			$d_{76}$ & $\lbrace \lbrace DACEB  \rbrace\rbrace$ & 1 & $d_{106}$ & $\lbrace \lbrace EBCDA  \rbrace\rbrace$ & 1 \\
			\hline
			$d_{77}$ & $\lbrace \lbrace DAEBC  \rbrace\rbrace$ & 1 & $d_{107}$ & $\lbrace \lbrace EBDAC  \rbrace\rbrace$ & 1 \\
			\hline
			$d_{78}$ & $\lbrace \lbrace DAECB \rbrace\rbrace$ & 1 & $d_{108}$ & $\lbrace \lbrace EBDCA \rbrace\rbrace$ & 1 \\
			\hline
			$d_{79}$ & $\lbrace \lbrace DBACE  \rbrace\rbrace$ & 1 & $d_{109}$ & $\lbrace \lbrace ECABD  \rbrace\rbrace$ & 1 \\
			\hline
			$d_{80}$ & $\lbrace \lbrace DBAEC  \rbrace\rbrace$ & 1 & $d_{110}$ & $\lbrace \lbrace ECADB  \rbrace\rbrace$ & 1 \\
			\hline
			$d_{81}$ & $\lbrace \lbrace DBCAE \rbrace\rbrace$ & 1 & $d_{111}$ & $\lbrace \lbrace ECBAD  \rbrace\rbrace$ & 1 \\
			\hline
			$d_{82}$ & $\lbrace \lbrace DBCEA  \rbrace\rbrace$ & 1 & $d_{112}$ & $\lbrace \lbrace ECBDA  \rbrace \rbrace$ & 1 \\
			\hline
			$d_{83}$ & $\lbrace \lbrace DBEAC  \rbrace\rbrace$ & 1 & $d_{113}$ & $\lbrace \lbrace ECDAB  \rbrace\rbrace$ & 1 \\
			\hline
			$d_{84}$ & $\lbrace \lbrace DBECA  \rbrace\rbrace$ & 1 & $d_{114}$ & $\lbrace \lbrace ECDBA  \rbrace\rbrace$ & 1 \\
			\hline
			$d_{85}$ & $\lbrace \lbrace DCABE  \rbrace\rbrace$ & 1 & $d_{115}$ & $\lbrace \lbrace EDABC  \rbrace\rbrace$ & 1 \\
			\hline
			$d_{86}$ & $\lbrace \lbrace DCAEB  \rbrace\rbrace$ & 1 & $d_{116}$ & $\lbrace \lbrace EDACB  \rbrace\rbrace$ & 1 \\
			\hline
			$d_{87}$ & $\lbrace \lbrace DCBAE  \rbrace\rbrace$ & 1 & $d_{117}$ & $\lbrace \lbrace EDBAC  \rbrace\rbrace$ & 1 \\
			\hline
			$d_{88}$ & $\lbrace \lbrace DCBEA  \rbrace\rbrace$ & 1 & $d_{118}$ & $\lbrace \lbrace EDBCA  \rbrace\rbrace$ & 1 \\
			\hline
			$d_{89}$ & $\lbrace \lbrace DCEAB  \rbrace\rbrace$ & 1 & $d_{119}$ & $\lbrace \lbrace EDCAB  \rbrace\rbrace$ & 1 \\
			\hline
			$d_{90}$ & $\lbrace \lbrace DCEBA  \rbrace\rbrace$ & 1 & $d_{120}$ & $\lbrace \lbrace EDCBA  \rbrace\rbrace$ & 1 \\
			\hline
	\end{tabular}	}
\end{minipage} 
\captionof{table}{ The sequences and $s$-factors of last sixty diagrams for the Cweb $W^{(5)}_{6} (1, 1, 1, 1, 1, 5)$ as shown in fig.~\ref{Diag:22}. }
\label{Table:Diag22-2}

\vspace{0.5cm}

The diagonal matrix $ D $ for this Cweb is
\begin{align}
	D=(\mathbf{1}_{24},0).
\end{align}
This Cweb has twenty-four exponentiated colour factors and they are given below
\begin{align}
	(YC)_1=      & - f^{abq}f^{cnt}f^{edn}f^{qtj} \tj{1} \ta{2} \tb{3} \tc{4} \td{5} \te{6}             + f^{ato}f^{cnt}f^{edn}f^{obk} \tkk{1} \ta{2} \tb{3} \tc{4} \td{5} \te{6}\, , \nonumber \\
	&=\;{\cal{B}}_{4}-{\cal{B}}_{15}\, .\nn\\
	(YC)_2=     & +f^{ato}f^{cnt}f^{edn}f^{obk} \tkk{1} \ta{2} \tb{3} \tc{4} \td{5} \te{6}\, , \nonumber \\
	&=\;-{\cal{B}}_{15} \, .\nn\\
	(YC)_3=     &  - f^{abq}f^{cnt}f^{edn}f^{qtj} \tj{1} \ta{2} \tb{3} \tc{4} \td{5} \te{6} 
	+  f^{ato}f^{cnt}f^{edn}f^{obk} \tkk{1} \ta{2} \tb{3} \tc{4} \td{5} \te{6}\nonumber \\
	&+  f^{auv}f^{cbs}f^{edn}f^{nsu} T_1^v \ta{2} \tb{3} \tc{4} \td{5} \te{6}\, , \nonumber \\
	&=\;{\cal{B}}_{4}-{\cal{B}}_{15}+{\cal{B}}_{29}\, .\nn\\
	(YC)_4=      & - f^{abq}f^{cnt}f^{edn}f^{qtj} \tj{1} \ta{2} \tb{3} \tc{4} \td{5} \te{6} 
	+  f^{anp}f^{cwj}f^{edn}f^{pbw} \tj{1} \ta{2} \tb{3} \tc{4} \td{5} \te{6}\nonumber \\
	&+  f^{abq}f^{cqk}f^{edn}f^{knj} \tj{1} \ta{2} \tb{3} \tc{4} \td{5} \te{6}\, ,\nonumber \\
	&=\;-{\cal{B}}_{2}-{\cal{B}}_{4}-{\cal{B}}_{19}\, .\nn\\
	(YC)_5=   & +f^{anp}f^{bmj}f^{edn}f^{pcm} \tj{1} \ta{2} \tb{3} \tc{4} \td{5} \te{6}\, , \nonumber \\
	&=\;-{\cal{B}}_{17}\, .\nn\\
	(YC)_6=  & + f^{anp}f^{cwj}f^{edn}f^{pbw} \tj{1} \ta{2} \tb{3} \tc{4} \td{5} \te{6} \, ,    \nonumber \\
	&=\;-{\cal{B}}_{19}\, .\nn\\
	(YC)_7=  & + f^{adp}f^{ecr}f^{pru}f^{ubj} \tj{1} \ta{2} \tb{3} \tc{4} \td{5} \te{6} 
	-  f^{abq}f^{ecr}f^{rds}f^{sqj} \tj{1} \ta{2} \tb{3} \tc{4} \td{5} \te{6} \nonumber \\
	&+  f^{aro}f^{dow}f^{ecr}f^{wbj} \tj{1} \ta{2} \tb{3} \tc{4} \td{5} \te{6}\, ,\nonumber \\
	&=\;{\cal{B}}_{4}-{\cal{B}}_{16}-{\cal{B}}_{27}+{\cal{B}}_{23}\, .\nn\\
	(YC)_8=   & + f^{adp}f^{ecr}f^{pru}f^{ubj} \tj{1} \ta{2} \tb{3} \tc{4} \td{5} \te{6} 
	+  f^{aro}f^{dow}f^{ecr}f^{wbj} \tj{1} \ta{2} \tb{3} \tc{4} \td{5} \te{6},\nonumber  \\
	&=\;-{\cal{B}}_{16}+{\cal{B}}_{27}\, .\nn\\
	(YC)_{9}=  & + f^{ajn}f^{bqj}f^{ceq}f^{dnt} \ttt{1} \ta{2} \tb{3} \tc{4} \td{5} \te{6} 
	+ f^{adp}f^{bqj}f^{ceq}f^{pju} T_1^u \ta{2} \tb{3} \tc{4} \td{5} \te{6}\, ,\nonumber \\
	&=\;-{\cal{B}}_{8}+{\cal{B}}_{18}\, .\nn\\
	(YC)_{10}=  & - f^{avj}f^{cbt}f^{edn}f^{ntv} \tj{1} \ta{2} \tb{3} \tc{4} \td{5} \te{6} 
	- f^{awy}f^{ebr}f^{rcw}f^{ydj} \tj{1} \ta{2} \tb{3} \tc{4} \td{5} \te{6} \,  \\
	&+  f^{adp}f^{ebr}f^{pwj}f^{rcw} \tj{1} \ta{2} \tb{3} \tc{4} \td{5} \te{6}
	-  f^{akj}f^{cbt}f^{euk}f^{tdu} \tj{1} \ta{2} \tb{3} \tc{4} \td{5} \te{6}\, ,\nonumber \\
	&=\;{\cal{B}}_{1}+{\cal{B}}_{8}+{\cal{B}}_{18}-2{\cal{B}}_{29}-2{\cal{B}}_{20}\, .\nn\\
	(YC)_{11}=     & - f^{asj}f^{cbt}f^{eus}f^{tdu} \tj{1} \ta{2} \tb{3} \tc{4} \td{5} \te{6}           +  f^{awy}f^{dyj}f^{ebr}f^{rcw} \tj{1} \ta{2} \tb{3} \tc{4} \td{5}    \te{6}\nonumber \\
	&+  f^{adp}f^{cbt}f^{etq}f^{pqj} \tj{1} \ta{2} \tb{3} \tc{4} \td{5} \te{6}
	-  f^{avj}f^{cbt}f^{edn}f^{ntv} \tj{1} \ta{2} \tb{3} \tc{4} \td{5} \te{6}\, ,\nonumber \\
	&=\;-2{\cal{B}}_{1}-2{\cal{B}}_{8}-{\cal{B}}_{18}\, .\nn\\
	(YC)_{12}=     & +f^{aro}f^{dow}f^{ecr}f^{wbj} \tj{1} \ta{2} \tb{3} \tc{4} \td{5}    \te{6}\, ,\nonumber \\
	&=\;{\cal{B}}_{16}\, .\nn\\
	(YC)_{13}=   & +f^{ank}f^{bdp}f^{ecn}f^{pkj} \tj{1} \ta{2} \tb{3} \tc{4} \td{5} \te{6}
	+ f^{anl}f^{bkj}f^{ecn}f^{ldk} \tj{1} \ta{2} \tb{3} \tc{4} \td{5} \te{6}\, ,\nonumber \\
	&=\;-{\cal{B}}_{5}+{\cal{B}}_{16}-{\cal{B}}_{22}\, .\nn\\
	(YC)_{14}=     &+ f^{akj}f^{cmk}f^{ebr}f^{rdm} \tj{1} \ta{2} \tb{3} \tc{4} \td{5}    \te{6}\, ,\nonumber \\
	&=\;{\cal{B}}_{10}\, .\nn\\
	(YC)_{15}=     &+ f^{amp}f^{cpj}f^{ebr}f^{rdm} \tj{1} \ta{2} \tb{3} \tc{4} \td{5}    \te{6}\, ,\nonumber \\
	&=\;{\cal{B}}_{14}\, . \nn
\end{align}

\begin{align}
	(YC)_{16}=     &+ f^{asj}f^{dws}f^{ebr}f^{rcw} \tj{1} \ta{2} \tb{3} \tc{4} \td{5}    \te{6}\, ,\nonumber  \\
	&=\;{\cal{B}}_{8}+{\cal{B}}_{18}-{\cal{B}}_{20}\, .\nn\\
	(YC)_{17}=     &+ f^{awy}f^{dyj}f^{ebr}f^{rcw} \tj{1} \ta{2} \tb{3} \tc{4} \td{5}    \te{6}\, ,\nonumber \\
	&=\;-{\cal{B}}_{1}-{\cal{B}}_{8}-{\cal{B}}_{18}+{\cal{B}}_{20}\, .\nn\\
	(YC)_{18}=     & + f^{ebr}f^{ldk}f^{ral}f^{kcj} \tj{1} \ta{2} \tb{3} \tc{4} \td{5}    \te{6}v ,\nonumber \\
	&=\;{\cal{B}}_{12} \, .\nn\\
	(YC)_{19}=   & + f^{aeg}f^{bko}f^{cdj}f^{jgk} \too{1} \ta{2} \tb{3} \tc{4} \td{5} \te{6} 
	-  f^{aeq}f^{bmq}f^{cgm}f^{qdr} \trr{1} \ta{2} \tb{3} \tc{4} \td{5} \te{6}\nonumber \\
	&+  f^{aeg}f^{bdn}f^{cgm}f^{nms} \ts{1} \ta{2} \tb{3} \tc{4} \td{5} \te{6}\, ,\nonumber \\
	&=\;{\cal{B}}_{9}+{\cal{B}}_{21}+{\cal{B}}_{22} \, .\nn\\
	(YC)_{20}=  & - f^{aev}f^{bdr}f^{cpj}f^{vrp} \tj{1} \ta{2} \tb{3} \tc{4} \td{5} \te{6} 
	+ f^{aev}f^{bvk}f^{kdq}f^{qcj} \tj{1} \ta{2} \tb{3} \tc{4} \td{5} \te{6} \, ,\nonumber \\
	&=\;{\cal{B}}_{28}+{\cal{B}}_{24}\, .\nn\\
	(YC)_{21}=     &+ f^{aev}f^{bkj}f^{cvq}f^{dqk} \tj{1} \ta{2} \tb{3} \tc{4} \td{5}    \te{6}\, ,\nonumber \\
	&=\;-{\cal{B}}_{9}-{\cal{B}}_{22}\, .\nn\\
	(YC)_{22}=     &- f^{aes}f^{btu}f^{cst}f^{udj} \tj{1} \ta{2} \tb{3} \tc{4} \td{5}    \te{6}\, ,\nonumber \\
	&=\;{\cal{B}}_{9} \, .\nn\\
	(YC)_{23}=     &- f^{aeq}f^{bqu}f^{cwj}f^{udw} \tj{1} \ta{2} \tb{3} \tc{4} \td{5}    \te{6}\, ,\nonumber \\
	&=\;{\cal{B}}_{24} \, .\nn\\
	(YC)_{24}=  & - f^{arv}f^{cvt}f^{dtj}f^{ebr} \tj{1} \ta{2} \tb{3} \tc{4} \td{5}    \te{6}-  f^{abq}f^{cus}f^{equ}f^{sdj} \tj{1} \ta{2} \tb{3} \tc{4} \td{5}   \te{6}\, ,\nonumber \\
	&=\;-{\cal{B}}_{3}+{\cal{B}}_{13}-{\cal{B}}_{25} \, .	\nn
\end{align}


\subsection{Orphan Cwebs} \label{sec:appdxOrphan}

\subsection*{(22)  \ $\mathbf{W^{(3,1)}_{6} (2, 2, 2, 1, 1, 1 )}$ } 
\label{C44}
This Cweb is unique at five loops and six lines as this Cweb does not belong to any previously known family neither it is a new basis Cweb. A classification of all the Cwebs present at five loops and six lines is given in section~\ref{sec:UNQ}, tables~\ref{tab:UNQ} and~\ref{tab:BASES}.
This Cweb again has eight diagrams because there are two attachments on each lines-1, 2 and 3. One of the diagrams is shown in fig.~\ref{Diag:12}. The sequences of diagrams and their corresponding $s$-factors are provided in table~\ref{Table:Diag12}.

\vskip0.5cm
\begin{minipage}{0.45\textwidth}
	\hspace{1.5cm}	\includegraphics[scale=0.5]{C44N}	
	\captionof{figure}{$W^{(3,1)}_{6} (2, 2, 2, 1, 1, 1)$}
	\label{Diag:12}
\end{minipage}
\begin{minipage}{0.45\textwidth}
	\footnotesize{
		\begin{tabular}{ | c | c | c |}
			\hline
			\textbf{Diagrams} & \textbf{Sequences} & \textbf{$s$-factors} \\ \hline
			$d_1$ & $\lbrace \lbrace AB \rbrace,  \lbrace CD \rbrace, \lbrace FE \rbrace\rbrace$ & 6 \\ 
			\hline
			$d_2$ & $\lbrace \lbrace AB \rbrace,  \lbrace CD \rbrace, \lbrace FE \rbrace\rbrace$ & 2 \\ 
			\hline	
			$d_3$ & $\lbrace \lbrace AB \rbrace,  \lbrace DC \rbrace, \lbrace EF \rbrace\rbrace$ & 2 \\ 
			\hline	
			$d_4$ & $\lbrace \lbrace AB \rbrace,  \lbrace DC \rbrace, \lbrace FE \rbrace\rbrace$ & 2 \\ 
			\hline
			$d_5$ & $\lbrace \lbrace BA \rbrace,  \lbrace CD \rbrace, \lbrace EF \rbrace\rbrace$ & 2 \\ 
			\hline
			$d_6$ & $\lbrace \lbrace BA \rbrace,  \lbrace CD \rbrace, \lbrace FE \rbrace\rbrace$ & 2 \\ 
			\hline
			$d_7$ & $\lbrace \lbrace BA \rbrace,  \lbrace DC \rbrace, \lbrace EF \rbrace\rbrace$ & 2 \\ 
			\hline
			$d_8$ & $\lbrace \lbrace BA \rbrace,  \lbrace DC \rbrace, \lbrace FE \rbrace\rbrace$ &6 \\ 
			\hline
		\end{tabular}
		\captionof{table}{Sequences and $s$-factors}
		\label{Table:Diag12}	}
\end{minipage} 

\vspace{0.5cm}
\noindent  The mixing matrix $R$, the diagonalizing matrix $Y$ and the diagonal matrix $D$ for the Cweb mentioned above are
\begin{equation} 
	R=\frac{1}{6} \left( \,\,\,
	\begin{array}{cccccccc}
		0 &\hspace{0.20cm} 0 &\hspace{0.20cm} 0 &\hspace{0.20cm} 0 &\hspace{0.20cm} 0 &\hspace{0.20cm} 0 &\hspace{0.20cm} 0 &\hspace{0.20cm} 0 \\
		\hspace{-0.25cm}-1 &\hspace{0.20cm} 1 &\hspace{0.20cm} 1 &\hspace{0.20cm} \hspace{-0.25cm}-1 &\hspace{0.20cm} 1 &\hspace{0.20cm} \hspace{-0.25cm}-1 &\hspace{0.20cm} \hspace{-0.25cm}-1 &\hspace{0.20cm} 1 \\
		\hspace{-0.25cm}-1 &\hspace{0.20cm} 1 &\hspace{0.20cm} 1 &\hspace{0.20cm} \hspace{-0.25cm}-1 &\hspace{0.20cm} 1 &\hspace{0.20cm} \hspace{-0.25cm}-1 &\hspace{0.20cm} \hspace{-0.25cm}-1 &\hspace{0.20cm} 1 \\
		1 &\hspace{0.20cm} \hspace{-0.25cm}-1 &\hspace{0.20cm} \hspace{-0.25cm}-1 &\hspace{0.20cm} 1 &\hspace{0.20cm} \hspace{-0.25cm}-1 &\hspace{0.20cm} 1 &\hspace{0.20cm} 1 &\hspace{0.20cm} \hspace{-0.25cm}-1 \\
		\hspace{-0.25cm}-1 &\hspace{0.20cm} 1 &\hspace{0.20cm} 1 &\hspace{0.20cm} \hspace{-0.25cm}-1 &\hspace{0.20cm} 1 &\hspace{0.20cm} \hspace{-0.25cm}-1 &\hspace{0.20cm} \hspace{-0.25cm}-1 &\hspace{0.20cm} 1 \\
		1 &\hspace{0.20cm} \hspace{-0.25cm}-1 &\hspace{0.20cm} \hspace{-0.25cm}-1 &\hspace{0.20cm} 1 &\hspace{0.20cm} \hspace{-0.25cm}-1 &\hspace{0.20cm} 1 &\hspace{0.20cm} 1 &\hspace{0.20cm} \hspace{-0.25cm}-1 \\
		1 &\hspace{0.20cm} \hspace{-0.25cm}-1 &\hspace{0.20cm} \hspace{-0.25cm}-1 &\hspace{0.20cm} 1 &\hspace{0.20cm} \hspace{-0.25cm}-1 &\hspace{0.20cm} 1 &\hspace{0.20cm} 1 &\hspace{0.20cm} \hspace{-0.25cm}-1 \\
		0 &\hspace{0.20cm} 0 &\hspace{0.20cm} 0 &\hspace{0.20cm} 0 &\hspace{0.20cm} 0 &\hspace{0.20cm} 0 &\hspace{0.20cm} 0 &\hspace{0.20cm} 0 \\
	\end{array}
	\right),\quad Y=\left( \,\,\,
	\begin{array}{cccccccc}
		\hspace{-0.25cm}-1 &\hspace{0.20cm} 1 &\hspace{0.20cm} 1 &\hspace{0.20cm} \hspace{-0.25cm}-1 &\hspace{0.20cm} 1 &\hspace{0.20cm} \hspace{-0.25cm}-1 &\hspace{0.20cm} \hspace{-0.25cm}-1 &\hspace{0.20cm} 1 \\
		0 &\hspace{0.20cm} 0 &\hspace{0.20cm} 0 &\hspace{0.20cm} 0 &\hspace{0.20cm} 0 &\hspace{0.20cm} 0 &\hspace{0.20cm} 0 &\hspace{0.20cm} 1 \\
		0 &\hspace{0.20cm} 1 &\hspace{0.20cm} 0 &\hspace{0.20cm} 0 &\hspace{0.20cm} 0 &\hspace{0.20cm} 0 &\hspace{0.20cm} 1 &\hspace{0.20cm} 0 \\
		0 &\hspace{0.20cm} 1 &\hspace{0.20cm} 0 &\hspace{0.20cm} 0 &\hspace{0.20cm} 0 &\hspace{0.20cm} 1 &\hspace{0.20cm} 0 &\hspace{0.20cm} 0 \\
		0 &\hspace{0.20cm} \hspace{-0.25cm}-1 &\hspace{0.20cm} 0 &\hspace{0.20cm} 0 &\hspace{0.20cm} 1 &\hspace{0.20cm} 0 &\hspace{0.20cm} 0 &\hspace{0.20cm} 0 \\
		0 &\hspace{0.20cm} 1 &\hspace{0.20cm} 0 &\hspace{0.20cm} 1 &\hspace{0.20cm} 0 &\hspace{0.20cm} 0 &\hspace{0.20cm} 0 &\hspace{0.20cm} 0 \\
		0 &\hspace{0.20cm} \hspace{-0.25cm}-1 &\hspace{0.20cm} 1 &\hspace{0.20cm} 0 &\hspace{0.20cm} 0 &\hspace{0.20cm} 0 &\hspace{0.20cm} 0 &\hspace{0.20cm} 0 \\
		1 &\hspace{0.20cm} 0 &\hspace{0.20cm} 0 &\hspace{0.20cm} 0 &\hspace{0.20cm} 0 &\hspace{0.20cm} 0 &\hspace{0.20cm} 0 &\hspace{0.20cm} 0 \\
	\end{array}
	\right), \nonumber
	\label{D44}
\end{equation}
\begin{align}
	D=(\mathbf{1}_1,0)\,.
\end{align}

The only exponentiated colour factor of this Cweb is
\begin{align}
	(YC)_1=&  f^{abc}f^{aep}f^{dcr}f^{gbq} \tp{1} \tq{2} \trr{3} \te{4} \tg{5} \td{6}\,,\nn\\
	=&\;-{\cal{B}}_{12}-{\cal{B}}_{14}\,.
\end{align}


This completes our listing of all Cwebs with a perturbative expansion starting at $\mathcal{O}(g^{10})$,
and connecting six  Wilson lines.

\bibliographystyle{JHEP}
\bibliography{boom}

\end{document}